\documentclass[11pt]{report}
\usepackage{setspace}
\usepackage[latin1]{inputenc}
\usepackage{epsfig}
\usepackage{color}
\usepackage{graphics}
\usepackage{graphicx}

\usepackage{amsthm}
\usepackage{amssymb}
\usepackage{amsmath}
\usepackage{amsfonts}

\numberwithin{equation}{section}
\input xy
\xyoption{all}

\begin{document}

\newcommand{\df}{\stackrel{\rm def}{=}}
\newcommand{\co}{{\scriptstyle \circ}}
\newcommand{\lb}{\lbrack}
\newcommand{\rb}{\rbrack}
\newcommand{\rn}[1]{\romannumeral #1}
\newcommand{\msc}[1]{\mbox{\scriptsize #1}}
\newcommand{\dsp}{\displaystyle}
\newcommand{\scs}[1]{{\scriptstyle #1}}

\newcommand{\ket}[1]{| #1 \rangle}
\newcommand{\bra}[1]{| #1 \langle}
\newcommand{\vac}{| \mbox{vac} \rangle }

\newcommand{\e}{\mbox{{\bf e}}}
\newcommand{\va}{\mbox{{\bf a}}}
\newcommand{\bc}{\mbox{{\bf C}}}

\newcommand{\com}{C\!\!\!\!|}

\newcommand{\br}{\mbox{{\bf R}}}
\newcommand{\bz}{\mbox{{\bf Z}}}
\newcommand{\bq}{\mbox{{\bf Q}}}
\newcommand{\bn}{\mbox{{\bf N}}}
\newcommand {\eqn}[1]{(\ref{#1})}

\newcommand{\cp}{\mbox{{\bf P}}^1}
\newcommand{\n}{\mbox{{\bf n}}}
\newcommand{\sbz}{\msc{{\bf Z}}}
\newcommand{\sn}{\msc{{\bf n}}}

\newcommand{\be}{\begin{equation}}\newcommand{\ee}{\end{equation}}
\newcommand{\bea}{\begin{eqnarray}} \newcommand{\eea}{\end{eqnarray}}
\newcommand{\ba}[1]{\begin{array}{#1}} \newcommand{\ea}{\end{array}}

\newcommand{\cleqn}{\setcounter{equation}{0}}

\makeatletter

\@addtoreset{equation}{section}

\def\theequation{\thesection.\arabic{equation}}
\makeatother

\def\oa{\bigcirc\!\!\!\! a}
\def\ob{\bigcirc\!\!\!\! b}
\def\oc{\bigcirc\!\!\!\! c}
\def\oi{\bigcirc\!\!\!\! i}
\def\oj{\bigcirc\!\!\!\! j}
\def\ok{\bigcirc\!\!\!\! k}
\def\ve{\vec e}\def\vk{\vec k}\def\vn{\vec n}\def\vp{\vec p}
\def\vr{\vec r}\def\vs{\vec s}\def\vt{\vec t}\def\vu{\vec u}
\def\vv{\vec v}\def\vx{\vec x}\def\vy{\vec y}\def\vz{\vec z}

\pagestyle{empty} 

\begin{figure}
\centering
\includegraphics[width=4.5cm]{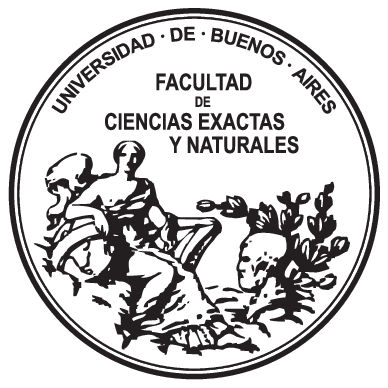}
\end{figure}

\begin{center}

\LARGE{\bf UNIVERSIDAD DE BUENOS AIRES}\\

\Large{Facultad de Ciencias Exactas y Naturales\\
Departamento de Física\\[8mm]}

\Huge{Cuerdas y $D$-branas en\\ espacio-tiempos curvos}\\[9mm] 

\large{Tesis presentada para optar por el título de Doctor de la\\ Universidad de Buenos
Aires en el área Ciencias Físicas}\\[1cm]

\large{\bf  Walter Helmut Baron}\\[1cm]
\end{center}
\large{Director de tesis: Dra. Carmen Alicia Nuñez\\[4mm]
Consejero de estudios: Dr. Gustavo Sergio Lozano\\[4mm]
Lugar de Trabajo: IAFE (CONICET/UBA)}\\[4mm]
Buenos Aires, 2012

\normalsize

\newpage

\pagestyle{empty} 
~~~~~~\\

\newpage
\pagestyle{plain}
\pagenumbering{roman}
\setcounter{page}{1}

\chapter*{\centerline{Cuerdas y $D$-branas en espacio-} \centerline{tiempos curvos}}

En esta tesis estudiamos el modelo $AdS_3$ Wess-Zumino-Novikov-Witten. Calculamos la Expansión en Producto de Operadores de campos primarios y de sus 
imágenes bajo el automorfismo de flujo espectral en todos los sectores del modelo considerado como una rotación de Wick del modelo coset $H_3^+$. 
Argumentamos que las simetrías afines del álgebra requieren un truncado que determina la clausura de las reglas de fusión del espacio de Hilbert. 
Estos resultados son luego utilizados para discutir la factorización de las funciones de cuatro puntos con la ayuda del formalismo conocido como 
bootstrap.

También realizamos un estudio de las propiedades modulares del modelo. Los caracteres sobre el toro Euclídeo divergen de una manera poco controlable. 
La regularización propuesta en la literatura es poco satisfactoria pues elimina información del espectro y se pierde así la relación uno a uno entre 
caracteres y representaciones del álgebra de simetría que forman el espectro. Proponemos estudiar entonces los caracteres definidos sobre el toro 
Lorentziano los cuales están perfectamente definidos sobre el espacio de funcionales lineales, recuperando así la biyección entre caracteres y 
representaciones. Luego obtenemos las transformaciones modulares generalizadas y las utilizamos para estudiar la conexión con los 
correladores que determinan los acoplamientos a las branas simétricas en tal espacio de fondo, obteniendo que en los casos particulares de branas 
puntuales o $dS_2$ branas se recuperan resultados típicos de Teorías de Campos Conformes Racionales como soluciones tipo Cardy o fórmulas tipo Verlinde.
\bigskip

Palabras claves: {\it teoría de cuerdas, teorías conformes no racionales, $D$-branas, $AdS_3$, reglas de fusión, transformaciones modulares}.

\chapter*{\centerline{Strings and $D$-branes in curved} \centerline{space-time}}

In this thesis we study the $AdS_3$ Wess-Zumino-Novikov-Witten model. We compute the Operator Product Expansion of primary fields as well as their images under the spectral flow automorphism in all sectors of the model by considering it as a Wick rotation of the $H_3^+$ coset model. We argue that the symmetries of the affine algebra require a truncation which establishes the closure of the fusion rules on the Hilbert space of the theory. These results are then used to discuss the factorization of four point functions by applying the bootstrap approach. 

We also study the modular properties of the model. Although the Euclidean partition function is modular invariant, the characters on the Euclidean torus diverge
and the regularization proposed in the literature removes information on the spectrum, so that the usual one to one map between characters and representations of rational models is lost. Reconsidering the characters defined on the Lorentzian torus and focusing on their structure as distributions, we obtain expressions that recover those properties. We then study their generalized modular properties and use them to discuss the relation between modular data and one point functions associated to symmetric D-branes, generalizing some results from Rational Conformal Field Theories in the particular cases of point like and $dS_2$ branes, such as Cardy type solutions or Verlinde like formulas.
\bigskip

Keywords: {\it string theory, non rational conformal field theories, $D$-branes, $AdS_3$, fusion rules, modular transformations}.

\newpage

\chapter*{Acknowledgments} 

Es el momento de agradecer apropiadamente a todas las personas e instituciones que de una u otra forma contribuyeron a la realización de esta tesis. 

Desde luego voy empezar agradeciendo a mi hija Juanita y a mi mujer Gisela por el apoyo y por haberme dado tantos gratos momentos que tantas fuerzas me dieron a lo largo de estos años. A mis padres y mis suegros también por su continuo apoyo. A mis hermanos y a mis amigos por la compañia. Quiero agradecer muy especialmente a mi directora, Carmen por haberme guiado y escuchado tantas veces y con tanta paciencia durante este tiempo. A diversos colegas que tanto me han enseñado en valiosas e innumerables discusiones durante la elaboración de mis investigaciones: Adrian Lugo, Alejandro Rosabal, Carlos Cardona, Diego Marqués, Eduardo Andrés, Gerardo Aldazabal, Jan Troost, Jörg Teschner, Jorge Russo, Juan Martín Maldacena, Mariana Graña, Pablo Minces, Robert Coquereaux, Sergio Iguri, Sylvain Ribault, Victor Penas, Volker Schomerus y Yuji Satoh. 

Quiero agradecer al Instituto de Astronomía y Física del Espacio (IAFE) por haberme dado no sólo un lugar físico y multiples facilidades para llevar a cabo la tarea de investigación sino también por saber generar una apropiada atmósfera de trabajo. Al Consejo Nacional de Investigaciones Científicas y Técnicas (CONICET) por haber permitido iniciarme en la inverstigación gracias a su financiamiento. A las diversas instituciones que me recibieron a lo largo de estos años y me permitieron discutir con diversos colegas expertos durante la elaboración de mis trabajos: el Instituto Balseiro (IB), Instituto de Física de La Plata (IFLP), Institut de Physique Théorique (IPTh, CEA Saclay), Laboratoire Charles Coulomb (LCC, Universidad de Montpellier II), Deutsches Elektronen-Synchrotron (DESY, Universidad de Hamburgo) , Institució Catalana de Recerca i Estudis Avançats (ICREA, Universidad de Barcelona), Scuola Internazionale Superiore di Studi Avanzati di Trieste (SISSA) y el Abdus Salam International Centre for Theoretical Physics (ICTP).

\newpage

\bigskip
\bigskip
\bigskip
\bigskip
\bigskip
\begin{center}
{\it  A Juanita y a Gisela por su incansable apoyo.}
\end{center}
\newpage

\tableofcontents

\newpage

\pagenumbering{arabic}

\chapter{Introduction}

String theory is one of the most ambitious projects in the field of high energy physics. Even though it was originally conceived to explain quark confinement, its development forced the opening of different roads and the extension of its original aims. Shortly after its beginning, people realized every consistent string theory had a massless spin two particle in the spectrum, a graviton. This opened a new possibility to reach a consistent theory of quantum gravity, with extended objects to quantize instead of point-like particles, and so to reach the dream of every theoretical physicist, $i.e.$ a unified theory for all known interactions or a theory of everything. 

This theory, yet under construction, is the best candidate to give a unified description of all interactions, but also has the beauty of having offered important results going beyond the theory itself. It gave rise to an amazing development of Conformal Field Theories (CFTs) providing several tools to deal with two dimensional statistical models. The consistency of the theory led to the idea of supersymmetry. The gauge/gravity duality, whose validity is not restricted to that of string theory, has offered the first realization of the holographic principle, leading not only to a theoretical frame relating gravitational theories with gauge theories in flat space, but a powerful tool allowing to study non perturbative regions of superconformal, but also of non conformal field theories or models with less supersymmetry or in different space-time dimensions \cite{magoo} than those proposed in the first example of this duality \cite{malda,wittenhol}, including holographic renormalization group flows \cite{NS,ST,GNS}, rotating strings \cite{BMN,GKP,S}, applications to the study of baryonic symmetry breaking \cite{GHK,S2}, applications to cosmology \cite{HHR,AEN}, applications to holographic QCD \cite{EKSS,RP,HHMS,HHMS2}, to electroweak symmetry breaking \cite{HMS}, and non relativistic quantum-mechanical systems \cite{MMT,ABM,HRM,S3}, among others. 

Many years have passed since its birth and it appears frustrating not having found yet a solid confirmation of the theory. There is not known way to reduce the theory to one which exactly reproduces all aspects of the Standard Model, or which solves the cosmological constant problem, among other drawbacks. But these puzzles must not be seen as a failure of the theory, but as a consequence of the complexity of the problems to be solved. The knowledge of the theory has increased considerably. There is a good understanding of the perturbative regime, and the discovery of $D-$branes, extended dynamical objects on which the ends of open strings live, has been crucial in the development of dualities that allow to explore non perturbative regions of the theory.          

One of the major challenges of the theory is to find a mechanism univocally leading to an effective reduced theory with a solution phenomenologically contrastable with the real world. Instead, one finds a large number of vacua, roughly $10^{500}$, a problem frequently named the $landscape$ of string theory \cite{landscape}.

There is no doubt that D-branes play a fundamental role in the resolution of these problems and so a fundamental step consists in the study of branes in non trivial curved backgrounds.   

A very powerful approach to explore string theories on non trivial curved backgrounds is to consider vacua with a lot of symmetry. This was successfully reached by considering the space-time to be given by the group manifold of a continuous compact group, $G$. In such a case the worldsheet theory is a Rational Conformal Field Theory (RCFT), $i.e.$ it contains a finite number of primary states. To date these theories can be solved exactly by using only algebraic tools \cite{dif,PZ,SA}. The situation is very different in the non compact case, as a consequence of the presence of a continuous spectrum of states (see for instance \cite{SNRat} for a review). This thesis is devoted to the study of the worldsheet of string theory on a three dimensional anti de Sitter ($AdS_3$) space time.  This theory is very interesting because it is one of the simplest models to test string theory in curved non compact backgrounds and with a non trivial timelike direction, it can be used to learn more on the not well known non RCFT, but also because of its relevance in the AdS/CFT correspondence.

Most of the studies regarding the AdS/CFT duality were explored only within the supergravity approximation. 
An example where the correspondence was successfully explored beyond the supergravity approximation is string theory in PP waves backgrounds \cite{M,MT} with RR fields, obtained by taking the Penrose limit of AdS. Another accessible background is  $AdS_3$ with Neveu Schwarz (NS) antisymmetric field, appearing within backgrounds like AdS$_3\times S^3\times M^4$, with $M^4= T^4$ or $K^3$ (obtained as the near horizon limit of the $D1-D5$ brane setup in the background $R^6\times M^4$) \cite{KG,DP,SP,Taylor,cn,ck}. 

The worldsheet of the bosonic string propagating on $AdS_3$ is described by the Wess Zumino Novikov Witten (WZNW) model associated to the universal cover of the $SL(2,{\mathbb R})$ group, $\widetilde{ SL(2,{\mathbb R})}$ but for short we will refer to this as the $AdS_3$ WZNW model in order to avoid confusion with the WZNW model on the single cover of $SL(2,{\mathbb R})$. So it is expected to be exactly solvable. The studies on this model started within the seminal work of O`Raifeartaigh, Balog, Forgacs and Wipf \cite{BRFW} on the $SU(1,1)$ WZNW model more than twenty years ago, but it was necessary to wait for more than ten years until the work of Maldacena and Ooguri \cite{mo1} correctly defined the spectrum, which is considerably more involved that those of RCFT. It consists of long strings with continuous energy spectrum arising from the principal continuous representation of $sl(2)$ and its spectral flow images, and short strings with discrete physical spectrum resulting from the highest-weight discrete representations and their spectral flow images.

By extending the result of Evans, Gaberdiel and Perry \cite{EGP} to nontrivial spectral flow sectors, a no ghost theorem for this spectrum was proved in \cite{mo1} and verified in \cite{mo2} through  the computation of the one-loop partition function on a Euclidean $AdS_3$ background at finite temperature. Amplitudes of string theory in $AdS_3$ on the sphere were computed in \cite{mo3}, analytically continuing the expressions obtained for the Euclidean H$^+_3=\frac{SL(2,\mathbb C)}{SU(2)}$ (gauged) WZNW model in \cite{tesch1,tesch2}. Some subtleties of the analytic continuation relating the H$^+_3$ and $AdS_3$ models were clarified in \cite{mo3} and this allowed to construct, in particular, the four-point functions of unflowed short strings. Integrating over the moduli space of the worldsheet, it was shown that the string amplitude can be expressed as a sum of products of three-point functions with intermediate physical states, $i.e.$ the structure of the factorization agrees with the Hilbert space of the theory.

A step up towards a proof of consistency and unitarity of the theory involves the construction of  four-point functions including states in different representations and the verification that only unitary states corresponding to long and short strings in agreement with the spectral flow selection rules  are produced in the intermediate channels. To achieve this goal, the analytic and algebraic structure of the AdS$_3$ WZNW model should be explored further. 

Most of the important progress achieved is based on the better understood Euclidean $H_3^+$ model. The absence of singular vectors and the lack of chiral factorization in the relevant current algebra representations obstruct the use of the powerful techniques from rational conformal field theories. Nevertheless, a  generalized conformal bootstrap approach was successfully applied in \cite{tesch1, tesch2} to the $H_3^+$ model on the punctured sphere, allowing to discuss the factorization of four-point functions. In principle, this method offers the possibility to unambiguously determine any $n>3-$point function in terms of two- and three-point functions once the operator product expansions of two operators and the structure constants are known.

We carried out some initial steps along the development of this thesis \cite{wc}, by examining the role of the spectral flow symmetry on the analytic continuation of the operator product expansion from $H_3^+$ to the relevant representations of $SL(2,\mathbb R)$ and on the factorization  properties of four-point functions. These results give fusion rules establishing the closure of the Hilbert space and the unitarity of the full interacting string theory. 
 
In RCFT, a practical derivation of the fusion rules ($i.e.$ of the representations contained in the Operator algebra) can be performed through the Verlinde theorem \cite{verlinde}, often formulated as the statement that the $S$ matrix of modular transformations diagonalizes the fusion rules. Moreover, besides leading to a Verlinde formula, the $S$ matrix allows a classification of modular invariants and a systematic study of boundary states for symmetric branes. It is interesting to explore whether analogous of these properties can be found in the $AdS_3$ WZNW  model. However, the relations among fusion algebra, boundary states and modular transformations are difficult to identify and have not been very convenient in non compact models \cite{TJ}. In general, the characters have an intricate behavior under the modular group \cite{stt}-\cite{t} and, as is often the case in theories with discrete and continuous representations, these mix under $S$ transformations. In the forthcoming chapters we will discuss these subjects based on our previous results \cite{wc2}.
 
This thesis is organized as follows, in Chapter \ref{chapGeo} we review the geometry, symmetries, and give some basis functions of $AdS_3$ and related spaces which are the basic objects in the minisuperspace limit. Chapter \ref{chapAdS} is devoted to introduce the WZNW models. We begin with a short introduction on general WZNW models and then present the $AdS_3$ WZNW model, and related ones like $H_3^+$ and the $\frac{SL(2,{\mathbb R})_k/U(1)\times U(1)_{-k}}{{\mathbb Z}_{N_k}}$ models. In Chapter \ref{chapOPE} we discuss interactions in the model. We consider two and three point functions of the Euclidean $H_3^+$ model, and assuming they are related to those of the $AdS_3$ model \cite{mo3}, we compute the Operator Product Expansion (OPE) for primary fields as well as for their images under spectral flows in all the sectors of the theory. After discussing the extension to descendant fields, we show that the spectral flow symmetry requires a truncation of the fusion rules determining the closure of the operator algebra on the Hilbert space of the theory. In Chapter \ref{chapFact} we consider the factorization of  four-point functions and study some of its properties. Chapter \ref{chapChar} is devoted to the characters of the relevant representations of the $AdS_3$ WZNW model. Since the standard Euclidean characters diverge and lack good modular properties, extended characters were originally introduced in \cite{hhrs} (see also \cite{HS})\footnote{Similar problems in non-compact coset models have also been considered in \cite{fnp}-\cite{bf}}. A different approach was followed in \cite{mo1} where the standard characters were computed on the Lorentzian torus and it was shown that the modular invariant partition function of the  $H_3^+$ model obtained in \cite{gawe} is recovered after performing analytic continuation and discarding contact terms. However, this trivial regularization removes information on the spectrum and the usual one to one map between characters and representations of rational models is lost. With the aim of overcoming these problems, we review (and redefine) the characters on the Lorentzian torus, focusing on their structure as distributions and compute the full set of generalized modular transformations in Chapter \ref{chapMP}. 

In order to explore the properties of the modular S matrix, in Chapter \ref{chapDbranes} we consider the maximally symmetric D-branes of the model. We 
explicitly construct the Ishibashi states and show that the coefficients of the boundary states turn out to be determined from the 
generalized $S$ matrix, suggesting that a Verlinde-like formula could give some information on the spectrum of open strings attached to 
certain D-branes. Furthermore, we show that a generalized Verlinde formula reproduces the fusion rules of the finite dimensional  degenerate 
representations of $sl(2, \mathbb R$) appearing in the boundary spectrum of the point-like D-branes.

In chapter \ref{chapConcl} we give a summary of the thesis and discuss the actual status and future challenges and perspectives regarding open problems. We 
also list the original contributions to the subject presented along the thesis. 

Some basic facts about CFTs are reviewed in appendix \ref{A:bfcft}, some technical details of the calculations are included in appendices \ref{A:asw}, \ref{A:mb} and \ref{A:gvf} and a discussion of the moduli space of the Lorentzian torus is found in appendix \ref{A:tlt}.

\chapter{Geometric aspects of maximally symmetric spaces}\label{chapGeo}

Before introducing the $AdS_3$ WZNW model, it is instructive to spend some pages reviewing some aspects of the geometry of maximally symmetric spaces and the minisuperspace limit in $AdS_3$ and related models. 

\section{Geometry and symmetries}

Along the bulk of the thesis the reader will find discussions concerning different types of geometries such as hyperbolic, Anti de Sitter or de Sitter spaces, so a good point to begin with is by defining all these geometries.
  
\subsection{Maximally symmetric spaces}

Maximally symmetric spaces are defined as those metric spaces with maximal number of isometries in a given spacetime dimension\footnote{In the concrete case of D spacetime dimensions these spaces admit $D(D+1)/2$ linear independent Killing vectors.}. Due to this important property such geometries were extensively studied in the literature. Here we will give a short $breviatum$. For a deeper study of the subject we redirect the reader to \cite{SH}. 

Every D dimensional maximally symmetric space has constant curvature and can be realized as a pseudosphere embedded in a D+1 dimensional flat space.

To be more precise, let $\left(X^0,X^1,\dots,X^D\right)$ represent the Cartesian coordinates of a particular point in such a flat space, the symmetric spaces can be realized as the hypersurfaces constrained by  
\bea
\epsilon R^2=X^\mu X_\mu, 
\eea
where the indices are lowered with the background metric
\bea
\eta_{\mu\nu}=diag\left(\epsilon_0,\epsilon_1,\dots,\epsilon_D\right)
\eea 
and the $\epsilon$'s are signs. $R$ is frecuently called the radius of the space because of the similarity with the radius of a sphere, but it must not be confused with the Ricci curvature scalar, ${\mathcal R}$, which is given by 
\bea
{\cal R}= \frac{\epsilon D(D-1)}{R^2}.
\eea
 
We are specially interested in Euclidean or Lorentzian (one timelike direction) D dimensional spaces, so we fix $\epsilon_1=\dots=\epsilon_{D-1}=1$. 

The cases of interest for us are anti de Sitter, Hyperbolic and de Sitter spaces. 

\subsubsection{ Anti de Sitter space: AdS$_D$}
This geometry corresponds to the case where $\epsilon_0=\epsilon_D=\epsilon=-1$. The space has Lorentzian signature and $SO(D-1,2)$ isometry group. The time like direction is compact. 

The particular case $D=3$ will be of special interest for us. 

Notice that the topology of $AdS_3$ coincides with that of the $SL(2,\mathbb R)$ group manifold as can be checked from the following parametrization of this group
\bea
g=R^{-1}\left(\begin{matrix}X^0+X^1&X^2+X^3\cr X^2-X^3&X^0-X^1\end{matrix}\right),~~~~X^\mu\in{\mathbb R}.\label{sl(2,r)}
\eea 

This relation is intimately linked to the fact that the $SO(2,2)$ isometry group is locally isomorphic to $SL(2)\otimes SL(2)$. 

For obvious reasons we will decompactify the time-like direction when discussing physical applications. And from now on we will denote this space as $\emph{AdS}_3$, which is nothing but the group manifold of the universal covering of $SL(2,{\mathbb R})$, $i.e.$ $\widetilde{SL(2,{\mathbb R})}$. 

Another useful coordinate system, frequently used in the literature, is the so called global or cylindrical coordinate system $(\rho,\theta,\tau)$, related to the previous one via
\bea
X^0+iX^3&=&e^{i\tau}\cosh\rho,\cr
X^1+iX^2&=&e^{i\theta}\sinh\rho.\label{cylcoord}
\eea

\subsubsection{Hyperbolic space: H$_D$}
This geometry is realized when $\epsilon=\epsilon_0=-\epsilon_D=-1$. It has Euclidean signature, $SO(D,1)$ isometry and decomposes in two disconnected branches, the upper sheet ($X^0\ge R$) denoted by  H$^{+}_D$ and the lower one ($X^0\le -R$) denoted by H$^-_D$. 

For the special case of $D=3$, this space has the topology of the group coset manifold of the subset of hermitian matrices of unit determinant in $SL(2,\mathbb C)$, $i.e.$ $SL(2,{\mathbb C})/SU(2)$. The subspace $H_3^+$ can be interpreted as a Euclidean Wick rotation of $AdS_3$, in fact it can be parametrized with the analytic continuation $\tau\rightarrow i\tau$ of the cylindrical coordinate system of $AdS_3$ so breaking the periodicity of the time-like direction. 

Another useful coordinate system is the one defined by the complex variables $(\phi,\gamma,\bar \gamma)$, where
\bea
\gamma     &=& e^{\tau+i\theta}\tanh\rho,\cr
\bar\gamma &=& e^{\tau-i\theta}\tanh\rho,\cr
e^{\phi}   &=& e^{-\tau}\cosh\rho.
\eea
In terms of these coordinate, the elements of the coset are parametrized by
\bea
h=\left(\begin{matrix}     e^{\phi}      &  e^{\phi}~\bar\gamma  \cr
                       e^{\phi}~\gamma   &  e^{\phi}\gamma\bar\gamma+e^{-\phi}
\end{matrix}\right)
\eea
and the coset structure is manifest when $h$ is written as $h=v\,v^\dagger$, with
\bea
v=\left(\begin{matrix}     e^{\phi/2}      &  0        \cr
                       e^{\phi/2}~\gamma   &  e^{-\phi/2}
\end{matrix}\right).
\eea
Clearly $h$ is invariant under $v\rightarrow v\,u$, with $u\in SU(2)$ and this explicit realization of the coset structure is the reason why these coordinates are implemented when constructing the $H_3^+$ model by gauging the $SU(2)$ subgroup of the $SL(2,{\mathbb C})$ WZNW model.
  
\subsubsection{de Sitter space: dS$_D$}
In this case only $\epsilon_0=-1$. The space has Lorentzian signature, $SO(D,1)$ isometry group and its topology coincides with the group manifold of unimodular antihermitian matrices 
\bea
g=R^{-1}\left(\begin{matrix}i\left(X^0+X^1\right)&\left(X^2+iX^3\right)\cr -\left(X^2-iX^3\right)&i\left(X^0-X^1\right)\end{matrix}\right),~~~~X^\mu\in{\mathbb R}.
\eea 
\medskip
Other geometries (not discussed in the thesis) are the (Euclidean) D-dimensional sphere with SO(D+1) isometry group ($\epsilon_\mu=1,\forall \mu$) and the ``two time" pseudosphere with $SO(D-1,2)$ isometry corresponding to $\epsilon_0=\epsilon_D=-\epsilon=-1$. 


\section{Basis functions}\label{sectBF}

In this section we describe some of the basis functions for $H_3^+, SL(2, {\mathbb R})$ and $AdS_3$, where by abuse of notation we refer to $SL(2,{\mathbb R})$ as the group manifold of the single cover of $SL(2,{\mathbb R})$. The fact that both $H_3^+$ and $SL(2,{\mathbb R})$ admit matricial representations makes the study of functions over these spaces much easier than in the $AdS_3$ case where such representations are not present.

Let us begin with $SL(2,{\mathbb R})$, and the parametrization (\ref{cylcoord}), then the elements of the group, $g$ are written as 
\bea
g(\rho,\theta,\tau)=\left(\begin{matrix}\cosh\rho \cos\tau + \sinh\rho \cos\theta & \cosh\rho \sin\tau + \sinh\rho \sin\theta \cr
                                        \sinh\rho \sin\theta - \cosh\rho \sin\theta & \cosh\rho \cos\tau - \sinh\rho \cos\theta \end{matrix}\right)\label{g}
\eea 

As we commented in the previous section, $AdS_3$ is obtained by decompactifying $\tau$. If we write $\tau=2\pi(q+\lambda)$, where $q\in {\mathbb Z},~\lambda\in[0,1)$, the elements $G(\rho,\theta,\tau)$ of $AdS_3$ admit a parametrization in terms of $SL(2,{\mathbb R})$
\bea
G=(g,q),~~~~g\in SL(2,{\mathbb R}),~q\in{\mathbb Z}
\eea  
The $AdS_3$ product is built from the one in the single cover as
\bea
GG'=(g,q)(g',q')=(gg',q+q'+F(g)+F(g')-F(gg'))\label{AdS3product}
\eea

Notice that both $SL(2,{\mathbb R})$ and the universal cover carry a natural left and right multiplication by themselves. For instance, for $AdS_3$ we have $(G_L,G_R)\cdot G=G_LGG_R$ and this will be the symmetry of the model under consideration (see section \ref{sectionWZNW}).

The geometric symmetry group of $SL(2,{\mathbb R})$ is $\frac{SL(2,{\mathbb R})\times SL(2,{\mathbb R})}{{\mathbb Z}_2}$, where ${\mathbb Z}_2$ is the center of the group. In the case of $AdS_3$ the symmetry group is $\frac{AdS_3\times AdS_3}{{\mathbb Z}}$, where the center is isomorphic to ${\mathbb Z}$. Indeed, it can be easily checked, from (\ref{AdS3product}), that it is given by the subgroup $\{(\pm id, q),~q\in{\mathbb Z}\}$, which is the subgroup freely generated by the element $(-id,0)$. 

A useful parametrization for $H_3^+$, different from the ones considered in the previous section is given by
\bea
h(\rho,\theta,\tilde\tau)=\left(\begin{matrix} e^{\tilde\tau}\cosh\rho       &   e^{i\theta}\sinh\rho \cr
                                         e^{-i\theta}\sinh\rho &   e^{-\tilde\tau}\cosh\rho \end{matrix}\right)\label{h}
\eea

Notice that the product of two matrices in $H_3^+$, $h_1$ and $h_2$ with $\theta_1\neq\theta_2$ is outside of the hyperbolic space. This is a not surprising because $H_3^+$ is not a group, but the group coset $SL(2,{\mathbb C})/SU(2)$. In this case the left and right action take the element out of the space and the geometric symmetry group acts as $k\cdot h = k h k^\dagger$, with $k\in SL(2,{\mathbb C})$. So it is given by $\frac{SL(2,{\mathbb C})}{{\mathbb Z}_2}$. 

\subsection{Continuous basis}

As we commented at the beginning of this section, the fact that $H_3^+$ and $SL(2,{\mathbb R})$ admit matrix representations simplifies matters.

A useful basis for $H_3^+$ is the well known $x$-$basis$, 
\bea
\phi^{j}(x,\bar x|h)=\frac{2j+1}{\pi}\left|(\bar x~1)h \left(\stackrel{x}{1}\right)\right|^{2j},
\eea
whit $x,\bar x\in {\mathbb C}$. These functions have simple behavior under symmetry transformations. If $k$ is parametrized by $k^\dagger=\left(\begin{matrix}a&b\cr c&d\end{matrix}\right)$ then
\bea
\phi^{j}(x,\bar x|k h k^\dagger)= |cx+d|^{4j}\phi^{j}(k\cdot x, \bar x\cdot k^{\dagger}|h),
\eea
where 
\bea
k\cdot x= \frac{ax+b}{cx+d}
\eea

It was shown \cite{teschmssl} that a complete basis of functions in the Hyperbolic space is generated by $\phi^j(x,\bar x|h)$, with $j\in{-\frac12+i{\mathbb R}^+}$.

Functions with $j$ and $j'=-j-1$ are related by the reflection symmetry 
\bea
\phi^{-j-1}(x,\bar x|h)=\frac{R(-1-j)}{\pi}\int d^2 y |x-y|^{4j}\phi^j(y,\bar y|h)\label{refsymm}
\eea
where the reflection $R(j)$ is such that 
\bea
R(j) R(-j-1)=-(2j+1)^2.
\eea
\medskip

A continuous basis for $SL(2,{\mathbb R})$ is given by the so called $t$-$basis$
\bea
\phi^{j,\eta}(t_L,t_R|g)=\frac{2j+1}{\pi}\left|(1~-t_L)g \left(\stackrel{t_R}{1}\right)\right|^{2j}
                           sgn^{2\eta}\left[(1~-t_L)g \left(\stackrel{t_R}{1}\right)\right]
\eea
The parity $\eta\in\{0,\frac12\}$, and the symmetry group acts as
\bea
\phi^{j,\eta}(t_L,t_R|g_L^{-1} g g_R)&=&\left|(c_R t_R+d_R)(c_L t_L+d_L)\right|^{2j}
                           sgn^{2\eta}\left[(c_R t_R+d_R)(c_L t_L+d_L)\right] \cr
                           &&\times ~~\phi^{j,\eta}(g_L\cdot t_L,g_R\cdot t_R|g),
\eea
where 
\bea
g\cdot t= \frac{a t+b}{c t + d}.
\eea

A complete basis of functions is known to be $\{\phi^{j,\eta}(t_L,t_R|g); j\in-\frac12+i{\mathbb R}^+,t_L,t_R\in{\mathbb R}\,,$ $\eta\in\{0,\frac12\}~\}\cup
\{\phi^{j,\eta}(t_L,t_R|g); j\in-1-\frac12 {\mathbb N},t_L,t_R\in{\mathbb R},\eta=j$ mod $1\}$

The situation is subtler in $AdS_3$ where the correct behavior under symmetry transformations was found in \cite{hosrib} 
\bea
\phi^{j,\alpha}(t_L,t_R|G_L^{-1} G G_R)&=&\left|(c_R t_R+d_R)(c_L t_L+d_L)\right|^{2j}
                           e^{2\pi i\alpha\left[N(G_L|t_L)-N(G_R|t_R)\right]}\cr
                           &&\times ~~\phi^{j,\alpha}(G_L\cdot t_L,G_R\cdot t_R|G),\label{adsfunct}
\eea   
with $\alpha\in [0,1)$, $G\cdot t\equiv g\cdot t$, $g$ being the $SL(2,{\mathbb R})$ projection of $G\in AdS_3$. $N(G|t)$ is a function with the following properties
\bea
N(G' G|t)&=&N(G'|G t) + N(G|t)\cr
N(\,(id, q)\,|t)&=& q.
\eea   

For instance $N(G|t)$ can be taken as the number of times $G'\cdot t$ crosses infinity when $G'$ moves from $(id,0)$ to $G$. A function satisfying (\ref{adsfunct}) was found in \cite{rib} and is given by
\bea
\phi^{j,\alpha}(t_L,t_R|G)= \frac{2j+1}{\pi} e^{2\pi i\alpha\; n (G|t_L,t_R)}\left| (1,-t_L) g \left(\stackrel{t_R}{1}\right)\right|^{2j},
\eea
where $n$ is the function defined such that $n(G|t_L,t_R)-\frac12 sgn(t_L,t_R)$ gives the number of times $G'\cdot t$ crosses $t_R$ as $G'$ moves from $(id,0)$ to $G$.
 
\subsection{Discrete basis}\label{sectDB}

The continuous $x$ and $t$ basis have simple properties under symmetry transformations. But as $g$ in (\ref{g}) and $h$ in (\ref{h}) are not related by Wick rotation, neither the continuous functions defined above are related by Wick rotation. 

So it is useful to introduce another type of basis, the so called ``$m$-$basis$''. These are sets of functions parametrized by discrete parameters $m,\bar m$, which even though not having a simple behavior under symmetry transformations have a simple connection via Wick rotation.  

The $m$-$basis$ of $H_3^+$ can be defined as a kind of Fourier transformation of the $x$-$basis$
\bea
\phi^j_{m\bar m}(h)\equiv \int d^2x~x^{j+m}\bar x^{j+\bar m}\phi^j(x,\bar x|h).\label{mtransf}
\eea
The combinations $m-\bar m$ and $m+\bar m$ are proportional to the momentum numbers along the compact $\theta$-direction and the non-compact $\tau$ direction respectively, which implies the decomposition
\bea
    m  &=&  \frac{n+p}2\cr
\bar m &=&  \frac{-n+p}2,\label{mh3+}
\eea
with $n\in {\mathbb Z},~ p\in i{\mathbb R}$. Notice that $m-\bar m\in{\mathbb Z}$ ensures the monodromy in (\ref{mtransf}).

The explicit computation of (\ref{mtransf}) gives 
\bea
\phi^j_{m\bar m} &=& -4~ \frac{\Gamma(1+j+\frac{|n|+p}2) \Gamma(1+j+\frac{|n|-p}2)}{\Gamma(|n|+1) \Gamma(1+2j)}e^{p\tilde \tau-in\theta} 
\sinh^{|n|}\rho \cosh^{-p}\rho\cr
&&~~~~~ \times ~~F(1+j+\frac{|n|-p}2,\; -j+\frac{|n|-p}2, |n|+1;-\sinh^2\rho)
\eea

The reflection property (\ref{refsymm}) translates in the $m$-$basis$ to
\bea
\phi^j_{m \bar m}=R^j_{m\bar m}~ \phi^{-j-1}_{m \bar m},\label{reflsymm}
\eea
where
\bea
R^j_{m,\bar m}=\frac {\Gamma(-2j-1)\Gamma(j+1+m)\Gamma(j+1-\bar m)}{\Gamma(2j+1)\Gamma(-j+m)\Gamma(-j-\bar m)}
\eea
\medskip

We take the $m$-$basis$ of the Lorentzian models as the Wick rotation $\tilde \tau\rightarrow i\tau$ of the basis above and by abuse of notation we use the same name, $i.e.$ $\phi^j_{m,\bar m}$. The first difference of the Wick rotated functions is that $p$ must be real in order to ensure the (delta function) normalization. 


If we introduce the parameter $\alpha\in[0,1)$ such that $m,\,\bar m\in\alpha+{\mathbb Z}$, the change of basis between $m$ and $t$-basis is
\bea
\phi^j_{m,\bar m}(G)=c^{j,\alpha}\int_{-\infty}^{\infty} dt_L (1+t_L^2)^{j}\left(\frac{1+it_L}{1-it_L} \right)^m \times \cr
\int_{-\infty}^{\infty} dt_R (1+t_R^2)^{j} \left(\frac{1-it_R}{1+it_R}\right)^{\bar m} \phi^{-1-j,\alpha}(t_L,t_R|G),\label{m.to.t}
\eea
where 
\bea
c^{j,\alpha}=-\frac{4^{-2-2j}\sin2\pi j}{\sin \pi(j-\alpha)\sin \pi(j+\alpha)}.
\eea

All the functions introduced in this section have a relevant role in the field theory description as they describe what is known as the minisuperspace limit of the WZNW models associated with $SL(2,{\mathbb R}), AdS_3$ or their Euclidean rotation, the $H_3^+$ space. In such a limit these functions represent the zero mode contribution of the primary fields. A thorough study of the minisuperspace limit of the $H_3^+$ coset model and the $AdS_3$ WZNW model was presented in \cite{teschmssl} and \cite{rib} respectively. The basis functions discussed in this section were proved to be a complete set of functions in each space. These functions not necesarily belong to the squared integrable set but form an orthogonal basis in the same sense that $\{e^{ik\,x}|k\in{\mathbb R} \}$ is a complete basis over the space of real functions. The completeness of the continuous and discrete basis of functions on $H_3^+$ was proved in \cite{teschmssl}, and as commented in $\cite{rib}$ the proof of the completeness of the discrete basis of $AdS_3$ follows from the results of \cite{basu} where a plancherel formula for $AdS_3$ was proved. The completeness of the continuous basis is a consequence of the integral relation (\ref{m.to.t}). And finally the completeness of the basis functions for $SL(2,{\mathbb R})$ follows from the observation that these are nothing but the functions of $AdS_3$ with $2\pi$ periodicity on $\tau$.

\chapter{$AdS_3$ WZNW Model}\label{chapAdS}

In this chapter we present the $AdS_3$ WZNW model and other coset models related to the former by Wick rotations. In next section we present a brief introduction on nongauged and gauged WZNW models, with the aim of introducing some basics tools, setting the notation and to present the formulae we will be using along the thesis. Then we turn to a description of the $AdS_3$ WZNW model, the $H_3^+$ or $SL(2,{\mathbb C})/SU(2)$ Coset model and the Coset $\frac{SL(2,{\mathbb R})_k/U(1)\times U(1)_{-k}}{{\mathbb Z}_{Nk}}$ model.

\section{WZNW models}\label{sectionWZNW}

WZNW models are CFTs with a Lie group symmetry, where the spectrum is built over representations of the affine algebra. These theories have the peculiarity of being defined with an action, a feature that usually does not occur in CFT's.

Sigma models defined with semisimple group manifolds as target space constitute a natural starting point to construct a theory with the properties mentioned above. Nevertheless, even though these theories are classically scale invariant, the $\beta$ function of the renormalization group is nonzero and so the effective theory becomes massive and a scale anomaly emerges at the quantum level.  

This observation is sufficient to realize that this is not the theory we are looking for. Another indication follows from the fact that the conserved currents do not satisfy the factorization property of CFTs, $i.e.$ they do not factorize in a holomorphic current and an antiholomorphic one.

The requested theory is obtained when the sigma model action is supplemented with a Wess-Zumino term. The WZNW action is \cite{Novikov,Witten1,GW}
\bea
S^{{\tiny  WZNW}} &=& S^{{\tiny sigma}}+S^{{\tiny WZ}}\cr
         &=& \frac{-k}{16\pi}\int d^2 x~ Tr'\left(\partial^\mu g^{-1}\partial_\mu g\right) \cr
         &+&\frac{i k}{24\pi} \int_{B^3} d^3y~ \epsilon_{\alpha\beta\gamma}~ Tr'\left(\tilde g^{-1} \partial^\alpha \tilde g~\tilde g^{-1} 
         \partial^\beta \tilde g~\tilde g^{-1} \partial^\gamma \tilde g~ \right)   
\eea 
where $B^3$ is the space whose boundary is the compactification of the space on which we defined the sigma model. The prime in the trace means a normalization in the trace such that in any representation the generators of the Lie algebra satisfy 
\bea
Tr'\left(t^a~ t^b\right)=2\delta_{ab}.
\eea
The field $g(x)$ lives in a unitary representation of the semisimple group G in order to ensure the sigma model action be real. $\tilde g(y)$ is the extension to the three-dimensional space $B^3$. The coupling constant $k$, usually referred to as the level, must be quantized because $B^3$ has the topology of a sphere. 

Even though the Wess-Zumino term is an integral over a three dimensional space, its variation being a divergence can be written as an integral over the two dimensional boundary and the solution to the Euler-Lagrange equation of the WZNW action is, after the change of variable $z=x^0+ix^1,\,\bar z=x^0-ix^1$, $g(z,\bar z)=f(z)\bar f(\bar z)$ with independent holomorphic and antiholomorphic functions. The conserved currents are 
\bea
J(z)&=&k~\partial g~ g^{-1}, \cr
\bar J(\bar z)&=& -k~g^{-1}\bar \partial g,\label{WZNWcurrent}
\eea 
where the notation $\partial\equiv\partial_z,~ \bar\partial\equiv \partial_{\bar z}$ was used. They are associated with the following invariance of the action
\bea
g(z,\bar z)\rightarrow\Omega(z)~g(z,\bar z)~\bar \Omega^{-1}(\bar z),\label{WZNWsymm} 
\eea
with $\Omega(z),~\bar \Omega(\bar z)$ two arbitrary functions living on G so that the global symmetry $G\times G$ of the sigma model was lifted to a local one  by adding the Wess-Zumino term. The transformation law of the currents is easily read out from (\ref{WZNWsymm}) and (\ref{WZNWcurrent}) and the {\it current algebra} can be determined using the Ward identities. It is found to be 
\bea
J^{a}(z_1) J^{b}(z_2)\sim \frac{-k\delta_{ab}}{(z_1-z_2)^2}+\sum_c if_{abc} \frac{J^c(z_2)}{z_1-z_2},
\eea
where $\sim$ means equal up to regular terms and $J^a$ are the components of $J$ in the $t^a$ basis. So defining the Laurent modes as
\bea
J^a(z)=\sum_{n\in {\mathbb Z}}z^{-n-1} J_n^a, 
\eea
the {\it current algebra} leads to the desired affine Lie algebra $\hat g$
\bea
[J_m^a, J_n^b]=\sum_{c}if_{abc}J^c_{m+n}- kn\delta_{ab}\delta_{m+n,0}
\eea
and similarly for the antiholomorphic sector. The OPE of holomorphic and antiholomorphic currents has no singular terms implying that the modes commute with each other.

The classical energy momentum tensor is obtained from varying the action with respect to the metric\footnote{ From now we will work only with the holomorphic sector, the antiholomorphic sector being analogous.}.
\bea
T(z)_{classic}=-\frac1{2k}\sum_a J^a(z) J^a(z).\label{sugclass}
\eea

Fields are not free, so that this expression will be corrected at quantum level. If the product of currents in (\ref{sugclass}) is replaced by a normal ordered product, namely
\bea
:A(z_1)B(z_2): =\frac{1}{2\pi i} \oint_{z_2} dz_1 \frac{A(z_1)B(z_2)}{z-w},
\eea
and the coefficient is left as a free parameter to be fixed by the requirement that the OPE between two energy momentum tensors be as required by a CFT, $i.e$  
\bea
T(z_1)T(z_2)\sim \frac{c/2}{(z_1-z_2)^4}+\frac{2T(z_2)}{(z_1-z_2)^2}+ \frac{\partial T(z_2)}{z_1-z_2},
\eea
one finds the quantum corrected energy momentum tensor
\bea
T(z)=\frac{-1}{2(k-{\mathfrak g}_c)}\sum_a:J^a J^a:(z),
\eea
where ${\mathfrak g}_c$ is the dual coxeter number and  
the central charge is found to be
\bea
c=\frac{k~ dim\, g}{k-{\mathfrak g}_c}.
\eea

It is bounded from below
\bea
c\geq Rank~ g,\label{ccbound}
\eea
so that $c\geq1$.

This realization of the energy momentum tensor in terms of the currents is usually named in the literature as the {\it Sugawara construction}.

After expanding $T$ according to 
\bea
T(z)=\sum_{n\in{\mathbb Z}}z^{-n-2}L_n
\eea
the Virasoro algebra is realized, namely
\bea
[L_m,L_n]=(n-m)L_{m+n}+ \frac{c}{12}n(n^2-1)\delta_{m+n,0}.
\eea

The commutator between Virasoro and current modes is 
\bea
[L_m,J_n^a]=-nJ_{m+n}^a.
\eea

Of course holomorphic and antiholomorphic modes commute with each other. 
\bigskip

There is much more to say about WZNW models, like discussing the Knizhnik-Zamolodchikov (KZ) equation, free field representations, the modular data, fusion rules and many other matters. We will only discuss the properties we need and at the appropriate time. We end this short review with a few comments on primary fields. 

In conformal field theory one defines primary fields, $\Phi(z,\,\bar z)$, as those transforming covariantly with respect to scale transformations and satisfying 
\bea
T(z_1)\phi(z_2,\bar z_2)\sim\frac{\Delta}{(z_1-z_2)^2}\Phi(z_2,\bar z_2)+\frac1{z_1-z_2}\partial_{z_2}\Phi(z_2,\bar z_2),
\eea 
where $\Delta$ is the conformal dimension of $\Phi$. So that
\bea
[L_n,\Phi(z,\bar z)] &=& \frac{1}{2\pi i}\oint_z dw~ w^{n+1}T(w)\Phi(z,\bar z)\cr
                     &=& \Delta(n+1)z^n~\Phi(z,\bar z) + z^{n+1} \partial\Phi(z,\bar z)\label{cVirPhi}, ~~~~~~~n\geq-1
\eea

On the other hand, in the case of WZNW models primary fields are those transforming covariantly under $G(z)\times G(\bar z)$ and so satisfying the following OPE with the current
\bea
J^a(z_1)~\Phi_{\mu,\bar \mu}(z_2,\bar z_2)\sim \frac{-t^a_\mu ~\Phi_{\mu,\bar \mu}(z_2,\bar z_2)}{z_1-z_2},
\eea
where $\mu, \bar \mu$ denote the holomorphic and antiholomorphic representations of the field, and $t^a_\mu$ is the realization of the generator $t^a$ in such representation. These conditions translate into
\bea
J_0^a|\Phi_{\mu,\bar \mu}>&=&-t^a_{\mu}|\Phi_{\mu,\bar \mu}>,\cr
J_n^a|\Phi_{\mu,\bar \mu}>&=&~0~,~~~~~n>0,
\eea
where $|\Phi_{\mu,\bar \mu}>$ represents the primary state $\Phi_{\mu,\bar \mu}(0)|0>$. 

As a consequence of the realization of the conformal symmetry via the {\it Sugawara construction}, WZNW primaries are also conformal primaries. They satisfy
\bea
L_n |\Phi_{\mu,\bar \mu}>&=&0~,~~~~~n>0,\cr
L_0 |\Phi_{\mu,\bar \mu}>&=&\Delta_{\mu}|\Phi_{\mu,\bar \mu}>,
\eea
where the conformal weight $\Delta_{\mu}$ is 
\bea
\Delta_{\mu}=\frac{-\,t^a_{\mu}t^a_{\mu}}{2(k-{\mathfrak g}_c)}, 
\eea
and $t^a_{\mu}t^a_{\mu}$ is the quadratic Casimir.

But the reader has to bear in mind that the inverse is not always true. A Virasoro primary can be a WZNW descendant.

\subsection{Gauged WZNW models}

We now consider the construction of Coset or Gauged WZNW models. These are constructed from two WZNW models where the first group is a subgroup of the second one. Contrary to what happens in WZNW models (see (\ref{ccbound})), these theories are less restrictive as there are no bounds for the central charge\footnote{The central charge of the Coset is the difference of the central charges of both WZNW models.}. Moreover it is expected that this framework provides a full classification of all RCFT \cite{dif}.

As we saw above a WZNW model with group G is invariant under $G(z)\times G(\bar z)$, thus it has a global symmetry $G\times G$. Given two subgroups $H_\pm\in G$ it is sometimes possible to obtain a theory with local $H_-(z,\bar z)\times H_{+}(z,\bar z)$ symmetry. 

The gauged action was obtained in \cite{Witten2}. This is given by 
\bea
S(g,{\cal A})=S^{sigma}(g,{\cal A})+S^{WZ}(g,{\cal A}),\label{gaugeWZNW}
\eea
where
\bea
S^{sigma}(g,{\cal A})&=&\frac{-k}{16\pi }\int Tr'\left( g^{-1}Dg\wedge*g^{-1} D g \right),\cr
S^{WZ}(g,{\cal A})&=& S^{WZ}(g)+\frac{ik}{16\pi}\int_\Sigma Tr'\left( {\cal A}_-\wedge dg g^{-1}
+ {\cal A}_+\wedge g^{-1} dg +{\cal A}_+ g^{-1}\wedge {\cal A}-L g\right),~~~~~~~~~~~
\eea
$D$ denotes the covariant derivative $D g=d g + {\cal A}_- g - g {\cal A}_+$ and ${\cal A}_\pm$ are the gauge fields associated to $H_\pm$ respectively. It was found that (\ref{gaugeWZNW}) is invariant under local transformation ($\xi_{L/R}=\xi^a(z,\bar z) t_{a,L/R}$)
\bea
\delta g&=&\xi_- g-g\xi_+
\eea
when the gauge fields transform as
\bea
\delta{\cal A}_{\pm}&=&-d\xi_{\pm}+[\xi_{\pm},{\cal A}_{\pm}],
\eea
and $H_{\pm}$ are anomaly free subgroups, $i.e.$ their Lie algebra generators ($t_{a,\pm}$) satisfy \cite{Witten2} 
\bea
Tr\left(t_{a,-} t_{b,-}- t_{a,+} t_{b,+}\right)=0
\eea 
The origin of this unusual restriction can be traced back to the fact that the $S^{WZ}$ is a three dimensional term which defines a two dimensional theory and the standard machinery implemented to gauge a field theory fails.\footnote{Of course the standard approach perfectly works for $S^{sigma}$, $i.e.$ replacing a derivative with a covariant derivative gives a gauge invariant expression.}

\section{$AdS_3$ WZNW and related models}

\subsection{The spectrum}

The spectrum of the $AdS_3$ WZNW model is built with representations of the affine $\hat sl(2)$ algebra, but which are the representations to consider is a subtle question.

The representations of the affine algebra are generated from those of the global $sl(2)$ algebra by freely acting with the current modes $J^a_n,~\bar J^a_n,~a=3, \pm,\,n\in{\mathbb Z}$, obeying the following commutation relations
\bea
[J_n^3,J_m^3]&=&-\frac k2 n\delta_{n+m,0}\, ,\cr
[ J_n^3, J_m^\pm ]&=&\pm J^\pm_{n+m}\, ,\cr
[J_n^+,J_m^-]&=& - 2J^3_{n+m}+kn\delta_{n+m,0}\, ,\label{sl2alg}
\eea
with level $k\in {\mathbb R}_{>2}$.
 
In the first attempts to define a consistent spectrum for the worldsheet theory of string theory on $SL(2,{\mathbb R})$ and $AdS_3$ \cite{BRFW,EGP}\cite{Mohammedi}-\cite{Petropoulos} only representations with $L_0$ bounded from below were considered. These decompose into direct products of the normalizable continuous, highest and lowest weight discrete representations. 
 
The principal continuous representations ${\mathcal C}_j^\alpha \times {\mathcal C}_j^\alpha$ contain the states  $|j,\alpha, m, \bar m>$ with, $\alpha\in[0,1)$, $m,\bar m\in \alpha+{\mathbb Z}$ and they are unitary as long as $j\in -\frac12+i{\mathbb R}^+$. The lowest weight principal discrete representations ${\mathcal D}_j^+ \times {\mathcal D}_j^+$ contain the states $|j,m,\bar m>$ with $j\in{\mathbb R}$, $m ,\bar m\in -j+{\mathbb Z}_{\ge 0}$. The highest weight principal discrete representations ${\mathcal D}_j^- \times {\mathcal D}_j^-$ contain the states $|j,m,\bar m>$ with $j\in{\mathbb R}$, $m ,\bar m\in j-{\mathbb Z}_{\ge 0}$. Both highest and lowest weight representations are unitary when the spin $j$ is constrained as $j<0$ .

The spectrum is supplemented by adding the affine descendants of the global representations defined above, $i.e.$ $\hat{\mathcal D}_j^\pm  \times \hat{\mathcal D}_j^\pm$ and $\hat{\mathcal C}_j^\alpha \times \hat{\mathcal C}_j^\alpha$. Even though these representations are not unitary for $j<0$, a no ghost theorem has been proved in \cite{EGP} for $-\frac k2<j<0$ guaranteeing the spectrum is unitary when the theory is supplemented with a unitary CFT such that the full central charge be $c=26$ and the Virasoro constraint is imposed. 
\medskip

This spectrum raised the problem that it leads to a non-modular invariant partition function. And it also raises two puzzles: an upper limit in the string mass spectrum and the absence of states corresponding to the long strings, which were expected to exist from the results of \cite{MMS,SW}.  

These problems were all solved when it was realized in \cite{mo1} that the spectrum must be extended to include representations with $L_0$ not bounded from below. These representations are nothing but the spectral flow images of the representations considered above. The no ghost theorem for the new representations requires the new bounds $-\frac{k-1}2<j<-\frac12$ to hold \cite{mo1}.\footnote{The bound $j<-\frac12$ already appears in the semiclassical limit. Indeed it is not hard to see that the wave functions in $AdS_3$ are squared integrable iff $j<-\frac12$.} 

The spectral flow transformation is generated by the operators $U_w, \bar U_{\bar w}$, defined by their action on the SL(2,$\mathbb R)$ currents $J^3, J^\pm$ as 
\bea
\left\{\begin{array}{lcr}
\tilde J^3(z) = U_{w} J^3(z) U_{-w}&=&J^3(z)-\frac k2 \frac wz~,\cr
&&\cr
\tilde J^{\pm}(z) = U_{w} J^\pm(z) U_{-w}&=&z^{\pm w}J^\pm(z)~,
\end{array}\right.\label{wtrans}
\eea 
where $U_{-w}=U_w^{-1}$ and similarly for the antiholomorphic sector. 

It is not hard to see that such a transformation leaves the algebra (\ref{sl2alg}) invariant. The right and left spectral flow numbers $w,\bar w$ are independent in the single cover of SL(2,${\mathbb R})$ where $\bar w-w$ is the winding number around the compact closed timelike direction. But in the universal covering, they are forced  to be equal, $w= \bar w \in {\mathbb Z}$. Using the {\it Sugawara construction}, the action of $U_w, \bar U_{\bar w}$ on the Virasoro generators is found to be
\bea
\tilde L_n = U_{-w} L_n U_{w}=L_n+wJ_n^3-\frac k4 w^2\delta_{n,0}~. \label{wl0}
\eea

The conformal weights are easily read as 
\bea
\Delta_j=\tilde \Delta_j+ wm-\frac k4 w^2\delta_{n,0}~, ~~~~~~\tilde\Delta_j=-\frac{j(j+1)}{k-2}.\label{confweight}
\eea
 
We define the nontrivial spectral flow representations as the discrete and continuous representations considered above but with respect to $\tilde J^a$. So that even though $\tilde L_0$ is bounded from below, this is clearly not the case for $L_0$ (see (\ref{wl0})). This implies that unlike in the compact $SU(2)$ case, different amounts of spectral flow give inequivalent representations of the current algebra of $SL(2,\mathbb R)$.

We will use the notation $|j,w,m,\bar m>=U_w \bar U_w|j,m,\bar m>$. The action of the currents on the spectral flowed states is
\bea
J^3_0|j,w,m,\bar m> &=& (m+\frac k2 w)|j,w,m,\bar m>\cr
J^3_n|j,w,m,\bar m> &=& 0~,~~~~~~~n=1,2,\dots\cr
J^+_n|j,w,m=-j,\bar m> &=& 0~,~~~~~~n=w,w+1,\dots\cr
J^-_n|j,w,m=j,\bar m> &=& 0~,~~~~~~n=-w,-w+1,\dots
\eea 

Let us end this presentation about the spectrum by noting that there is an overcounting when considering the spectral flow images of the representations above. In fact the spectral flow image of highest and lowest weight representations are related via the identification $\hat{\cal D}_j^{+,w} \equiv \hat{\cal D}_{-\frac k2-j}^{-,w+1}$. The reader can easily check from (\ref{wtrans}) and (\ref{wl0}) that they have the same spectrum. So, in what follows we will only consider continuous and lowest weight discrete representations and their images under spectral flow. 
\bigskip 

The spectrum of the $H_3^+$ model is simpler. It was determined in \cite{gawe} and it is built from the principal continuous representations, but contrary to the $AdS_3$ case, it does not factorize between holomorphic and antiholomorphic representations\footnote{It is important to note that even though the representations of the spectrum, and so the fields, do not factorize in holomorphic-antiholomorphic sectors, the correlation functions do.}. The Virasoro primary states are parametrized by $|j,m,\bar m>$, with $j\in-\frac12+i{\mathbb R}^+$ and $m,\bar m$ are given by (\ref{mh3+}). 

\subsection{Primary fields}\label{sectPF}

The interpretation of the Lorentzian field theory is subtle and it is instructive to begin with a discussion on the sigma model whose target space is a Euclidean rotation of $AdS_3$, with a non-zero NS-NS 2 form field $B_{\mu\nu}$. This is the $H_3^+$ model, constructed by gauging the $SL(2,{\mathbb C})$ WZNW model with an $SU(2)$ right action \cite{gawe}. A thorough study of the model was presented in \cite{tesch1, tesch2}. 
The 
Lagrangian formulation was  developed in \cite{gawe} and it
follows from
\begin{equation}
{\cal
  L}=\frac k\pi(\partial\phi\overline\partial\phi+e^{2\phi}\overline\partial\gamma
\partial\overline\gamma)\, .\label{h3lagrangian}
\end{equation}

The relevance of this model relies in that it is used to compute the correlation functions of the Lorentzian $AdS_3$ model. The Euclidean action (\ref{h3lagrangian}) is real valued and positive definite and normalizable fields have positive conformal weights, so that Euclidean path integrals are expected to be well defined. As a consequence of the B field, the Euclidean action is not invariant under Euclidean time inversion and so the Lorentzian action is not real and the theory is not unitary.     

Normalizable operators in the $H_3^+$ model, $\Phi_j(x,\overline x; z,\overline z)$, $x,z \in \mathbb C$, are labeled by the spin $j$ of a principal continuous representation of $SL(2,\mathbb C)$. These are primary operators with respect to the Virasoro generators (but not with respect to the $sl(2)$ ones) and can be semiclassically identified with the expression 
\bea
\Phi_j(x,\overline x; z,\overline z)
=\frac{2j+1}{\pi}\left ((\gamma-x)(\overline\gamma-\overline
x)e^\phi+e^{-\phi}
\right )^{2j}\, . \label{phij}
\eea
The variables $x,\bar x$ allow to build a continuous representation where the generators of the $sl(2)$ algebra are differential operators. But as is discussed in \cite{BORT}, these variables have a very important interpretation in string theory as the coordinates of the operators in the dual CFT living in the boundary of $H_3^+$. 

These fields satisfy the following OPE with the holomorphic $SL(2,\mathbb C)$ currents
\bea
J^a(z)\Phi_j(x,\overline x;z',\overline z')\sim\frac{D^a\Phi_j
(x,\overline x;z',\overline z')}{z-z'}\, ,\qquad a=\pm, 3\, ,
\eea
where $D^-=\partial_x\, ,  D^3=x\partial_x-j\, ,D^+=x^2\partial_x-2jx$ and similarly with the antiholomorphic modes, and they have conformal weight $\widetilde \Delta_j$ as defined in (\ref{confweight}). The asymptotic  $\phi\rightarrow\infty$ expansion, given by 
\bea
\Phi_j(x,\overline x| z,\overline z) \sim :e^{2(-1-j)\phi(z)}:\delta^2\left (\gamma(z)-x\right
)+B(j):e^{2j\phi(z)}: |\gamma(z)-x|^{4j}\, ,\label{asym}
\eea
fixes a normalization and determines the relation between $\Phi_j$ and $\Phi_{-1-j}$ as
\bea
\Phi_j(x,\overline x| z,\overline z)= B(j)\int_{\mathbb C}d^2x'
|x-x'|^{4j}\Phi_{-1-j}(x',\overline x';z,\overline z)\, ,
\eea
which explicitly realizes the equivalence between representations with ``spin'' $j$ and $j'=-1-j$. The reflection coefficient $B(j)$ is given by
\bea
B(j) = \frac{k-2}{\pi} \,\frac{\nu^{1+2j}}{\gamma\left (-\frac{1+2j}{k-2}\right )}\, , \quad
\nu=\pi\frac{\Gamma\left (1-\frac 1{k-2}\right )}{\Gamma\left (1+\frac
  1{k-2}\right )}\, ,\quad \gamma(x)=\frac{\Gamma(x)}{\Gamma(1-x)} \, .\label{2pf}
\eea

As the path integral of the Lorentzian theory is ill defined, the correlation functions are defined via analytic continuation from those of the $H_3^+$ model. Even those involving fields in non trivial spectral flow sectors, absent in the Euclidean model, can be obtained from the latter by certain formal manipulations \cite{mo3}. Then, as it is clear from section \ref{sectBF} we will need to Fourier transform these expressions to the $m$-$basis$ in order to obtain fields which can be interpreted as fields of the $AdS_3$ model after analytic continuation. This transformation is exactly the same carried out for the basis functions of section \ref{sectDB}, $i.e.$
\bea
\Phi^j_{m,\overline m}(z,\overline z)=\displaystyle\int d^2x ~ x^{j+m} ~\overline
x^{j+\overline m}~\Phi_{-1-j}(x,\overline x; z,\overline z)\, ,\label{phim}
\eea
where $m=\frac {n+is}2,~ \overline m=\frac{-n+is}2,~ n\in {\mathbb Z}, s\in \mathbb R$. The fields $\Phi^j_{m,\overline m}$ have the following OPE with the chiral currents
\bea
J^\pm(z)\Phi^j_{m,\overline m}(z',\overline z')&\sim&\frac{\mp j + m}{z-z'} \Phi^j_{m\pm 1,\overline m}(z',\overline z')\, ,\cr\cr
 J^3(z)\Phi^j_{m,\overline m}(z',\overline z') &\sim& \frac m{z-z'}\Phi^j_{m,\overline m}(z',\overline z')\, ,~~~~
\eea
and the relation between $\Phi^j_{m,\overline m}$ and $\Phi^{-1-j}_{m,\overline m}$ is given by 
\bea
\Phi^j_{m,\overline m}(z,\overline z)=B(-1-j) c^{-1-j}_{m,\overline m}\Phi^{-1-j}_{m,\overline m}
(z,\overline z) \label{gnm}
\eea
which generalized the reflection relation of the superspace limit (\ref{reflsymm}). The reflection coefficient $c^j_{m,\bar m}$ is defined as
\bea
c^j_{m,\bar m}=\frac{-\pi~}{1+2j}\frac{\Gamma(1+2j)~\Gamma(-j+m)~\Gamma(-j-\overline m) }{\Gamma(-1-2j)
\Gamma(1+j+m)\Gamma(1+j-\overline m)}\, .\label{cjmm}
\eea

We take the fields of the $AdS_3$ WZNW model as the Wick rotated from those of the $H_3^+$ coset but we will use the same notation, with the difference that now the quantum numbers are in agreement with the representations appearing in the former model.
 
An affine primary state in the unflowed sector is mapped by the automorphism (\ref{wtrans}) to a highest/lowest-weight state of the global $sl(2)$ algebra.
We denote these fields in the spectral flow sector $w$ as $\Phi^{j,w}_{m,\overline m}$. Their explicit expressions will not be needed below. It is only necessary to know that they verify the following OPE with the currents:
\bea
J^3(z)\Phi^{j,w}_{m,\overline m}(z',\overline z')&\sim&\frac{m+\frac k2w}{z-z'}\Phi^{j,w}_{m,\overline m}(z',\overline z')\, ,\cr\cr
  J^\pm(z)\Phi^{j,w}_{m,\overline m}(z',\overline z')&\sim&\frac{\mp j+m}{(z-z')^{\pm w}}\Phi^{j,w}_{m\pm 1,\overline m}(z',\overline z')+\cdots\, \label{opew}
\eea
$m-\overline m \in \mathbb Z$, $m+\overline m\in \mathbb R$ and dots denote we only write down the leading singular term in the OPE. Its conformal weight is given by $\Delta_j$ as defined in (\ref{confweight}).

These fields satisfy the same reflection relation as those of the $w=0$ sector (\ref{gnm}),
\begin{equation}
\Phi^{j,w}_{m,\overline m}(z,\overline z)=B(-1-j) c^{-1-j}_{m,\overline
  m}\Phi^{-1-j,w}_{m,\overline m}(z,\overline z)\,. \label{rr}
\end{equation}
\bigskip

Let us end this section presenting a model constructed from the axial coset $SL(2,{\mathbb R})/U(1)_A$, which also admits an analytic continuation to the $AdS_3$ model. It was shown that such a model has some advantages with respect to the $H_3^+$ model in that the spectrum is similar to that of $AdS_3$ and it was shown that the correlation functions \cite{mo3} and the partition function \cite{mo1} obtained from the hyperbolic model can be exactly reproduced with this new construction \cite{kounnas} using path integral techniques. It was also used to compute one point functions associated with symmetric and symmetry breaking $D$-branes ( see \cite{israel} and section \ref{sect1pf} for the case of maximally symmetric $D$-branes).
 
The construction rests on the observation that, after doing a T duality in the timelike 
direction, the 
$N$-th cover of SL(2,${\mathbb R})$, $i.e.$
SL(2,${\mathbb R})^N_k$, is given by the orbifold
\bea
\frac{SL(2,{\mathbb R})_k/U(1)\times U(1)_{-k}}{{\mathbb Z}_{Nk}}\, .
\eea

Because now the timelike direction is a free compact boson, 
the analytic continuation to  Euclidean space 
is simply obtained by replacing $U(1)_{-k} \rightarrow U(1)_{R^2k}$. 
Thus, one can construct arbitrary correlation functions in 
$AdS_3$ from those in the cigar and  the free compact boson theories,
after taking 
the limits 
$N\rightarrow\infty$, $R^2\rightarrow-1$. The effect of the orbifold is to
 produce new (twisted) sectors. 
 These can be read in the following modification of the left and right 
momentum modes in the coset 
and the free boson models, respectively,  
\bea
\frac{\left(n+k\omega,n-k\omega\right)}{\sqrt{2k}}&\longrightarrow& 
\frac{\left(n+k\omega-\frac{\gamma}{N},n-k\omega+\frac{\gamma}{N} \right)}
{\sqrt{2k}},~~ \gamma\in{\mathbb Z}_{kN}\, ,\cr
\frac{\left(\tilde n+R^2k\tilde\omega,\tilde n-R^2k\tilde \omega\right)}
{R\sqrt{2k}}&\longrightarrow& \frac{\left(n+kNp+R^2k\tilde 
\omega+\frac{R^2\gamma}{N},n+kNp-R^2k\tilde \omega-\frac{R^2\gamma}{N} 
\right)}{R\sqrt{2k}}
\, ,\cr&&
\eea
with $p\in{\mathbb Z}$ and $\omega, \tilde\omega$ being the 
winding numbers in the cigar and U(1) respectively.
In the $N$-th cover, 
$k$ has to be an integer, but in the universal 
covering,
the theory can be defined for arbitrary real level $k>2$ \cite{israel}.

The vertex operators for the orbifold theory are the product of the 
vertices in each space, namely
\bea
V^j_{n\omega\gamma p\tilde\omega}(z,\bar z)=\Phi^{sl(2)/u(1)}_{j,n,
\omega-\frac{\gamma}{kN}}
(z,\bar z)~~\Phi^{u(1)}_{n+kNp,\tilde\omega+\frac{\gamma}{kN}}(z,\bar z)\, .
\label{ver}
\eea
In the universal covering, 
the discrete momentum $\frac{\gamma}{kN}$ becomes a continuous parameter 
$\lambda\in [0,1)$,
the $J_0^3,\bar J_0^3$ quantum numbers read
\bea
M=-\frac n2+\frac k2(\tilde\omega+\lambda),~~~~ \bar M=\frac n2 +\frac 
k2(\tilde\omega+\lambda),
\eea 
and the spectral flow number is given by 
\bea
w=\omega+\tilde\omega.
\eea

\chapter{Operator algebra}\label{chapOPE}

The OPE in non RCFTs, as for instance in the case of WZNW (or gauged WZNW) models with non compact groups, is much subtler than those of RCFT. The structure constants can be determined by the usage of null states as commented in section \ref{bootapp} and generalizing the bootstrap approach of \cite{BPZ} complementing this with the KZ equation (in the case of WZNW) models. So that these new approaches are analytic rather than algebraic as in the RCFT case. 

We begin this chapter presenting a brief review of the OPE in the $H_3^+$ Coset model. Then we turn to the construction of the Operator Algebra in $AdS_3$ WZNW model by an appropriate Wick rotation of that of the $H_3^+$ model and by adding the non preserving spectral flow structure constants.

\section{Operator algebra in $H_3^+$}

In the case of $H_3^+$ the degenerate (reducible) representations of $sl(2)$ are not included in the spectrum and this is the reason why the conjecture that the correlation functions are analytic in their parameters is so important. Two and three point functions of the hyperbolic model was determined by Teschner in \cite{tesch1,tesch2}. 

The following operator product expansion for any product $\Phi_{j_1}\Phi_{j_2}$ in the $H_3^+$ model was determined in {\it loc. cit.}:
\bea
\Phi_{j_2}(x_2|z_2)\Phi_{j_1}(x_1|z_1)&=&\displaystyle\int_{{\cal P}^+}dj_3~ C(-j_1,-j_2,-j_3)
~|z_2-z_1|^{-\widetilde\Delta_{12}}\int_{\mathbb C}d^2x_3|x_1-x_2|^{2j_{12}}\nonumber\\
&&\times |x_1-x_3|^{2j_{13}} |x_2-x_3|^{2j_{23}}\Phi_{-1-j_3}(x_3|z_1)+{\rm descendants}.\label{ope}
\eea
Here, the integration contour is ${\cal P}^+=-\frac 12+i{\mathbb R}_+$, the structure constants $C(j_1,j_2,j_3)$ are given by

\bea
C(j_1,j_2,j_3) = - \frac{G(1- j_1-j_2-j_3)G(-j_{12})G(-j_{13})G(-j_{23})}{2 \pi^2 \nu^{j_1+j_2+j_3-1} 
\gamma\left(\frac{k-1}{k-2}\right) G(-1) G(1-2j_1) G(1-2j_2)  G(1-2j_3)}\,, \label{sc}
\eea
with $G(j)=(k-2)^{\frac{j(1-j-k)}{2(k-2)}} \, \Gamma_2(-j |1,k-2 ) \, \Gamma_2(k-1+j |1,k-2 ),$ $\Gamma_2(x|1,w)$ being the Barnes double Gamma function,
$\widetilde\Delta_{12}=\widetilde\Delta(j_1)+\widetilde\Delta(j_2)-\widetilde\Delta(j_3)$ and $j_{12}=j_1+j_2-j_3$, etc.

The OPE (\ref{ope}) holds for a range of values of $j_1, j_2$ given by 
\begin{equation}
|{\rm Re}(j_{21}^\pm)|<\frac 12\, ,\qquad j_{21}^+=j_2+j_1+1\, ,\qquad
 j_{21}^-=j_2-j_1\, .\label{range}
\end{equation}
This is the maximal region in which $j_1, j_2$ may vary such that none of the poles of the integrand hits the contour of integration over $j_3$. However, as long as the imaginary parts of $j_{21}^\pm$ do not vanish, J. Teschner \cite{tesch2} showed that (\ref{ope}) admits an analytic continuation to generic complex values of $j_1, j_2$, defined by deforming the contour ${\cal P}^+$. The deformed contour is given by the sum of the original one plus a finite number of circles around the poles leading to a finite sum of residue contributions to the OPE. When $j_{21}^\pm$ are real one can give them a small imaginary part which is sent to zero after deforming the contour. 

\section{Operator algebra in $AdS_3$}

\subsection{Correlation functions in $AdS_3$}

Following \cite{mo3},\cite{hs}-\cite{gk2} we assume that correlation functions of primary fields in the $AdS_3$ WZNW model are those of $H_3^+$ with $j_i, m_i,\overline m_i$ taking values in representations of $\widetilde{SL(2,\mathbb R)}$. The  spectral flow operation is straightforwardly performed in the $m-$basis where the only change in the $w-$conserving expectation values of  fields $\Phi_{m,\overline m}^{j,w}$ in different $w$ sectors is in the powers of the coordinates $z_i, \overline z_i$. Correlation functions may violate $w-$conservation according to the following spectral flow selection rules established in \cite{mo3}
\bea
-N_t+2 ~ \le ~ \sum_{i=1}^{N_t} &w_i&  \le ~ N_c-2\, ,\qquad {\rm {at ~least~ one ~
state ~in}}~ \widehat{\cal C}^{\alpha, w}_j \otimes\widehat{\cal C}^{\alpha, w}_j , \label{cc}\\
-N_d+1~\le ~ \sum_{i=1}^{N_t} &w_i & \le ~ -1\, ,\qquad\qquad {\rm {all~ states ~in}}~
\widehat{\cal D}_j^{+,w} \otimes\widehat{\cal D}_j^{+,w},\label{dd}
\eea
with $N_t=N_c+N_d$ and $N_c, N_d$ are the total numbers of operators in $\widehat{\cal C}^{\alpha, w}_j\otimes\widehat{\cal C}^{\alpha, w}_j$ and $\widehat{\cal D}^{+,w}_j\otimes\widehat{\cal D}^{+,w}_j$, respectively.

The spectral flow preserving two-point function is given by
\begin{eqnarray}
\label{2point}
 \langle \Phi^{j,w}_{m,\overline m}(z,\overline z)\Phi^{j',-w}_{m',\overline
  m'}(z', \overline z')\rangle &=&  \delta^2(m+m') \, (z-z')^{-2\Delta(j)}
  (\overline{z}-\overline{z}')^{-2\overline{\Delta}(j)}  \nonumber \\
&& \times ~ \left[ \delta(j+j'+1) +  B(-1-j) c^{-1-j}_{m,\overline{m}} \, \delta(j-j')\right], \label{218}
\end{eqnarray}
where $\Delta(j) =\widetilde\Delta(j)-wm-\frac{k}{4} w^2 = -\frac{j(j+1)}{k-2}-wm-\frac{k}{4} w^2$, $B(j)$ and $c^j_{m,\bar m}$ were defined in (\ref{2pf}) and (\ref{cjmm}) respectively

For states in discrete series it is convenient to work with spectral flow images of both lowest- and highest-weight representations related by the identification $\widehat{\cal D}_j^{+, w}\equiv \widehat{\cal D}_{-\frac k2-j}^{-, w+ 1}$, which determines the range of values for the spin 
\begin{equation}
-\frac{k-1}2<j<-\frac 12\, , \label{unitary}
\end{equation}
and allows to obtain the ($\pm 1$) unit spectral flow two-point functions from (\ref{218}).

Spectral flow conserving three-point functions are the following:
\begin{eqnarray}
\label{threepoint}
&& \left\langle \prod_{i=1}^3 \Phi^{j_i,w_i}_{m_i,\overline m_i}(z_i,\overline z_i)\right\rangle 
= \delta^2({\mbox{$\sum m_i$}}) \, C(1+j_i) W\left[\begin{matrix} j_1\,,\,j_2\,,\,j_3\cr 
m_1,m_2,m_3\cr\end{matrix}\right] \prod_{i<j} z_{ij}^{-\Delta_{ij}} \overline z_{ij}^{-\overline \Delta_{ij}} ,
\end{eqnarray}
where $z_{ij}=z_i-z_j$ and $C(j_i)$ is given by (\ref{sc}). The function $W$ is 
\begin{eqnarray}
&& W\left[\begin{matrix}
j_1\,,\,j_2\,,\,j_3\cr m_1,m_2,m_3\cr\end{matrix}\right] =\int d^2x_1\, d^2x_2\, x_1^{j_1+m_1}
\overline{x}_1^{j_1+\overline{m}_1} x_2^{j_2+m_2} \overline{x}_2^{j_2+\overline{m}_2} \nonumber \\
&& ~~~~~~~~~~~~~~~~~~~~ \qquad\qquad\times ~ |1-x_1|^{-2j_{13}-2} |1-x_2|^{-2j_{23}-2}|x_1-x_2|^{-2j_{12}-2}\, ,\label{w1i}
\end{eqnarray}
and we omit the obvious $\overline m-$dependence in the arguments to lighten the notation. This integral was computed in \cite{fukuda}.

The one unit spectral flow three-point function  \cite{mo3} is given by \footnote{For an independent calculation of three-point functions using the free field approach see \cite{in}}
\begin{eqnarray}
\label{3pointwind}
\left\langle \prod_{i=1}^3 \Phi^{j_i,w_i}_{m_i,\overline{m}_i}(z_i,\overline z_i)\right\rangle 
= \delta^2(\mbox{$\sum m_i$}\pm\frac k2) \, \frac{\widetilde{C}(1+j_i) \widetilde{W}
\left[\begin{matrix}j_1~~,~j_2~,~~j_3\cr \pm m_1,\pm m_2,\pm m_3\cr\end{matrix}\right]}
{\gamma(j_1+j_2+j_3+3-\frac k2)} \, \prod_{i<j} z_{ij}^{-\Delta_{ij}}\overline{z}_{ij}^{-\overline{\Delta}_{ij}} ,
\end{eqnarray}
where $\sum_iw_i=\pm 1$, the $\pm$ signs corresponding to the $\pm$ signs in the r.h.s.,
\begin{eqnarray}
\widetilde{C}(j_i) \sim B(-j_1) C\left (\frac k2-j_1,j_2,j_3 \right ),
\end{eqnarray}
up to $k-$dependent, $j-$independent factors and
\begin{eqnarray}
\widetilde{W}\left[\begin{matrix} j_1\,,\,j_2\,,\,j_3\cr  m_1, m_2, m_3\cr\end{matrix}\right]
=\frac{\Gamma(1+j_1+m_1)} {\Gamma(-j_1-\overline{m}_1)}\frac{\Gamma(1+j_2+\overline m_2)}{\Gamma(-j_2-m_2)}
\frac{\Gamma(1+j_3+\overline{m}_3)} {\Gamma(-j_3-m_3)}.
\end{eqnarray}

For discrete states, this expression is related to the $\sum_iw_i=\pm 2$ three-point function through $\widehat{\cal D}_j^{+, w}\equiv \widehat{\cal D}_{-\frac k2-j}^{-, w+ 1}$.

\subsection{Operator Product Expansion in $AdS_3$}\label{opeads3}

A non-trivial check on
the   OPE (\ref{ope})
and structure constants (\ref{sc}) of
the $H_3^+$ WZNW model is that 
the well-known
fusion rules of degenerate representations \cite{ay} 
are exactly recovered by analytically continuing  $j_i, i=1,2$  \cite{tesch1}. 
 On the other hand, it was argued in \cite{mo3}-\cite{tesch2}, \cite{hs}-\cite{gk2} that correlation
functions  in the $H_3^+$ and $AdS_3$ WZNW models are related
by analytic continuation and moreover, the
$k\rightarrow \infty$ limit of the
OPE of  unflowed fields computed along these lines in
\cite{hs, satoh} exhibits complete agreement with the classical tensor
products of representations of $SL(2,\mathbb R)$ \cite{holman}.
It seems then natural to conjecture that the 
 OPE of all fields in the spectrum of the $AdS_3$ WZNW
model can be obtained from
(\ref{ope}) 
analytically continuing $j_1, j_2$ from the range
 $(\ref{range})$.

 However, the spectral flowed fields do not belong to the spectrum of the $H_3^+$ model and moreover, the spectral flow symmetry transforms primaries into descendants. Thus, a better knowledge of these representations seems necessary in order to obtain the fusion rules in the $AdS_3$ model and these cannot be simply obtained by a straight forward analytic continuation. Nevertheless, we will show that the OPE and fusion rules obtained from the $H_3^+$ model by analytic continuation and by taking into account the $w-$violating structure constants in addition to (\ref{sc}) require a truncation imposed by the spectral flow symmetry and once it is carried out the OPE and fusion rules close in the spectrum of the theory and satisfy many consistency checks. For instance, the selection rules of arbitrary correlation functions with fields in any representation of the model and the appropriate semiclassical limit are reproduced. In this section we explore this possibility in order to get the OPE of primary fields and their spectral flow images in the $AdS_3$ WZNW model.

To deal with highest/lowest-weight and spectral flow representations it is convenient to work in the $m-$basis.  We have to keep in mind that when $j$ is real, new divergences appear in the transformation from the $x-$basis and this must be performed for certain values of $m_i,\overline m_i$, $i=1,2$. Indeed, to transform the OPE (\ref{ope}) to the $m-$basis using (\ref{phim}), the integrals over $x_1, x_2$ in the r.h.s. must be interchanged with the integral over $j_3$ and this process does not commute in general if there are divergences. However, restricting $j_1, j_2$ to the range (\ref{range}), one can check that the integrals commute and are regular when  $|m_i|<\frac 12$ and $|\overline m_i|<\frac12$, $i=1,2,3$, where $m_3=m_1+m_2, \overline m_3=\overline m_1+\overline m_2$. For other values of $m_i, \overline m_i$ the OPE must be defined, as usual, by analytic continuation of the parameters. Therefore, after performing the $x_1, x_2$ integrals, the
OPE (\ref{ope}) in the $m-$basis is found to be 
\bea
\left.\Phi^{j_1}_{m_1,\overline m_1}(z_1, \overline z_1)\Phi^{j_2}_{m_2,\overline m_2}(z_2, \overline z_2)\right|_{w=0}
&=& \displaystyle\int_{\cal P} dj_3~|z_{12}|^{-2\widetilde\Delta_{12}}~Q^{w=0}\left[\begin{matrix}j_1\,,j_2,\,j_3\cr 
m_1, m_2, m_3\cr\end{matrix}\right] \Phi^{j_3}_{m_3,\overline m_3}(z_1,\overline z_2)\cr 
&&~~~~~~~+~{\rm descendants} ,\label{opetm}
\eea
where we have defined
\begin{equation}
Q^{w=0}\left[
\begin{matrix}
j_1\,,j_2,\,j_3\cr  m_1, m_2, m_3\cr\end{matrix}\right]=C(1+j_1,1+j_2,-j_3)W
\left[
\begin{matrix}
j_1,j_2,-1-j_3\cr  m_1, m_2, -m_3\cr\end{matrix}\right]\, .\label{q0}
\end{equation}

It is easy to see that the integrand is symmetric under
$j_3\rightarrow-1-j_3$ using the identity \cite{satoh}
\begin{eqnarray}
\frac{W\left[\begin{matrix}
j_1\,,\,j_2\,,\,j_3\cr  m_1, m_2, m_3\cr\end{matrix}\right]}{W\left[\begin{matrix}
j_1,j_2,-1-j_3\cr  m_1\,,\, m_2\,,\, m_3\cr\end{matrix}\right]}
=\frac{C(1+j_1,1+j_2,-j_3)}
{C(1+j_1,1+j_2,1+j_3)}B(-1-j_3)c^{-1-j_3}_{m_3,\overline m_3}\, ,\label{eqsat}
\end{eqnarray}
and as a consequence of (\ref{gnm}). 
In the $x-$basis, every pole in (\ref{ope}) appears duplicated, one over
 the real axis
and another one below, and
the 
 $j_3\rightarrow-1-j_3$ symmetry implies that the integral may be equivalently
 performed either over ${\rm Im} ~j_3>0$ or
over ${\rm Im}~ j_3<0$  \cite{tesch2}.
 In the
 $m-$basis, the $(j_1,j_2)-$dependent poles are also duplicated but
the $m-$dependent poles  are not. The 
 $j_3\rightarrow-1-j_3$ symmetry is still
present, as we discussed above, because of poles and zeros in the
normalization of $\Phi^j_{m,\overline m}$.
The integral must be extended to the full axis
${\cal P}=-\frac12+i\mathbb R$ before
performing the analytic continuation in $m_1,\, m_2$ because the $m-$dependent
poles fall on the real axis.

Since the $w-$conserving structure constants of  operators $\Phi_{m,\overline m}^{j,w}\in {\cal C}_{j}^{\alpha, w}$ or ${\cal D}_{j}^{+,w}$ in different $w$ sectors do not change in the $m-$basis \footnote{~ We denote the series containing the highest/lowest-weight states obtained by spectral flowing primaries as ${\cal C}_{j}^{\alpha, w},{\cal D}_{j}^{+,w}$.}, the OPE (\ref{opetm}) should also hold  for fields obtained by spectral flowing primaries to arbitrary $w$ sectors, as long as they satisfy $w_1+w_2=w_3$. But this OPE only reproduces the $w$-preserving three-point functions, which are the ones directly obtained from $H_3^+$. The full OPE requires to additionally take into account the spectral flow non-preserving structure constants and to consider the following OPE \footnote{A similar expression was proposed in  \cite{ribault} and some supporting evidence was presented from the relation between the $H_3^+$ model and Liouville theory.}
\begin{eqnarray}
\Phi^{j_1,w_1}_{m_1,\overline m_1}(z_1,\overline z_1)\Phi^{j_2,w_2}_{m_2,\overline
  m_2}(z_2, \overline z_2) = \sum_{w=-1}^1 \displaystyle\int_{{\cal P}} dj_3~Q^w
 z_{12}^{-\Delta_{12}}\overline z_{12}^{-\overline\Delta_{12}} \Phi^{j_3,w_3}_{m_3,\overline
m_3}(z_2,\overline z_2) +\cdots\, , \label{opec}
\end{eqnarray}
with $w=w_3-w_1-w_2$, $m_3=m_1+m_2-\frac k2w$, $\overline m_3=\overline m_1+\overline m_2-\frac k2w$, and

\begin{eqnarray}
Q^{w=\pm 1}(j_i;m_i,\overline m_i)  &=& \widetilde W\left[\begin{matrix} j_1\,,\,j_2\,,\,j_3\cr
\mp m_1, \mp m_2, \pm m_3\cr\end{matrix}\right]\frac{\widetilde C(j_i+1)} {B(-1-j_3) c_{m_3,\overline m_3}^{-j_3-1}
\gamma(j_1+j_2+j_3+3-\frac{k}{2})}\nonumber\\
&\sim & \frac{\Gamma(\pm\overline m_3-j_3)}{\Gamma(1+j_3\mp m_3)} \prod_{a=1}^2\frac{\Gamma(1+j_a\mp m_a)}
{\Gamma(-j_a\pm\overline m_a)} \frac{C(\frac{k}{2}-1-j_1,1+j_2,1+j_3)} {\gamma(j_1+j_2+j_3+3-\frac{k}{2})}\, .\nonumber\\
\label{qw}
\end{eqnarray}
For completeness, according to the spectral flow selection rules (\ref{dd}), we should also include terms with $w=\pm 2$ in the sum. However, we shall show in the next section that these do not affect the results of the OPE. The integrand is also symmetric under $j_3\rightarrow -1-j_3$. This follows from (\ref{eqsat}) and the analogous identity
\begin{eqnarray}
\frac{\widetilde {W}\left[\begin{matrix} j_1\,,\,j_2\,,\,j_3\cr  m_1, m_2, m_3\cr\end{matrix}\right]}{\widetilde {W}
\left[\begin{matrix} j_1,j_2,-1-j_3\cr  m_1\,,\, m_2\,,\, m_3\cr\end{matrix}\right]}
=\frac{\widetilde {C}(1+j_1,1+j_2,-j_3)\gamma(j_1+j_2+j_3+3-\frac k2)} {\widetilde {C}(1+j_1,1+j_2,1+j_3)
\gamma(j_1+j_2-j_3+2-\frac k2)}B(-1-j_3)c^{-1-j_3}_{m_3,\overline m_3}\, ,~~~~~~~\label{eqsat2}
\end{eqnarray}
together with the reflection relation (\ref{rr}). The dots in (\ref{opec}) stand for spectral flow images of current algebra descendants with the same $J_0^3$ eigenvalues $m_3,\overline m_3$. This expression is valid for $j_1, j_2$ in the range (\ref{range}) and the restrictions on $m_1,m_2$ depend on $Q^w$. The maximal regions in which they may vary such that none of the poles hit the contour of integration are, $min\{m_1+m_2,\overline m_1+\overline m_2\}<\frac 12$ and $max\{m_1+m_2,\overline m_1+\overline m_2\}>-\frac 12$ for $Q^{w=0}$, $min\left\{m_1+m_2,\overline m_1+\overline m_2\right\}<-\frac{k-1}{2}$ for $Q^{w=-1}$ and $max\left\{m_1+m_2,\overline m_1+\overline m_2\right\}>\frac{k-1}{2}$  for $Q^{w=+1}$. So that the bounds for $Q^{w=1}$ and $Q^{w=-1}$ ensure the convergence for the contributions from $Q^{w=0}$. For other values of $j_1,j_2$ and $m_1,m_2$ the OPE must be defined by analytic continuation. In the rest of this section we perform this continuation.

To specifically display the contributions to (\ref{opec}) we have
to study the analytic structure of $Q^w$. We first consider the
simpler case
$w=\pm 1$ and we refer to the terms proportional to
$Q^{w=\pm 1}$ as {\it spectral flow non-preserving contributions} to the OPE. Then,
 we investigate $Q^{w=0}$ and obtain the {\it spectral flow preserving contributions}.

\subsection{Spectral flow non-preserving contributions}

Let us study the analytic structure of
$Q^{w=\pm 1}$ in (\ref{qw}).
The $m$-independent poles arising from the last
factor  are
 the same for both $w=\pm 1$ sectors
and
are explicitly given by
\begin{eqnarray}
     j_3 = \pm j^-_{21}+\frac k2 -1+p+q(k-2)\, ,\qquad  &&
j_3=\pm j^-_{21}-\frac k2 -p-q(k-2) \, ,\nonumber\\
        j_3=\pm j^+_{21}+\frac k2-1+p+q(k-2)\, ,\qquad &&
    j_3=\pm j^+_{21}-\frac k2-p-q(k-2)
\, ,\label{polos}
\end{eqnarray}
with $p,q = 0,1,2,\dots$
The $m$-dependent poles, instead, vary according to the spectral flow
sector.
However they
are connected through
 $(m,\overline m)\leftrightarrow(-m,-\overline m)$ and thus
going from
 $w= -1$ to  $w=+1$  involves the
change
${\cal D}^{-,\;w_i}_{j_i}
\otimes{\cal D}^{-,\;w_i}_{j_i}
\leftrightarrow
{\cal D}^{+,\;w_i}_{j_i}\otimes{\cal D}^{+,\;w_i}_{j_i}$.
Therefore we concentrate on the contributions from
$w=-1$.

 By abuse of
notation, from now on
we denote the states by the representations they belong to and we write only the
holomorphic sector for short, $e.g.$ when $\Phi_{m_i,\overline m_i}^{j_i,w_i}\in
{\cal D}^{+,w_i}_{j_i}\otimes{\cal D}^{+,w_i}_{j_i},
\, i=1,2$, we
 write the set of all possible operator products
$\Phi_{m_1,\overline m_1}^{j_1,w_1}\Phi_{m_2,\overline
m_2}^{j_2,w_2}$ for generic quantum numbers within these
representations as $ {\cal D}^{+,w_1}_{j_1}\times{\cal
D}^{+,w_2}_{j_2}$. 

Let us study the OPE of fields in all different
combinations of representations. First consider  the case
$\Phi_{m_i,\overline m_i}^{w_i,j_i}\in {\cal
C}_{j_i}^{\alpha_i,w_i} \otimes {\cal C}_{j_i}^{\alpha_i,w_i},
i=1,2$, $i.e.$

\begin{itemize}
    \item ${\cal C}_{j_1}^{\alpha_1,\;w_1}\times\;
{\cal C}_{j_2}^{\alpha_2,\;w_2}$
\end{itemize}

The pole structure of $Q^{w=-1}$
 is represented in Figure 1.$a)$ for $min\left\{m_1+m_2,\,\overline m_1 +
\overline m_2\right\}<-\frac {k-1}{2}$. Recalling that
$m_3=m_1+m_2+\frac k2$, then
$min\left\{m_3,\overline m_3\right\}<\frac 12$, and therefore the poles
 from the factor $\frac{\Gamma(-j_3-\overline m_3)}{\Gamma(1+j_3+m_3)}$
are to the right of the  integration contour. Moreover, given that all
$m-$independent poles are to the right of the axis $\frac k2-1$
or to the left of $-\frac k2$, we conclude that
the OPE ${\cal
  C}_{j_1}^{\alpha_1,w_1}\times ~{\cal C}_{j_2}^{\alpha_2,w_2}$ receives no
 spectral flow
violating contributions from discrete representations when $min\{m_1+m_2,
\overline m_1+\overline m_2\}<-\frac{k-1}2$.

\bigskip

\medskip

\centerline{\psfig{figure=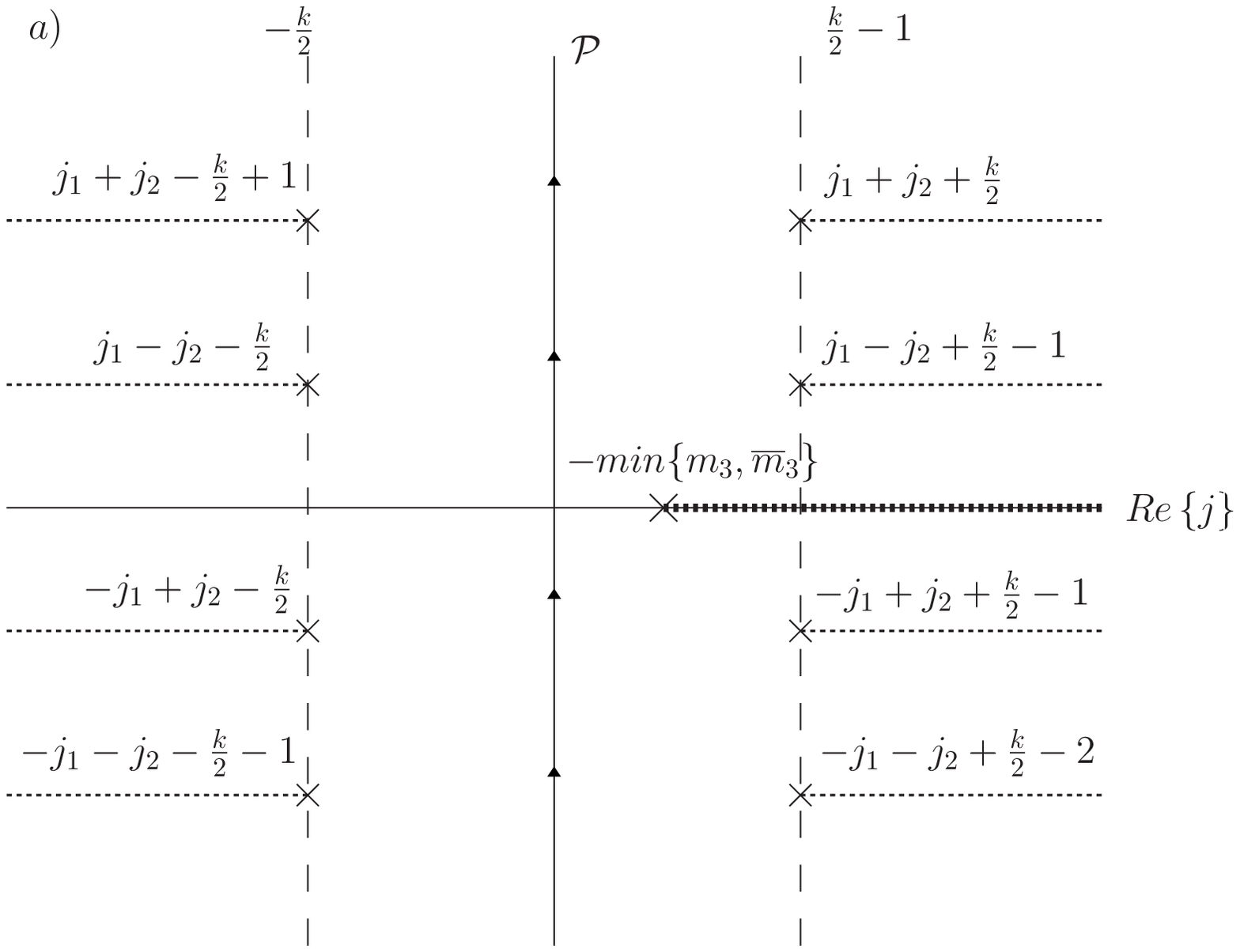,width=6.5cm}~~~~~~~~~~~~
\psfig{figure=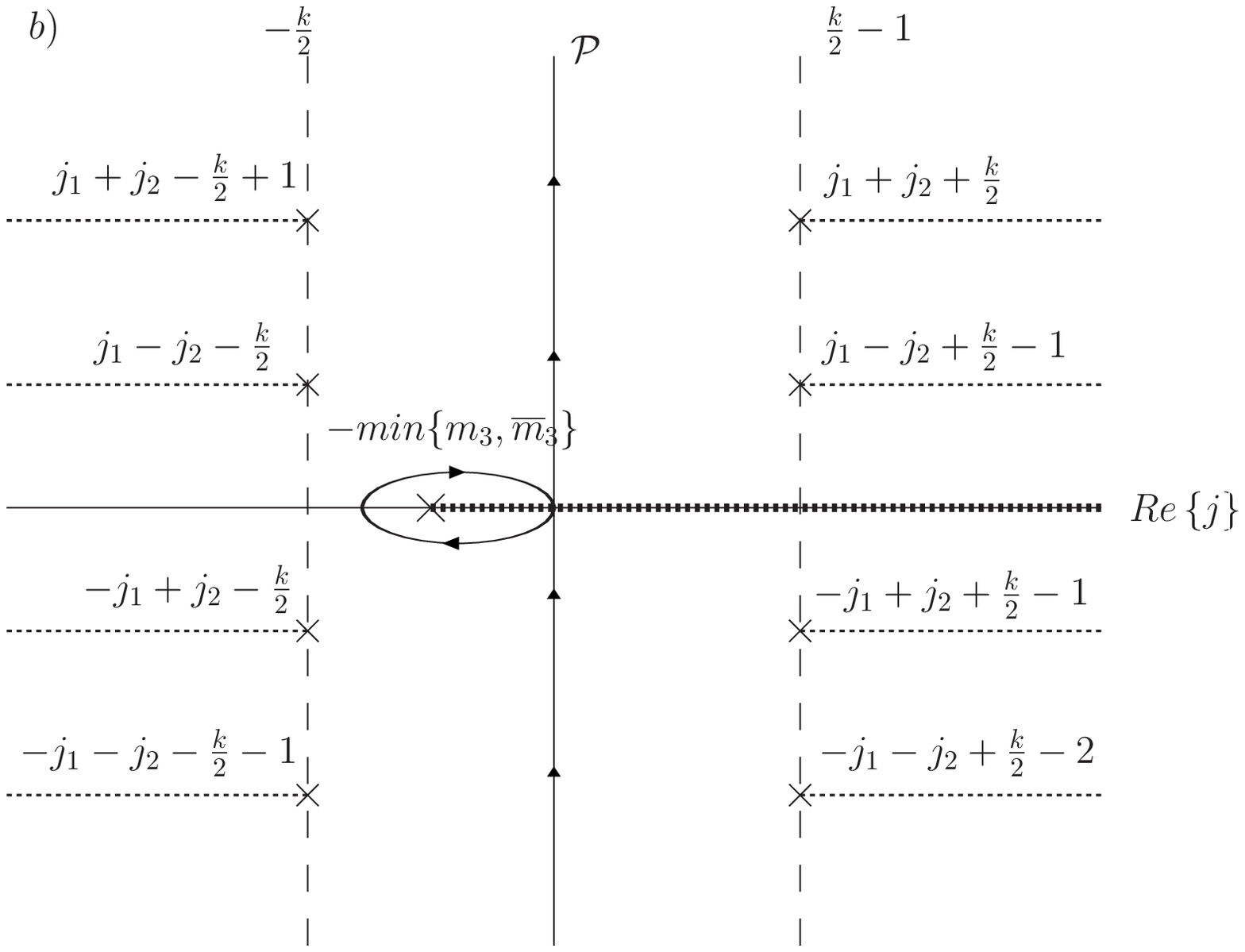,width=6.5cm}}
 {\footnotesize
{ Figure 1: Case ${\cal C}^{\alpha_1,w_1}_{j_1}\times{\cal
    C}^{\alpha_2,w_2}_{j_2}$.
The solid line indicates the integration
     contour ${\cal P}=-\frac 12+i{\mathbb R}$
in the $j_3$ complex plane. The dots above or below the real axis
     represent the ($j_1,j_2$)-dependent poles and those on the real axis
      correspond to  the $m-$dependent poles. The crosses are
the positions of the first poles in the series. $a)$
When
$m_1+m_2<-\frac{k-1}2$ or $\overline m_1+\overline
      m_2<-\frac{k-1}{2}$,
there are
no poles crossing the contour of integration. $b)$ When
$m_1+m_2>-\frac{k-1}2$ and $\overline m_1+\overline
      m_2>-\frac{k-1}{2}$,
 poles from the factor $\frac{\Gamma(-j_3-\overline m_3)}   {\Gamma(1+j_3+m_3)}$
cross the contour, indicating the contribution to
the OPE from states in discrete
     representations.}}
\bigskip

\medskip
Some poles cross the integration
contour when
$min\left\{m_1+m_2,\,\overline m_1 + \overline m_2\right\}>-\frac
{k-1}{2}$.
They are sketched in Figure $1.b)$ and  indicate contributions
from the
discrete series ${\cal D}_{j_3}^{+,w_3=w_1+w_2-1}$ with
$j_3=-min\left\{m_3,\overline m_3\right\}+n$,  $n=0,1,2,\dots$, and
such that $j_3<-\frac 12$. Since $Q^{w=\pm 1}$ does not vanish for
$j_3=-\frac 12 +i\mathbb R$ and $m_3$ not correlated with $j_3$,
 there are terms from ${\cal
  C}_{j_3}^{\alpha_3,w_3=w_1+w_2-1}$ in this OPE as well. Therefore we get

\bea
\left.{\cal C}^{\alpha_1,\;w_1}_{j_1}\times
{\cal C}^{\alpha_2,\;w_2}_{j_2}\right|_{|w|=1}
&=&\sum_{j_3<-\frac
  12} {\cal
    D}^{+,\;w_3=w_1+w_2-1}_{j_3}+\sum_{j_3<-\frac
  12} {\cal
    D}^{-,\;w_3=w_1+w_2+1}_{j_3}\nonumber\\
&&+\;\sum_{w=-1,1}
\int_{\mathcal{P}} dj_3\;{\cal
  C}^{\alpha_3,\;w_3=w_1+w_2+w}_{j_3}+\cdots ,\label{cc-1}
\eea
where  $|_{|w|= 1}$ denotes  that only spectral flow
non-preserving contributions are displayed in the right-hand side.

\begin{itemize}
    \item ${\cal C}_{j_1}^{\alpha_1,\;w_1}\times\;
{\cal D}_{j_2}^{\pm,\;w_2}$
\end{itemize}

To analyze this case, we need to perform the analytic continuation
for $j_2$ away from $-\frac 12+is_2$. When $is_2$ is continued to
the real interval $(-\frac{k-2}{2},\;0)$, the series of
$m-$independent poles changes as  shown in Figure
2. It is easy to see that these poles do not cross the contour of
integration. For instance, ${\rm
Re}\left\{j_1+j_2+\frac k2\right\}>0$, ${\rm
Re}\left\{j_1-j_2+\frac k2-1\right\}>\frac k2-1$, etc. Similarly as
in the previous case, only poles  from
$\frac{\Gamma(-j_3-\overline m_3)}{\Gamma(1+j_3+m_3)}$ can cross
the contour, but due to the factor
$\frac{\Gamma(1+j_2+\overline m_2)}{\Gamma(-j_2- m_2)}$ there are
contributions from the discrete series just for
$\Phi^{j_2,w_2}_{m_2,\overline m_2}\in {\cal D}^{-,w_2}_{j_2}\otimes
{\cal D}^{-,w_2}_{j_2}$. Therefore we get
\bea
\left.{\cal C}^{\alpha_1,\;w_1}_{j_1}
\times\;{\cal D}^{\pm,\;w_2}_{j_2}
\right|_{|w|=1}= \displaystyle  \int_{\mathcal{P}}
dj_3\;{\cal C}^{\alpha_3,\;w_3=w_1+w_2\pm1}_{j_3} + ~
\sum_{j_3<-\frac 12} {\cal D}^{\mp,\;w_3=w_1+w_2\pm1}_{j_3}+\cdots\,
.\label{d-c-1}
\eea

\centerline{\psfig{figure=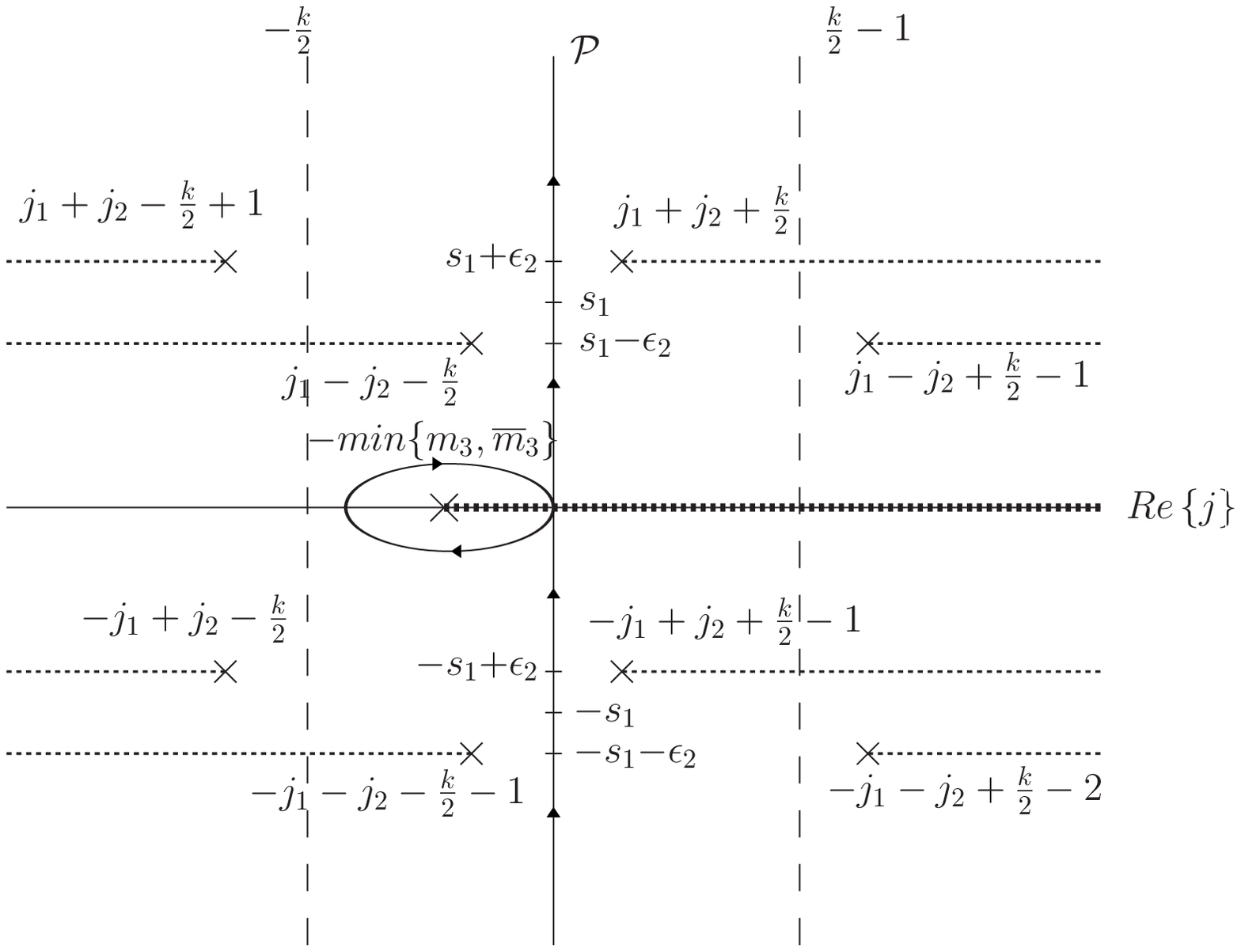,width=7cm}}
{\footnotesize Figure 2: Case ${\cal
      C}^{\alpha_1,w_1}_{j_1}\times{\cal D}^{\pm,w_2}_{j_2}$. Only $m-$dependent
      poles can cross
the contour of integration. This occurs when
 both $m_1+m_2$ and $\overline m_1+\overline m_2$ are
      larger
than $-\frac{k-1}{2}$. We have given $j_2$  an
infinitesimal imaginary part, $\epsilon_2$, to better display the
($j_1,j_2$)-dependent series of poles.}

\begin{itemize}
    \item ${\cal D}_{j_1}^{\pm,\;w_1}\times\;
{\cal D}_{j_2}^{\pm,\;w_2}$ and ${\cal D}_{j_1}^{\pm,\;w_1}\times\;
{\cal D}_{j_2}^{\mp,\;w_2}$

\end{itemize}

Let us first analytically continue both $j_1$ and $j_2$
to the interval $(-\frac{k-1}{2},\;-\frac 12)$, which is shown in Figure
3. The correct way to do this is to consider that both
$j_1$ and $j_2$
have an infinitesimal imaginary part, $\epsilon_1$ and $\epsilon_2$
respectively, which is sent  to zero after computing the integral.

The $m-$independent poles cross the contour of integration only when $j_1+j_2<-\frac{k+1}{2}$. However, due to the factors $\frac{\Gamma(1+j_1+m_1)} {\Gamma(-j_1-\overline m_1)}\frac{\Gamma(1+j_2+\overline m_2)}{\Gamma(-j_2- m_2)}$ in $Q^{w=-1}$, the contributions from these poles only survive when the quantum numbers of both $\Phi^{j_1,w_1}_{m_1,\overline m_1}$ and $\Phi^{j_2,w_2}_{m_2,\overline m_2}$ are in ${\cal D}_{j_i}^{-,w_i}\otimes{\cal D}_{j_i}^{-,w_i}, i=1,2$.
In this case, the poles at $j_3=j_1+j_2+\frac k2+n$ give contributions from ${\cal D}^{-,w_3=w_1+w_2-1}_{j_3}$. This may be seen noticing that $j_3=m_1+m_2+\frac k2 +n_3=\overline m_1+\overline m_2+\frac k2 + \overline n_3$, with $n_3=n+n_1+n_2$ and $\overline n_3=n+\overline n_1+\overline n_2$, or using $m_3=m_1+m_2+\frac k2,\,\overline m_3=\overline m_1+\overline m_2+\frac k2,$ so that $j_3=m_3+n_3=\overline m_3+\overline n_3$. Instead, the contributions from the poles at $j_3=-j_1-j_2-\frac k2-1-n$ seem to cancel due to the factor $\frac{\Gamma(-j_3-\overline m_3)}{\Gamma(1+j_3+m_3)}$. However, these zeros are canceled because the operator  diverges. In fact, using (\ref{rr}) and relabeling $j_3\rightarrow-1-j_3$, it is straightforward to recover exactly the same contribution from the
poles at $j_3=j_1+j_2+\frac k2+n$. Obviously, this was expected  as a consequence of the symmetry $j_3\leftrightarrow-1-j_3$ of the integrand in (\ref{opec}).

\bigskip

\centerline{
\psfig{figure=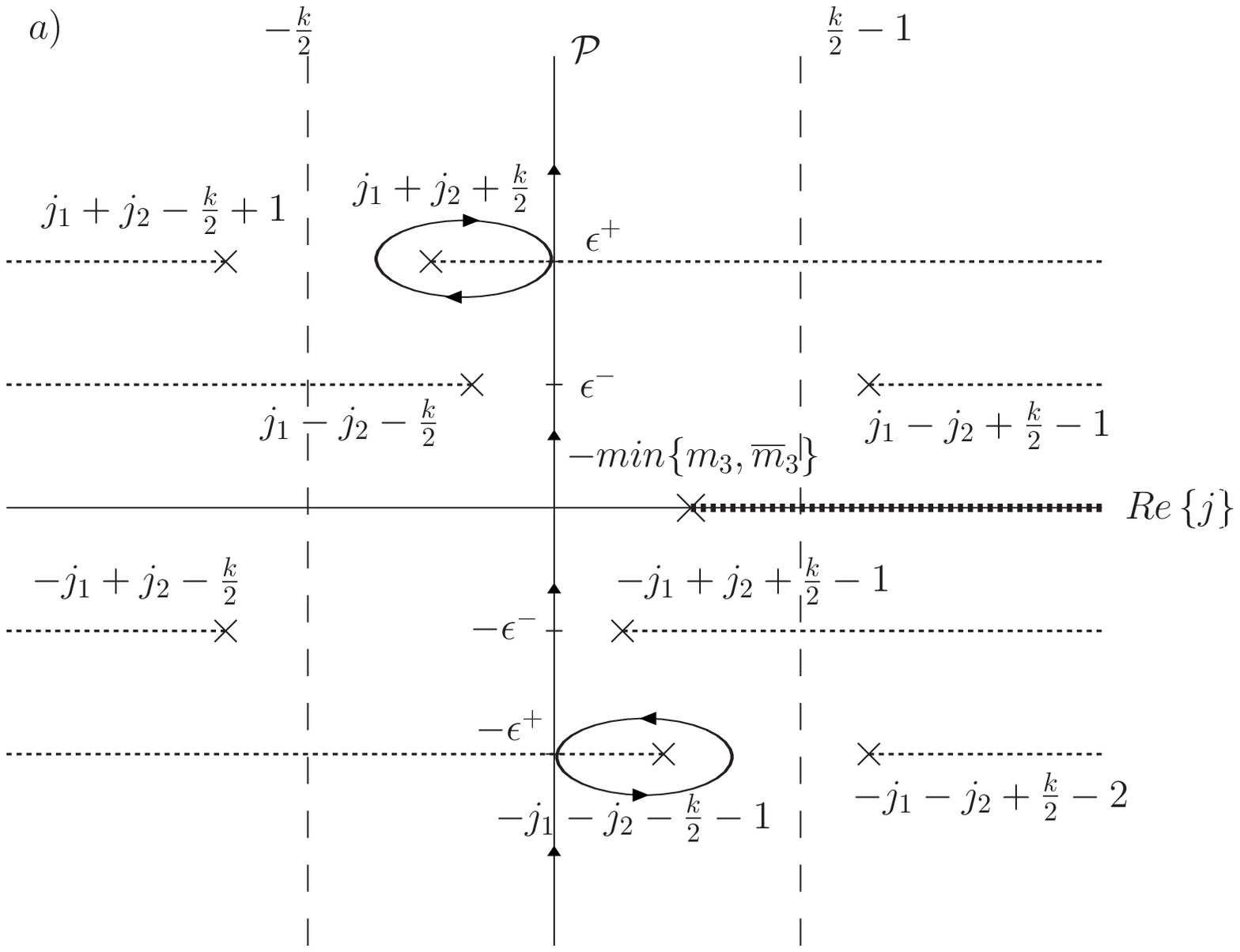,width=6.5cm}~~~~~~~~~~~~\psfig{figure=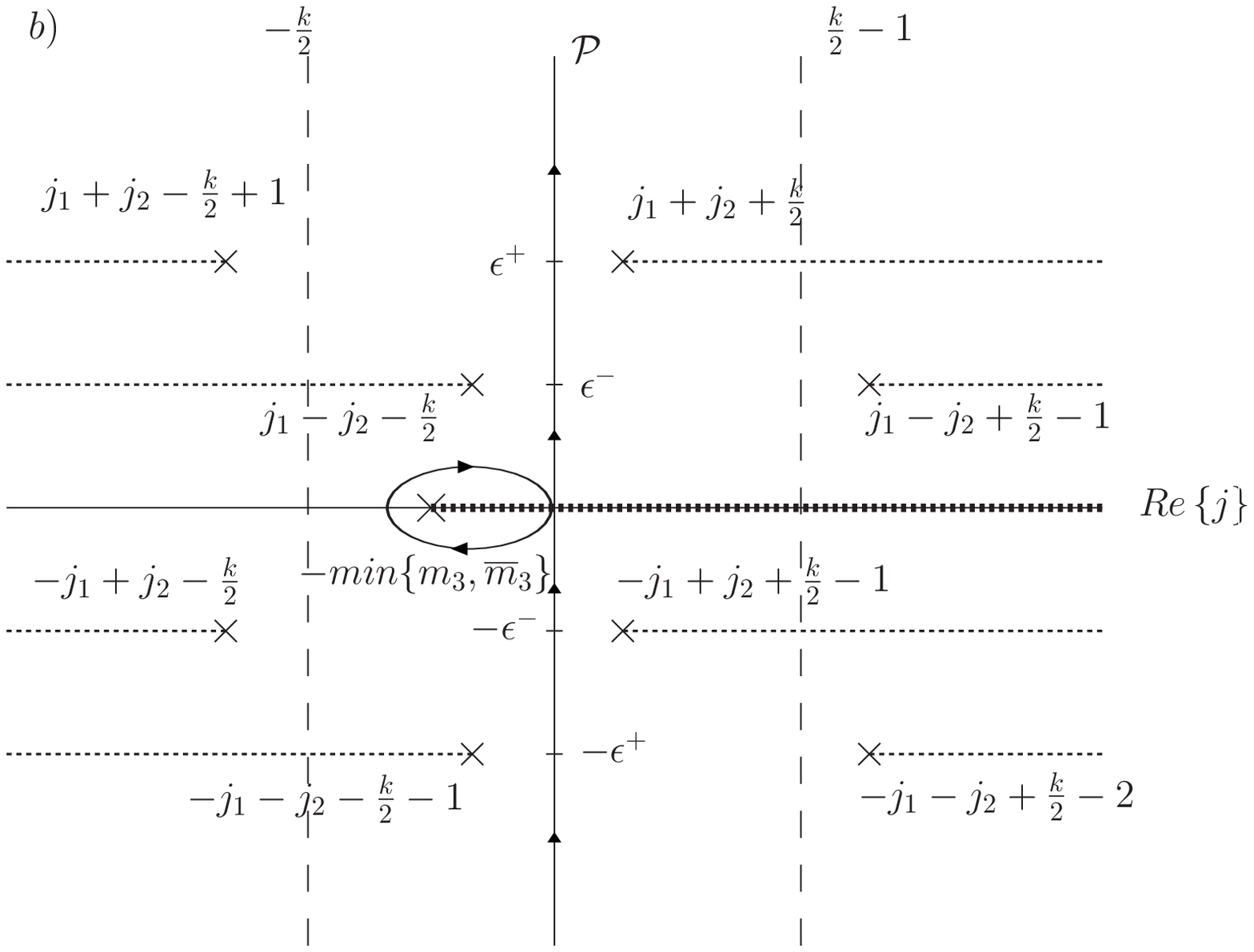,width=6.5cm}}
{\footnotesize Figure 3: Case
${\cal D}^{w_1}_{j_1}\times{\cal
       D}^{w_2}_{j_2}$.
Both $m-$dependent and
    $m-$independent poles can cross the contour of  integration. There
    are two
possibilities: 1)  ${\cal D}^{-,w_1}_{j_1}\times{\cal
       D}^{-,w_2}_{j_2}$.
When $j_1+j_2<-\frac{k+1}{2}$, only
$m-$independent poles can cross the contour, as shown in
Figure 3.$a$)
and when
$j_1+j_2>-\frac{k-1}{2}$, only $m-$dependent poles can  cross as shown in
Figure 3.$b$). 2) ${\cal D}^{\mp,w_1}_{j_1}\times{\cal
       D}^{\pm,w_2}_{j_2}$. 
Both
$m-$dependent and
$m-$independent poles can cross the contour  but only the former
survive after
taking the limit $\epsilon^+,\epsilon^-\rightarrow0$, where
$\epsilon^{\pm}=\epsilon_1\pm\epsilon_2$.}

\bigskip

Finally, the $m-$dependent poles give contributions from
${\cal D}^{+,w_3=w_1+w_2-1}_{j_3}$. Actually, when
$min\left\{m_3,\overline m_3\right\}>
\frac 12$
some of the $m-$dependent poles cross the  contour. Using $m-$conservation
it is not difficult to check that these contributions
fall inside the range (\ref{unitary}).

Let us continue the analysis, considering the OPE ${\cal D}^{\mp,w_1}_{j_1}\times{\cal
       D}^{\pm,w_2}_{j_2}$.
For instance, take the limiting case
$j_1=m_1+n_1+i\epsilon_1$ and $j_2=-m_2+n_2+i\epsilon_2$ with
$\epsilon_1,\;\epsilon_2\rightarrow0$. The factor
$\frac{\Gamma(1+j_2+\overline m_2)}{\Gamma(-j_2-m_2)}$ 
vanishes as a simple zero. However, some poles from the
series $j_3=j_2-j_1-\frac k2-n$ will overlap with
the $m-$dependent
poles. But because the $m-$independent simple poles are outside the
contour of integration, in the limit  $\epsilon_i\rightarrow0$ they may
cancel the simple zeros.
The way to compute this limit is determined by
the definition of
the three-point function. We
assume that a finite and nonzero term remains in the
limit \footnote{
In the limit $\epsilon_1,\epsilon_2\rightarrow0$,
$Res(Q^{w=-1})\sim\frac{\epsilon_2}{\epsilon_2-\epsilon_1}$.
The same ambiguity appears in the
three-point function including $\Phi^{j_1,w_1}_{m_1,\overline m_1}\in\mathcal
  D^{-,w_1}_{j_1}\otimes \mathcal D^{-,w_1}_{j_1}$,
$\Phi^{j_2,w_2}_{m_2,\overline m_2}\in \mathcal D^{+,w_2}_{j_2}\otimes
\mathcal D^{+,w_2}_{j_2}$, with
  $n_1\leq n_2$
such that $j_3=j_1-j_2-\frac k2-{\mathbb Z}_{n\ge 0}$.
The  resolution of this ambiguity requires an interpretation of the
divergences. 
The $w-$selection rules allow to
assume that a finite term survives  in the
limit. For instance, consider a generic three-point function 
$\langle{\cal D}_{j_1}^{-,w_1}{\cal D}^{+,w_2}_{j_2}
{\cal D}^{+,w_3}_{j_3}\rangle$ with $w_1+w_2+w_3=-1$.
According to (\ref{dd}) this is non-vanishing (for certain  values of
$j_i$, not determined from the $w-$selection rules). 
Indeed, the divergence from the $\delta^2(\sum_im_i-\frac k2)$ in
(\ref{3pointwind}) cancels the zero
from $\Gamma(-j_3-m_3)$ and then the pole in
$\widetilde C(1+j_i)\sim\frac{1}{\epsilon_2-\epsilon_1}$ must cancel
the zero from $\frac{\Gamma(1+j_2+\overline m_2)}{\Gamma(-j_2-\overline m_2)}
\sim\epsilon_2$, leaving a finite and non vanishing contribution.}.

Including the contributions
from continuous representations, we get the following  results:

\bea
\left.{\cal D}^{\pm,\;w_1}_{j_1}
\times ~
{\cal D}^{\pm,\;w_2}_{j_2}\right |_{|w|= 1}
&=& \int_{\mathcal{P}^+} dj_3 ~ {\cal C}^{\alpha_3,\;
w_3=w_1+w_2\pm1}_{j_3}
~ + ~\sum_{-j_1-j_2-\frac k2\le j_3< -\frac 12}
 {\cal D}^{\mp,\;w_3=w_1+w_2\pm 1}_{j_3}\nonumber\\
 && + ~
\sum_{j_1+j_2+\frac k2\le j_3< -\frac 12}
{\cal D}^{\pm,\;w_3=w_1+w_2\pm1}_{j_3}~ +\cdots .\label{lu}
\eea

\bea
\left .{\cal D}^{+,\;w_1}_{j_1}\times\; {\cal D}^{-,\;w_2}_{j_2}\right |_{|w|=1} &=&
\sum_{j_3<j_2-j_1-\frac k2}{\cal D}^{-,w_3=w_1+w_2+1}_{j_3}~ +
\sum_{j_3<j_1-j_2-\frac k2}{\cal D}^{+,w_3=w_1+w_2-1}_{j_3}~
+\cdots\nonumber\\
\label{d+td-}
\eea

\subsection{Spectral flow preserving contributions}

The analytic structure of  $Q^{w=0}(j_i;m_i,\overline
m_i)$ in (\ref{q0}) was studied in \cite{satoh}. Here we present the
analysis mainly to discuss some subtleties which are crucial to
perform the analytic continuation of $m_i, \overline m_i,
i=1,2$. Although our treatment of the $m-$dependent poles differs from
that followed in \cite{satoh}, we show in this section that the  results coincide.

The function $C(1+j_i)$ has zeros at $j_i=\frac{j-1}{2}$, $i=1,2,3$
and poles at $j=-j_1-j_2-j_3-2$,  $-1-j_1-j_2+j_3, ~ -1-j_1-j_3+j_2$, or $
-1-j_2-j_3+j_1$ where
$j:=p+q(k-2), -(p+1)-(q+1)(k-2)$, $p, q = 0, 1, 2, \cdots$.
To explore the behavior of the function $W$, we  use the
expression
\cite{satoh}
\begin{equation}
W\left[\begin{matrix}
j_1\,,\,j_2\,,\,j_3\cr  m_1, m_2,
m_3\cr\end{matrix}\right]=(i/2)^2\left [C^{12}
\overline P^{12}+C^{21}\overline P^{21}
\right ]\, ,\label{w}
\end{equation}
with $~~~~(i/2)^2P^{12}=s(j_1+m_1)s(j_2+m_2)C^{31}-s(j_2+m_2)s(m_1-j_2+j_3)C^{13}\,,$
\begin{eqnarray}
C^{12} &=&\frac{\Gamma(-N)\Gamma(1+j_3-m_3)}{\Gamma(-j_3-m_3)}
G\left[\begin{matrix}
    -m_3-j_3,\;-j_{13},\;1+m_2+j_2
\cr
                       -m_3-j_1+j_2+1,\;m_2-j_1-j_3\cr\end{matrix}\right]\, ,\nonumber\\
C^{31}&=&\frac{\Gamma(1+j_3+m_3)\Gamma(1+j_3-m_3)}{\Gamma(1+N)}
G\left[\begin{matrix}
    1+N,\;1+j_{1}+m_1,\;1-m_2+j_2
\cr
                       j_3+j_2+m_1+2,\;j_1+j_3-m_2+2\cr\end{matrix}\right]\, ,\nonumber\\
G\left[\begin{matrix} a,b,c\cr
    e,f\cr\end{matrix}\right]&=&\frac{\Gamma(a)\Gamma(b)
\Gamma(c)}
{\Gamma(e)\Gamma(f)}
F\left[\begin{matrix} a,b,c\cr e,f\cr\end{matrix}\right]
=\sum_{n=0}^{\infty}\frac 1{n!}\frac{\Gamma(a+n)\Gamma(b+n)
\Gamma(c+n)}
{\Gamma(e+n)\Gamma(f+n)\Gamma(n+1)}\,\label{Wfull} ,
\eea
and $N=1+j_1+j_2+j_3$, $s(x)=\sin(\pi x)$.
$\overline P^{ab}\,(\overline C^{ab})$ is obtained from $P^{ab}\,(C^{ab})$ by
 replacing $(m_i\rightarrow \overline m_i)$ and $P^{ba}\,(C^{ba})$ from
   $P^{ab}\, (C^{ab})$ by  changing $(j_1,m_1
\leftrightarrow j_2,m_2)$ and
$F\left[\begin{matrix} a,b,c\cr e,f\cr\end{matrix}\right]={}_3F_2(a,b,c;e,f;1)$.
An equivalent expression for $W$ which will be
useful below is the following \cite{satoh}
\bea
W\left[\begin{matrix}
j_1\,,\,j_2\,,\,j_3\cr m_1, m_2, m_3\cr\end{matrix}\right]=D_1C^{12}\overline C^{12}+
D_2C^{21}\overline C^{21}+D_3[C^{12}\overline C^{21}+C^{21}\overline
  C^{12}]\, ,
\label{ww}
\eea
where
\bea
D_1 &=& \frac{s(j_2+m_2)s(j_{13})}{s(j_1-m_1)s(j_2-m_2)s(j_3+m_3)}
[s(j_1+m_1)s(j_1-m_1)s(j_2+m_2)\nonumber\\
&&\; -
s(j_2-m_2)s(j_2-j_3-m_1)s(j_2+j_3-m_1)]\, ,\nonumber\\
D_2&=&D_1(j_1,m_1\leftrightarrow j_2,m_2)\, ,\nonumber\\
D_3&=&-\frac{s(j_{13})s(j_{23})s(j_1+m_1)s(j_2+m_2)s(j_1+j_2+m_3)}
{s(j_1-m_1)s(j_2-m_2)s(j_3+m_3)}\,\label{D} .
\eea

Studying the analytic structure of $Q^{w=0}$ is a difficult task as a
consequence of the complicated form
of  $W$.
The analysis greatly simplifies  when analytically continuing the
quantum numbers of one  operator to those of a
discrete representation. Indeed,
when
$j_1=-m_1+n_1=-\overline m_1+\overline n_1,$ and
$n_1,\overline n_1= 0,1,2,\cdots$,
 $W{\scriptsize\left[\begin{matrix}
j_1\,,\,j_2\,,\,j_3\cr  m_1, m_2, m_3\cr\end{matrix}\right]}$ reduces to
$W_1=D_1C^{12}\overline C^{12}$
\cite{satoh}, $i.e.$
\begin{equation}
W_1\left[\begin{matrix}
j_1\,,\,j_2\,,\,j_3\cr  m_1, m_2, m_3\cr\end{matrix}\right]=
\frac{(-)^{m_3-\overline m_3+\overline n_1}\pi^2 \gamma(-N)
}{\gamma(-2j_1)\gamma(1+j_{12})\gamma(1+j_{13})}\frac{\Gamma(1+j_3-m_3)
\Gamma(1+j_3-\overline
m_3)}{\Gamma(1+j_3-m_3-n_1)\Gamma(1+j_3-\overline
 m_3-\overline n_1)}\nonumber
\end{equation}
\begin{equation}
~~~~~~~~~~~~~
\times \prod_{i=2,3}\frac{\Gamma(1+j_i+m_i)}{\Gamma(-j_i-\overline m_i)}F\left
[\begin{matrix} -n_1,-j_{12},1+j_{23}\cr
-2j_1,1+j_3-m_3-n_1\cr\end{matrix}\right]
F\left[\begin{matrix} -\overline n_1,-j_{12},1+j_{23}\cr
-2j_1,1+j_3-\overline m_3-\overline
n_1\cr\end{matrix}\right]\label{w1}\, .
\end{equation}

It is easy to see that
\bea
& &\frac{\Gamma(1+j_3-m_3)}{\Gamma(1+j_3-m_3-n_1)}F\left[\begin{matrix}
-n_1,-j_{12},1+j_{23}\cr-2j_1,1+j_3-m_3-n_1\cr\end{matrix}\right]=\cr\cr
& &\sum_{n=0}^{n_1}\frac{(-)^n n_1!}{n!(n_1-n)!}\frac{\Gamma(n-j_{12})}
{\Gamma(-j_{12})} \frac{\Gamma(n+1+j_{23})}{\Gamma(1+j_{23})} \frac{\Gamma(-2j_1)}
{\Gamma(n-2j_1)}\frac{\Gamma(1+j_3-m_3)}{\Gamma(n+1+j_3-m_3-n_1)}\,
 .
\label{Fn1} \eea
Recall that the OPE
involves the function $W{\scriptsize\left[\begin{matrix}
j_1\,,\,j_2\,,\,j_3\cr  m_1, m_2, -m_3\cr\end{matrix}\right]}$ and
then the change $(m_3,\overline m_3)\rightarrow (-m_3,-\overline
m_3)$ is required in the above expressions to analyze $Q^{w=0}$.
 Thus, for generic $2j_i\notin\mathbb{Z}$, the
poles and zeros of $Q^{w=0}(j_i;m_i,\overline m_i)$ are contained
in

\begin{equation}
C(1+j_i) \frac{\gamma(-1-j_1-j_2-j_3)}
{\gamma(1+j_{12})\gamma(1+j_{13})}
\frac{\Gamma(1+m_2+j_2)\Gamma(-m_3-j_3)} {\Gamma(-\overline
m_2-j_2)\Gamma(1+\overline m_3+j_3)}\, , \label{q0a}
\end{equation}
plus possible additional zeros in (\ref{Fn1}) and its antiholomorphic
equivalent expression (see appendix \ref{A:asw}).
The $(j_1,j_2)-$dependent poles in (\ref{q0a}) are at
$j_3=\pm j_{21}^{\pm}+p+(q+1)(k-2)$, $\pm j_{21}^{\pm}-(p+1)-q(k-2)$,
$\mp j_{21}^{\pm}+p+q(k-2)$, $\mp j_{21}^{\pm}-(p+1)-(q+1)(k-2)$. There are also
 zeros at $1+2j_i=p+q(k-2),\;-(p+1)-(q+1)(k-2)$, $i=1,2,3$.

Let us first
consider  $\Phi_{m_1,\overline m_1}^{w_1,j_1}\in
{\cal D}_{j_1}^{+,w_1}
\otimes {\cal D}_{j_1}^{+,w_1}$
 and note that when
 $\Phi_{m_1,\overline m_1}^{w_1,j_1}\in
{\cal D}_{j_1}^{-,w_1}
\otimes {\cal D}_{j_1}^{-,w_1}$
the OPE follows directly
 using the symmetry of the spectral flow conserving two- and three-point
functions under
$(m_i,\overline m_i)\leftrightarrow(-m_i,-\overline m_i), \forall ~
i=1,2,3$.\footnote {
~This symmetry follows directly from the integral expression for
 $W{\scriptsize\left[\begin{matrix}
j_1\,,\,j_2\,,\,j_3\cr  m_1, m_2, m_3\cr\end{matrix}\right]}$
performing the change of variables ($x_i,\overline
x_i)\rightarrow(x_i^{-1},\overline x_i^{-1})$ in (\ref{w1i}).}

\begin{itemize}
    \item ${\cal D}_{j_1}^{\pm,\;w_1}\times\;
{\cal C}_{j_2}^{\alpha_2,\;w_2}$
\end{itemize}

Consider
$j_1=-m_1+n_1+i\epsilon_1$ with $n_i\in{\mathbb Z}_{\ge 0}$ and
 $\epsilon_1$ an infinitesimal
positive number, and $j_2=-\frac 12 +is_2$
not correlated with $m_2$.
In this case,
$W{\scriptsize\left[\begin{matrix}
j_1\,,\,j_2\,,\,j_3\cr  m_1, m_2, m_3\cr\end{matrix}\right]}\approx
W_1{\scriptsize\left[\begin{matrix}
j_1\,,\,j_2\,,\,j_3\cr  m_1, m_2, m_3\cr\end{matrix}\right]}$.

The $m-$independent poles are to the right or to the left
of the contour of integration as  sketched in Figure 4.$a$).
If $min\left\{m_3,\overline
  m_3\right\}<\frac12$,
none of the $m-$dependent poles cross the contour, implying that only continuous
  series contribute to
the spectral flow conserving terms of the OPE
$ {\cal D}^{+,w_1}_{j_1}
\times
{\cal C}^{\alpha_2,w_2}_{j_2}$.   On the
  other hand if
$min\left\{m_3,\overline m_3\right\}>\frac12$, this OPE also receives
contributions
from ${\cal  D}^{+,w_3=w_1+w_2}_{j_3}$.
 Note that when $j_1\approx m_1+n_1$, $W{\scriptsize\left[\begin{matrix}
j_1\,,\,j_2\,,\,j_3\cr  m_1, m_2, m_3\cr\end{matrix}\right]}\approx
W_1{\scriptsize\left[\begin{matrix}
j_1\,,\,j_2\,,\,j_3\cr  -m_1, -m_2, -m_3\cr\end{matrix}\right]}$,
which implies that the spectral flow conserving terms in the OPE
${\cal D}^{-,w_1}_{j_1}\times
{\cal C}^{\alpha_2, w_2}_{j_2}$ contain  contributions from
the continuous representations as well as from
${\cal D}^{-,w_3}_{j_3}$ when $max\left\{m_3,\overline
m_3\right\}<-\frac12$. So we find
\bea
\left.{\cal D}^{\pm,\;w_1}_{j_1}\times\;
{\cal C}^{\alpha_2,\;w_2}_{j_2}\right|_{w=0}=\;\int_{\mathcal{P}}
dj_3\, {\cal C}_{j_3}^{\alpha_3,\;w_3=w_1+w_2}\;+\;
\sum_{j_3<-1/2}{\cal D}^{\pm,\;w_3=w_1+w_2}_{j_3}+\cdots \, .\label{dpmc0}
\eea

\begin{itemize}
    \item ${\cal D}_{j_1}^{\pm,\;w_1}\otimes\;
{\cal D}_{j_2}^{\mp,\;w_2}$ and ${\cal D}_{j_1}^{\mp,\;w_1}\otimes\;
{\cal D}_{j_2}^{\mp,\;w_2}$
\end{itemize}

When $j_2$ is
continued to $(-\frac{k-1}{2}+i\epsilon_2,-\frac12+i\epsilon_2)$,
$\epsilon_2$ being an infinitesimal positive number, $W$ is again
well approximated by $W_1$ as long as $j_2\neq-m_2+n_2+i\epsilon_2,
-\overline m_2+\overline n_2+i\epsilon_2$.
Otherwise, one also
has to consider $W_2\equiv D_{2}C^{21}\overline C^{21}$, but the result
coincides exactly with the one obtained using $W_1$, so we
restrict to this.
Two $m-$independent series of poles may cross the
contour of integration: $j_3=j_1-j_2-1-p-q(k-2)$ and
$j_3=j_2-j_1+p+q(k-2)$, both with $q=0$. The former has
$j_3>-\frac12$ and the latter,  $j_3<-\frac12$.
  The $m-$dependent poles in $Q^{w=0}$ arise from
$\frac{\Gamma(-j_3-\overline m_3)}{\Gamma(1+j_3+m_3)}$.
When $j_2=-m_2+n_2+i\epsilon_2$, because of the factor
$\Gamma(-j_2-m_2)^{-1}$, only $m-$dependent
poles give contributions from discrete series. To see this, consider
the $m-$independent
poles at
$j_3=j_1+j_2-p-q(k-2)$.
These  are outside the contour of integration and in the limit
$\epsilon_1,\epsilon_2\rightarrow0$ some of them may overlap with the
$m-$dependent ones.
Again,  one may argue that this limit leaves a finite and
non-vanishing factor.
\medskip

\centerline{
\psfig{figure=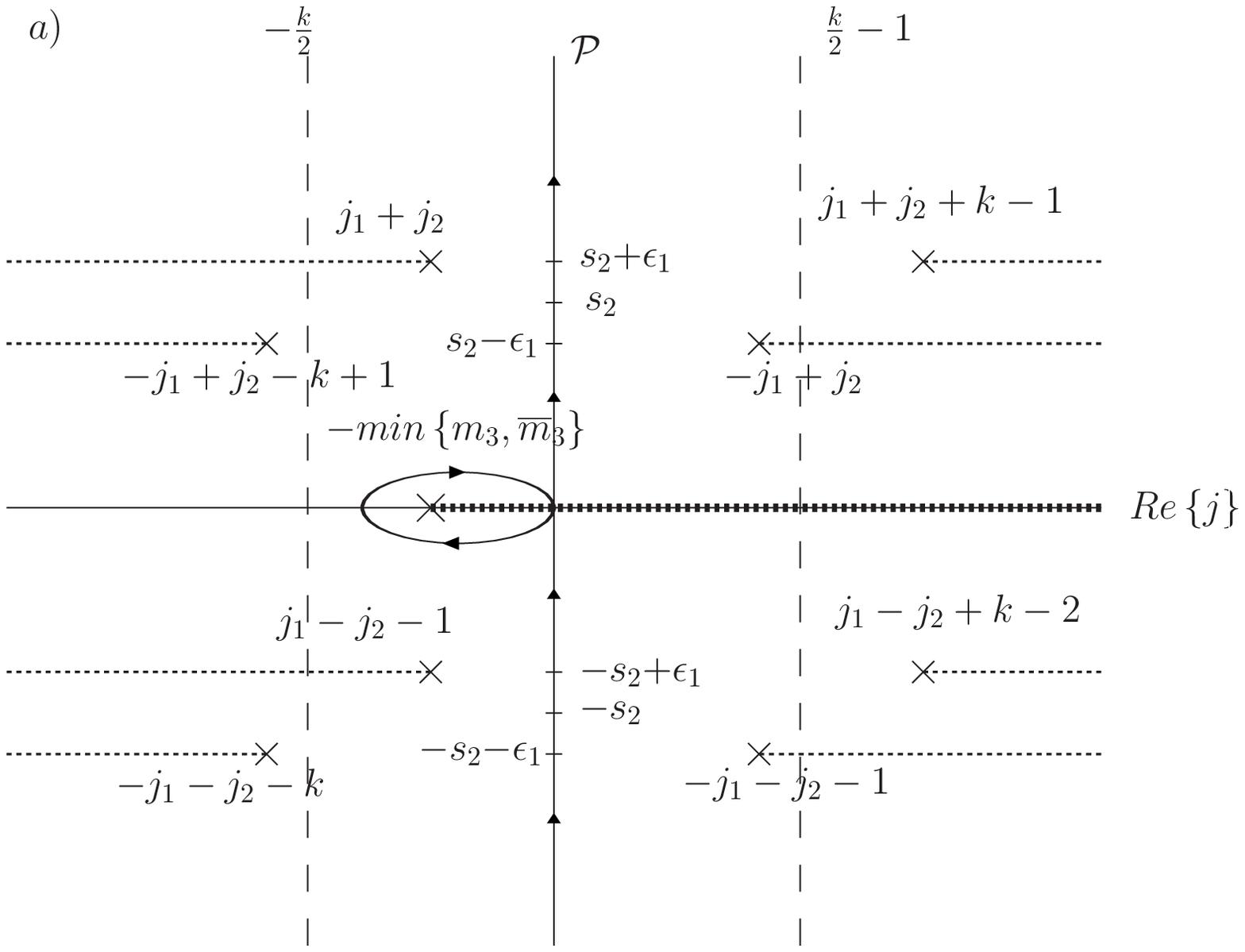,width=6.5cm}~~~~~~~~~~~~\psfig{figure=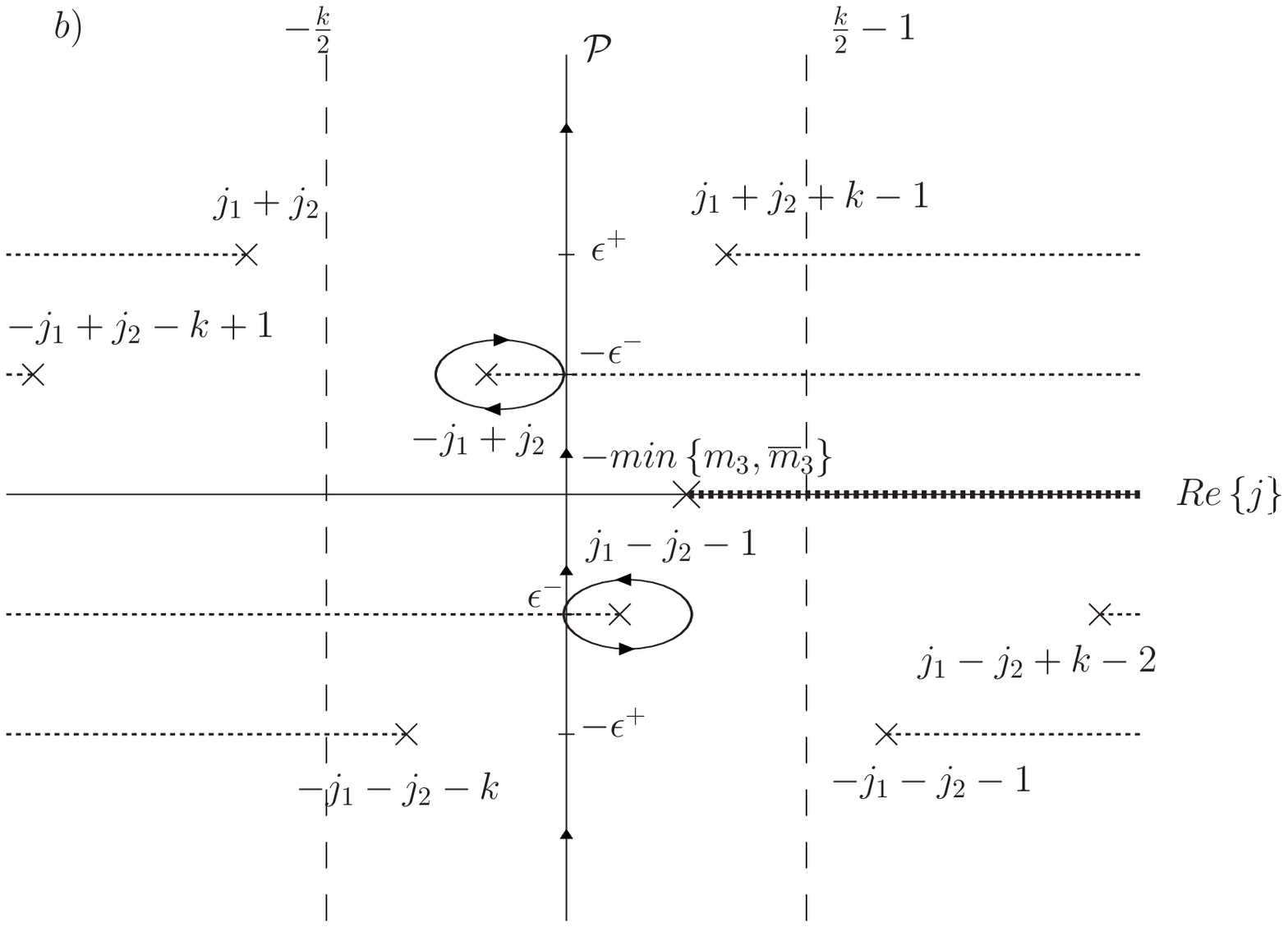,width=6.5cm}}

{\footnotesize Figure 4:
Analytic continuation of $Q^{w=0}$ for
($j_1,j_2$)-values away from the axis $-\frac 12+i\mathbb R$,
using $W_1$ instead of W.
 In  $4.a)$ $j_2=-\frac 12+is_2$
and only $m-$dependent poles can cross the contour of integration.
In $4.b)$ $-\frac{k-1}{2}<j_2<-\frac 12$ was considered. While
$m-$independent poles only cross the contour when $j_2<j_1$,
$m-$dependent poles can cross independently of the values of
$j_1,j_2$, but they are annihilated unless $j_2>j_1$. }
\bigskip

When $j_2=m_2+n_2+i\epsilon_2$, at first sight there are no
zeros.
 If $j_2-j_1<-\frac 12$, some poles with $q=0$ in the series
 $j_3=j_2-j_1+p+q(k-2)$ and
$j_3=j_1-j_2-1-p-q(k-2)$ cross the contour, as  shown in Figure
4.$b$). Using the relation between $j_i$ and $m_i$ and
$m-$conservation, it follows that the former poles
can be rewritten as $j_3=m_3+n_3=\overline m_3
+\overline n_3$, where $n_3=n_2-n_1+p$ and $\overline n_3=\overline
n_2-\overline n_1+p$. Obviously, if $n_2\geq n_1$ and $\overline
n_2\geq \overline n_1$ all the residues picked up by the contour
deformation
imply contributions to
the OPE from ${\cal D}^{-,w_3=w_1+w_2}_{j_3}$. When
$n_2<n_1$ or $\overline n_2<\overline n_1$,
only those values of $p$ for which both $n_3$ and $\overline n_3$
are non-negative integers remain after taking the limit
$\epsilon_1,\epsilon_2\rightarrow 0$. This is
because of extra zeros appearing in $W_1$ which are not explicit in
(\ref{w1})
(see appendix \ref{A:asw}).
 Using the results in the Appendix and the identity (\ref{rr}) it is
 straightforward to see that the latter  series of poles give the same contributions.

  The poles at
 $j_3=-min\left\{m_3,\overline m_3\right\}+n_3$ may cross the contour.
 If this happens they overlap with the $m-$independent poles. But
 there are
double zeros canceling these contributions.

 If $j_2-j_1>-\frac 12$, only $m-$dependent poles may cross the
 contour. But they give
contributions only if they do not overlap with the poles at
 $j_3=j_1-j_2-1-n$, again because of the presence of double zeros. 
Therefore, these contributions remain only for
 $j_3\geq j_1-j_2$.

Putting all together we get
\bea
 \left.{\cal D}^{+,\; w_1}_{j_1}\times\; {\cal D}^{-,\;w_2}_{j_2}\right|_{w=0} &=&
\int_{\mathcal{P}} dj_3~
{\cal C}_{j_3}^{\alpha_3,\;w_3=w_1+w_2}\;
+\displaystyle\sum_{j_2-j_1\leq j_3<-\frac 12
}{\cal D}^{-,\;w_3=w_1+w_2}_{j_3}\; \nonumber\\
&&~~ + ~
\displaystyle\sum_{ j_1-j_2\leq j_3<-\frac
12}{\cal D}^{+,\;w_3=w_1+w_2}_{j_3}+\cdots \, ,\label{d+1d-20}\\
    \left.{\cal D}^{\pm,\;w_1}_{j_1}\times\;{\cal D}^{\pm,\;w_2}_{j_2}\right|_{w=0}&=&
    \displaystyle\sum_{j_3\leq j_1+j_2} {\cal
    D}^{\pm,\;w_3=w_1+w_2}_{j_3}+\cdots\, .
\label{ddis}
\eea

\begin{itemize}
    \item
${\cal C}_{j_1}^{\alpha_1,\;w_1}\times\;
{\cal C}_{j_2}^{\alpha_2,\;w_2}$
\end{itemize}

The zero and pole structure of $Q^{w=0}$  is given by
\bea
Q^{w=0}(j_i;m_i,\overline m_i) &\sim & C(1+j_i)
\frac{\gamma(-N)}{s(\overline m_3+j_3)}
G\left[\begin{matrix}
    m_3-j_3,\;-j_{13},\;1+m_2+j_2
\cr
                       m_3-j_1+j_2+1,\;m_2-j_1-j_3\cr\end{matrix}\right]\cr\cr
&&\times ~
\left\{s(\overline m_1+j_1)G\left[\begin{matrix} 1+N,\;1+\overline
    m_1+j_1,\;1-
\overline m_2+j_2\cr
                       2+\overline m_1+j_2+j_3,\;2-\overline
               m_2+j_1+j_3\cr\end{matrix}\right]
\right.\cr\cr&& ~~~ - ~ \left.
 s(\overline m_1-j_2+j_3) G\left[\begin{matrix} 1+N,\;1+\overline
 m_2+j_2,\;1-\overline m_1+j_1\cr
                       2+\overline m_2+j_1+j_3,\;2-\overline
               m_1+j_2+j_3\cr\end{matrix}\right]
\right\}\nonumber\\
&& + ~
(j_1,m_1,\overline m_1)\leftrightarrow(j_2,m_2,\overline m_2)\, .\nonumber \eea

$G{\scriptsize\left[\begin{matrix} a,b,c\cr e,f\cr\end{matrix}\right]}$ has simple poles at $a,b,c=0,-1,-2,\dots$ as well as at $u=e+f-a-b-c=0,-1,-2,\dots$, if $a,b, c\ne 0,-1,-2,\cdots$. The pole structure of $Q^{w=0}$ is much subtler when $j_i=-\frac 12+is_i,i=1,2$ as the naive poles cancel because of the presence of hidden zeros. Actually, the correct behavior of $Q^{w=0}$, must be of the form \cite{wc}
\bea 
Q^{w=0}\sim \frac{\Gamma(-j_3-m_3)\Gamma(-j_3+\overline m_3)}{\Gamma(1+j_3-m_3)\Gamma(1+j_3+\overline m_3)}\, ,\label{wcxc3}
\eea
for generic $j_1,j_2$ and for $m_1,m_2$ not correlated with them, up to regular and non-vanishing contributions for $j_3=\pm m_3+n_3=\pm\overline m_3+\overline n_3$, with $n_3,\overline n_3\in \mathbb Z$. No other $m$-dependent pole series appears and the $m$-independent pole series do not cross the integration contour

We may now analyze the OPE ${\cal C}_{j_1}^{\alpha_1,\;w_1}\times\; {\cal C}_{j_2}^{\alpha_2,\;w_2}$. A sum over continuous representations appears because  $Q^{w=0}$ does not vanish for $j_3\in-\frac 12+i{\mathbb R}$. On the other hand, the expression (\ref{wcxc3}) shows that there are no contributions from discrete representations provided  $min\left\{m_1+m_2,\overline m_1+\overline m_2\right\}<\frac 12$ and $max\left\{m_1+m_2,\overline m_1+\overline m_2\right\}>-\frac 12$. Obviously both bounds cannot be violated at the same time. When the first one is violated, operators belonging to  spectral flow images of  lowest-weight representations contribute to the OPE. On the contrary, when the second bound is not satisfied, operators in spectral flow images of  highest-weight representations appear in the OPE.

Therefore, we conclude that the $w-$conserving contributions to the OPE of two continuous representations are the following:
\bea
\left.{\cal C}_{j_1}^{\alpha_1,\;w_1}\times\; {\cal C}_{j_2}^{\alpha_2,\;w_2}\right|_{w=0} \sim
\int_{\mathcal{P}} dj_3\, {\cal C}_{j_3}^{\alpha_3,\;w_3=w_1+w_2} &+& \sum_{j_3<-\frac12} {\cal D}^{+,\;w_3=w_1+w_2}_{j_3}
+\;\sum_{j_3<-\frac 12} {\cal D}^{-,\;w_3=w_1+w_2}_{j_3}\, ,\nonumber\\\label{cc0}
\eea
up to descendants. Note that, in a particular OPE with $m_i,\overline m_i$ fixed, only one of the discrete series contributes, depending on the signs of $m_3,\overline m_3$.

Collecting all the results, the OPE for primary fields and their spectral flow images in the spectrum of the $AdS_3$ WZNW model are the following:
\bea
{\cal D}^{\pm,\;w_1}_{j_1} \times ~ {\cal D}^{\pm,\;w_2}_{j_2}
&=&  \sum_{ j_3\le j_1+j_2}  {\cal D}^{\pm,\;w_3=w_1+w_2}_{j_3}+\sum_{-j_1-j_2-\frac k2\le j_3< -\frac 12}
     {\cal D}^{\mp,\;w_3=w_1+w_2\pm 1}_{j_3} \cr
&+&  \sum_{j_1+j_2+\frac k2\le j_3< -\frac 12 } {\cal D}^{\pm,\;w_3=w_1+w_2\pm 1}_{j_3}+ 
     \int_{\mathcal{P}} dj_3 ~ {\cal C}^{\alpha_3,\; w_3=w_1+w_2\pm 1}_{j_3}+\cdots . \label{1}\cr&&
\eea
\bea
{\cal D}^{+,\; w_1}_{j_1}\times\; {\cal D}^{-,\;w_2}_{j_2} 
&=& \displaystyle\sum_{ j_1-j_2\leq j_3<-\frac 12}{\cal D}^{+,\;w_3=w_1+w_2}_{j_3} +\displaystyle\sum_{j_2-j_1\leq j_3<-\frac 12
}{\cal D}^{-,\;w_3=w_1+w_2}_{j_3}+ \cr
&&+ \displaystyle\sum_{ j_3\le j_2-j_1-\frac k2} {\cal D}^{-,\;w_3=w_1+w_2+1}_{j_3}+\displaystyle\sum_{j_3\le j_1-j_2-\frac k2
}{\cal D}^{+,\;w_3=w_1+w_2-1}_{j_3}\cr 
&&+ ~\int_{\mathcal{P}} dj_3~ {\cal C}_{j_3}^{\alpha_3,\;w_3=w_1+w_2}\; +\cdots \, ,\label{d+d-0}
\eea
\bea
{\cal D}^{\pm,\;w_1}_{j_1}\times\; {\cal C}^{\alpha_2,\;w_2}_{j_2}
&=& \sum_{w=0}^1 \displaystyle  \int_{\mathcal{P}} dj_3\;{\cal C}^{\alpha_3,\;w_3=w_1+w_2\pm w}_{j_3} + \sum_{j_3<-\frac 12}
{\cal D}^{\pm,\;w_3=w_1+w_2}_{j_3}\cr
&&+ \sum_{j_3<-\frac 12}D^{\mp,w_3=w_1=w_2\pm 1}_{j_3}+\cdots \,,\label{dporc}
\eea
\bea
{\cal C}^{\alpha_1,\;w_1}_{j_1}\times {\cal C}^{\alpha_2,\;w_2}_{j_2}
&=& \sum_{w=0}^1\sum_{j_3<-\frac 12} \left ({\cal D}^{+,\;w_3=w_1+w_2-w}_{j_3} +\;  {\cal D}^{-,\;w_3=w_1+w_2+w}_{j_3}\right )\cr
&&+ \sum_{w=-1}^1 \int_{\mathcal{P}} dj_3\;{\cal C}^{\alpha_3,\;w_3=w_1+w_2+w}_{j_3}+\cdots .\label{u}
\eea
\medskip

\subsubsection{Satoh's prescription}

In order to analyze these results, let us first restrict to the spectral flow conserving contributions and for the particular case of $w_i=0,~ i=1,2.$ In this case, exactly the same results were obtained in \cite{satoh} using the following prescription for the OPE of $w=0$ primary fields $\Phi^{j_1}_{m_1,\overline m_1}$ $\Phi^{j_2}_{m_2,\overline m_2}$ \footnote{~ See \cite{hs} for previous work involving highest-weight representations.}:
\bea
\Phi^{ j_1}_{m_1,\overline m_1}(z_1,\overline z_1)\Phi^{j_2}_{m_2,\overline m_2} (z_2, \overline z_2)
{}^{~~\sim}_{z_1\rightarrow z_2} \displaystyle\sum_{j_3} |z_{12}|^{-2\widetilde\Delta_{12}} 
Q^{w=0}(j_i;m_i,\overline m_i)\Phi^{j_3}_{m_1+m_2,\overline m_1+\overline m_2}(z_2,\overline z_2),~~~~~~\label{opesatoh}
\eea
where $Q^{w=0}$ was obtained using the standard procedure, $i.e.$ multiplying both sides of (\ref{opesatoh}) by a fourth field in the $w=0$ sector and taking 
expectation values. The formal symbol $\sum_{j_3}$ denotes integration over ${\cal D}_{j_3}^{\pm}$ and ${\cal C}_{j_3}^{\alpha_3}$, namely
\bea\sum_{j_3}=\displaystyle\int_{{\cal P}^+} d j_3 + \delta_{{\cal D}_{j_3}^{\pm}}\displaystyle\oint_{\cal C}dj_3 \, .\label{formal}
\eea
The integration over ${\cal P}^+$ stands for  summation over $C_j^\alpha$. The contour integral along ${\cal C}$ encloses the poles from ${\cal D}^{\pm}_{j_3}$
and $\delta_{{\cal D}_{j_3}^{\pm}}$ means that $j_3$ is picked up from the poles in $Q^{w=0}$ by the contour ${\cal C}$ only when it belongs to a discrete representation. The range of $j_3$ is Re $j_3\le -\frac 12$ and Im $j_3\ge 0$, consistently with the argument which determined $Q^{w=0}$ because $\sum_{j_3}$ picks up only one term in (\ref{218}). This prescription to deal with the $j-$dependent $m-$independent poles was shown  to be compatible with the one suggested in \cite{tesch2} for the $H_3^+$ model. The strategy designed in (\ref{formal})  for the treatment of $m-$dependent poles,  which were absent in \cite{tesch2}, aimed to reproducing the classical tensor product of representations of $SL(2, \mathbb R)$ in the limit $k\rightarrow \infty$. This proposal for the OPE includes in addition the requirement that poles with divergent residues should not be picked up.

In this section, we have followed a different path. We have treated the $j-$ and $m-$dependent poles alike. However, although the equivalence between both prescriptions is not obvious a priori, we obtained the same results for the OPE of unflowed primary fields \footnote{ More generally, it can be shown that a
generalization of the $ansatz$ (\ref{opesatoh}) for fields $\Phi^{j_1,w_1}_{m_1,\overline m_1}$ $\Phi^{j_2,w_2}_{m_2,\overline m_2}$, by adding the contributions from terms proportional to $Q^{w=\pm 1}$ and replacing $\delta_{D_{j_3}^{\pm}}$ by $\delta_{D_{j_3}^{\pm,w_3}}$, leads to the same results (\ref{1})$-$(\ref{u}).}. 
Indeed, notice that poles in $Q^{w=0}$ at values of quantum numbers in $C_j^\alpha$ or ${\cal D}^{\pm}_{j_3}$ would not contribute to the OPE determined by (\ref{opec}) if they do not cross the contour ${\cal P}$, unlike to (\ref{opesatoh}). On the other hand, contributions from operators in other  representations, $i.e.$ neither in $C_j^\alpha$ nor  in ${\cal D}^{\pm}_{j_3}$, could have appeared in (\ref{1})$-$(\ref{u}), but they did not. Moreover, by a careful analysis of the analytic structure of $Q^{w=0}$ we have shown  that there are no double poles, so that the regularization proposed in \cite{satoh} is not really necessary. \footnote{ This is very important because the double poles discussed in \cite{satoh} would lead to inconsistencies in the analytic continuation of the OPE from $H_3^+$ that we have performed in this chapter. In particular, they would give divergent contributions to the OPE ${\cal D}^+_j\times{\cal D}^-_j$ and, in addition, this OPE would be incompatible with  ${\cal D}^-_j\times{\cal D}^+_j$, in contradiction with expectations from the symmetries of the function $W$.}.

In the case $w_1=w_2=0$, $k\rightarrow \infty$, the $w-$conserving contributions to the OPE of representations of the zero modes in (\ref{1})$-$(\ref{u}) reproduce the classical tensor products of representations of $SL(2,\mathbb R)$ obtained in \cite{holman}. Continuous series appear twice in the  product of two continuous representations due to the existence of two linearly independent Clebsh-Gordan coefficients. As noted in \cite{satoh}, this is in agreement with the fact that both terms $C^{12}$ and $C^{21}$ in (\ref{w}) contribute to $Q^{w=0}$ in the fusion  of two continuous series. Moreover, it was also observed that the analysis can be applied for finite $k$ without modifications. The results are given by replacing ${\cal D}_j^\pm, {\cal C}_j^{\alpha}$ in (\ref{1})$-$(\ref{u}) by the corresponding affine representations $\widehat{\cal D}_j^\pm, \widehat{\cal C}_j^\alpha$. It is easy to see that this OPE of unflowed fields in the spectrum of the $AdS_3$ WZNW model is not closed, $i.e.$ it gets contributions from discrete representations with $j_3<-\frac {k-1}2$. When  spectral flow is turned on, incorporating all the relevant representations of the theory and the complete set of structure constants as we have  done in this section, the OPE still does not close, namely there are contributions from discrete representations outside the range (\ref{unitary}).  In particular, this feature of the OPE of fields in discrete representations differs from the results in \cite{mo3} where the factorization limit of the four-point function of $w=0$ short strings was shown to be in accord with the Hilbert space of the string dual theory.

In the following section we will show how the spectral flow symmetry imposes a truncation of the OPE (\ref{1})$-$(\ref{u}) which guaranties the closure in the physical spectrum of the model.

\subsection{Truncation of the operator algebra and fusion rules}\label{secttrunc}

The analysis of the previous section involved primary operators and their spectral flow images. Then, the OPE (\ref{1})-(\ref{u}) explicitly includes some descendant fields. Assuming the appearance of spectral flow images of primary states in the fusion rules indicates that there are also contributions from descendants not obtained by spectral flowing primaries but descendants with the same $J_0^3$ eigenvalue. As we commented in section \ref{opesection}, the presence of a descendant in the OPE requires the presence of the Virasoro primary associated to it, so the inclusion of descendants implies the replacement of ${\cal D}_j^{\pm,w}, {\cal C}_j^{\alpha, w}$ by $\hat{\cal D}_j^{\pm,w}, \hat{\cal C}_j^{\alpha,w}$ in (\ref{1})-(\ref{u}), we will show that this observation implies some interesting conclusions.

For instance, consider the spectral flow non-preserving terms in
the OPE ${\cal D}^{-,w_1}_{j_1}\times{\cal
 D}_{j_2}^{-,w_2}$,  (\ref{1}). If they are extended to the affine series, using
 the spectral flow symmetry they may be identified as
\begin{equation}
\sum_{-\frac{k-1}2<\tilde j_3\leq
  j_1+j_2} \widehat{\cal D}^{+,\;w_3=w_1+w_2-1}_{-\frac k2-\tilde j_3}
\equiv\sum_{-\frac{k-1}2<j_3\leq
  j_1+j_2} \widehat{\cal D}^{-,\;w_3=w_1+w_2}_{j_3}\, .
\end{equation}
This reproduces the spectral flow conserving terms in the first sum in 
(\ref{1}).
However, there is an important difference:  here $j_3$ is
automatically restricted to
the region (\ref{unitary}).

Analogously, applying the spectral flow symmetry  to the discrete
series contributing to the OPE
$\displaystyle\left.{\cal D}^{+,w_1}_{j_1}\times{\cal D}_{j_2}^{-,w_2}\right|_{|w|= 1}$
in (\ref{d+td-}) leads to contributions from $\displaystyle
\sum_{j_2-j_1\leq j_3} {\widehat{\cal D}}^{-,\;w_3=w_1+w_2}_{j_3}$
 as well as from $\displaystyle\sum_{j_1-j_2\leq j_3} {\widehat{\cal D}}^
{+,\;w_3=w_1+w_2}_{j_3}$, which were found
among the spectral flow conserving terms
 with the extra condition $j_3<-\frac 12$.

In order to see further implications of the spectral flow symmetry on the OPE
 (\ref{1})-(\ref{u}), let us now consider operator products of descendants.
Take
 the OPE
 $\widehat{D}_{j_1}^{+,w_1=0}\otimes \widehat{D}_{j_2}^{-,w_2=1}$
\footnote{ We use the tensor product symbol $\otimes$ to denote the
 OPE of fields in representations of 
the current algebra, to distinguish it from that of
 highest/lowest-weight fields.}. Equation (\ref{d+d-0})
 gives  spectral flow conserving contributions from
$\widehat{\cal D}_{j_3}^{-,w_3=1}$, for certain $m_i,\overline m_i,
 i=1,2$, with $j_3$ verifying (\ref{unitary}).
Using the spectral flow symmetry,  one might infer that
the contributions from $\widehat{\cal D}^{+,w_3=0}_{j_3}$
to  the OPE $\widehat{\cal D}_{j_1}^{+,w_1=0}\otimes \widehat{\cal
 D}_{j_2}^{+,w_2=0}$ in (\ref{1}) would also be
 within the region (\ref{unitary}).
  On the contrary, we found 
 terms in $\widehat{\cal D}^{+,w_3=0}_{j_3}$ with $j_3<-\frac {k-1}{2}$.
Moreover, using the spectral flow symmetry again,
these terms can be identified with contributions from
 $\widehat{\cal D}^{-,w_3=1}_{j_3}$ with $j_3>-\frac
 12$ to the OPE $\widehat{\cal D}^{+,w_1=0}_{j_1}\otimes \widehat{\cal
 D}^{-,w_2=1}_{j_2}$, in contradiction with our previous result.

Similar puzzles are found identifying
 $\displaystyle\sum_{j_3<-\frac 12} \widehat{\cal
D}^{+,\;w_3=w_1+w_2-1}_{ j_3} =\displaystyle\sum_{-\frac{k-1}{2}<j_3} \widehat{\cal
D}^{-,\;w_3=w_1+w_2}_{ j_3}$  in (\ref{dporc}),
which gives some of the spectral flow conserving contributions.
 It is interesting to note that
only the states within the  region (\ref{unitary}) contribute in both cases,
explicitly $j_3=j_1+\alpha_2+n$, with $n\in \mathbb Z$ such
that $-\frac{k-1}2<j_3<-\frac 12$.
It is also important to stress the following observation. For given  $j_1,
m_1$ and $j_2,m_2$ the spectral flow conserving part of the OPE
 (\ref{dporc}) 
receives contributions
from states with $\tilde j_3, \tilde m_3$ verifying $\tilde j_3=\tilde
m_3+\tilde n_3$ with $\tilde n_3=0, 1,\cdots, \tilde n_{3}^{max}$,
$\tilde n_{3}^{max}$ being the maximum integer
such that $\tilde j_3<-\frac 12$.
On the other hand, the spectral flow non-conserving  terms
get contributions from  $j_3=- m_3+
n_3$
with $ n_3=0,1,2,\dots, n_{3}^{max}$ and here $ n_{3}^{max}$ is the
maximal non-negative integer such that $j_3<-\frac 12$. So,
identifying both series implies considering
 $\tilde j_3=-\frac k2- j_3$ and
now  $n_{3}^{max}$ (which is the same as before) has to be the
maximal non-negative integer for which $\tilde j_3>-\frac{k-1}{2}$.
There is just one operator appearing in both contributions to the OPE. It has
 $\tilde n_3=0$ in the former and
     $n_3=0$ in the latter.
This is a consequence of the relation
$\Phi_{m=\overline
m=-j}^{j,w=0}=\frac{\nu^{\frac k2-1}}{(k-2)}\frac{1}{B(-1-j')}
\Phi_{m'=\overline m'=j'}^{j',w'=1}$
with $j'=-\frac
k2-j$ \cite{mo3}. One
can check that the $w-$conserving three-point functions containing
$\Phi_{m=\overline m=-j}^{j,w=0}$ reduce to the $w-$non-conserving ones
involving $\Phi_{m'=\overline m'=j'}^{j',w'=1}$. This result can be
generalized for arbitrary $w$ sectors in the $m-$basis,
 $i.e.$ $\Phi_{m=\overline m=-j}^{j,w}\sim$
$\Phi_{m'=\overline m'=j'}^{j',w'=w+1}$ up to a regular
normalization for $j$ in the region (\ref{unitary}).
 For instance, one can reduce a spectral flow conserving three-point
function including $\Phi_{m=\overline m=-j}^{j,w}$ to a one unit
violating amplitude containing $\Phi_{m'=\overline m'=j'}^{j', w+1}$
using the identity
\bea
C(1+j_1,1+j_2,1+j_3)=\frac{\nu^{k-2}\gamma(k-2-j_{23})
\gamma(2-k-2j_1)C(k+j_1-1,1+j_2,1+j_3)}
{(k-2) \gamma(1+2j_1) \gamma(-N) \gamma(-j_{12}) \gamma(-j_{13})}\, ,
\label{Cidentity}
\eea
which is a consequence of the relation
$G(j)=(k-2)^{1+2j}\gamma(-j)G(j-k+2)$.\\

\bigskip

\centerline{\psfig{figure=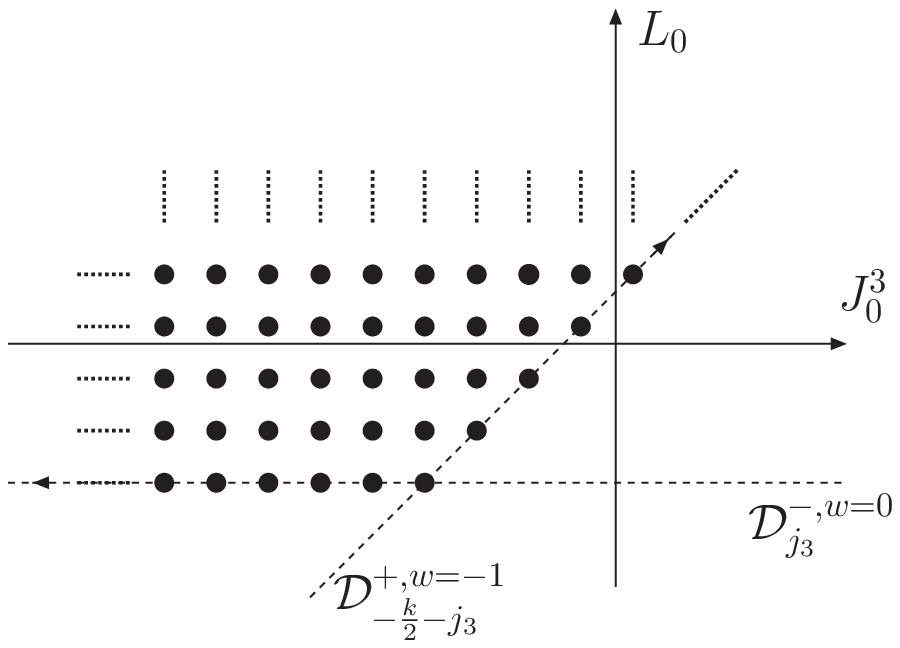,width=7cm}}
 {\footnotesize
{ Figure 5: Weight diagram of $\widehat{\cal D}^{-,w=0}_{j_3}$. The lines with arrows indicate the states in ${\cal D}_{j_3}^{-,w=0}$ and ${\cal D}_{-\frac k2-j_3}^{+,w=-1}$. Consider a state in $\widehat{\cal D}^{+,w=0}_{\tilde{j}}$, at level $\tilde{N}$ and weight $\tilde{m}=-\tilde{j}+\tilde{n}$. It follows from (\ref{wtrans}), (\ref{wl0}) that after spectral flowing by ($-1$) unit, this state maps to a state in $\widehat{\cal D}^{-,w=0}_{j}$, with  $j=-\frac k2-\tilde{j}$, level $N=\tilde{n}$ and weight $m=j-n$, with $n=\tilde{N}$. For instance primary states in $\widehat{\cal D}^{+,w=0}_{-\frac k2-j_3}$, denoted simply by ${\cal D}^{+,w=0}_{-\frac k2 -j_3}$, map to highest-weight states in $\widehat{\cal D}^{-,w=0}_{j_3}$. So, only one state in ${\cal D}^{+,w=-1}_{-\frac  k2-j_3}$ coincides with one in ${\cal D}^{-,w=0}_{j_3}$, namely that with $\tilde{n}=0$. }}
\bigskip

The OPE (\ref{dporc}) was obtained for states in ${\cal C}^{\alpha,w}_j$ and ${\cal D}^{\pm, w}_j$. When replacing operators in, say ${{\cal D}}^{-,w}_j$ by those in ${\widehat{\cal D}}^{-,w}_j$, the latter can be interpreted as having been obtained by performing $w$ units of spectral flow on primaries of ${\widehat{\cal D}}^{-,w=0}_j$ or $w-1$ units of spectral flow on primaries of ${\cal D}^{+}_{-\frac k2-j}$, that is $w$ units of spectral flow from ${\cal D}^{+,w=-1}_{-\frac
k2-j}$, which in turn may be thought of as the highest-weight field in ${\widehat{\cal D}}^{-,w=0}_j$ (see figure 5). Only the spectral flowed primary of highest-weight appears in both sets of contributions, $i.e.$ the one with $n_3 =\tilde n_3=0$. This behavior was observed in all other cases, namely, the same discrete series arising in the OPE from $Q^{w=0}$ can be also seen to arise from $Q^{w=1}$ or $Q^{w=-1}$, but only one operator appears in both simultaneously.  

Thus, even if the calculations involved operators in the series ${\cal D}_j^{\pm, w}$ and ${\cal C}_j^{\alpha, w}$, we collect here the results for the fusion rules \footnote{ Actually, the fusion rules for two representations determine the exact decomposition of their tensor products. These not only contain information on the conformal families appearing in the r.h.s of the OPE, but also on their multiplicities. We shall not attempt to determine the latter here.} assuming $\Phi^{j_i,w_i}_{m_i,\overline m_i}(z_i,\overline z_i)\in\widehat{\cal D}_{j_i}^{\pm,w_i}$ or $\widehat{\cal C}_{j_i}^{\alpha_i, w_i}$, $ i=1,2,3$.
Using the spectral flow symmetry to identify $\widehat{\cal D}_j^{-,w}= \widehat{\cal D}_{-\frac k2-j}^{+,w- 1}$, we obtain:

\noindent
\begin{enumerate}

\item ~~ $
\widehat{\cal D}^{+,\;w_1}_{j_1}
\otimes ~
\widehat{\cal D}^{+,\;w_2}_{j_2}
= \displaystyle\int_{\mathcal{P}} dj_3 ~ \widehat{\cal C}^{\alpha_3,\;
w_3=w_1+w_2+1}_{j_3}
 \oplus ~\sum_{-\frac{k-1}{2}<j_3\leq
  j_1+j_2} \widehat{\cal D}^{+,\;w_3=w_1+w_2}_{j_3}$ \label{ffr}

$~~~~~~~~~~~~~~~~~~~~~~~~~~~ \oplus ~\displaystyle
\sum_{j_1+j_2+\frac k2\leq j_3<-\frac
  12}  \widehat{\cal D}^{+,\;w_3=w_1+w_2+1}_{j_3},$

\noindent
\item $ ~~
\widehat{\cal D}^{+,\;w_1}_{j_1}\otimes ~
\widehat{\cal C}^{\alpha_2,\;w_2}_{j_2}
= \displaystyle
\sum_{-\frac{k-1}{2}<j_3<-\frac
  12} \widehat{\cal D}^{+,\;w_3=w_1+w_2}_{j_3}
\oplus ~
\sum_{w=0}^1\int_{\mathcal{P}} dj_3 ~
\widehat{\cal C}^{\alpha_3,\;w_3=w_1+
w_2+ w}_{j_3} ,$

\noindent
\item $ ~~ \widehat{\cal C}^{\alpha_1,\;w_1}_{j_1}
~ \otimes ~ \widehat{\cal C}^{\alpha_2,\;w_2}_{j_2}
= \displaystyle\sum_{w=-1}^0\sum_{-\frac{k-1}{2}<j_3<-\frac
  12} \widehat{\cal D}^{+,\;w_3=w_1+w_2+w}_{j_3}
\oplus ~~\displaystyle \sum_{w=-1}^1 \int_{\mathcal{P}} dj_3 ~
\widehat{\cal C}^{\alpha_3,\;w_3=w_1+w_2+w}_{j_3} .$

\end{enumerate}

We have truncated the spin of the contributions from discrete representations following the criterion that processes related through the identity $\widehat{\cal D}_j^{+, w}\equiv \widehat{\cal D}_{-\frac k2-j}^{-, w+ 1}$ must be equal, $i.e.$ equivalent operator products should get the same contributions.
Indeed, one finds contradictions unless the OPE is truncated to keep $j_3$ within the region (\ref{unitary}). As we have seen through some examples, extending the OPE (\ref{1})-(\ref{u}) to representations of the current algebra, discrepancies occur both when comparing $w-$conserving with non-conserving contributions
as well as when comparing $w-$conserving terms among themselves. So the truncation is imposed by self-consistency. 

A strong argument in support of the fusion rules 1.$-$3. is that only operators violating the bound (\ref{unitary}) must be discarded. Indeed, the cut  amounts to keeping just contributions from states in the spectrum\footnote{ It is important to stress that the truncation is not discarding contributions from the {\it microstates} associated to the $(j_1,j_2)-$dependent poles that were found in \cite{tesch2}. Only $m-$dependent poles  which are absent in the $x-$basis present 
inconsistences with the spectral flow symmetry.}, $i.e.$ it implies that the operator algebra is closed on the Hilbert space of the theory. However, the spectrum involves irreducible representations and there are no singular vectors to decouple states like in $SU(2)$ \cite{zf} \footnote{~The spectral flow operators $\Phi^{-\frac k2}_{\pm\frac k2,\pm\frac k2}$ have null descendants. Even though they are excluded from the range (\ref{unitary}) they are necessary auxiliary fields to construct the states in spectral flow representations. Although the  physical mechanism is not clear to us, these operators might play a role in the decoupling.}. So there is a yet to be discovered physical mechanism decoupling states. 

Nevertheless, the results listed in  items 1.$-$3. above are supported by several consistency checks. First, the limit $k\rightarrow \infty$ contains the classical tensor products of representations of $SL(2,\mathbb R)$ \cite{holman} when restricted to $w=0$ fields. Second, as mentioned in the previous paragraph, 
once the OPE is truncated  to keep only  contributions from the spectrum, one can verify full consistency. In particular, the OPE $\widehat{\cal D}_{j_1}^{+,w_1}\otimes \widehat{\cal D}^{+,w_2}_{j_2}$ is consistent with the results in \cite{mo3} (see the discussion in appendix \ref{rtmo1}).
Finally, based on the spectral flow selection rules (\ref{cc}) and (\ref{dd}), the following alternative analysis can be performed. Let us consider, for instance, the operator product $\widehat{\cal D}_{j_1}^{+,w_1}\otimes \widehat{\cal D}^{+,w_2}_{j_2}$. Applying equation (\ref{dd}) to correlators involving three discrete states in $\widehat{\cal D}_{j}^{+,w}$ requires either $i) ~w_3=-w_1-w_2-1$ or $ii) ~w_3=-w_1-w_2-2$. Therefore, together with $m$ conservation, $i)$ implies  that the three-point function $<\widehat{\cal D}_{j_1}^{+,w_1} \widehat{\cal D}^{+,w_2}_{j_2}\widehat{\cal  D}_{j_3}^{+,w_3=-w_1-w_2-1}> $ will not vanish
as long as the OPE $\widehat{\cal D}_{j_1}^{+,w_1}\otimes \widehat{\cal D}^{+,w_2}_{j_2}$ contains a state in $\widehat{\cal D}_{j_3}^{-,w=w_1+w_2+1}$, which is equivalent to $\widehat{\cal D}_{\tilde j_3}^{+,w=w_1+w_2}$. Indeed, this contribution appeared above. Similarly, $ii)$ implies that  in order for $<\widehat{\cal D}_{j_1}^{+,w_1} \widehat{\cal D}^{+,w_2}_{j_2}\widehat{\cal D}_{j_3}^{+,w_3=-w_1-w_2-2}>$ to be non-vanishing, the OPE $\widehat{\cal D}_{j_1}^{+,w_1}\otimes
\widehat{\cal D}^{+,w_2}_{j_2}$ must have contributions from $\widehat{\cal D}_{j_3}^{-,w_3=w_1+w_2+2}\equiv \widehat{\cal D}_{\tilde j_3}^{+, w_3=w_1+w_2+1}$,
which in fact were found. Finally, when the third state involved in the three-point function is in the series $\widehat{\cal C}_{j_3}^{\alpha_3,w_3}$, equation (\ref{cc}) leaves only one possibility, namely $w_3=-w_1-w_2-1$, and thus the OPE must include terms in $\widehat{\cal C}_{j_3}^{\alpha_3, w_3=w_1+w_2+1}$, which actually appear in the list above. Although this analysis based on the spectral flow selection rules does not allow to determine either the range of $j_3-$values or the OPE coefficients, it is easy to check that the series content in 1.$-$3. is indeed completely reproduced in this way.

As mentioned in the previous section, in principle $w=\pm 2$ 
three-point functions should have been considered. However, the
contributions from these terms are already contained in our
results. If they gave contributions from discrete representations
outside the spectrum, they should be truncated since the equivalent
terms listed above do not include them. Contributions from operators
in $\widehat{\cal D}_{j_3}^{-,w_3=w_1+w_2+2}$ can only appear in case
1., namely $\widehat{\cal D}_{j_1}^{+,w_1}\otimes
\widehat{\cal D}^{+,w_2}_{j_2}$,
 for $j_3=-k-j_1-j_2-n$. These correspond to the terms denoted as
Poles$_2$ in \cite{mo3}, where they could not be interpreted in terms
of physical string states and were then truncated. See section \ref{rtmo1} for a
detailed discussion.

In conclusion, the results presented in this section are in agreement with the spectral flow selection pattern (\ref{cc})-(\ref{dd}), they are consistent with the results in \cite{mo3} and determine the closure of the operator algebra when properly treating the spectral flow symmetry. The full consistency of the OPE should follow from a proof of factorization and crossing symmetry of the four-point functions, but closed expressions for these amplitudes are not known, even in the simpler $H_3^+$ model. In order to make some preliminary progress in this direction, in the next chapter we discuss certain properties of the factorization of
four-point amplitudes involving states in different representations of the $AdS_3$ WZNW model, constructed along the lines in \cite{tesch2}.

\chapter{Factorization of  four-point functions}\label{chapFact}

In this chapter we discuss the issue of the factorization of four point functions in $AdS_3$ and its consistency with the OPE found in previous chapter. After discussing the factorization in the Euclidean rotation of $AdS_3$, the $H_3^+$ model we turn to the Lorentzian model.

Although a complete description of the contributions of descendant operators is not available to complete the bootstrap program, in section \ref{sectfactAdS3} we display some interesting properties of the amplitudes that can be useful to achieve a resolution of the theory. We will display a very odd property of the factorization in the $AdS_3$ WZNW model which follows from the assumption that correlators  are related through analytic continuation to those in $H_3^+$: Both, spectral flow conserving and spectral flow non conserving channels seem to give the same numerical result. This observation was later explicitly confirmed in some particular examples in the Coulomb-Gas formalism \cite{in2}. Then, we perform a qualitative study of the contributions of primaries and flowed primaries in the intermediate channels of the amplitudes and finally, we discuss the consistency of the factorization with the spectral flow selection rules.

\section{Factorization in $H_3^+$}

A decomposition of the four-point function in the Euclidean $H_3^+$ model was worked out in \cite{tesch1, tesch2} using the OPE (\ref{ope}) for pairs of primary operators $\Phi_{j_1}\Phi_{j_2}$ and $\Phi_{j_3}\Phi_{j_4}$. The $s-$channel factorization was written as follows

\begin{eqnarray}
\left\langle \Phi_{j_1}(x_1|z_1)\Phi_{j_2}(x_2|z_2)
\Phi_{j_3}(x_3|z_3)\Phi_{j_4}(x_4|z_4)
\right\rangle =
|z_{34}|^{2(\tilde\Delta_2+\tilde\Delta_1-\tilde\Delta_4-\tilde\Delta_3)}
|z_{14}|^{2(\tilde\Delta_2+\tilde\Delta_3-
\tilde\Delta_4-\tilde\Delta_1)}\nonumber\\
\times~|z_{24}|^{-4\tilde\Delta_2}
~|z_{13}|^{2(\tilde\Delta_4-\tilde\Delta_1-
\tilde\Delta_2-\tilde\Delta_3)}
\int_{{\cal P}^+}dj ~
{\cal A}(j_i, j)~
{\cal G}_j(j_i,z,\overline
z,x_i,\overline x_i)~|z|^{2(
  \Delta_j-\Delta_{1}-\Delta_{2})} .
\label{4pt}
\end{eqnarray}
Here
\begin{equation}
{\cal A}(j_i,j)=C(-j_1,
-j_2,-j)B(-j-1)
C(-j,-j_3,-j_4)
\end{equation}
and
\begin{equation}
{\cal G}_j(j_i,z,\overline z,x_i,\overline x_i)=\sum_{n,\overline
  n =0}^{\infty}z^n\overline z^{\overline
  n}D^{(n)}_{x,j}(j_i,x_i)
\overline D^{\overline n}_{\overline x,
  j}(j_i,\overline x_i)G_j(j_i,x_i, \overline x_i)\, ,
\end{equation}
where $D^{(n)}_{x,j}(j_i,x_i)$ are differential operators containing
the contributions from intermediate descendant states and
\begin{eqnarray}
 G_j(j_i, x_i, \overline x_i) & = &|x_{12}|^{2(j_{1}+j_2-j)}
|x_{34}|^{2(j_{3}+j_4-j)}
\int  d^2x d^2x'
|x_{1}-x|^{2(j_{1}+j-j_2)}
|x_{2}-x|^{2(j_{2}+j-j_1)}\nonumber\\
&&\times ~
|x_3-x'|^{2(j_{3}+j-j_4)}|x_4-x'|^{2(j_{4}+j-j_3)}|x-x'|^{-4j-4}\, ,\label{gj}
\end{eqnarray}
which may be rewritten as
\begin{eqnarray}
G_j(j_i, x_i, \overline x_i) & = &\frac{\pi^2}{(2j+1)^2}|x_{34}|^{2(j_{4}+j_3-j_2-j_1)}
|x_{24}|^{4j_{2}}|x_{14}|^{2(j_{4}+j_1-j_2-j_3)}|x_{13}|^{2(j_{3}+j_2+j_1-j_4)}
\nonumber\\
&\times &\left\{|F_j(j_i,x)|^2+
\frac{\gamma(1+j+j_4-j_3)\gamma(1+j+j_3-j_4)}{\gamma(2j+1)
\gamma(j_1-j_2-j)\gamma(j_2-j_1-j)}
|F_{-1-j}(j_i,x)|^2\right\}\, ,\nonumber
\end{eqnarray}
with $F_j(j_i, x)\equiv x^{j_1+j_2-j} {}_{2}F_1(j_1-j_2-j,j_4-j_3-j;-2j;x)$
and $x=\frac {x_{12}x_{34}}{x_{13}x_{24}}$.

The properties of (\ref{4pt}) under $j\rightarrow -1-j$ allow to extend the integration contour from ${\cal P}^+$ to the full axis ${\cal P}=-\frac 12+i{\mathbb R}$ and rewrite it in a holomorphically factorized form. Crossing symmetry follows from similar properties of a five-point function in Liouville theory to which this model is closely relates and it amounts to establishing the consistency of the $H_3^+$ WZNW model  \cite{tesch3}.

\section{Analytic continuation to $AdS_3$}\label{sectfactAdS3}

Expression (\ref{4pt}) is valid for external states
$\Phi_{j_1},\Phi_{j_2}$
in the range (\ref{range}) and similarly
for $\Phi_{j_3}, \Phi_{j_4}$. In particular, it holds for operators
 in continuous representations of the $AdS_3$ WZNW model.
The analytic continuation to other values of
$j_i$ was performed in
\cite{mo3}. In this process, some poles in the integrand cross the
integration contour and the four-point function is defined as
(\ref{4pt}) plus the contributions of all these poles.
This procedure allowed to analyze the factorization of 
four-point functions of $w=0$ short
strings in the boundary conformal field theory, obtained from primary states in discrete
representations ${\cal D}_j^{ w=0}\otimes {\cal D}_j^{
  w=0}$, by integrating over the world-sheet moduli.
It is important to stress that the aim in \cite{mo3} was to study
the factorization in the boundary conformal field theory with
coordinates  $x_i, \overline x_i$, so
the $x-$basis was found convenient. The conformal blocks were expanded
in powers of the cross ratios $x, \overline x$ and then integrated
over the worldsheet coordinates
$z,\overline z$. 
To study the factorization in the $AdS_3$
WZNW model instead, we expand the conformal blocks in powers of $z,
\overline z$, and in order to consider the various sectors, we find
convenient to translate (\ref{4pt}) to the $m-$basis. 

To this purpose, one can
verify that the integral over $j$ commutes with the integrals over
$x_i,\overline x_i,\,i=1,\dots,4$ and that it is regular for
$j_{21}^{\pm}$ and $j_{43}^{\pm}$ in the range (\ref{range}) and for all of
$|m|,|\overline m|, |m_i|,|\overline m_i|<\frac12$, where we have
introduced $m=m_1+m_2=-m_3-m_4$, $\overline m=\overline
m_1+\overline m_2=-\overline m_3 -\overline m_4$. Integrating in
addition over $x$ and $x'$ in (\ref{gj}), we get
\begin{eqnarray}
\left\langle \Phi^{j_1}_{m_1,\overline m_1}\Phi^{j_2}_{m_2,\overline m_2}
\Phi^{j_3}_{m_3,\overline m_3}\Phi^{j_4}_{m_4,\overline m_4}
\right\rangle  =
|z_{34}|^{2(\tilde\Delta_2+\tilde\Delta_1-
\tilde\Delta_4-\tilde\Delta_3)}
|z_{14}|^{2(\tilde\Delta_2+\tilde\Delta_3-
\tilde\Delta_4-\tilde\Delta_1)}|z_{24}|^{-4\tilde\Delta_2}\nonumber\\
~~~~~~~~\times ~ |z_{13}|^{2(\tilde\Delta_4-\tilde\Delta_1-
\tilde\Delta_2-\tilde\Delta_3)}
\int_{{\cal P}^+}dj ~
{\mathbb A}_j^{w=0}(j_i;m_i,\overline m_i)~
|z|^{2(
  \tilde\Delta_j-\tilde\Delta_{1}-\tilde\Delta_{2})}+\cdots\, ,\label{4ptm}
\end{eqnarray} 
where
\bea 
{\mathbb A}_j^{w=0}(j_i;m_i,\overline m_i)& = &
\delta^{(2)}(m_1+\dots+m_4)~C(1+j_1,1+j_2,1+j)~W\left[\begin{matrix}
j_1\,,j_2,\,j\cr m_1,m_2,-m\cr\end{matrix}\right]\cr &&\times ~
\frac{1}{B(-1-j)~c^{-1-j}_{m,\overline m}} ~C(1+j_3,1+j_4,1+j)~ W\left[\begin{matrix}
j_3\,,j_4,\,j\cr m_3,m_4,m\cr\end{matrix}\right].\label{A4pt1}
\eea

An alternative representation of (\ref{A4pt1}) was found in
\cite{mn} in terms of higher generalized hypergeometric functions
$_4F_3$.
This new
identity among hypergeometric functions is an interesting
by-product of the present result.

The dots in (\ref{4ptm}) refer to higher
powers of $z, \overline z$ corresponding to the integration of terms
of the form
$\mathbb A_j^{N,w=0}|z|^{2(
  \Delta^{(N)}_j-\tilde\Delta_{1}-\tilde\Delta_{2})}$, where
$\mathbb A_j^{N,w=0},\;N=1,2,3,\dots$
  stand for  contributions from
descendant operators at level $N$ with conformal weights $
  \Delta^{(N)}_j=
  \tilde\Delta_j+N$.

Notice that the symmetry under $j\leftrightarrow-1-j$ in (\ref{A4pt1}),
which can be easily checked by using the identity (\ref{eqsat}), allows  to
extend the integral to the full axis ${\cal P} = -\frac{1}{2}+i\mathbb{R}$.

Given that correlation functions in the $AdS_3$ WZNW model in the $m-$basis depend on the sum of $w_i$ numbers, except for the powers of the coordinates $z_i, \overline z_i$, if the Lorentzian and Euclidean theories are simply related by analytic continuation, this result should hold, in particular, for states in continuous representations in arbitrary spectral flow sectors (with $|m_i|,|\overline m_i|, |m| <\frac 12$), as long as $\sum_iw_i=0$, $i.e.$
\begin{eqnarray}
\left\langle \Phi^{j_1, w_1}_{m_1,\overline m_1}\Phi^{j_2,w_2}_{m_2,\overline m_2}
\Phi^{j_3,w_3}_{m_3,\overline m_3}\Phi^{j_4,w_4}_{m_4,\overline m_4}\right\rangle _{\sum_{i=1}^4w_i=0} ~= ~
z_{34}^{\Delta_2+\Delta_1- \Delta_4-\Delta_3}z_{14}^{\Delta_2+\Delta_3-
\Delta_4-\Delta_1}z_{13}^{\Delta_4-\Delta_1- \Delta_2-\Delta_3}\nonumber\\
\times ~ z_{24}^{-2\Delta_2}\times ~c.c.~\times \int_{\cal P} dj ~ {\mathbb A}_j^{w=0}(j_i;m_i,\overline m_i)~
z^{\Delta_j-\Delta_{1}-\Delta_{2}}\overline z^{\overline \Delta_j-\overline\Delta_{1}-\overline\Delta_{2}}
+\cdots \, ,~~~~~\label{final4}
\end{eqnarray}
where $\Delta_j=-\frac {j(j+1)}{k-2}-m(w_1+w_2)-\frac k4(w_1+w_2)^2$
 and
 $c.c.$ stands for the obvious antiholomorphic $\overline
z_i-$dependence.
For other values of
$j_1, \cdots ,j_4$, $m_1,\dots,\overline m_4$ the integral
may diverge and must be
defined by analytic continuation.

That a generic $w-$conserving four-point function involving primaries or
highest/lowest-weight states in 
${\cal C}_j^{\alpha,w}$ or ${\cal
  D}_j^{\pm,w}$ 
should factorize
as in (\ref{final4}), if the amplitude with four $w=0$ states is 
given by (\ref{4ptm}), can be deduced from
the  relation \cite{ribault}:
\begin{equation}
\displaystyle\left\langle \prod_{i=1}^n \Phi_{m_i,\overline m_i}^{j_i,w_i}
(z_i,\overline z_i)\right\rangle_{\sum_{i=1}^n w_i=0}=\kappa
\overline{\kappa}\left\langle \prod_{i=1}^n
\Phi_{m_i,\overline m_i}^{j_i,\widetilde{w}_i=0}(z_i,\overline
z_i)\right\rangle,\label{kap}
\end{equation}
where $\displaystyle\kappa=\prod_{i<j}z_{ij}^{-w_i m_j-w_j m_i-\frac
  k2 w_i w_j}$,
$\displaystyle\overline\kappa=\prod_{i<j}z_{ij}^{-w_i \overline
m_j-w_j \overline m_i-\frac k2 w_i w_j}$,
after Taylor expanding around $z=0$ the r.h.s. of the following identity:
\bea
&&\kappa~~
z_{34}^{\tilde\Delta_2+\tilde\Delta_1-\tilde\Delta_4-\tilde\Delta_3}
z_{14}^{\tilde\Delta_2+\tilde\Delta_3-\tilde\Delta_4-\tilde\Delta_1}
z_{24}^{-2\tilde\Delta_2} z_{13}^{\tilde\Delta_4-\tilde\Delta_1-
\tilde\Delta_2-\tilde\Delta_3}z^{\tilde\Delta_j-\tilde\Delta_{1}-\tilde\Delta_{2}}=\cr
&& z_{34}^{\Delta_2+\Delta_1-\Delta_4-\Delta_3}
z_{14}^{\Delta_2+\Delta_3-\Delta_4-\Delta_1}
z_{13}^{\Delta_4-\Delta_1-
\Delta_2-\Delta_3}z_{24}^{-2\Delta_2}
z^{\Delta_j-\Delta_{1}-\Delta_{2}}\left(1-z\right)^{-m_2w_3-m_3w_2-
\frac k2w_2w_3}.\label{kappa}\nonumber\\
\eea

The conclusion is that, if the $H_3^+$ and $AdS_3$ models are simply related by
analytic continuation, then 
(\ref{final4}) and its analytic continuation should hold for
generic $w-$conserving four-point functions of fields in 
${\cal C}_j^{\alpha,w}$ or ${\cal
  D}_j^{\pm,w}$ \footnote{See section \ref{kz} for an alternative discussion
directly in the $m-$basis, independent of the $x-$basis.}.
However, expression  (\ref{final4}) appears to be in 
contradiction with the factorization $ansatz$ 
and the OPE found in section \ref{opeads3} for the $AdS_3$ WZNW model, 
because it seems to contain
 just $w-$conserving channels.
Actually, directly applying the
factorization $ansatz$ based on the OPE (\ref{opec}) would give the
following expression for both $w-$conserving and violating four-point functions:
\begin{equation}
\left\langle \Phi^{j_1, w_1}_{m_1,\overline m_1}\Phi^{j_2,w_2}_{m_2,\overline m_2}
\Phi^{j_3,w_3}_{m_3,\overline m_3}\Phi^{j_4,w_4}_{m_4,\overline m_4}
\right\rangle  ~\sim ~
z_{34}^{\Delta_2+\Delta_1-
\Delta_4-\Delta_3}z_{14}^{\Delta_2+\Delta_3-
\Delta_4-\Delta_1}z_{13}^{\Delta_4-\Delta_1-
\Delta_2-\Delta_3} z_{24}^{-2\Delta_2}\times c.c.
\nonumber
\end{equation}
\begin{equation}
~~\times~\delta^2(\sum_{i=1}^4m_i+\frac k2w_i) \sum_{w=-1}^1\int_{\cal P} dj ~
Q^wQ^{-w-\sum_{i=1}^4w_i}B(-1-j)c^{-1-j}_{m,\overline m}
z^{\Delta_j-\Delta_{1}-\Delta_{2}}\overline z^{\overline
\Delta_j-\overline\Delta_{1}-\overline\Delta_{2}}
+\cdots
\label{f4pt}
\end{equation}
with $m=m_1+m_2-\frac k2w = -m_3-m_4-\frac k2w$,
$\overline m=\overline m_1+\overline m_2-\frac k2w = -\overline
m_3-\overline m_4-\frac k2w$ and $\Delta_j=-\frac
{j(j+1)}{k-2}-m(w_1+w_2+w)-\frac k4(w_1+w_2+w)^2$  (similarly for
$\overline \Delta_j$).
Actually,
in the $m-$basis, the starting point for the $w-$conserving four-point
function would have been 
(\ref{final4}) plus an analogous contribution involving one unit spectral 
flow three-point functions, $i.e.$
(\ref{final4}) rewritten in terms of ${\mathbb A}_j^{w=1}$
or ${\mathbb A}_j^{w=-1}$ instead of ${\mathbb A}_j^{w=0}$, where
\bea
{\mathbb A}_j^{w=\pm 1}(j_i;m_i,\overline m_i) &=&
\delta^{(2)}(\sum_{i=1}^4m_i)~\frac{\widetilde
C(1+j_1,1+j_2,1+j)}{\gamma(j_1+j_2+j+3-
\frac k2)}
\widetilde
W\left[\begin{matrix}
j_1~,~~j_2~~,~j\cr \mp m_1, \mp m_2, \pm
m\cr\end{matrix}\right]\cr
&\times& \frac{1}{B(-1-j)c^{-1-j}_{m,\overline m}}
\frac{\widetilde C(1+j_3,1+j_4,1+j)}{\gamma(j_3+j_4+j+3-\frac k2)}
\widetilde W\left[\begin{matrix}
j_3~,~~j_4~~,~j\cr \pm m_3, \pm m_4, \pm m \cr\end{matrix}\right]
.\label{Aw1}
\eea
But if correlation functions in this model
 are to be obtained from those in the $H_3^+$ model
\cite{mo3}-\cite{tesch2}, \cite{hs}-\cite{gk2},
spectral flow conserving and non-conserving channels should give the same
result for the $w-$conserving four-point functions. 
This does not imply that $\mathbb A_j^{w=0}$ and $\mathbb A_j^{w=\pm1}$
carry the same amount of information \footnote{ In other words, both
 expressions seem to give the same contribution in $w-$conserving four-point
functions. However one cannot always use either one of them. In
particular, this is not expected to hold for $w-$violating amplitudes.}. In
general, if both expressions for the four-point functions 
were equivalent, one would expect that part of the
information in $\mathbb A_j^{w=0}$ were contained in $\mathbb
A_j^{w=\pm1}$ and the rest in the
contributions from descendants in $\mathbb A_j^{N,w=\pm1}$.

A proof of this statement would require making  explicit  the
higher order terms and possibly some contour manipulations,
 which we shall not attempt. Nevertheless there are several
 indications supporting this claim.
A similar proposition
was advanced in \cite{ribault} for the $H_3^+$ model and some
evidence was given that these possibilities might not be exclusive,
depending on which correlator the OPE is inserted in.
Furthermore,
$w=1$ long strings were
found in the $s-$channel factorization of the four-point amplitude
of $w=0$ short strings in \cite{mo3} starting from the holomorphically
factorized expression for (\ref{4pt}), rewriting the integrand and
moving the integration contour. Moreover,
in the $m-$basis, spectral flow non-conserving channels can be seen to
appear naturally from
 (\ref{final4}) in certain special cases, as we now show.

Identities
among different expansions of four-point functions containing at least one field
in discrete representations can be generated using the spectral flow symmetry.
In particular,  $w-$conserving four-point functions
 involving the fields $\Phi_{m_1=\overline m_1=-j_1}^{j_1,w_1}$ and  
$\Phi_{m_3=\overline m_3=j_3}^{j_3,w_3}$ coincide (up to
 $B(j_1),B(j_3)$ factors) with the 
 $w-$conserving amplitudes involving
$\Phi_{m'_1=\overline m'_1=j'_1}^{j'_1=-\frac k2-j_1,w'_1=w_1+1}$ and  
$\Phi_{m'_3=\overline m'_3=-j'_3}^{j'_3=-\frac k2-j_3,w'_3=w_3-1}$
\footnote{ This is a consequence of the identities discussed in the
  paragraph containing equation (\ref{Cidentity}) in the previous section.}. This allows
to expand the four-point amplitude in two alternative ways, namely
\bea
\int_{{\cal P}}dj ~ {\mathbb
A}_j^{w=0}(j_1,j_2,j_3,j_4;m_1,\dots,\overline m_3,\overline m_4)~ z^{
  \Delta(j)-\Delta(j_1)-\Delta(j_2)} \overline z^{\overline
  \Delta(j)-\overline\Delta(j_1)-\overline\Delta(j_2)}
+\cdots\label{factAw0}
\eea
or
\bea
\beta_{1,3}\int_{{\cal P}}dj ~
{\mathbb A}_j^{w=0}(j'_1,j_2,j'_3,j_4;m'_1,\dots,\overline m'_3,\overline m_4)~
  z^{\Delta'(j)-\Delta(j'_1)-\Delta(j_2)}
\overline z^{\overline\Delta'(j)-\overline\Delta(j'_1)-\overline\Delta(j_2)}
+ \cdots ,  \label{A'}
\eea 
where $\beta_{1,3}\equiv\frac{B(-1-j_3)}{B(-1-j'_1)}$ and the dots refer  
to  contributions from descendants and, in addition, to
residues at poles in $\mathbb A_j^{w=0}$ crossing ${\cal P}$
after analytic continuation of $j_i$ ($i=1,3$ and eventually $2,4$) to the
region (\ref{unitary}). Explicitly, ${\mathbb
A}_j^{w=0}(j'_1,j_2,j'_3,j_4;m'_1,\dots,\overline m'_3,\overline m_4)$ is 
given by 
\bea 
&&C(1+j'_1,1+j_2,1+j)C(1+j'_3,1+j_4,1+j)
\frac{\pi^3\gamma(2+2j)}{B(-1-j)}
\frac{\gamma(j-j'_1-j_2)\gamma(j_2-j'_1-j)
}{\gamma(2+j'_1+j_2+j)\gamma(-2j'_1)}\cr\cr &&\times
\frac{\gamma(j-j'_3-j_4)\gamma(j_4-j'_3-j)
}{\gamma(2+j'_3+j_4+j)\gamma(-2j'_3)}
\frac{\Gamma(1+j_2-m_2)\Gamma(1+j_4+\overline 
m_4)}{\Gamma(-j_2+\overline m_2) \Gamma(-j_4-m_4)}
\frac{\Gamma(-j-\overline m)
\Gamma(1+j-m)}{\Gamma(1+j+m)\Gamma(-j+\overline m)} .\nonumber
\eea

Using  (\ref{Cidentity}) and rewriting this expression in terms of $j_i,m_i$, the
following equivalence can be shown
\bea
(\ref{A'})=\int_{{\cal P}}dj ~
{\mathbb A}_j^{w=1}(j_1,j_2,j_3,j_4;m_1,\dots,\overline m_3,\overline m_4)~
z^{\Delta(j)-\Delta(j_1)-\Delta(j_2)}\overline 
z^{\overline\Delta(j)-\overline\Delta(j_1)-\overline \Delta(j_2)}+
\cdots.\label{factAw1}
\eea

Notice that not only the coefficient $\mathbb A_j^{w=1}$  but also the
$z_i,\overline z_i$ dependence are as expected.
In fact, $\Delta(j'_1)=\tilde\Delta(j'_1)-m'_1 w'_1-\frac k4 w'_1{}^2=$
$\tilde\Delta(j_1)-m_1w_1-\frac k4w_1^2=\Delta(j_1)$ and
$\Delta'(j)=\tilde\Delta(j)-(m'_1+m_2)(w'_1+w_2)-\frac
k4(w'_1+w_2)^2=$
$\tilde\Delta(j_1)-mw-\frac k4w^2=\Delta(j_1)$,
where $m=m_1+m_2-\frac k2$ and $w=w_1+w_2+1$. Therefore, we have seen
in a particular example that spectral flow conserving and violating
channels can give the same result for four-point functions.
This is a nontrivial result showing that 
the spectral flow 
symmetry allows to exhibit
$w-$non-conserving channels that are not equivalent to other $w-$conserving 
ones in expressions constructed as sums over
 $w-$conserving exchanges. 

In section \ref{kz} we  show that the terms explicitly displayed in both
(\ref{final4}) and (\ref{factAw1}) are 
solutions of the Knizhnik-Zamolodchikov (KZ) equations. 
However, these equations do not give 
 enough information  to 
confirm that the full expressions  (\ref{final4}) and (\ref{factAw1}) 
are equivalent.

The factorization of  four-point functions 
reproduces  the field
content of the OPE. 
Therefore, the truncation imposed on the operator algebra
by the spectral flow symmetry must be realized in physical amplitudes. Again,
to confirm this would
require  more information on the contributions from descendant fields
and studying
crossing symmetry.
Here, we just illustrate this point with one example.
Take for instance the following four-point function \footnote{ Here,
  as in the previous chapter, we
denote the states by the representations they belong to and we omit
the antiholomorphic part for short.}:
\begin{equation}
\left\langle{\cal D}^{+,w_1=0}_{j_1}{\cal D}^{+,w_2=-1}_{j_2}
{\cal D}^{-,w_3=0}_{j_3}{\cal D}^{-,w_4=-1}_{j_4}
\right\rangle\, ,\label{4pte}
\end{equation}
in the particular case with $n_i=0,\forall i$ (where $m_i=\pm j_i\mp
n_i$) and
$j_1+j_2=j_3+j_4<-\frac{k-1}{2}$. The OPE (\ref{1}) implies one
intermediate state
in the $s-$channel in ${\cal D}^{+,w=-1}_j$, with $j=j_1+j_2=-m$ as
well as exchanges of
states in
${\cal D}^{+,w=0}_j$  if $j_1+j_2=j_3+j_4<-\frac{k+1}2$ with
$j=j_1+j_2+\frac k2+n$, $n=0,1,2,\dots$ such that $j<-\frac12$, and
also of continuous states in ${\cal C}_j^{\alpha, w=0}$. The unique state found in ${\cal
   D}^{+,w=-1}_j$
is equivalent to the highest-weight state in ${\cal D}^{-,w=0}_{\tilde j}$
with
$\tilde j=-\frac k2-j>-\frac 12$.

This four-point
function must
coincide with the following one:
\begin{equation}
\left\langle{\cal D}^{+,w_1=0}_{j_1}{\cal D}^{-,w_2=0}_{\tilde j_2}
{\cal D}^{-,w_3=0}_{j_3}{\cal D}^{+,w_4=0}_{\tilde j_4}
\right\rangle\, ,
\end{equation}
where as usual $\tilde j_i=-\frac k2-j_i$ (notice that this holds
without ``hats"
because $n_i=0, \forall i$). Now $\tilde j_2-j_1=\tilde
 j_4-j_3>-\frac 12$.
Therefore, (\ref{d+d-0}) implies that only states from ${\cal C}_j^{\alpha, w=0}$
as well as  from
${\cal D}^{+,w=0}_j$ with $j=j_1-\tilde j_2+n=j_1+j_2+\frac k2+n$
 propagate in the
intermediate $s-$channel, the latter  requiring the extra condition
 $\tilde j_2-j_1=\tilde
j_4-j_3>\frac 12$, $i.e.$
 $j_1+j_2=j_3+j_4<-\frac{k+1}{2}$. The
important remark is that no intermediate states from ${\cal D}^{-,w=0}_{\tilde
 j}$ appear in the
factorization.
 This behavior was discussed in the previous chapter
when studying the consequences of the spectral flow symmetry on the OPE. However, we
 have considered
this case carefully here because it explicitly displays the fact that
 the same
four-point function
factorizes in two different ways and the unique difference is an extra state
 violating the bounds (\ref{unitary}). Recall that we are only
 considering primaries and their spectral flow images.
We expect that some consistency requirements, such as
crossing symmetry, will 
automatically realize
the OPE displayed in the previous chapter in 
physical amplitudes.

An indication in favor of the factorization of this non-rational CFT is that the expressions reproduce the spectral flow selection rules (\ref{cc}) and (\ref{dd}) for four-point functions in different sectors. Indeed, let us analyze this feature in a four-point function involving only external discrete states
or their spectral flow images. The  bounds (\ref{dd}) require $-3\le \sum_{i=1}^4 w_i \le -1$, in agreement  with the  factorization of this amplitude in any channel. Indeed, consider for instance
\begin{equation}
\left\langle{\widehat{\cal D}}^{+,w_1}_{j_1}{\widehat{\cal D}}^{+,w_2}_{j_2}
{\widehat{\cal D}}^{+,w_3}_{j_3}{\widehat{\cal D}}^{+,w_4}_{j_4}
\right\rangle\, .\label{4d}
\end{equation}
The OPE ${\widehat{\cal D}}^{+,w_1}_{j_1}\otimes{\widehat{\cal
 D}}^{+,w_2}_{j_2}$
 computed
in the previous chapter (and similarly for $j_3,j_4$) requires
either
$w_1+w_2=-w_3-w_4-1$ or $w_1+w_2=-w_3-w_4- 2$ or
$w_1+w_2=-w_3-w_4-3$ for discrete intermediate states and
 $w_1+w_2=-w_3-w_4- 2$
for continuous intermediate states. And similarly in the other channels.

Repeating this analysis for four-point functions involving fields in
different representations, it is straightforward to conclude that the spectral
flow selection rules  for four-point functions in different sectors
can be obtained from those
 for two- and three-point functions, or equivalently from the OPE found
in chapter \ref{chapOPE}.

\subsection{ Relation to \cite{mo3}}\label{rtmo1}

This section contains some comments about the relation
between our work and \cite{mo3}. For simplicity, we use the
conventions of the latter,
related  to ours by $j\rightarrow-j$ in the $x-$basis, up to normalizations. 
The range of $j$ for discrete representations is now $\frac 12<j<\frac{k-1}{2}$
and for continuous representations, $j=\frac 12+i\mathbb R$.

One of the aims of \cite{mo3} was
to study the factorization of four-point functions involving $w=0$ short
strings in the boundary conformal field theory.
The $x-$basis seems  
appropriate for this purpose  since $x_i,\overline x_i$ 
can be interpreted as the coordinates of the boundary.
Naturally, both the OPE and the factorization
 look very different in the $m-$ and $x-$basis. 
For instance, it is not obvious how
discrete series would 
appear in the
OPE or factorization of fields in continuous representations if they are to be obtained from the analogous expressions in the $H_3^+$ model in the $x-$basis.
However, when discrete representations are involved, there are certain similarities. Actually, in agreement with the fusion rules $\hat{\cal D}_{j_1}^{+,w_1}\otimes\hat{\cal D}_{j_2}^{+,w_2}$ obtained in the previous chapter, $w=1$ long strings and $w=0$ short strings were found in the factorization studied in \cite{mo3}. Conversely, it was interpreted that $w=1$ short strings  do not
  propagate in the intermediate channels, while we found spectral flow
  non-preserving contributions of discrete representations in the OPE.
In this section we analyze this issue. We reexamine 
the three-point functions involving two $w=0$ strings and one $w=1$ short string 
and certain divergences in the  four-point functions
of $w=0$ short strings, namely the so-called
Poles$_2$, which seem to break the factorization.

\begin{itemize}
    \item {\it Three-point functions involving one $w=1$ short string and
    two $w=0$ strings}
\end{itemize}

The $w-$conserving two-point functions 
of short strings in the target space ($w\ge 0$) are given by
\bea
\langle\Phi^{\omega,j}_{J,\bar J}(x_1, \overline x_1)\Phi^{\omega,j}_{J,\bar
J}(x_2, \overline x_2)\rangle\sim
|2j-1\pm(k-2)\omega|\frac{\Gamma(2j+p)\Gamma(2j+\bar
p)}{\Gamma(2j)^2p!\bar p!}\frac{{\cal B}(j)}{x_{12}^{2J}\bar
x_{12}^{2\bar J}}\,,\label{tpf}
\eea
where ${\cal B}(j)=B(-j)$ and the upper (lower) sign holds for $J=j+p+\frac k2w$ 
($J=-j-p+\frac k2w$), $p, \overline p$ being non-negative integers. 
Three-point functions of $w=0$ string states are 
\bea
\left\langle \Phi_{j_1}(x_1, \overline x_1) \Phi_{j_2}(x_2, \overline x_2)
\Phi_{j_3}(x_3, \overline x_3)\right\rangle= C(j_1,j_2,j_3)
\prod_{i>j}|x_{ij}|^{-2j_{ij}}\, ,\label{3ptxw=0}
\eea
and for one $w=1$  short string
and two $w=0$ strings they are given by
(we omit the $x, \overline x-$dependence)
\bea
\langle\Phi^{j_1,\omega=1}_{J_1,\bar
J_1}(x_1, \overline x_1)\Phi_{j_2}(x_2, \overline x_2)
\Phi_{j_3}(x_3, \overline x_3)\rangle\sim\frac{1}{\Gamma(0)}{\cal B}(j_1)
C\left(\frac k2-j_1,j_2,j_3\right)\times~~~~~~~~~~~~~\cr 
~~~~~~~~~~~~~~~~~\frac{\Gamma(j_2+j_3-J_1)}{\Gamma(1-j_2-j_3+\bar J_1)}
\frac{\Gamma(j_1+J_1-\frac k2)}{\Gamma(1-j_1-\bar J_1+\frac k2)}
\frac{1}{\gamma(j_1+j_2+j_3-\frac k2)}\, .\label{3ptx}
\eea
The $\Gamma(0)^{-1}$ factor is absent when the $w=1$ operator is a
long string state. This three-point function was obtained in
\cite{mo3} from an equivalent expression in the $m-$basis.  $J_1,
\overline J_1$  label the  global
$AdS_3$ representations and can be written in terms of 
parameters $m_1, \overline m_1$ as
$J_1=\mp m_1+\frac k2$,
$\overline J_1=\mp\overline m_1+\frac k2$, depending if the correlator
involved the field $\Phi_{m_1,\overline m_1}^{j_1, w_1=\mp 1}$.

As observed in \cite{mo3}, when $J_1=\frac
k2-j_1-p$, $\overline J_1=\frac k2-j_1-\overline p$, 
the factor $\frac{\Gamma(j_1+J_1-\frac
k2)}{\Gamma(1-j_1-\overline J_1+\frac k2)}$ cancels the $\Gamma(0)$ and
the three-point function is finite and
 can be  interpreted as a $w-$conserving amplitude. 
To see this, recall that if it was obtained from a $w=-1$
three-point function in the $m-$basis and
$m_1=j_1+p$, then
\bea
\langle\Phi^{j_1,w=1}_{J_1,\bar
J_1}(x_1,\overline x_1)\Phi_{j_2}(x_2,\overline x_2)\Phi_{j_3}(x_3,
\overline x_3)\rangle\sim  (-)^{p+\bar
p} {\cal B}(j_1)C\left(\frac k2-j_1,j_2,j_3\right)\times~~~~~~~~~~~~~~~~~\cr 
\frac{\Gamma(j_2+j_3+j_1-\frac k2+p)}{p!
\Gamma(j_2+j_3+j_1-\frac k2)}\frac{\Gamma(j_2+j_3+j_1-\frac
k2+\bar p)}{\bar p! \Gamma(j_2+j_3+j_1-\frac k2)}~~~~~\label{3ptw1}
\eea
 reduces to (\ref{3ptxw=0}) when
$p=\overline p=0$ and $j_1\rightarrow \frac k2-j_1$, as expected from
  spectral flow symmetry. Similarly, if $w=+1$ and $m_1=-j_1-p$, the same
 interpretation holds.

On the contrary, for  $w=-1$ ($w=+1$)
and $m_1=-j_1-p$ ($m_1=j_1+p$),  the $\Gamma(j_1+J_1-\frac k2)$ does not cancel
the factor $\Gamma(0)^{-1}$ and then, it was
concluded in \cite{mo3} that the three-point function 
 vanishes in this case.

However, notice that if
$J_1=\frac
k2+j_1+n=j_2+j_3+p$, $\overline J_1=\frac k2+j_1+\overline
n=j_2+j_3+\overline p$, $n,\overline n\in {\mathbb Z}_{\ge0}$, 
the r.h.s. of (\ref{3ptx}) can also be rewritten as
the r.h.s. of (\ref{3ptw1}), but now
this non-vanishing amplitude corresponds to
a $w=1$ three-point function which is not equivalent to a $w-$conserving one.
Indeed,  (\ref{3ptw1}) is regular as long as $n<p$ ($\overline
n<\overline p$) and when
$n\ge p$ ($\overline n\ge \overline p$)  there are 
divergences in $C(\frac k2-j_1,j_2,j_3)$  at $j_1=j_2+j_3-\frac k2-q$ 
with $q=0,1,2,\cdots$.
Using the spectral flow symmetry, the $w=1$ short string
can be identified with a $w=2$ short string with  $\tilde j_1=\frac
k2-j_1=k-j_2-j_3+q$, which correspond to the
Poles$_2$ in \cite{mo3}.

\begin{itemize}
    \item {\it Factorization of four-point functions of
    $w=0$ short strings}
\end{itemize}

The four-point amplitude of $w=0$ short strings
was  extensively studied in \cite{mo3}. 
The conformal blocks were rearranged  as sums of products of
positive powers of $x$ times  functions of $u=z/x$. In order to perform the
integral over the worldsheet before the $j-$integral, it was
necessary to change the $j-$integration contour from $\frac 12+i\mathbb R$ to
$\frac{k-1}{2}+i\mathbb R$, and
in this process two types of sequences of poles
were picked up, namely \bea &&{\rm Poles}_1:~~j_3=j_1+j_2+n,\cr
&&{\rm Poles}_2:~~j_3=k-j_1-j_2+n, \nonumber\eea where $n=0,1,2,\dots$.
Only values of $n$ for which $j_3<\frac{k-1}{2}$
 contribute to the factorization, so Poles$_1$ appear when
$j_1+j_2<\frac{k-1}{2}$ and Poles$_2$ when
$j_1+j_2>\frac{k+1}{2}$. The contributions from
Poles$_1$ were identified as two
particle states of short strings in the boundary conformal field
theory,
but no interpretation was found
for Poles$_2$ as $s-$channel exchange. 

Recall that we found Poles$_1$ among
the $w-$conserving discrete contributions to the OPE  
$\mathcal
D^{+,w_i}_{j_i}\times \mathcal
D^{+,w_i}_{j_i}$ (see (\ref{1}))
 and Poles$_2$ in the $w-$violating terms with
$\tilde j_3=\frac k2-j_3=j_1+j_2-\frac k2 -n$.
Therefore, it seems tempting to consider
  Poles$_2$ as
 two particle states of $w=1$ short strings in the boundary conformal
 field theory. However, neither the powers of $x, \overline x$ nor the
 residues of the poles in the four-point function studied in
 \cite{mo3} allow this interpretation and thus the Poles$_2$ had to be
 truncated. Clearly,  more work is necessary to determine the
 four-point function and understand
 the factorization.

\section{ Knizhnik-Zamolodchikov equation}\label{kz}

\subsection{KZ equations in the $m-$basis and the factorization {\it ansatz}}

In this section we show
some consistency conditions of the expressions used in the previous section.

Let us start by considering the KZ equation 
for $w-$conserving $n-$point functions in the $m-$basis, namely \cite{ribault}
\bea
\mathcal{E}_i~\kappa^{-1}\left\langle \prod_{\ell=1}^n
\Phi_{m_\ell,\overline{m}_\ell}^{j_\ell,w_\ell}(z_\ell,\overline{z}_\ell)
\right\rangle=0,\label{KZm}
\eea
where
\bea
\mathcal E_i\equiv (k-2)\frac{\partial}{\partial z_i}+\sum_{j\neq i}
\frac{Q_{ij}}{z_{ji}},~~~~Q_{ij}=-2t_i^3t_j^3+t_i^-t_j^++t_i^+t_j^-,
\eea
$t^a$ are defined by $\tilde J_0^a|j,m, \overline m, w>=
-t^a|j,m,\overline m,w>$, $|j,m,\overline m, w>$ 
being the state corresponding to the field
$\Phi_{m,\overline m}^{j,w}$ and $\kappa$ was introduced in (\ref{kap}). 

Since  a generic
$w-$conserving four-point function can be obtained from the expression
involving four $w=0$ fields,  we concentrate on
\bea
\left\langle \prod_{i=1}^4\Phi_{m_i,\overline{m}_i}^{j_i,w_i=0}
(z_i,\overline{z}_i)\right\rangle&=&|z_{34}|^{2(\tilde\Delta_2+\tilde\Delta_1-
\tilde\Delta_4-\tilde\Delta_3)}|z_{14}|^{2(\tilde\Delta_2+\tilde\Delta_3-
\tilde
\Delta_4-\tilde\Delta_1)}
|z_{13}|^{2(\tilde\Delta_4-\tilde\Delta_1-\tilde\Delta_2-\tilde\Delta_3)}
\nonumber\\
&&\times~ |z_{24}|^{-4\tilde
\Delta_2}~\mathcal F_j(z,\overline z)\, ,\nonumber
\eea
$\mathcal F_j(z,\overline z)$ being a function of the cross ratios $z,
\overline z$, not
determined by conformal symmetry. The KZ equation (\ref{KZm})
implies the following constraint
\bea
\frac{\partial\mathcal
F_j(z,\overline z)}{\partial z}=\frac{1}{k-2}
\left[\frac{Q_{21}}{z}+\frac{Q_{23}}{z-1}\right] \mathcal
F_j(z,\overline z)\, .\label{KZF}
\eea

Assuming that $\mathcal F_j(z,\overline z)$ has the following form
\bea
\mathcal F_j(z,\overline z)=\sum_{N,\overline
N=0}^\infty\int dj \left\{A_j^{(N,\overline N)}\left[\begin{matrix}
j_1\,,\,j_2\,,\,j_3\,,\,j_4\cr m_1,m_2,\dots,\overline
m_4\cr\end{matrix}\right]
z^{\Delta_j-\tilde\Delta_1-\tilde\Delta_2+N}\overline
z^{\Delta_j-\tilde\Delta_1-\tilde\Delta_2+ \overline
N}\right\}\, ,\label{factansm}
\eea
inserting it into
(\ref{KZF}) with
$\Delta_j=\tilde\Delta_j\equiv-\frac{j(1+j)}{k-2}$, then
$A_j^{(0,0)}\left[\begin{matrix}
j_1\,,\,j_2\,,\,j_3\,,\,j_4\cr m_1,m_2,\dots,\overline
m_4\cr\end{matrix}\right]$ satisfies
\bea
&&\left\{2m_1m_2-j(1+j)+j_1(1+j_1)+j_2(1+j_2)\right\}
A_j^{(0,0)}\left[\begin{matrix} j_1\,,\,j_2\,,\,j_3\,,\,j_4\cr
m_1,m_2,\dots,\overline m_4\cr\end{matrix}\right]=~~~~~~~~~~~~~~~~~~~~\nonumber\\
&&~~~~~~~~~~~~~~~~~~~~~~~~~~~~~~~=~(m_1-j_1)(m_2+j_2)A_j^{(0,0)}\left[\begin{matrix}
j_1~,~j_2~,~j_3~,~j_4\cr m_1+1,m_2-1,\dots,\overline
m_4\cr\end{matrix}\right]\nonumber\\
&&~~~~~~~~~~~~~~~~~~~~~~~~~~~~~~~+~(m_1+j_1)(m_2-j_2)A_j^{(0,0)}\left[\begin{matrix}
j_1~,~j_2~,~j_3~,~j_4\cr m_1-1,m_2+1,\dots,\overline
m_4\cr\end{matrix}\right]\, .\label{KZA}
\eea

The equations relating coefficients  
$A_j^{(N,\overline N)}$ with $N, \overline N\ne 0$,
are much more
complicated because they mix terms with different values of $m_i,
\overline m_i$ with terms at different levels $N, \overline N$.

This equation does not have enough information to determine
$A_j^{(0,0)}$ completely. So we just check that
the expression found in (\ref{4ptm}) is consistent with an analysis
performed directly in the
$m-$basis. Inserting 
$A_j^{(0,0)}
\left[\begin{matrix} j_1\,,\,j_2\,,\,j_3\,,\,j_4\cr
m_1,m_2,\dots,\overline m_4\cr\end{matrix}\right]= {\mathbb
A}_j^{w=0}(j_1,\dots,j_4;m_1,\dots,\overline m_4)$ 
into (\ref{KZA}) reproduces the same equation
 with $A_j^{(0,0)}$ replaced by $W(j_1,j_2,j;m_1,m_2,m)$.
Because of the complicated expressions known for $W$, we  focus
on the case in which one of the fields in the four-point function
is a discrete primary, namely $\Phi_{m_1,\overline
m_1}^{j_1,w_1=0}\in\mathcal D_{j_1}^{+,w=0}$. In this case, using
(\ref{w1}) one
can show that
(\ref{KZA}) is equivalent to 
\bea
0&=&\sum_{n=0}^{n_1-1}(-)^n\left(\begin{matrix} n_1\cr
n\cr\end{matrix}\right)\left[j-m+\frac{(m_1-j_1)(1+j_1+m_1)}{n_1+1-n}
+\frac{(m_2-j_2)(1+j_2+m_2)(n_1-n)} {n+1+j+m-n_1}\right]\cr
&&~~~\times \frac{\Gamma(n-j_{1}-j_2+j)}{\Gamma(-j_{1}-j_2+j)}
\frac{\Gamma(n+1+j+j_{2}-j_1)}{\Gamma(1+j+j_{2}-j_1)}
\frac{\Gamma(-2j_1)}{\Gamma(n-2j_1)}
\frac{\Gamma(1+j+m)}{\Gamma(n-n_1+1+j+m)}\cr
~~&-&(-)^{n_1}\left[m_1(1-m_1)+j_1(1+j_1)\right]
\frac{\Gamma(n_1-j_{1}-j_2+j)}{\Gamma(-j_{1}-j_2+j)}
\frac{\Gamma(n_1+1+j+j_{2}-j_1)}{\Gamma(1+j+j_{2}-j_1)}
\frac{\Gamma(-2j_1)}{\Gamma(n_1-2j_1)} \nonumber
\eea 
where $n_1=m_1+j_1$ and
$m=m_1+m_2$.
Using $m-$conservation this can be rewritten as 

\bea
0&=&\sum_{n=0}^{n_1-1}(-)^n\left(\begin{matrix} n_1\cr
n\cr\end{matrix}\right)\left[-n\frac{1-n+2j_1}{n_1+1-n}
+\frac{(n-j_{1}-j_2+j)(n+1+j_2+j-j_1)} {n+1+j+m-n_1}\right]\cr\cr
&& ~~~~~\times~ \frac{\Gamma(n-j_{1}-j_2+j)}{\Gamma(-j_{1}-j_2+j)}
\frac{\Gamma(n+1+j_{2}+j-j_1)}{\Gamma(1+j_{2}+j-j_1)}
\frac{\Gamma(-2j_1)}{\Gamma(n-2j_1)}
\frac{\Gamma(1+j+m)}{\Gamma(n-n_1+1+j+m)}\cr\cr
&-& (-)^{n_1}\left[m_1(1-m_1)+j_1(1+j_1)\right]
\frac{\Gamma(n_1-j_{1}-j_2+j)}{\Gamma(-j_{1}-j_2+j)}
\frac{\Gamma(n_1+1+j_{2}+j-j_1)}{\Gamma(1+j_{2}+j-j_1)}
\frac{\Gamma(-2j_1)}{\Gamma(n_1-2j_1)}\, .\nonumber
\eea 
To see
that this vanishes, it is sufficient to note that 
\bea
&&\sum_{n=0}^{n_1-1}(-)^n\left(\begin{matrix} n_1\cr
n\cr\end{matrix}\right)\left[-n\frac{1-n+2j_1}{n_1+1-n}
\right]\frac{\Gamma(n-j_{1}-j_2+j)}{\Gamma(-j_{1}-j_2+j)}
\frac{\Gamma(n+1+j_{2}+j-j_1)}{\Gamma(1+j_{2}+j-j_1)}
\frac{\Gamma(-2j_1)}{\Gamma(n-2j_1)}\nonumber\\
&&~~~~~~~~~~~~~~~~~\times~\frac{\Gamma(1+j+m)}{\Gamma(n-n_1+1+j+m)}\nonumber\\
&&-(-)^{n_1}\left[m_1(1-m_1)+j_1(1+j_1)\right]
\frac{\Gamma(n_1-j_{1}-j_2+j)}{\Gamma(-j_{1}-j_2+j)}
\frac{\Gamma(n_1+1+j_{2}+j-j_1)}{\Gamma(1+j_{2}+j-j_1)}\frac{\Gamma(-2j_1)}
{\Gamma(n_1-2j_1)}\nonumber\\
&&=-\sum_{\tilde n=0}^{n_1-1}(-)^{\tilde n}
\left(\begin{matrix} n_1\cr
\tilde n\cr\end{matrix}\right)\left[\frac{(\tilde n-j_{1}-j_2+j)
(\tilde n+1+j_{2}+j-j_1)}{\tilde n+1+j+m-n_1}\right]
\frac{\Gamma(\tilde n-j_{1}-j_2+j)}{\Gamma(-j_{1}-j_2+j)}
\nonumber\\
&&~~~~~~~~~~~~~~~~~\times~~\frac{\Gamma(\tilde n+1+j_{2}+j-j_1)}{\Gamma(1+j_{2}+j-j_1)}
\frac{\Gamma(-2j_1)}{\Gamma(\tilde n-2j_1)}
\frac{\Gamma(1+j+m)}{\Gamma(\tilde n-n_1+1+j+m)}\, ,
\nonumber
\eea
where $\tilde n=n-1$.

Let us now discuss the other possible $ansatz$, namely (\ref{Aw1}).
To see that ${\mathbb A}_j^{w=1}$ also verifies the KZ equation,
consider $\Delta_j= 
-\frac{j(1+j)}{k-2}-m-\frac k4 $ and $m=m_1+m_2-\frac k2$ in (\ref{KZF}). 
In this case, the equation to be satisfied by $A_j^{(0,0)}$, obtained
by replacing 
(\ref{factansm}) into
(\ref{KZF}), is the following:
\bea
\left\{2m_1m_2-j(1+j)+j_1(1+j_1)+j_2(1+j_2)-(k-2)(m_1+m_2-\frac k4)\right\}
A_j^{(0,0)}\left[\begin{matrix} j_1\,,\,j_2\,,\,j_3\,,\,j_4\cr
m_1,m_2,\dots,\overline m_4\cr\end{matrix}\right]\nonumber
\eea
\bea
&=&(m_1-j_1)(m_2+j_2)A_j^{(0,0)}\left[\begin{matrix}
j_1~,~j_2~,~j_3~,~j_4\cr m_1+1,m_2-1,\dots,\overline
m_4\cr\end{matrix}\right]\nonumber\\
&&+~ (m_1+j_1)(m_2-j_2)A_j^{(0,0)}\left[\begin{matrix}
j_1~,~j_2~,~j_3~,~j_4\cr m_1-1,m_2+1,\dots,\overline
m_4\cr\end{matrix}\right]\nonumber\\
&&-~(m_2-j_2)(m_3+j_3)A_j^{(0,0)}\left[\begin{matrix}
j_1~,~j_2~,~j_3~,~j_4\cr m_1,m_2+1,m_3-1,\dots,\overline
m_4\cr\end{matrix}\right]\, .
\eea

It is not difficult to check that $A_j^{(0,0)}
\left[\begin{matrix} j_1\,,\,j_2\,,\,j_3\,,\,j_4\cr
m_1,m_2,\dots,\overline m_4\cr\end{matrix}\right]=
{\mathbb A}_j^{w=1}(j_1,\dots,j_4;m_1,\dots,\overline m_4)$ 
is a solution of this equation. 

Obviously, $A_j^{w=-1}$ is also a solution of (\ref{KZF})
 when $\Delta_j= 
-\frac{j(1+j)}{k-2}+m-\frac k4 $ and $m=m_1+m_2+\frac k2$.  

Here, we have considered the simple case of four $w=0$ fields.
However, these results can be generalized for arbitrary  
$w-$conserving correlators using the identity (\ref{kappa}).

\bigskip

\chapter{Characters on the Lorentzian torus}\label{chapChar}

The spectrum of a quantum model is built up by appropriately choosing a subset of the representation spaces of the symmetry group. There are two quantities storing this information, the characters associated to each representation and the partition function storing the information of the full spectrum. Such a subset of representations is chosen by constraining the spectrum with physical conditions, $e.g.$ unitarity. 

A consistent CFT must be well defined independently of the boundary conditions imposed on the fields, which means that the theory can be defined in any Riemannian surface. Consistency is guarantied by imposing modular invariance on the surface. The particular case with periodic boundary conditions in the two directions gives place to a CFT defined on a torus topology and a well defined CFT on this surface requires the partition function to be invariant under $SL(2,{\mathbb Z})$, the modular group of the torus. This is the reason why a standard method, used in CFTs theories to determine the appropriate representations contained in the spectrum of the model is looking for modular invariant quantities, which are then interpreted as the partition function. In this chapter we will concentrate on the definition of characters of the Lorentzian $AdS_3$ WZNW model. The next chapter is devoted to discuss the issue of modular trnasformations.

The partition function of the $AdS_3$ WZNW model was computed 
on the Lorentzian torus in  \cite{mo1} because it diverges
on the Euclidean signature torus,
and it was shown that  a modular invariant expression is obtained
after analytic continuation of the modular parameters. 
In this chapter we 
rederive the characters of the relevant representations and
stress some important issues 
related to the regions of convergence of the expressions involved,
focussing on their structure as distributions.

The characters on the Lorentzian signature torus  
are defined 
from the standard
expressions as
\bea
\chi_{{\cal V}_L}(\theta_-,\tau_-,u_-)={\rm Tr}_{{\cal V}_L} e^{2\pi 
i\tau_-(L_0-\frac{c}{24})}
e^{2\pi i\theta_- J_0^3}e^{\pi iu_- K}\, ,\nonumber\\
\chi_{{\cal V}_R}(\theta_+,\tau_+,u_+)={\rm Tr}_{{\cal V}_R} e^{2\pi 
i\tau_+(\bar L_0-\frac{\bar c}{24})}
e^{2\pi i\theta_+ \bar J_0^3}e^{\pi iu_+ K}\, ,\label{char}
\eea
where 
 $\tau_\pm, \theta_\pm, u_\pm$  are independent real parameters, 
  $c=\bar c$ are the left- and right-moving central charges
and
$K$ is the central element of the affine algebra. 
The traces 
are taken over the left and
right representation
modules of the
Hilbert space of the theory, ${\cal V}_L$ and  ${\cal V}_R$, respectively.
The Euclidean version of (\ref{char}) is obtained replacing
the real parameters by complex ones.
For completeness, a
 description of the moduli space of the Lorentzian torus is presented in 
appendix \ref{A:tlt}.

In the remaining of this chapter we compute the complete set of characters of the relevant representations making up the spectrum of the bulk $AdS_3$ conformal field theory and of the finite dimensional representations appearing in the open string spectrum of some brane solutions.  

To lighten notation, from now on  $\tau , \theta, u$ will denote the real parameters $\tau_-, \theta_-, u_-$ and the following compact notation will be used: 
$\chi_j^{\pm,w}:=\chi_{\hat{\mathcal D}_j^{\pm ,w}}$, $\chi_j^{\alpha,w}:= \chi_{\hat{\mathcal C}_j^{\alpha,w}}$.

\section{Discrete representations}

The naive computation of the characters (\ref{char}) for the
 discrete representations
leads to $\theta$ and $\tau$ dependent 
divergences. This is not a problem because the characters
 are typically not functions but distributions. Indeed,
similarly as the characters of the continuous representations, which contain
 a series of delta functions \cite{mo1},
those of the discrete representations need also be interpreted as 
distributions. 

Let us consider the distributions constructed from
 the series defining the characters of the discrete 
representations. Shifting  $\tau\rightarrow\tau+i\xi_1$ and
 $\theta\rightarrow\theta+i\xi_2^w$ in (\ref{char}), 
where $\xi_1,\xi_2^w$ are two real non vanishing parameters,
 a regular distribution can be defined.
Indeed,
the deformed characters of  discrete representations
in an arbitrary spectral flow
sector $w$ can be written in terms 
of those of unflowed representations as
\bea
&&\chi_{j,\xi_2^w,\xi_1}^{+,w}
(\theta, \tau,u)=
e^{i\pi k u}\sum_{\rm n}
\epsilon_{\rm n} <{\rm n}|U_{-w}e^{2\pi i 
(\tau+i\xi_1)(L_0-\frac{c}{24})} e^{2\pi i(\theta+i\xi_2^w) J_0^3} U_{w}|{\rm
n}> \,
,
\nonumber
\eea
where $|{\rm n}>$ 
is a complete orthonormal basis in $\hat{\mathcal D}_j^{+,0}$,
 with norm 
$\epsilon_{\rm n}=\pm1$ (remember that
 this model is not unitary unless the Virasoro constraint is imposed). 
Since $U_w$ is unitary, 
$U_w|{\rm n}>$ defines an orthonormal basis in 
$\hat{\mathcal D}_j^{+,w}$ and from (\ref{wtrans}) one can rewrite 
\bea
\chi_{j,\xi_2^w,\xi_1}^{+,w}=
e^{i\pi k u} e^{-2\pi i\tau\frac k4w^2}e^{2\pi i\theta \frac k2 w}
\sum_{\rm n}\epsilon_{\rm n}<{\rm n}|e^{2\pi i 
(\tau+i\xi_1)(L_0-\frac{c}{24})} e^{2\pi i(\theta-w\tau+ 
i(\xi_2^w-w\xi_1))J_0^3}|{\rm n}>.~
\label{charfromw0}
\eea
Choosing an orthonormal basis of eigenvectors of $L_0$ and $J_0^3$, the
following behavior of the sum is easy to see
\bea
\chi_{j,\xi_2^w,\xi_1}^{+,w}\sim
 \sum_{N,n=0}^\infty \rho(n,N) ~e^{2\pi i [(1+w)\tau-
\theta +i\left((1+w)\xi_1-\xi_2^w\right)]N} 
e^{2\pi i[\theta-w\tau+i(\xi_2^w-w\xi_1)]n}\, ,\nonumber
\eea
where $\rho(n,N)$ gives the degeneracy of states. 
This expression requires the necessary condition
\bea
\left\{\begin{array}{lcr}
~~~~~~~~~~~\xi_1>0\,,\cr
(1+w)\xi_1>\xi_2^w>w\xi_1 \, ,
\end{array}\right.\label{epsiloncond}
\eea
 and it gives 
\bea
\chi_{j,\xi_2^w,\xi_1}^{+,w}
 =
e^{i\pi k u} e^{-2\pi i(\tau+i\xi_1)\frac k4w^2}e^{2\pi i(\theta+i\xi_2^w) 
\frac k2 w} 
\frac{e^{-\frac{2\pi i (\tau+i\xi_1)}{k-2}(j+\frac12)^2}
e^{-2\pi i(\theta+i\xi_2^w-
w(\tau+i\xi_1))(j+\frac12)}}
{i\vartheta_{11}(\theta+i\xi_2^w-w(\tau+i\xi_1),\tau+i\xi_1)}\, .~~ ~~\label{chireg}
\eea

Analyzing  expression (\ref{chireg}) it is found that region (\ref{epsiloncond}) is free of poles and that the nearest poles are located when the inequalities saturate. So that (\ref{epsiloncond}) are not only necessary but also sufficient conditions.
 
This character defines a regular distribution and,
given that the series 
of regular 
distributions are continuous with respect to the weak limit, 
this implies 
\bea
\chi_j^{+,w}(\theta,\tau,u)&=&
e^{i\pi k u}~\frac{e^{-\frac{2\pi i \tau}{k-2}(j+\frac12-w\frac{k-2}{2})^2} 
e^{-2\pi i\theta(j+\frac12-w\frac{k-2}{2})}}{i\vartheta_{11} 
(\theta+i\epsilon_2^w,\tau+i\epsilon_1)}\, ,\label{chidiscw}
\eea
where we have used the identity
\bea
\vartheta_{11}\left(\theta+i\epsilon_2^w -w(\tau +i\epsilon_1),
\tau+i\epsilon_1\right)=(-)^w e^{-\pi i \tau w^2+2\pi i \theta w}
\vartheta_{11}(\theta +i\epsilon_2^w, \tau+i\epsilon_1)
\eea
and the $i\epsilon$'s denote the usual $i0$ prescriptions, but constrained as the corresponding finite parameters in (\ref{epsiloncond}),
which dictate how to avoid the poles of $\vartheta_{11}^{-1}$ 
at $n\tau\in {\mathbb Z},~m\tau+\theta\in{\mathbb Z}$, 
for $n\in{\mathbb N},m\in{\mathbb Z}$.
These poles are easily seen in the following alternative expression for the 
elliptic theta function
\bea
\frac{1}{\vartheta_{11}(\theta+i\epsilon_2^w,\tau+i\epsilon_1)} &=&
\frac{-e^{-i\frac\pi4 \tau}}{\sin\left[\pi\left(\theta+i\epsilon_2^w\right)
\right]}\frac 1{
\prod_{n=1}^{\infty}\left[1-e^{2\pi in(\tau+i\epsilon_1)
}\right] } 
\nonumber\\
&& \times ~\frac 1{\prod_{n=1}^\infty \left
[1-e^{2\pi i(n\tau-\theta+i\epsilon_3^{n,w}})
\right]
\left[1-e^{2\pi i(n\tau+\theta+i\epsilon_4^{n,w})}\right]}\, ,\label{theta11b}
\eea
with 
\bea 
\ \ \ \ \ \ \ 
\left\{\begin{array}{lcr}
\epsilon_3^{n,w}=n\epsilon_1-\epsilon_2^w\cr
\epsilon_4^{n,w}=n\epsilon_1+\epsilon_2^w
\end{array}\right.\, ,\label{epsilon34}
\eea
$i.e.$, $\epsilon_3^{n,w}>0~(<0)$ for $n\geq1+w ~(n\leq w)$ 
and $\epsilon_4^{n,w}>0~(<0)$ for $n\geq-w~ (n\leq -1-w)$.

Notice that, in the weak limit, one can take
 $\epsilon_1,\,\epsilon_2^w=0$ in the arguments of the exponential 
terms in (\ref{chidiscw}) because they are perfectly regular.

It is useful to rewrite (\ref{chidiscw}) using the identity 
(\ref{identdeltas}), which allows  to change the signs of  $\epsilon_2^w,\,
\epsilon_3^{n,w}$ and $\epsilon_4^{n,w}$,
in order to get the following expressions in terms of
only one parameter, say $\epsilon_2^{w'}$, with arbitrary $w'$:
\bea
\chi_j^{+,w < w'}(\theta,\tau,u)&=&(-)^{w} e^{i\pi ku}\frac{
e^{-\frac{2\pi i 
\tau}{k-2}(j+\frac12-w\frac{k-2}{2})^2}e^{-2\pi i\theta(j+\frac12-w\frac{k-2}
{2})}}{i\vartheta_{11}(\theta+i\epsilon_2^{w'},\tau+i\epsilon_1)}\cr\cr
&&-~(-)^{w} e^{i\pi ku}\frac{
e^{-\frac{2\pi i \tau}{k-2}(j+\frac12-w\frac{k-2}{2})^2} 
e^{-2\pi i\theta(j+\frac12-w\frac{k-2}{2})}}{\eta^3(\tau+i\epsilon_1)}\cr
&&\times ~\sum_{n=1+w}^{w'} 
(-)^ne^{2i\pi \tau \frac{n^2}{2}}\sum_{m=-\infty}^{\infty}(-)^m
\delta(\theta-n\tau+m)\, \label{chiw<w'}
\eea  
and
\bea
\chi_j^{+,w > w'}(\theta,\tau,0)&=&(-)^{w} e^{i\pi k u}
\frac{e^{-\frac{2\pi i \tau}{k-2}(j+\frac12-w\frac{k-2}{2})^2}
e^{-2\pi i\theta(j+\frac12-w\frac{k-2}{2})}}{i\vartheta_{11}
(\theta+i\epsilon_2^{w'},\tau+i\epsilon_1)}\cr\cr
&& +~(-)^{w} e^{i\pi ku}
\frac{e^{-\frac{2\pi i \tau}{k-2}(j+\frac12-w\frac{k-2}{2})^2}
e^{-2\pi i\theta(j+\frac12-w\frac{k-2}{2})}}{\eta^3(\tau+i\epsilon_1)} \cr
&&\times ~ \sum_{n=1+w'}^{w} 
(-)^n e^{2i\pi \tau \frac{n^2}{2}}\sum_{m=-\infty}^{\infty}(-)^m
\delta(\theta-n\tau+m)\, .\label{chiw>w'}
\eea

These expressions are in perfect agreement with the spectral flow symmetry, 
which implies $\chi_{j}^{+,w}(-\theta,\tau,u)=\,\chi_{-\frac k2-j}^{+,-w-1}
(\theta,\tau,u)$.   
They lead to the following contribution to the partition function  
\bea
Z_{\cal D}^{AdS_3}= \sqrt{
\frac{k-2}{2i(\tau_- -\tau_+)}}
\frac{e^{i\pi k(u_--u_+)}
e^{2\pi i\frac{k-2}{4}\frac{(\theta_--\theta_+)^2}{\tau_- -\tau_+}}
}{
\vartheta_{11}(\theta_-+i\epsilon^0_2,\tau_-+i\epsilon_1)
\vartheta_{11}^*(\theta_+-i\epsilon^0_2,\tau_+-i\epsilon_1)}
~+~\dots\, , \label{parf}
\eea
where the
ellipses stand for the contributions of the contact terms. 
This expression differs formally from the equivalent one in \cite{mo1}, 
where no $\epsilon$ prescription or contact terms were considered. 
Nevertheless, 
the ultimate goal in \cite{mo1}
was to 
 reproduce the Euclidean partition function
continuing the modular parameters 
away from the real axes and discarding
contact terms such as those of the 
characters of the continuous representations.

\section{Continuous representations}

A similar analysis can be performed for the characters of the
continuous representations. 
Using (\ref{charfromw0}), one can compute these characters in terms of 
those of the unflowed continuous representations. The result is  
\bea
\chi_{j}^{\alpha,w}&=&e^{i\pi k u}
\frac{-2\sin[\pi (\theta-w\tau)]e^{-2\pi i\tau\frac k4w^2}
e^{2\pi i\theta \frac k2 w} 
e^{-\frac{2\pi i \tau}{k-2}(j+\frac12)^2}e^{2\pi i(\theta-w\tau) \alpha} }
{\vartheta_{11}
(\theta-w\tau,\tau+i\epsilon_1)} 
\sum_{n=-\infty}^{\infty}e^{2\pi i(\theta-w\tau) n}\cr\cr
&=&e^{i\pi ku}\frac{e^{2\pi i \tau\left(\frac{s^2}{k-2}+\frac{k}{4}w^2\right)}}
{\eta^3(\tau+i\epsilon_1)}  \sum_{m=-\infty}^{\infty}
e^{-2\pi im\left(\alpha+\frac k2 w\right)} 
\delta\left(\theta-w\tau+m\right)\, ,\label{chiwcon}
\eea
where the following identity was used
\bea
\sum_{n=-\infty}^{\infty}e^{2\pi ixn}=
\sum_{m=-\infty}^{\infty}\delta\left(x+m\right)\, .
\eea

In previous attempts to find the modular $S$-transformation of the principal continuous series (see for instance \cite{TJ,GKS}) the $\theta$ variable was turned off and so the factor $\displaystyle\sum_m e^{-2\pi i m\alpha}\delta(\theta+m)\sim \delta(\theta)$ was interpreted as the infinite volume of the target space and was factorized out in the transformation. It is clear that the situation is much more involved because such a term has a non trivial behavior under $SL(2,{\mathbb Z})$. After a modular $S$ transformation one finds $\delta\left(\frac\theta\tau\right)=|\tau|\delta(\theta)$, which prevents one from simply taking the limit $\theta=0$ discarding the $\delta$ function. The modular transformation will differ from the $\theta\neq0$ case, and so it will not give the correct modular $S$ matrix (which must not depend on  $\theta$).

In this case, the characters are defined as the weak limit 
$\epsilon_1,\,\epsilon_2^w\rightarrow0$, with the constraints
\bea
\left\{\begin{array}{lcr}
~~~~\epsilon_1>0\,,\cr
\epsilon_2^w-w\epsilon_1=0\, ,
\end{array}\right.
\eea
and they give  the
following contribution to the partition function:
\bea
Z_{\cal C}^{AdS_3}&=&
\sqrt{\frac{2-k}{8i(\tau_--\tau_+)}}\frac{e^{i\pi k(u_--u_+)}}
{\eta^3(\tau_-+i\epsilon_1)\eta^*{}^3(\tau_+-i\epsilon_1)}\cr
&&\times ~ \sum_{m,w=-\infty}^{\infty}e^{-2\pi
i\frac  k4 w(\theta_- -\theta_+)}\delta(\theta_- -w\tau_-+m)\delta(\theta_+
-w\tau_+ +m)\, .~~\label{pfcon}
\eea

\section{Degenerate representations}

Degenerate representations  are not contained
 in the spectrum 
of the $AdS_3$ WZNW model but 
they  play an important role in the description of the boundary CFT. 
Indeed, using worldsheet duality, it was argued that they 
make up the Hilbert 
space of open string 
excitations of S$^2$ branes in the $H_3^+$ model \cite{GKS,PST}. 
For the analysis that we shall perform in the forthcoming chapters, 
 it is useful to note the relation among their characters 
and those of 
discrete  and continuous representations of the universal cover 
of SL(2,${\mathbb R})$ discussed above.

The finite dimensional degenerate representations 
 are labeled by the spin $j_{rs}^\pm$ defined 
by $1+2j^\pm_{rs}=
\pm \left(r+ s(k-2)\right)$, with $r,s+1=1,2,3,\dots$ for the 
upper sign 
and $r,s=1,2,3,\dots$ for the lower one. 
Here we consider 
$J=j^+_{r0}$, with characters  given by
\bea
\chi_J(\theta,\tau,u)=-\frac{2 e^{i\pi ku} e^{-2\pi i 
\tau\frac{(2J+1)^2}{4(k-2)}} 
\sin\left[\pi\theta (2J+1)\right]}{\vartheta_{11}(\theta+i\epsilon_2,\tau+i\epsilon_1)}\,,
\label{cdr}
\eea 
where the $\epsilon$'s are restricted to 
\bea
\left\{\begin{array}{lcr}
~\,\epsilon_1>0\,,\cr
|\epsilon_2|<\epsilon_1\,.
\end{array}\right.
\eea

Extrapolating the values of the 
spins in the expressions obtained in the previous sections, 
(\ref{cdr}) can be rewritten as 
\bea
\chi_J(\theta,\tau,u)
&=&\chi_{J}^{+,w=0}(\theta,\tau,u)+\chi_{-\frac k2-J}^{+,w=-1}(\theta,\tau,u)
-\chi_{J}^{\alpha=\{J\},w=0}(\theta,\tau,u)\,,\label{deg-dc}
\eea
where $\{J\}$ is the sawtooth function. 
Actually, this relation could have been guessed from a 
simple inspection of the spectrum (see Figure 6). This can be seen as a non trivial check of the characters defined above and,
 simultaneously, it shows the important role played by the $i0$ prescription in the definition of the characters of discrete representations. A naive computation of these characters, ignoring the $i0'$s, would yield the (wrong) conclusion 
$\chi_J=\chi^{+,w=0}_J+\chi^{+,w=-1}_{-\frac k2-J}$.

\bigskip

\centerline{\psfig{figure=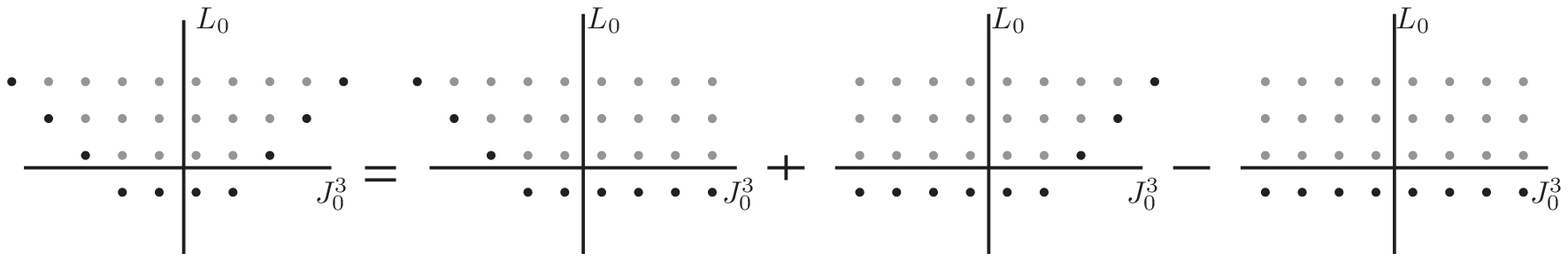,width=16.0cm}}
{\footnotesize{Figure 6: The weight diagram of the degenerate representations
with spin $J=j^+_{r0}=\frac{r-1}2$, $r=1,2,3...$ can be decomposed as the
sum of the weight diagrams of the lowest and highest weight unflowed discrete
representations  minus that of the continuous representation
of spin $J$.}}

\chapter{Modular properties }\label{chapMP}

In this chapter we discuss the modular properties of the Lorentzian characters defined in the previous chapter. Even though the Lorentzian torus is not modular invariant, the characters transform as pseudovectors and the full modular $S$ and $T$ matrices can be rigorously defined. It is shown that these satisfy the expected relations $S^2=(ST)^3=C$, $C$ being the charge conjugation matrix. In the next chapter we will explore if these modular matrices have a similar role as the modular $S$ matrices in the microscopic description of boundary CFTs.

\section{Modular group of the torus}

The modular transformation $\tau\rightarrow\frac{a\tau + b}{c\tau + d}$, 
with integer parameters $a,b,c,d$ such that $ad-bc=1$, can be easily extended to include
$\theta, u$. Characters generating a representation space of the modular group
transform as \cite{dif}
\bea
\chi_{\mu} \left(\frac{\theta}{c\,\tau+d},\frac{a\,\tau+b}{c\,\tau+d},
u+\frac{c~\theta^2}{2(c\,\tau+d)}\right)=\sum_{\nu}M_{\mu}{}^{\nu} 
\chi_{\nu}(\theta,\tau,u),\label{modultrans}
\eea 
$M$ being the matrix associated to the group element. Insofar as $\tau$ and $u$ are concerned, the sign of all the parameters $a,b,c,d$ may be simultaneously changed without affecting the transformation. In models where the representations are self conjugate ($e.g.$ the $SU(2)$ WZNW model), the invariance $\theta\leftrightarrow-\theta$ allows to put $\theta=0$ and so the modular group is simply PSL(2, ${\mathbb Z})= \frac{SL(2, {\mathbb Z})}{{\mathbb Z}_2}$. When this is not the case, the characters form a representation of $SL(2,{\mathbb Z})$ which is freely generated by $T=\left(\begin{matrix}1&1\cr 0&1\end{matrix}\right)$ and $S=\left(\begin{matrix}0&-1\cr 1&0\end{matrix}\right)$, where $S^2$ is no longer the identity (as in $SU(2)$) but the charge conjugation matrix. In fact, $S^2$ produces time and parity inversion on the torus geometry and, by CPT invariance, it transforms a character into its conjugate.

\section{The $S$ matrix} 

Below we will find explicit expressions for generalized $S$ transformations of the characters introduced in the previous chapter, setting $u=0$ for short, as\footnote{Some authors use the $\tilde S$ matrix generating $\chi_{\mu}\left(-\frac{\theta}{\tau},-\frac{1}{\tau},u+\frac{\theta^2} {2\tau}\right)$. This is given by $\tilde S_\mu{}^\nu = S_\mu {}^{\nu^ +}$, where $\nu ^+$ labels the conjugate $\nu$-representation.}
\bea
\chi_{\mu}(\frac{\theta}{\tau},-\frac{1}{\tau},0)=e^{-2\pi i 
\frac k4\frac{\theta^2}{\tau}} \sum_{\nu} S_{\mu}{}^{\nu}
\chi_{\nu}(\theta,\tau,0)\, ,\label{Stransf}
\eea
and we will show that, unlike standard expressions, 
they contain a sign of $\tau$ factor. This result can already
be inferred from the $S$ modular transformation of the partition 
function. Indeed,  ignoring the $\epsilon$'s and the
contact terms, one
 finds for the contributions from discrete representations\footnote{ 
$\tilde Z^{AdS_3}_{\mathcal D}$ is  the contribution to the partition 
function 
for $\theta$ and $\tau$  far from $\theta+n\tau\in{\mathbb Z}$, 
$\forall n\in{\mathbb Z}$.}
\bea
\tilde Z^{AdS_3}_{\cal D}
(\tau '_-,
\theta '_-,u '_- ;\tau '_+,
\theta '_+,u '_+)
~=~sgn(\tau_-\,\tau_+)~~
\tilde Z^{AdS_3}_{\cal D}
(\tau_-,\theta_-,u_-;\tau_+,\theta_+,u_+)\, ,
\ \ \ \ \  \ 
\label{ZDmod}
\eea
while the contributions from the continuous series verify
\bea
Z^{AdS_3}_{\cal C}(\tau '_-,
\theta '_-,u '_- ;\tau '_+,
\theta '_+,u '_+)~=~
 sgn(\tau_-\,\tau_+)~~
Z^{AdS_3}_{\cal C}(\tau_-,\theta_-,u_-;\tau_+,\theta_+,u_+) \, , \ \ \ \ \ \
\label{ZCmod}
\eea 
where the primes denote the $S$ modular transformed parameters.
This suggests that the block $S_{d_i}{}^{d_j}$, 
$d_i$ labeling discrete representations, is given by 
$sgn(\tau)~{\cal S}_{d_i}{}^{d_j}$ with
${\cal S}_{d_i}{}^{d_j}$ being unitary. Moreover, since  
 the characters of the continuous representations
 contain purely contact terms, 
one expects that 
 they close
among themselves. 
This together with (\ref{ZCmod}) suggest that the block $S_{c_i}{}^{c_j}$, 
$c_i$ labeling continuous representations, is given by 
$sgn(\tau)~{\cal S}_{c_i}{}^{c_j}$ with
${\cal S}_{c_i}{}^{c_j}$ being unitary. 
We will explicitly show these features of the generalized modular 
transformations in the next section. 
In this sense, the characters of the $AdS_3$ model on the Lorentzian torus
are pseudovectors with respect to the standard modular $S$ transformations.

A naive treatment of the Lorentzian partition function as a Wick rotation of the Euclidean path integral, would suggest the appearance of this sign after an $S$ transformation from the measure, when one takes into account the change in the metric  (see appendix \ref{A:tlt}). However, it will be clear from the results
of section \ref{secSdiscete}, that the failure in the modular invariance of $Z_{\cal D}^{AdS_3}$ is subtler than just the sign appearing in (\ref{ZDmod}). Of course this is not a problem, the Lorentzian partition functions are not modular invariant, the $S$ transformation exchanges time and space directions and this is generically not a symmetry. The relevant question is how to perform the Wick rotation to a finite modular invariant quantity without lost on the information of the spectrum.

\subsection{The strategy}

The modular $S$ matrix is an interesting object in itself and proving its existence is a fundamental step to ensure the consistency of a given CFT, but it also plays an important role in the microscopic description of a string theory, $e.g.$ in RCFT worldsheet it was proved the $S$ matrix defines the coupling (one point functions) to maximally symmetric $D$-branes and determines the fusion rules of the theory. We will come back to these issues in the next chapter.

In order to present and to motivate the relevance of the modular transformation of the characters defined in the previous chapter let us concentrate for a moment in the simplest Lorentzian model, where the target space is the $D$-dimensional Minkowski spacetime. 

Contrary to the $AdS_3$ case, string theory in flat space can be consistently defined with a Lorentzian target space and a Euclidean worldsheet. The representations building up the worldsheet spectrum are labeled by the $D$-dimensional momentum $k^{\mu}$ and the characters on the Euclidean torus are defined as (we set $\alpha'=1$)
\bea
\chi_{\bf k} (\tau)=V\prod_{\mu=0}^{D-1}\chi_{\mu}(\tau),
\eea 
where $V$ is the volume of the space time and the normalized characters, $\chi_{\mu}$ are defined by
\bea
\chi_0(\tau)&=&\frac1{\eta(\tau)}e^{-2\pi i \tau \frac{k_0^2}{2}}\,,\cr
\chi_j(\tau)&=&\,\frac1{\eta(\tau)}\,e^{2\pi i \tau \frac{k_j^2}{2}}~,~j=1,\dots,D-1.\label{flatchar}
\eea 
The partition function and the modular transformations are ill defined, so one makes a Wick rotation to the Euclidean target space, where $k_0\rightarrow ik_0$. The rotated character, $\chi^E_{\bf k}$ has a well defined modular $S$ transformation given by
\bea
\chi_{\bf k}^E(-\frac1\tau)=\int d^{D}k'~ S_{\bf k, k'}~ \chi_{\bf k'}^E(\tau)~,~~~~~~~~
S_{\bf k, k'}=  e^{2\pi i {\bf k}\,\cdot\, {\bf k'}}.
\eea 

The partition function is modular invariant and the indices in the modular $S$ matrix are in a one to one relation with the representations of the original Lorentzian model. 

The situation is completely different in the $AdS_3$ model. 
Here the representations in the Lorentzian and Euclidean models are completely different. The characters of $H_3^+$ do not even factorize in holomorphic and antiholomorphic factors and their Wick rotation has no information on the $AdS_3$ spectrum so that there is no reason to expect that the $H_3^+$ modular $S$ matrix has some relation with the microscopic description of $AdS_3$.  
  
An interesting observation is that the $S$ matrix of the Minkowski space can be obtained without invoking the Euclidean rotation. In fact the characters on the Lorentzian torus are given by (\ref{flatchar}) where now $\tau$ is a real parameter and the Dedekind function is now interpreted as the distribution $\frac1{\eta(\tau+i0^+)}$. Even though these characters are not in a vector representation of the modular group, their transformations are perfectly well defined (in a distributional sense) and do not require Wick rotation or other regularization. They transform as pseudovectors as their transformations introduces a $sgn(\tau)$ factor. In fact by using  
\bea
\frac1{\eta(-\frac1\tau+i0^+)}=\frac1{\eta(-\frac1{\tau+i0^+})}=\frac{e^{sgn(\tau)~\frac{i\pi}4}}{\sqrt{|\tau|}~\eta(\tau+i0^+)},\label{etatransf}\\ \cr
e^{- 2\pi i\frac1\tau \frac{\lambda^2}2}=e^{-sgn(\tau)\frac{i\pi}4}\sqrt{|\tau|}\int d\lambda' 
~e^{2\pi i \lambda \lambda'}e^{2\pi i\tau\frac{\lambda'{}^2}2},\label{gaussint}
\eea
it is found that
\bea
\chi_{\bf k}(-\frac1\tau)=sgn(\tau)\int d^{D}k'~ {\cal S}_{\bf k, k'}~ \chi_{\bf k'}(\tau)~,~~~~~~~~
{\cal S}_{\bf k, k'}=  i\, e^{2\pi i {\bf k}\,\cdot\, {\bf k'}}.
\eea 

So, up to the $sign(\tau)$ factor and the $i$ phase which can be interpreted as coming from the Euclidean rotation $dk^0\rightarrow i\,dk^0$, we have obtained the modular $S$ matrix of the model without any reference to the Euclidean theory. 
 
In the rest of the chapter we will generalize this procedure to the more involved $AdS_3$ model and we will find that the characters found in the previous chapter transform as pseudovectors with respect to the modular group.

\subsection{Continuous representations }

The $S$ transformed characters of continuous representations can be written as:
\bea
\chi_j^{\alpha,w}(\frac{\theta}{\tau},-\frac{1}{\tau},0)=
\frac{e^{-2\pi i\left(\frac{s^2}{k-2}+\frac k4 w^2\right)\frac1\tau}}
{(-i\tau)^{\frac32}\eta^3(\tau+i\epsilon_1)} \sum_{m=-\infty}^{\infty} 
e^{2\pi im\left(\alpha+\frac k2 w\right)}\delta\left(\frac{\theta}{\tau}+\frac{w}{\tau}-m\right),
\eea
where (\ref{etatransf}) was used. After inserting equation (\ref{gaussint}) with the appropriate relabeling we find
\bea
\chi_j^{\alpha,w}(\frac{\theta}{\tau},-\frac{1}{\tau},0)&=&\frac{e^{-2\pi i
\frac k4\frac{\theta^2}{\tau}}}{\tau}\int_{-\infty}^{+\infty}ds'\; 
\tilde{\cal S}_s{}^{s'} \frac{e^{\frac{2\pi i}{k-2}\tau s'{}^{2}}}{{\eta^3(\tau+i\epsilon_1)} }\nonumber\\
&&\times~
\sum_{m=-\infty}^{\infty} e^{2\pi i\frac k4 \tau m^2} e^{2\pi im\alpha}~~
\delta\left(\frac{\theta}{\tau}+\frac{w}{\tau}-m\right)\, ,
\eea
with $~\displaystyle\tilde{\cal S}_s{}^{s'}=i\sqrt{\frac{2}
{k-2}}e^{-4\pi i\frac{ss'}{k-2}}$.

From $\delta\left(\frac{\theta}{\tau}+\frac{w}{\tau}-m\right)=
|\tau|~ \delta\left(\theta+w-m\tau\right)$ and renaming variables, one gets
\bea
\chi_j^{\alpha,w}(\frac{\theta}{\tau},-\frac{1}{\tau},0)&=&e^{-2\pi i
\frac k4\frac{\theta^2}{\tau}}sgn(\tau)\nonumber\\
&\times &\sum_{w'=-\infty}^{\infty} 
\int_{-\infty}^{+\infty}ds' \tilde{\cal S}_s{}^{s'} \frac{e^{2\pi i\tau
\left(\frac{ s'^{2}}{k-2}+\frac k4 w'^{2}\right)}}{\eta^3(\tau+i\epsilon_1)}  
e^{2\pi iw'\alpha}\delta\left(\theta-w'\tau+w\right) .~~~~
\label{chiconttrans}
\eea
In order to reconstruct the character  $\chi_{j'}^{\alpha',w'}$ 
in the $r.h.s.$, we use the identity
\bea
\delta\left(\theta-w'\tau+w\right)=
\sum_{m'=-\infty}^\infty\int_0^1 d\alpha' e^{2\pi i \left(w\alpha'+\frac k2 
ww'\right)}
e^{-2\pi im'\left(\alpha'+\frac k2 w'\right)}
\delta\left(\theta-w'\tau+m'\right)\, ,\label{conttrick}
\eea
and exchanging summation and integration\footnote{Here, summation and 
integration can be exchanged because, for a fixed $w'$,
 the series always reduces to a finite sum when it is considered
as a distribution acting on a test function.}, 
(\ref{chiconttrans}) can be rewritten as
\bea
\chi_j^{\alpha,w}(\frac{\theta}{\tau},-\frac{1}{\tau},0)=e^{-2\pi i
\frac k4\frac{\theta^2}{\tau}}sgn(\tau)\sum_{w'=-\infty}^{\infty} 
\int_{0}^{+\infty}ds'\; \int_0^1 d\alpha'~~
{\cal S}_{s,\alpha,w}{}^{s',\alpha',w'} \chi_{j'=-\frac12+is'}^{\alpha',w'}
(\theta,\tau,0)\, ,\nonumber
\eea
with
\bea
{\cal S}_{s,\alpha,w}{}^{s',\alpha',w'}= 2i\sqrt{\frac{2}{k-2}}~~
\cos\left(4\pi 
\frac{ss'}{k-2}\right)~~ e^{2\pi i \left(w\alpha'+w'\alpha+\frac k2
  ww'\right)}\, ,
\eea
which is symmetric and, 
as expected from (\ref{ZCmod}), unitary, $i.e.$
\bea
\sum_{w'=-\infty}^{\infty}\int_{0}^\infty ds'\int_0^1 d\alpha' {\cal S}_{s_1,
\alpha_1,w_1}{}^{s',\alpha',w'}
{\cal S}^{\dagger}_{s',\alpha',w'}{}^{s_2,\alpha_2,w_2}=\delta(s_1-s_2)
\delta(\alpha_1-\alpha_2)\delta_{w_1,w_2}\, .
\eea

\subsection{Discrete representations}\label{secSdiscete}

The structure of the characters of the discrete representations
 is more involved than that of the continuous ones. A priori, we expect
 that characters of
both discrete and continuous representations
appear in the generalized modular transformations. So, generically we can
 assume
\bea
\chi_j^{+,w}(\frac\theta\tau,-\frac1\tau,0)&=&
e^{-2\pi i
\frac k4\frac{\theta^2}{\tau}}sgn(\tau) \sum_{w'=-\infty}^{\infty}\left\{ 
\int_{-\frac{k-1}{2}}^{-\frac12}{\cal S}_{j,w}{}^{j',w'}~\chi_{j'}^{+,w'}(\theta,\tau,0)\right.\cr\cr
&& ~~~~~~\qquad ~~~~~~~~
+~
\left.\int_0^1 d\alpha'\int_0^\infty ds'~{\cal S}_{j,w}{}^{s',\alpha',w'}~\chi_{j'=-\frac12+is'}^{\alpha',w'}(\theta,\tau,0) \right\}\, .\nonumber
\eea
Fortunately, it is easy to separate the contributions from
 discrete and continuous representations. If one considers generic
values of $\theta$ and $\tau$ far from $\theta+n\tau\in{\mathbb Z}$ 
for $n\in{\mathbb Z}$, the contributions of the continuous series
 in the $r.h.s.$ can be neglected as well as all contact terms and 
$\epsilon$'s. On the other hand, if 
$\theta+n\tau\notin{\mathbb Z},\forall n\in {\mathbb Z}$ then
 $\frac\theta\tau-p\frac1\tau\notin{\mathbb Z},\forall p\in {\mathbb Z}$ 
and  all contact terms and $\epsilon$'s of the $l.h.s.$ can be neglected 
too. Thus, we obtain
\bea
\chi_j^{+,w}(\frac\theta\tau,-\frac1\tau,0)&=&
\frac{(-)^we^{\frac{2\pi i}
{k-2}\frac1\tau \left(j+\frac12-w\frac{k-2}{2}\right)^2}e^{-2\pi i\frac\theta
\tau \left(j+\frac12 -w\frac{k-2}{2}\right)}} {i\vartheta_{11}(\frac\theta
\tau,-\frac1\tau)}\nonumber\\
&=&(-)^{w+1}
\frac{
e^{\frac{2\pi i}{k-2}\frac1\tau \left(j+\frac12-(w+\theta)
\frac{k-2}{2}\right)^2}
e^{-2\pi i\frac k4\frac{\theta^2}{\tau}}} 
{i\sqrt{i\tau}\vartheta_{11}(\theta,\tau)}\, ,\label{tmd}
\eea
where the following identity was  used for 
$\tau\in \mathbb R$:
 \bea
\vartheta_{11}(\frac\theta\tau,-\frac1\tau)=
sgn(\tau)~ e^{\pi i \frac{\theta^2}{\tau}}e^{sgn(\tau)~ i\frac\pi4 }\sqrt{|\tau|}\;
\vartheta_{11}(\theta,\tau)\,.\label{thetataureal}
\eea

Inserting
\bea
e^{\frac{2\pi i}{k-2}\frac1\tau \left(j+\frac12-(w+\theta)\frac{k-2}{2}
\right)^2}= e^{sgn(\tau)~ i\frac\pi4}\sqrt{\frac{2|\tau|}{k-2}}
\int_{-\infty}^{+\infty}
d\lambda'e^{\frac{4\pi i}{k-2} 
\lambda'\left(j+\frac12-(w+\theta)\frac{k-2}{2}\right)} 
e^{-\frac{2\pi i}{k-2}\tau\lambda^{'2}} \label{gaussian}
\eea
into (\ref{tmd}), changing the integration variable 
to $j'+\frac 12-w'\frac {k-2}2$ 
and using (\ref{thetataureal}), we get
\bea
\chi_j^{+,w}(\frac\theta\tau,-\frac1\tau,0)=e^{-2\pi i
\frac k4\frac{\theta^2}{\tau}} sgn(\tau)
\sum_{w'=-\infty}^{\infty}\int^{-\frac{1}{2}}_{-\frac{k-1}{2}}dj' 
{\cal S}_{j,w}{}^{j',w'}\chi_{j'}^{+,w'}(\theta,\tau,0), \label{chidis}
\eea
with
\bea
{\cal S}_{j,w}{}^{j',w'}=(-)^{w+w'+1}~\sqrt{\frac{2}{k-2}}~e^{\frac{4\pi i}
{k-2} 
\left(j'+\frac12-w'\frac{k-2}{2}\right)\left(j+\frac12-w\frac{k-2}{2}\right)}
\, .
\label{SD}
\eea
Notice that this block of the  ${\cal S}$ matrix is symmetric and, again as expected from (\ref{ZDmod}), unitary\footnote{Changing $e^{\pm i\frac\pi4}\sqrt{\tau}$ by $\sqrt{i\tau}$, the validity of (\ref{gaussian}) can be extended to the full lower half plane and that of (\ref{thetataureal}) can be extended to the upper half plane, giving 
\bea
\vartheta_{11}(\frac\theta\tau,-\frac1\tau)=
- e^{\pi i \frac{\theta^2}{\tau}}\sqrt{i\tau}\;
\vartheta_{11}(\theta,\tau)\, .\label{stheta}
\eea
Nevertheless, one cannot cancel the $\sqrt{i\tau}$ terms and ignore the sign factor due to the different branches.}.

While the identity (\ref{gaussian}), which 
is essential to reconstruct the discrete characters in the
$r.h.s.$ of (\ref{chidis}), only makes sense 
for Im $\tau\leq0$, the characters are only
well defined  for Im $\tau\geq0$. Therefore, to determine 
 the generalized $S$ transformation, it is  crucial 
that $\tau\in {\mathbb R}$. 

\medskip

Finding the block 
${\mathcal S}_{j,w}{}^{s',\alpha',w'}$ 
mixing discrete with continuous representations 
is a much more technical issue, which we discuss in appendix \ref{A:mb}. 
Here we simply display the result, namely
\bea
{\cal S}_{j,w}{}^{s',\alpha',w'}=-i\sqrt{\frac{2}{k-2}}e^{-2\pi i
\left(w'j-w\alpha'-ww'\frac k2\right)} \left[\frac{e^{\frac{4\pi}{k-2}s'
\left(j+\frac12\right)}}{1+e^{-2\pi i(\alpha'-is')}} + 
\frac{e^{-\frac{4\pi}{k-2}s'\left(j+\frac12\right)}}
{1+e^{-2\pi i(\alpha'+is')}}\right]. \label{degcont}~~~~
\eea

This block prevents the full ${\mathcal S}$ matrix 
from being unitary. Instead, we find ${\mathcal S}^*{\mathcal S}=id$. 
This implies that the full partition function defined from the product
of characters is not modular invariant, not only due to the 
sign of the modular 
parameters. Actually,
after a modular transformation, the mixing block introduces terms where the 
left modes are in discrete representations
and the right ones in continuous series,
and vice versa, as well as new terms containing left and right continuous 
representations.

In section 
\ref{STproperties}, we explicitly check that the blocks of the 
$S$ matrix determined here have the correct properties.

\subsection{Degenerate representations} 
\label{Sdeg}

The modular properties discussed 
above can be used to write the 
$S$ transformation of the characters of the
degenerate representations 
with $1+2J\in{\mathbb N}$
as:
\bea
\chi_J(\frac\theta\tau,-\frac1\tau,0)&=&e^{-2\pi i\frac k4 \frac{\theta^2}
{\tau}}
sgn(\tau)\sum_{w=-\infty}^\infty \left\{\int^{-\frac12}_{-\frac{k-1}{2}} dj~
{\cal S}_J{}^{j,w}\chi_j^{+,w}(\theta,\tau,0)\right.\cr 
&&~~~~~~~~~~~~~~~~~~~~~~~~~~
+~\left.\int_0^1 d\alpha\int_{-\frac{k-1}{2}}^{-\frac12} ds ~
{\cal S}_{J}{}^{s,\alpha,w}\chi_{j=-\frac12+is}^{\alpha,w}(\theta,\tau,0)
\right\},\nonumber
\eea
where
\bea
{\cal S}_J{}^{j,w}=2i\sqrt{\frac{2}{k-2}}(-)^{w+1} \sin
\left[\frac{\pi}{k-2}\left(1+2j-w
(k-2) \right)\left(2J+1\right)\right]\,,
\eea
and
\bea
{\cal S}_J{}^{s,\alpha,w}&=&-i(-)^{2Jw}\sqrt{\frac{2}{k-2}}e^{\frac{4\pi}
{k-2}s(J+\frac 12)}\left(1+\frac{1}{1+e^{-2\pi i (\alpha-is)}}+\frac{1}{1+
e^{2\pi i (\alpha+is)}}\right) \nonumber\\
&&+~~(s\leftrightarrow -s) \, .
\label{degs}
\eea

\section{The $T$ matrix }
Together with the $S$ matrix, the  $T$ matrix defines a basis over the space 
of modular transformations.
Using
\bea
\vartheta_{11}(\theta,\tau+1)=e^{\frac{\pi i}{4}}\vartheta_{11}(\theta,\tau),
\qquad
\eta(\tau+1)=e^{\frac{\pi i}{12}}\eta(\tau)\, ,
\eea
the characters of the 
discrete and continuous representations transform respectively with 
\bea
T_{j,w}{}^{j',w'}=\delta_{w,w'}\delta(j-j')e^{-\frac{2\pi i}{k-2}
\left(j'+\frac12-w'\frac{k-2}{2}\right)^2-\frac{\pi i}{4}}
\eea
and 
\bea
T_{s,\alpha,w}{}^{s',\alpha',w'}=\delta_{w,w'}\delta(\alpha-\alpha')
\delta(s-s')e^{2\pi i\left(\frac{ s^2}{k-2}-\frac{k}{4}w^2-w\alpha-
\frac 18\right)}\, ,
\eea
while the $T$
 transformation of the characters of the degenerate representations is given by
\bea
\chi_J(\theta,\tau+1,0)=e^{-\frac{2\pi i}{k-2}\left(J+\frac12\right)^2}
e^{-\frac{\pi i}{4}}\chi_J(\theta,\tau,0)\, .
\eea

\section{Properties of the $S$ and $T$ matrices}\label{STproperties}
 
The expressions $(ST)^3$ and $S^2$ must give the conjugation matrix, $C$. We have found above that the characters of the $AdS_3$  model do not expand a representation space for the modular group since the generators depend on the sign of $\tau$. Nevertheless, in terms of the $\tau$ independent part of $S$, that we have denoted ${\mathcal S}$, these identities read $C=(ST)^3=sgn(\tau+1)$ $sgn(\frac{\tau}{\tau+1})sgn(-\frac1\tau)({\mathcal S}T)^3=-({\mathcal S}T)^3$ and $C=S^2=sgn(\tau)sgn(-\frac1\tau){\mathcal S}^2=-{\mathcal S}^2$.

As a consistency check on the  expressions found above for $S$ and $T$, an explicit computation gives 
\bea
-({\cal S}T)^3{}_{j_1,w_1}{}^{j_2,w_2}=-{\cal S}^2{}_{j_1,w_1}{}^{j_2,w_2}= 
\delta_{w_1+w_2+1,\,0}~\delta\left(j_1+j_2+\frac k2\right)\, ,
\eea
which corresponds to the conjugation matrix restricted to the discrete 
sector, since
$\hat {\mathcal D}_j^{+,w}$ is the conjugate representation of
$\hat {\mathcal D}_j^{-,-w}$, which in turn can be identified with
$\hat {\mathcal D}_{-\frac k2-j}^{+,-w-1}$ using the spectral flow symmetry.
Similarly,
for the block of continuous representations we get
\bea
-({\cal S}T)^3_{s_1,\alpha_1,w_1}{}^{s_2,\alpha_2,w_2}=
-{\cal S}^2{}_{s_1,\alpha_1,w_1}{}^{s_2,\alpha_2,w_2}= 
\delta_{w_1,-w_2}\delta(s_1-s_2)\delta(\alpha_1+\alpha_2-1)\, ,\label{STcontprop}
\eea
which is
again the charge conjugation matrix, since $\hat {\mathcal C}_j^{1-\alpha,-w}$
is the conjugate representation of
$\hat{\mathcal C}_j^{\alpha,w}$. 

Of course, one also needs to show that the non diagonal terms vanish. 
The equalities
$({\mathcal S}T)^3{}_{s_1,\alpha_1,w_1}{}^{j_2,w_2}=
{\mathcal S}^2{}_{s_1,\alpha_1,w_1}{}^{j_2,w_2}=0$ are trivially 
satisfied as a consequence of 
${\mathcal S}_{s_1,\alpha_1,w_1}{}^{j_2,w_2}=0$. One can also show that
$({\mathcal S}T)^3{}_{j_1,w_1}{}^{s_2,\alpha_2,w_2}=
{\mathcal S}^2{}_{j_1,w_1}{}^{s_2,\alpha_2,w_2}$=0, but this computation
is more involved, so
the details are left to appendix \ref{A:mb}.

\chapter{D-branes in $AdS_3$}\label{chapDbranes}

D-branes can be characterized by the one-point functions of the states in the bulk, living on the upper half plane. In RCFT, these one-point functions can be determined from the entries of the $S$ matrix, a property that we will call a {\it Cardy structure}. This property is closely related to the Verlinde formula and, {\it a priori}, there is no reason for it to hold in non RCFT. In this chapter we explore the scope of this connexion for the $AdS_3$ model.

D-branes in $AdS_3$ and related models have been studied in several works (see for instance \cite{GKS}-\cite{AS} and references therein). Here, we shall restrict to the maximally symmetric D-branes discussed in \cite{BP}. Because the Lorentzian $AdS_3$ geometry is obtained by sewing an infinite number of  SL(2,${\mathbb R})$ group manifolds, these D-brane solutions can be trivially obtained from those of SL(2,${\mathbb R})$. Their geometry was considered semiclassically in \cite{BP}, where it was found that solutions of the Dirac Born Infeld  action stand for regular and twined conjugacy classes of SL(2,${\mathbb R})$. The model also has symmetry breaking D-brane solutions, but in this case, the open string spectrum is not a sum of $sl(2)$ representations and then the one-point functions cannot be determined by the $S$ matrix.

We begin this chapter with a short introduction to Boundary Conformal Field Theories (BCFT) and the geometry of D-branes in $AdS_3$. A very comprehensive study about the (twined) conjugacy classes of SL(2,${\mathbb R})$ and a semiclassical analysis of branes can be found in \cite{BP} and \cite{Stanciu}. Both can be easily extended to the universal covering. 
Here, we review the analysis of the conjugacy classes in order to make the discussion self contained and discuss the extension to the universal covering with the aim of obtaining the key relation (\ref{pardS2}) obtained in \cite{wc2}. 

Third section is devoted to give a brief review of the one point functions computed in \cite{israel} and to translate them to our conventions. Then we turn to the explicit construction of the Ishibashi states for regular and twisted boundary gluing conditions which give rise to the maximally symmetric D-branes. These conditions were solved in the past for the single cover of SL(2,${\mathbb R})$ (see \cite{RR} for twisted gluing conditions) with different amounts of spectral flow in the left and right sectors, namely $w_L=-w_R$, and therefore, these solutions are not contained in the spectrum of the $AdS_3$ model (with the obvious exception of $w=0$ discrete and $w=0$, $\alpha=0,\frac12$  continuous representations).

We will show that the one-point functions of states in discrete representations coupled to point-like and $H_2$ branes exhibit a {\it Cardy structure} and we present a generalized Verlinde formula giving the fusion rules of the degenerate representations with $1+2J\in {\mathbb N}$. 

\section{Boundary CFT}\label{sectBCFT}

In this section we present a brief review on CFTs with boundaries in order to develop the machinery to deal with $D$-branes on Conformal theories.

\subsection{Closed String Sector}

Bulk fields constitute the field content of the worldsheet theory describing closed strings. These can be defined in CFT with and without boundaries. In both cases there is a correspondence between states of the Hilbert space and fields of the CFT. Even though the fields differ in each case the Hilbert space (${\cal H}_C=\oplus_{m\overline m}{\cal V}_m\otimes\overline{\cal V}_{\overline m}$ \footnote{We assume the spectrum factorizes as a sum of representations, ${\cal V}_m$ and $\overline{\cal V}_{\overline m}$ of certain symmetry algebra generated by chiral currents $J(z)$ and $\bar J(\bar z)$ respectively.}) is the same in both, the theory over the full plane (P) and the one over the upper half plane (H). So that for a given state $|m\,\overline m>\in {\cal H}_C$ the state operator correspondence is
\bea
\varphi_{m,\overline m}(z,\bar z)&=&\Phi^{(P)}(|m\,\overline m>;z,\bar z),\cr
\phi_{m,\overline m}(z,\bar z)&=&\Phi^{(H)}\left(|m\,\overline m>;z,\bar z\right),
\eea
where $\varphi(z,\bar z)$ are well defined over the full complex plane and $\phi(z,\bar z)$ is only required to be well defined for Im $z>0$. The condition of introducing a $D$-brane in the background defined by the Bulk theory is to require that Bulk fields of the BCFT and those of the CFT without boundaries (Bulk CFT) are equivalent, where equivalence means that both spectra coincide but also that the OPE is the same in both theories.

The conformal symmetry requires 
\bea
T(z)=\overline T(\bar z),~~~~~~~~z=\bar z.\label{Tgluing}
\eea

It is important to stress that a given Bulk CFT may be connected with several BCFTs and that there is no systematic approach to dealing with all possibilities.

In this short review we will restrict to the case of maximally symmetric $D$-branes, that is we will consider boundary conditions preserving the maximal amount of symmetry of the parent Bulk CFT. Assuming that the chiral fields $J(z),\bar J(\bar z)$ can be analytically continued to the real line we will look for an automorphism $\Omega$, that we will denote as {\it gluing map}, such that 
\bea
J(z)=\Omega\bar J(\bar z),~~~~~~~~z=\bar z\label{gluingz}
\eea 

Notice that this map, $\Omega$, induces a map over the different sectors, $j\rightarrow w(j)$.\footnote{For instance in a theory with $U(1)$ symmetry and Dirichlet gluing map $\Omega(J)=-J,~~\Rightarrow\Omega(\alpha_n)=-\alpha_n$. Then, taking into account that $\alpha_0|k>=\sqrt{\alpha'}k|k>$
\bea
\Omega(\alpha_0)|k>=-\alpha_0|k>=-\sqrt{\alpha'}k|k>=\alpha_0|-k>,\nonumber
\eea
which induces the map $w(k)=-k$.}    
\bigskip

\subsubsection{One point functions}

Ward identities of the chiral currents and the OPE of the bulk fields can be exploited to reduce the computation of correlation functions with arbitrary number of points of bulk fields on the disk to the calculus of one point functions.

The transformation properties of the bulk fields in the BCFT, $\phi_{m\bar m}$ (the indices label left and right representations) with the modes $L_n,~n=\pm1,0$ and the zero modes of the currents, $J_0$
\bea
&&[J_0,\phi_{m\overline m}(z,\bar z)]=X^m_{J}\phi_{m\overline m}(z,\bar z)-\phi_{m \overline m}(z,\bar z)X^{\overline m}_{\Omega\bar{ J}}~,\cr
&&[L_n,\phi_{m\overline m}(z,\bar z)]=z^n[z\partial+\Delta_m(n+1)]~\phi_{m\overline m}(z,\bar z)+\bar z^n [\bar z\bar\partial + \bar\Delta_{\overline m}(n+1)]
~\phi_{m \overline m}(z,\bar z)~,~~~~~~~~
\eea 
determine the structure of the one point functions on the disk to be
\bea
\left\langle \phi_{i\bar i}(z,\bar z)\right\rangle_\alpha=\frac{{\mathcal A}^\alpha_{i\bar i}}{|z-\bar z|^{\Delta_i+\bar \Delta_{\bar i}}},
\eea
where ${\mathcal A}^\alpha_{m \overline m}:{\cal V}_{\overline m}^0\rightarrow {\cal V}_{m}^0$, obey $X^m_{ J}{\mathcal A}^\alpha_{m \overline m}={\mathcal A}^\alpha_{m \overline m}X^{\overline m}_{\Omega\bar{J}}$, which implies $\overline m=w(m^+)\equiv m^w$ is a necessary condition for the one point function to be non vanishing. $m^+$ is the conjugate representation of $m$ and $X^{m}_{J}$ is the action of the zero mode $J_0$ of the affine algebra over the primaries $i.e.$ 
\bea
X^{m}_{{J}}:=\left.{J}_0\right|_{{\cal V}_m^0}: {\cal V}_m^0\rightarrow {\cal V}_m^0,
\eea

The zero modes of the currents are irreducible in the subspaces $ {\cal V}_m^0$, thus Schur's lemma implies
\bea
{\mathcal A}^\alpha_{m\overline m}=A^\alpha_m~\delta_{\overline m,m^w}~{\cal U}_{m\overline m},
\eea
where ${\mathcal U}_{m \overline m}$ intertwines between the zero modes of the algebra $X^m_{J}$ and $X^{\overline m}_{\Omega\bar J}$ and is normalized such that ${\mathcal U}_{m\overline m}^*{\mathcal U}_{m \overline m}=1$. $\alpha$ indexes different Boundary Theories appearing for a given {\it gluing map} $\Omega$. 

We have found that one point functions for different boundary conditions associated to a given Bulk Theory and the same {\it gluing map} $\Omega$ may differ just in a set of scalars $A_m^\alpha$. Once this set is known we will have completely solved the Boundary Theory (including the open string sector that we have not considered yet).

\subsubsection{Boundary states}

It is possible to store all the information of the couplings $A_m^\alpha$ in a unique object, the so called {\it boundary state}. 

The boundary state is a linear combination of coherent states known as Ishibashi states \cite{Ishibashi} and the coefficients of this expansion are essentially the couplings $A_m^\alpha$. 

A way to introduce {\it boundary states} is by equaling Bulk fields in the upper half plane of the Boundary theory and in the exterior of the unit disk in the Bulk theory.

Let $z,\bar z$ be coordinates in the upper half and $\zeta,\bar \zeta$ those of the exterior unit disk. They are related through
\bea
\zeta=\frac{1-iz}{1+iz}~~~~;~~~~~~~~\bar\zeta=\frac{1+i\bar z}{1-i\bar z}. 
\eea 
Thus if $|0>$ is the vacuum of the Bulk CFT, the {\it boundary state} $|\alpha>$ is defined via the identity
\bea
\left\langle \Phi^{(H)}(|\varphi>;z,\bar z)\right\rangle_\alpha=\left(\frac{d\zeta}{dz}\right)^{h} \left(\frac{d\bar\zeta}{d\bar z}\right)^{\bar h}<0|\Phi^{(P)}(|\varphi>;\zeta,\bar \zeta)|\alpha>\label{defbs}
\eea   
The boundary condition (\ref{gluingz}) of the BCFT at $z=\bar z$ translates in the Bulk CFT as the condition
\bea
\left[J(\zeta)-\bar\zeta^{2\Delta_J}(-)^{\Delta_J}\Omega\bar J(\bar \zeta)\right]|\alpha>=0,~~~~~~~~\zeta\bar\zeta=1, \label{gluingzeta}
\eea 
where $\Delta_J$ is the conformal weight of $J$. The constraint above can be rewritten in terms of the Laurent modes as
\bea
\left[J_n-(-)^{\Delta_J}\Omega\bar J_{-n}\right]|\alpha>_{\Omega}=0,\label{constraintWn}
\eea
where the label $\Omega$ was written to make the {\it gluing map} considered explicit.

These constraints are linear and leave each representation invariant. Given an automorphism $\Omega$ there exists a unique solution in each representation of the form $(m,w(m^+))$ \cite{Ishibashi} and this is given by a coherent state (or Ishibashi state) $|m\gg_\Omega$ univocally defined up to a normalization which can be fixed such that 
\bea
{}_\Omega\ll m|~\tilde q^{L_0^{(P)}-\frac{c}{24}}~|n\gg_{\Omega}=\delta_{mn}~\chi(\tilde q).\label{condIsh}
\eea

From (\ref{bs}), (\ref{defbs}) and (\ref{condIsh}) one can prove that the {\it boundary states}, $|\alpha>_\Omega$ are
\bea
|\alpha>_\Omega=\sum_m A_{m^+}^\alpha~|m\gg_\Omega\label{bs}
\eea

\subsubsection{Open String Sector}

World sheet duality changes open string channels and closed string channels when time and space coordinates are exchanged. Thus the boundary partition function, $Z_{\alpha\beta}(q),~q=e^{2\pi i\tau}$, associated with the Boundary spectrum with $\alpha$ and $\beta$ boundary conditions\footnote{This is interpreted in string theory as the spectrum of open strings with one end in the brane $\alpha$ and the other one in the brane $\beta$.} ($\equiv$ One-loop vacuum open string amplitude) must agree with the tree level amplitude of a boundary state emitted in the brane $\alpha$ and reabsorbed in the brane $\beta$, $i.e.$ it is interpreted as the probability that the boundary state be emitted and reabsorbed in a time interval $\tilde \tau=-1/\tau~$ \footnote{Remember that the modular $S$ matrix exchanges cycles $a$ and $b$ in the torus and so exchanges time and space coordinates}
\bea
Z_{\alpha\beta}(q)=<\Theta\beta|\tilde q^{H^{(p)}}|\alpha>,
\eea
where $\Theta$ denotes the worldsheet CPT operator, defined such that
\bea
\Theta A_{m^+}^\beta|m\gg=(A_{m^+}^\beta)^*|m^+\gg.
\eea

Using that the Hamiltonian in the Bulk CFT is $H^{(P)}=\frac12(L_0+\bar L_0)-\frac{c}{24}$ and the gluing condition of the energy momentum tensor (\ref{Tgluing}) imposes in the Bulk CFT the equation $[L_n-\bar L_{-n}]|\alpha>=0$ it is found that
\bea
&&<\Theta\beta|\tilde q^{H^{(p)}}|\alpha>=\sum_{m,n}\left(\Theta A_{m^+}^\beta\right)^*\ll m| \tilde q^{L_0-\frac{c}{24}}|n\gg A_{n^+}^\alpha\cr
&=&\sum_{m,n} A_{m^+}^\beta A_{n^+}^\alpha \ll m^+| \tilde q^{L_0-\frac{c}{24}}|n\gg=\sum_m  A_{m^+}^\beta A_{m^+}^\alpha \chi_{m^+}(\tilde q).
\eea

Then 
\bea
Z_{\alpha\beta}(q)=\sum_{m,n} A_{m^+}^\beta A_{m^+}^\alpha S_{m^+}{}^{n}\chi_{n}( q),
\eea
with $S_{m^+ n}$ being the modular $S$ matrix. 

Boundary conditions discussed here preserve part of the chiral symmetry, thus the boundary spectrum must be decomposed as a sum over certain representations ${\cal V}_m$. Let $N_{\alpha\beta}{}^m$ be its spectrum degeneracy, thus the {\it Cardy condition} \cite{Cardy} 
\bea
N_{\alpha\beta}{}^n=\sum_m A_{m^+}^\beta~ A_m^{\alpha}~S_{m^+}{}^{n}\in {\mathbb Z}_{\geq0},\label{Cardycond}
\eea
must be satisfied.

\subsubsection{ Cardy Solution}

Let us consider a Bulk RCFT with modular invariant partition function
\bea
Z(q,\bar q)=\sum_m\chi_m(q)\chi_{\overline m}(\bar q),
\eea
with $m$ taking values in some given set ${\mathcal M}$ and $\overline m$ being paired with some given $m$, assuming that\footnote{Remember that the condition $m^w=\overline m$ was required in order to get non vanishing one point functions.}
\bea
m^w\equiv m(j)^+=\overline m.
\eea

Cardy claimed \cite{Cardy} that the number of boundary theories coincide with the number of representations in the boundary spectrum. In other words that there is a one to one map between the labels $\alpha$ and the indices $m\in{\cal M}$.

Then Cardy proposed that the one point function coefficients $A_m^n$ (before $A_m^\alpha$) are determined by the entries of the modular $S$ matrix
\bea
A_m^n=\frac{S_n{}^m}{\sqrt{S_0{}^m}},\label{cardycoeff}
\eea
with 0 indexing the representation containing the identity field. The disk one point functions of primary fields are
\bea
\left\langle \phi_{m,m^w}(z\bar z)\right\rangle_n=\frac{S_{n}{}^{m}}{\sqrt{S_{0}{}^{m}}}\frac{{\cal U}_{m m^w}}{|z-\bar z|^{2\Delta_m}},\label{cardysol}
\eea   
with ${\cal U}_{jj^w}$ the unitary intertwiner operator previously defined. 

The {\it Cardy ansatz} follows from the observation that such a solution naturally solves the {\it Cardy condition}. Indeed, after inserting (\ref{cardycoeff}) in (\ref{Cardycond}) one obtains
\bea
\sum_q A_{q^+}^n~A_q^m~S_{q^+}{}^{k}=\sum_q \frac{S_{n}{}^{q^+}~S_{m}{}^{q}~S_{q^+}{}^{k}}{S_0}{}^{q}=N_{mn^+}{}^k\in {\mathbb Z}_{\ge0}
\eea 
where the last equality is nothing but the famous {\it Verline formula} determining the fusion rules ($N_{mn^+}^k$) in RCFTs. It guaranties the ansatz satisfies the {\it Cardy condition} but does not prove that it is a solution. It may happen that fusion rules do not reproduce the spectral decomposition of the Boundary CFT or it may occur that Verlinde formula does not works, as in most Non RCFTs. Nevertheless it has shown to be a powerful framework to solve Rational BCFTs.

\section{Conjugacy classes in $AdS_3$}\label{conjclass}

As is well known \cite{AS}, the world-volume of a symmetric D-brane on the $G$ group manifold is given by the (twined) conjugacy classes 
\bea
{\cal W}_g^\omega=\left\{\omega (h)gh^{-1},\forall h\in G\right\}\, ,
\eea
where $\omega$ determines the gluing condition connecting left and right moving currents, $\omega(g)=\omega^{-1}g\,\omega$. When $\omega$ is an inner automorphism, ${\mathcal W}_g^\omega$ can be seen as left group translations of the regular conjugacy class (of the element $\omega g$). So, one can restrict 
attention to the case $\omega=id.$, and in the case of $SL(2,{\mathbb R})$ the conjugacy classes are simply given by the solution to (see (\ref{sl(2,r)}))
\bea
tr~ g=2\frac{X_0}{\ell}=2\tilde C\, .
\eea  

The 
geometry of the world-volume is then parametrized by the constant $\tilde C$ as
\bea
-X_1{}^2-X_2{}^2+X_3{}^2=\ell^2\left(1-\tilde C^2\right)\, .
\eea
Different geometries can be distinguished for $\tilde C^2$  bigger,
 equal or smaller than one. 
The former gives rise to a two dimensional de Sitter space, $dS_2$,
the latter to a two dimensional hyperbolic 
space, $H_2$, and the case $|\tilde C|=1$ splits into
 three different geometries: the apex,
the future and the past of a light-cone.

 A more convenient way to parametrize these solutions is given 
by the redefinition   
\bea
\tilde C= \cos\sigma\, .
\eea   

For $|\tilde C|>1$, $\sigma=ir+\pi v,~r\in{\mathbb R^+},~v\in{\mathbb Z_2}$.
The world-volumes are given by 
\bea
\cosh\rho \cos t=\pm\cosh r\, .
\eea
Each circular D-string is emitted and absorbed at the boundary in a time 
interval of width $\pi$ but does not reach the origin 
unless $r=0$. Their lifetime is determined by $v$.
 
For $|\tilde C|<1$, $\sigma$ is real and   
\bea
\cosh\rho \cos t=\cos \sigma\, .
\eea
If one restricts $\sigma\in(0,\pi)$, there are two different solutions for each $\sigma$, for instance one with $t\in(-\frac\pi2,-\sigma]$ and another one
with $t\in[\sigma,\frac{\pi}{2})$. To distinguish between these two solutions we 
can take
$\sigma=\lambda+\pi v,~\lambda\in\left(-\pi,0\right),~v\in{\mathbb Z_2}$, 
 such that $t={\rm arcos}(\cos\sigma/\cosh\rho)$, taking the branch 
where $t=\sigma$ when it crosses over the origin.
Because these solutions have Euclidean signature, they are identified 
as instantons in $AdS_3$. In fact, they represent constant time slices in hyperbolic coordinates.

For $|\tilde C|=1$, $\sigma=0$ or $\pi$ and
\bea
\cosh\rho\cos t=\pm1\, .
\eea
For example, for 
$\tilde C=1$, this corresponds to a circular D-string at the boundary at
$t=-\pi/2$ collapsing 
to the instantonic solution in $\rho=0$ at $t=0$, and
 then expanding again to a D-string reaching the boundary at $t=\pi/2$.

All of these solutions are restricted to the single covering of 
SL(2,${\mathbb R}$). In the universal covering, $t$ is decompactified 
and  the picture is periodically repeated. 
The general solutions can be parametrized by a pair $(\sigma,q)$, 
$q\in{\mathbb Z}$, or equivalently, the range of $\sigma$ can be extended to
 $\sigma=ir+q\pi$ for $dS_2$ branes, 
$\sigma=\lambda+q\pi$ for $H_2$ branes or $\sigma=q\pi$
 for point-like and light-cone branes.

Preparing for
the discussions on  one-point functions and  $Cardy~structure$, it 
is interesting to note that these parameters can be naturally identified 
with representations 
of the model. For instance, one can label the 
D-brane solutions 
as
\bea
\sigma=\frac{2\pi}{k-2}\left(j+\frac12-w\frac{k-2}{2}\right)
\, ,\label{pardS2} 
\eea
with $j=-\frac12+is,~s\in{\mathbb R^+},~w\in{\mathbb Z}$ for $dS_2$ branes, $j\in\left(-\frac{k-1}{2},-\frac12\right),~w\in{\mathbb Z}$ for $H_2$ branes and finally $\sigma=n\pi,~n\in{\mathbb Z}$ for the point-like and light-cone D-brane solutions.

The appearance of
the level $k$ in a classical regime could seem awkward.
However, it is useful to  recall
that $\sigma$ is just a parameter labeling the conjugacy classes, and
the factor $k-2$ can be eliminated  by simply redefining 
 $j$ through a change of variables. 
The important observation is that this suggests 
 $\sigma$ labels the exact solutions, $e.g$ the one-point 
functions at finite $k$ will be found to be parametrized
exactly by (\ref{pardS2}) and in fact, in the semiclassical 
regime $k\rightarrow\infty$, the domain of $\sigma$ does not change at all.
\bigskip

When $\omega$ is an outer automorphism, 
one can take 
$\omega=\left(\begin{matrix}0&1\cr1&0\end{matrix}\right)$
up to group translations.  
In this case, the twined conjugacy classes are given by
\bea
tr~ \omega g=2\frac{X_2}{\ell}=2C\, .
\eea
The world-volume geometry now describes an $AdS_2$ space for all $C$ since
\bea
X_0^2-X_1^2+X_3^2=\ell^2(1+C^2)\, .
\eea
These are static open D-strings with endpoints fixed at the boundary. This is obvious in cylindrical coordinates, $i.e.$
\begin{equation}
\sinh\rho\sin\theta=\sinh r\, ,
\end{equation}
where we have renamed $C=\sinh r$. So, 
after decompactifying the time-like direction $t$, there is no 
need to extend the domain of 
$r$. 

Let us end this brief review with a word of caution. 
In this section we have reviewed the 
twined conjugacy classes and, although branes wrap conjugacy classes, 
extra restrictions appear when studying the semiclassical or 
exact solutions. In particular,
it was found  in \cite{BP} that 
 $r$ becomes a positive quantized parameter at the semiclassical
 level.

\section{One-point functions}\label{sect1pf}

In this section we summarize
 the results for one-point functions in
 maximally symmetric D-branes, obtained by applying the method of 
\cite{israel}. The solution for 
 one-point functions in $H_2$ D-branes found in {\it loc. cit.} 
holds for integer level $k$. 
Here we work with an alternative expression, equivalent to the one obtained 
in \cite{israel}, but with a different extension for generic $k\in{\mathbb R}$.

\subsection{One-point functions for point-like instanton branes}\label{plb}

To obtain the one-point functions for the point-like branes, 
we simply take the ${\mathbb Z}_{kN}$ orbifold 
action on the product of the one-point functions  associated 
to D0  
branes in the cigar
\cite{RS} 
and to Neumann boundary conditions in the U(1) theories, 
respectively
\bea
&&\left\langle \Phi_{j,n,\omega}^{sl(2)/u(1)}(z,\bar z)\right\rangle_{\bf s}
^{D0} =
\frac{\delta_{n,0}~(-)^{r\omega}}
{\left|z-\bar z\right|^{h^j_{nr}+\bar h^j_{nr}}} 
\frac{\Gamma\left(-j+\frac k2 \omega\right)
\Gamma\left(-j-\frac k2 \omega\right)}{\Gamma\left(-2j-1\right)}\cr\cr
&&  \ \ \ \ \ \ \ \ \ \ \ 
\times~\left(\frac{k}{k-2}\right)^{\frac14}\left(\frac{ \sin[\pi b^2]}
{4\pi}\right)^{\frac12}\frac{\sin[{\bf s}(2j+1)]}{\sin[\pi b^2(2j+1)]}
\frac{\Gamma\left(1+b^2\right)~\nu^{1+j}}{\Gamma\left(1-b^2(2j+1)\right)}
\, ,
\label{D0}
\eea 
and
$\ \ \ \ \ \ \ \ \ \ \ \ \ \ \ \ \ \ 
\displaystyle
\left\langle \Phi^{u(1)}_{\tilde n,\tilde \omega}(z,\bar z)\right
\rangle_{x_0}^{\cal N} ~= ~
\frac{\delta_{\tilde n,0}e^{i\tilde \omega x_0}
\left(\sqrt{k/2}R\right)^{\frac12}}{\left|z-\bar z\right|^{\frac k2 
\tilde\omega^2}}\, .
$

\medskip

Here ${\bf s}
=\pi rb^2,~r\in{\mathbb N}$, $b^2=\frac{1}{k-2}$, $\nu$ was defined in (\ref{2pf}), ${\mathcal N}$ refers to Neumann boundary conditions\footnote{In section \ref{sectcohstat} when explicitly constructing the coherent states we will consider Dirichlet gluing conditions. Here we take Neumann boundary conditions because this is the T dual version in the time direction.} and $x_0$ is the position of the D0 brane in the timelike direction. In the single covering of SL(2,${\mathbb R})$, the only possibilities are $x_0=0$ and $\pi$, which represent the center of the group ${\mathbb Z}_2$ (see \cite{Stanciu}). But in the universal covering, one can take $x_0=q\pi$ with $q\in {\mathbb Z}$ (see section \ref{conjclass}).   

To compare these one-point functions with those obtained 
in section \ref{sectcohstat}, it is convenient to 
consider the conventions introduced in section \ref{sectPF} but with a different normalization in order to explicitly realize the relation between the spectral flow image of highest and lowest weight representations. The fields $\tilde\Phi_{m,\bar m}^{j,w}$ represent the spectral flow images of the primary fields
$\tilde\Phi_{m,\bar m}^{j,0}$, $i.e$ they are  in correspondence with  highest or lowest weight states depending if $w<0$ or $w>0$, and have $J_0^3,~\bar J_0^3$ 
eigenvalues $M=m+\frac k2 w,~\bar M=\bar m+\frac k2 w$. They are related to the vertex operators (\ref{ver}) as
\bea
\tilde\Phi_{m,\bar m}^{j,w}(z,\bar z)=(-)^w \sqrt{\nu^{-\frac12-j}B(j)}~~V^{-1-j}_{n\omega\gamma p\tilde \omega}
(z,\bar z)\,.
\eea 

When looking for $w=0$ solutions, $i.e.$ $\omega=-\tilde\omega$, one expects 
to reproduce the  one-point functions of point-like D-branes
in the $H_3^+$ model, which forces $x_0=r\pi$. So,
\bea
\left\langle \tilde\Phi_{m,\bar m}^{j,w}(z,\bar z)\right\rangle_{\bf s} &=& 
\frac{\delta_{m,\bar m}}{\left|z-\bar z\right|^{\Delta_j+\bar\Delta_j}}
~\frac{\Gamma\left(1+j-m\right) \Gamma\left(1+j+m\right)}{\Gamma
\left(2j+1\right)}\cr\cr
&&\times ~\frac{i\sqrt k~(-)^{w+1}}{2^{\frac54}}~ \frac{\sin[{\bf s}\left((2j+1)-w(k-2)\right)]}
{\sqrt{\sin[\frac{\pi}{k-2}\left(2j+1\right)]}}\label{final}~,
\eea
with the parameter ${\bf s}$ labeling the positions of the instanton 
solutions. 

\subsection{One point-functions for $H_2$, $dS_2$ and light-cone branes} 
 
All of the $H_2$, $dS_2$ and light-cone  branes can be constructed from a D2-brane in the cigar and taking Neumann boundary conditions in the U(1). They are simply related to each other by analytic continuation of a parameter labeling the scale of the branes. Here, we discuss in detail the case of the one-point functions of fields in discrete representations on $H_2$ branes in order to prepare the discussion for section \ref{sectcohstat}. The approach developed in this section to construct the one-point functions gives the correlators associated to $H_2$ branes placed at the surface $X^3=cons$ rather than that at $X^0=cons$ presented in section \ref{conjclass}, so we have to translate these solutions before comparing with the results of the previous section.

The one point-functions for the D2-branes in the cigar are given by \cite{RS} 
\bea
\left\langle \Phi_{j,n,\omega}^{sl(2)/u(1)}(z,\bar z)
\right\rangle_{\tilde\sigma}^{D2} &=& 
\frac{\frac12 \delta_{n,0} (-)^\omega e^{-i\tilde\sigma\omega(k-2)}
\left(\frac{k-2}{k}\right)^{\frac14}}{\left|z-\bar z\right|^{h^j_{nr}+
\bar h^j_{nr}}} 
\Gamma\left(1+2j\right)\Gamma\left(1+\frac{1+2j}{k-2}\right)
\nu^{\frac12+j}\cr\cr
&&\times ~\left(\frac{\Gamma\left(-j+\frac k2 \omega\right)}{\Gamma\left(1+j+
\frac k2\omega\right)}e^{i\tilde\sigma(1+2j)}+ 
\frac{\Gamma\left(-j-\frac k2 \omega\right)}
{\Gamma\left(1+j-\frac k2\omega\right)}e^{-i\tilde\sigma(1+2j)}\right)\, .
\nonumber
\eea 
Notice that this differs from the result
 in \cite{RS} by the $\omega$ 
dependent phase 
$(-)^\omega~e^{-i\tilde\sigma\omega(k-2)}$.\footnote{ This phase 
that we added by hand is required by the spectral flow symmetry, 
when used to construct the one-point functions for $H_2$ branes, 
which demands
$\left\langle 
\tilde\Phi_{j,j}^{j,w}\right\rangle^{H_2}=\left\langle 
\tilde\Phi_{\frac k2+j,\frac k2+j}^{-\frac k2-j,w-1}\right\rangle^{H_2}$,
 in our conventions.
 The one-point function for D2 branes 
was constructed in \cite{RS}
beginning from the parent 
$H_3^+$ model  and was found to have some sign 
problems. We claim this phase cannot be deduced  from the $H_3^+$ model 
because of the absence of spectral flowed states. It would be interesting to 
investigate the implications of this modification in the sign. 
Unfortunately, this information cannot be obtained from the $w$
 independent
 semiclassical limit of the one-point functions.}
The position of the D-brane over the $U(1)$ is again fixed 
by the one-point function of 
the $H_3^+$ model. We find
\bea
\left\langle \tilde\Phi_{m,\bar m}^{j,w}(z,\bar z)\right\rangle_{\tilde\sigma}^{H_2,X^3} &=& 
\frac{\delta_{m,\bar m}}{\left|z-\bar z\right|^{\Delta_j+\bar 
\Delta_j}}\frac{-1}{2^{\frac54}~\sqrt{i} (k-2)^{\frac14}}\frac{\pi ~e^{-i\tilde\sigma w(k-2)}}{\sqrt{\sin\left[\frac{\pi}{k-2}(2j+1)\right]}}\cr\cr
&\times & \left(\frac{\Gamma\left(1+j-m\right)}{\Gamma\left(-j-m\right)}
e^{-i\tilde\sigma(1+2j)}+ 
\frac{\Gamma\left(1+j+m\right)}{\Gamma\left(-j+m\right)}
e^{i\tilde\sigma(1+2j)}\right)\, .~~~~
\label{1ptH2}
\eea

For fields in discrete representations with $m=-j+{\mathbb Z}_{\geq0}$ and $j\notin{\mathbb Z}$, only one factor survives in the  last line. Here $\tilde\sigma$ is a real parameter, determining the embedding of the brane in $AdS_3$ as $X^3=\cosh\rho \sin t=\sin\tilde\sigma$. So, in order to compare with the solutions discussed in section \ref{conjclass}, the identification $\tilde\sigma=\sigma+ \frac\pi2$ and the global shift in the time-like coordinate on the cylinder, namely $t\rightarrow t+\frac\pi2$, must be perfomed. The latter simply adds a phase $e^{i\frac\pi2(M+\bar M)}$ \footnote{In fact, $J_0^3+\bar J_0^3$ gives the energy in $AdS_3$ and so this combination is the generator of $t$ translations.}. 

From the analysis of conjugacy classes, it is natural to relabel $\sigma=\frac{\pi}{k-2}(2j'+1)-w'\pi$, with $j'\in (-\frac{k-1}{2},-\frac12)$, $w'\in{\mathbb Z}$ \footnote{The one-point functions for $dS_2$ branes are given by (\ref{1ptH2}) with $j'\in\left\{-\frac12+i{\mathbb R^+}\right\}$ and for light-cone branes, they are given by $\sigma=n\pi,\,n\in{\mathbb Z}$.}, and
\bea
\left\langle \tilde \Phi_{m,\bar m}^{j,w}(z,\bar z)\right\rangle_{\sigma(j',w')}^{H_2,X^0} 
&=& \frac{\delta_{m,\bar m}}{\left|z-\bar z\right|^{\Delta_j+\bar \Delta_j}} ~ 
\frac{\Gamma\left(1+j+m\right)\Gamma\left(1+j-m\right)}{\Gamma\left(1+2j\right)}\cr\cr
&&\times ~ \frac{-\pi \sqrt{-i}}{2^{\frac54}(k-2)^{\frac14}}\frac{(-)^w e^{\frac{4\pi i}{k-2}
(j'+\frac12-w'\frac{k-2}{2})(j+\frac12-w\frac{k-2}{2})}}{\sqrt{\sin\left[\frac{\pi}{k-2}(2j+1)\right]}}\,.\label{H2dis}
\eea 

\subsection{One-point functions for $AdS_2$ branes}

For completeness, we display here the one-point functions for $AdS_2$ branes obtained in \cite{israel}, in our conventions. These are constructed by gluing two one-point functions: one for a  D1-brane in the coset model and another one with Dirichlet boundary conditions in the U(1) model. The result is
\bea
\left\langle \tilde \Phi^{j,w}_{m,\bar m}(z,\bar z)\right\rangle^{AdS_2}_r 
&=& \frac{\delta_{w,0}\delta_{m,-\bar m} e^{-i\frac\pi4}e^{in(\theta_0+x_0)}
\left(\frac{k-2}{2}\right)^{\frac14}}{\left|z-\bar z\right|^{\Delta_j+\bar \Delta_j}}
\frac{\Gamma\left(-1-2j\right)}{\Gamma\left(-j-m\right)\Gamma \left(-j+m\right)}\cr\cr
&&\times ~\cos\left(ir(j+\frac12)+m\pi\right) \Gamma\left(1-\frac{1+2j}{k-2}\right)\nu^{-\frac12-j}\, ,\label{gam}
\eea
where $\theta_0$ is related to the angles (in cylindrical coordinates) 
to which the branes asymptote when they get close to the boundary of $AdS_3$,
 $x_0$ is the location of the brane and $r$ determines their scale. 
From the geometrical point of view, $r$ seems to be an arbitrary real number, 
but as  shown in \cite{BP}, it becomes quantized at the semiclassical level.

\section{Coherent states and relation to modular data}\label{sectcohstat}

Boundary states play a fundamental role in understanding boundary conformal field theories. They store all the information about possible D-brane solutions and their couplings to bulk states. Following the ideas developed in section \ref{sectBCFT}, one can study different gluing conditions for the left and right current modes, 
consistent with the affine algebra \cite{KO} as well as with the conformal symmetry via the Sugawara construction \cite{Ishibashi}.

In the case of $AdS_3$, much of the progress reached
in this direction is based on the analytic continuation from 
$H_3^+$ \cite{PST}.
Gluing conditions were imposed
as differential equations 
applied directly to find, with 
the help of certain {\it sewing contraints}, the one-point 
functions of maximally symmetric D-branes. It would 
be interesting to get the one-point functions of the $AdS_3$ 
model without reference to other models, but the approach used so far
 cannot 
be easily extended. In the first place, it was developed in
 the $x$-basis  of the $H_3^+$ model, which is not a good 
basis for the representations of the universal 
covering of SL(2,${\mathbb R})$. 
 Suitable bases instead
are the $m$-
or $t$-basis  \cite{mo1, rib}. Moreover,
 there are still some open questions about the fusion rules of the $AdS_3$
model \cite{wc} 
which deserve further attention before analyzing the 
{\it sewing constraints}. Therefore, we will not compute the one-point 
functions in this way, but will give the first step in this 
direction by finding the explicit expressions for the Ishibashi states in 
the $m$-basis for all 
the representations of the Hilbert space of the bulk theory.

\subsection{Coherent states for regular gluing conditions} \label{coherentstates}             

Boundary states associated to $dS_2$, H$_2,$ light-cone and point-like D-branes in $AdS_3$ must satisfy the following regular gluing conditions \cite{RR}
\bea
&&\left(J_n^3-\bar J_{-n}^3\right)\left|{\bf s}\right\rangle=0,\cr
&&\left(J_n^{\pm}+\bar J_{-n}^{\mp}\right)\left|{\bf s}\right\rangle=0,\label{boundstatecond}
\eea
where ${\bf s}$ labels the members of the family of branes allowed by the gluing conditions.

These constraints are linear and leave  each representation invariant, so that the boundary states must be expanded as a sum of solutions in each representation. The solutions are coherent states, usually called Ishibashi states \cite{Ishibashi}. 

Let us begin introducing the following notation which will be useful in the subsequent discussions. Let
\bea
\left|j,w,\alpha;{\rm n, m}\right\rangle=\left|j,w,\alpha\right\rangle
\{\left|{\rm n}\right\rangle\otimes\overline{\left|{\rm m}\right\rangle}\},~~
\left|j,w,+;{\rm n, m}\right\rangle=\left|j,w,+\right
\rangle\{\left|{\rm n}\right
\rangle\otimes\overline{\left| {\rm m}\right\rangle}\},
\eea
denote  orthonormal bases for 
$\hat{\mathcal C}_j^{\alpha,w}\otimes\hat{\mathcal C}_j^{\alpha,w}$ 
and $\hat{\mathcal D}_j^{+,w}\otimes\hat{\mathcal D}_j^{+,w}$, 
respectively. They 
satisfy\footnote{The separation between 
$|j,w,\alpha\rangle$ or $|j,w,+\rangle$ 
and $|{\rm n}\rangle, \overline{|{\rm m}\rangle}$
 in different kets is simply a matter of 
 useful notation for calculus and does not denote tensor product.}
\bea
\langle j,w,\alpha;{\rm n, m}|j',w',\alpha';{\rm n',m'}\rangle &=&~~~~~
\langle j,w,\alpha|j,w,\alpha\rangle~\times ~ 
\langle {\rm n}|{\rm n}'\rangle\times \overline{\langle {\rm m}}
|\overline{{\rm m}'\rangle}\cr
&=&\delta(s-s')\delta_{w,w'}\delta(\alpha-\alpha')~\epsilon_{\rm n}
\delta_{{\rm n,n}'}~
\epsilon_{\rm m} \delta_{{\rm m,m}'},\cr\cr
\left\langle j,w,+;{\rm n,m}|j',w',+;{\rm n',m'}\right\rangle &=&\left\langle j,w,+
|j,w,+\right\rangle \left\langle {\rm n}|{\rm n}'\right\rangle \overline{\left
\langle {\rm m}\right.}|\overline{\left.{\rm m}'\right\rangle}\cr
&=&~ ~  \delta(j-j')\delta_{w,w'}~\epsilon_n\delta_{{\rm n,n}'}~
\epsilon_{\rm m} 
\delta_{{\rm m,m}'},
\eea
$\left\{|{\rm n}>\right\}$ 
is an orthonormal basis in $\hat{\mathcal C}_j^{\alpha,w}$ 
(or $\hat{\mathcal D}_j^{+,w}$) for which the expectation values of 
$J^3_n, J^{\pm}_n$ are real numbers and $\epsilon_{\rm n}=\pm1$ is its 
norm squared.
It is constructed by the action of the affine currents over the ket 
$|j,w,m=\alpha>=U_w|j,m=\alpha>$  ($|j,w,m=-j>$).

The Ishibashi states for continuous and discrete representations 
are found to be 
\bea
\left|j,w,\alpha\right.\gg= \sum_{\rm n} \epsilon_{\rm n}\bar 
V \left|j,w,\alpha;{\rm n,n}
\right\rangle~~~ 
{\rm and}\quad ~
\left|j,w,+\right.\gg= \sum_{\rm n} \epsilon_{\rm n}\bar V\left|j,w,+;{\rm n,n}
\right\rangle,~~~\label{Ishdef}
\eea
respectively, where $V$ is defined as the linear operator satisfying 
\bea
V\prod_I J_{n_I}^{a_I}\left|j,w,m=-j\right\rangle &=& \prod_I \eta_{a_I b_I}
J^{b_I}_{n_I}\left|j,w,m=-j\right\rangle,\cr
V\prod_I J_{n_I}^{a_I}\left|j,w,m=\alpha\right\rangle & =& \prod_I 
\eta_{a_I b_I}J^{b_I}_{n_I}\left|j,w,m=\alpha\right\rangle,\label{Vdef}
\eea 
with $a=1,2,3$, $\eta_{ab}=diag(-1,-1,1)$
and the bar  denotes 
 action restricted to the antiholomorphic 
sector. It is easy to see that this defines a	unitary operator. 
The proof that 
they are solutions to (\ref{boundstatecond}) follows similar lines as 
those of \cite{Ishibashi}. 
As an example, let us consider an arbitrary base state 
$|j',w',\alpha';{\rm n',m'}>$:
\bea
&&<j',w',\alpha';{\rm n',m'}|J_r^3-\bar J_{-r}^3|j,w,\alpha\gg =\cr
&& ~~~
\delta(s-s')\delta_{w,w'}\delta(\alpha-\alpha')\sum_{\rm n}\epsilon_{\rm n} 
\left\langle {\rm n'}
\right|J_n^3\left|{\rm n}\right\rangle\overline{\left\langle 
{\rm m'}\right|}\bar V\overline{\left|{\rm n}\right\rangle}-\epsilon_{\rm n} 
\left\langle {\rm n'}\right|\left.{\rm n}\right\rangle \overline{\left\langle 
{\rm m'}
\right|}\bar J_{-n}^3 \bar V\overline{\left|{\rm n}\right\rangle}=\cr
&&  ~
\delta(s-s')\delta_{w,w'}\delta(\alpha-\alpha')\sum_{\rm n}\epsilon_{\rm n} 
\left\langle {\rm n'}\right|J_n^3\left|{\rm n} \right\rangle \left\langle 
{\rm n}\right|
V\left|{\rm m'}\right\rangle-\epsilon_{\rm n} \left\langle {\rm n'}
\right|\left.{\rm n}
\right\rangle \left\langle {\rm n}\right|VJ_{n}^3 \left|{\rm m'}
\right\rangle=0\, .
\nonumber
\eea

The normalization fixed above for the Ishibashi states implies
\bea
\ll j,w,\alpha|e^{\pi i\tau\left(L_0+\bar L_0-\frac{c}{12}\right)}
e^{\pi i\theta(J_0^3+\bar J_0^3)}|j',w',\alpha'\gg &=& \delta (s-s')
~\delta_{w,w'}~\delta(\alpha-\alpha')~\chi_j^{\alpha,w}(\tau,\theta),\cr\cr
\ll j,w,+|e^{\pi i\tau\left(L_0+\bar L_0-\frac{c}{12}\right)}
e^{\pi i\theta(J_0^3+\bar J_0^3)}|j',w',+\gg &=&
 \delta(j-j')~\delta_{w,w'}~\chi_j^{+,w}(\tau,\theta).\label{IshNorm}
\eea

\subsection{{\it Cardy structure} and one-point functions for point-like branes}\label{pointbranes}

Assuming that after Wick rotation the open string partition function in $AdS_3$ reproduces that of the $H_3^+$ model and a generalized Verlinde formula, we show in this section that the one-point functions on localized branes in $AdS_3$ previously found in \cite{israel} (and reviewed in section \ref{sect1pf}) can be recovered. 
We also verify that the  one-point 
functions on  point-like and $H_2$ D-branes
exhibit a {\it Cardy structure}. 
Usually, this structure is accompanied by a Verlinde formula for 
the representations appearing in the boundary
 spectrum. In fact, the {\it Cardy structure} is a natural 
solution to the {\it Cardy condition} when the
Verlinde theorem holds. However, as we shall discuss, the latter 
does not hold in the $AdS_3$ WZNW model. 
The generalized 
Verlinde formula proposed in  appendix \ref{A:gvf}  
reproduces the fusion rules of the degenerate representations, but it gives
contributions to the fusion rules of the discrete representations with 
an arbitrary amount of spectral flow, thus contradicting the 
selection rules determined in \cite{mo3}. Nevertheless, we find a
{\it Cardy structure}. 
\bigskip

\subsubsection{Boundary states}
\medskip

Worldsheet duality allows to write the one loop partition function 
for open strings ending on point-like branes labeled by ${\bf s}_1$ and 
${\bf s}_2$ as
\bea
e^{-2\pi i\frac k4\frac{\theta^2}{\tau}}~Z_{{\bf s}_1{\bf s}_2}^{AdS_3}(\theta,\tau,0) &=&
\left
\langle \left.\Theta {\bf s}_1\right|\tilde q^{H^{(P)}}\tilde z^{J_0^3}
 \left|{\bf s}_2
\right.\right\rangle ~~\cr &=& \sum_{w=-\infty}^{\infty} 
\int^{-\frac12}_{-\frac{k-1}{2}}dj~{\cal A}_{(j,w)}^{{\bf s}_1}
{\cal A}_{(j^+,w^+)}^{{\bf s}_2}\chi_j^{+,w}(\tilde \theta,\tilde \tau,0)~
+~
ccr\, ,\nonumber
\eea
where $\Theta$ denotes the worldsheet CPT operator in the bulk theory, 
$\tilde q=e^{2\pi i\tilde 
\tau}$, $\tilde z=e^{2\pi i\tilde \theta}$, $\tilde\tau=-1/\tau$,
$\tilde\theta=\theta/\tau$, $(j^+,w^+)$ refer to the labels of the 
$(j,w)$-conjugate representations, $ccr$ denotes the contributions of
continuous representations
 and 
${\cal A}_{(j,w)}^{{\bf s}}$ 
are the Ishibashi coefficients of the boundary states.

The open string partition function for the ``spherical branes'' 
of the $H_3^+$  model was found in \cite{GKS} for $\theta=0$ and extended
to the case $\theta\neq0$ in \cite{RS}. It reads
\bea
Z_{{\bf s}_1{\bf s}_2}^{H_3^+}(\theta,\tau,0)=\sum_{J_3=\left|J_1-J_2\right|}^{J_1+J_2}~~
\chi_{J_3}(\theta,\tau,0)\, ,\label{Verldeg}
\eea
where ${\bf s}_i=\frac{\pi}{k-2}(1+2J_i)$ and $1+2J_i\in{\mathbb N}$. This
 reveals an open string spectrum of discrete degenerate representations.

The Lorentzian partition function is expected to reproduce 
that of the $H_3^+$ model after analytic continuation in $\theta$ and $\tau$.
Then, if we concentrate on the one-point functions of fields in discrete 
representations, we only need to consider the case 
$\theta+n\tau\not\in{\mathbb Z}$. 
Thus, using the generalized Verlinde formula 
(see appendix \ref{A:gvf} for details), namely 
\bea
\sum_{J_3=\left|J_1-J_2\right|}^{J_1+J_2}~~
\chi_{J_3}(\theta,\tau,0)=\sum_{w=-\infty}^{\infty}
\int^{-\frac12}_{-\frac{k-1}{2}}
dj~~
\frac{S_{J_1}{}^{j,w}S_{J_2}{}^{j,w}}{S_{0}{}^{j,w}}~~
e^{2\pi i\frac k4 \frac{\theta^2}{\tau}}\chi_j^{+,w}(\frac\theta\tau,
-\frac1\tau,0)\, ,\label{verdeg}
\eea
we obtain the following expression for the coefficients of the boundary states:
\bea
{\cal A}_{(j,w)}^{{\bf s}}
=f(j,w)~(-)^w\sqrt{\frac{ 2}{ i}}\left(\frac{2}{k-2}
\right)^{\frac14}
\frac{\sin\left[{\bf s}\left(1+2j-w(k-2)\right)\right]}{\sqrt{\sin\left[\frac{\pi}{k-2}\left(1+2j\right)\right]}}\, ,
\label{A}
\eea
defined up to a function $f(j,w)$ satisfying 
\bea
f(j,w)~f(-\frac k2-j,-w-1)=1.\label{fconst}
\eea

\subsubsection{One-point functions}

To find the one-point functions associated to these point-like branes, let us make use of the definition (\ref{defbs}) of the boundary states (see also 
\cite{schomerus})\footnote{Strictly speaking, this identity is valid on a Euclidean worldsheet. However, it is appropriate to use it here since we want to explore the relation of our results with those of the Euclidean model defined in \cite{israel} where the coefficients of the one-point functions are assumed to coincide with those of the Lorentzian $AdS_3$.}:
\bea
\left\langle \Phi^{(H)}\left(\left|j,w,m,\bar m\right\rangle;z,
\bar z\right)\right\rangle_{\bf s}=\left(\frac{d\xi}{dz}\right)^{\Delta_j} 
\left(\frac{d\bar\xi}{d\bar z}\right)^{\bar\Delta_j}
\left\langle 0\right|\Phi^{(P)}
\left(\left|j,w,m,\bar m\right\rangle;\xi,\bar \xi\right)
\left|{\bf s}\right\rangle\, ,\label{1pf}
\eea
where $\Phi^{(H)}\left(\left|j,w,m,\bar m\right\rangle;z,\bar z\right)$ 
($\Phi^{(P)}\left(\left|j,w,m,\bar m\right\rangle;\xi,\bar \xi\right)$) 
is the bulk field of the boundary (bulk) CFT
 corresponding to the state inside the 
brackets\footnote{Here $|j,w,m,\bar m>$ is a shorthand notation
 for $|j,w,m>\otimes|j,w,\bar m>$ and it must be distinguished from
 the orthonormal basis introduced in section \ref{coherentstates}.},
$z,\bar z$ denote the coordinates of the upper half plane and 
$\xi,\bar \xi$ those of the exterior of the unit disc. 

Conformal invariance forces the $l.h.s.$ of (\ref{1pf}) to be 
\bea
\left\langle \Phi^{(H)}\left(\left|j,w,m,\bar m\right\rangle;z,
\bar z\right)\right\rangle_{\bf s}=\frac{{\mathcal B}
({\bf s})_{m,\bar m}^{j,w}}
{|z-\bar z|^{\Delta_j+\bar\Delta_j}}\, ,
\eea
where the $z$-independent factor ${\cal B}({\bf s})_{m,\bar m}^{j,w}$ 
is not fixed by the conformal symmetry. The solution (\ref{Ishdef}), (\ref{Vdef}) implies
\bea
{\mathcal B}({\bf s})_{m,\bar m}^{j,w}=(-)^{j+m}\delta_{m,\bar m}
{\mathcal A}_{j,w}^{\bf s}~~,\label{resul}
\eea
from which the spectral flow symmetry determines $f=1$. \footnote{In fact, the identification $<\Phi^{(H)}\left(\left|j,w,m=\bar m=-j\right\rangle;z,
\bar z\right)>$ $=<\Phi^{(H)}\left(\left|j',w',m'=\bar m'=j'\right\rangle;z,\bar z\right)>$ where $j'=-\frac k2-j,~w'=w+1$ together with (\ref{resul}) requires $f(j,w)=f(-\frac k2-j, w+1)$ which implies
\bea
f(j,w)=f(j,w+2)
\eea
So it is sufficient to find $f(j,0)$ and $f(j,1)$. It is immediate from this constraint and (\ref{fconst}) that $f(j,0)^2=f(j,1)^2=1$. But as $f$ must be continuous in $j$, it must be $j$ independent, $f(j,w)=f(w)=\pm1$ ($i.e.$ the same sign for every $w$). To choose the upper or the lower sign is simply a matter of convention.}  
\medskip

It is important to note that the normalization used here differs from 
the one usually considered in the literature. Our normalization is such that 
the spectral flow image of the primary operator
corresponding to the state 
 $|j,w,m,\bar m>$
is normalized to 1. 
In particular, it implies the following operator product expansions
\bea
J^3(\zeta)\Phi^{(P)}\left(\left|j,w,m,\bar m\right\rangle;\xi,\bar \xi\right)
&=& 
\frac{m+\frac k2 w}{\zeta-\xi}\Phi^{(P)}\left(\left|j,w,m,\bar m\right\rangle;
\xi,\bar \xi\right)+\dots\cr\cr
J^{\pm}(\zeta)\Phi^{(P)}\left(\left|j,w,m,\bar m\right\rangle;\xi,\bar \xi
\right)&=&
 \frac{\sqrt{-j(1+j)+m(m\pm1)}}{\left(\zeta-\xi\right)^{1\pm w}}\Phi^{(P)}
\left(\left|j,w,m\pm1,\bar m\right\rangle;\xi,\bar \xi\right)\nonumber\\
&&+\dots~~~~\label{ope(p)}
\eea 

Comparing with the OPE of $\tilde \Phi^{j,w}_{m,\bar m}\left(\xi,\bar \xi\right)$ which coincides with (\ref{opew}) we obtain the following relation, 
valid for $m=\bar m\in-j+{\mathbb Z}_{\geq0}$,
\bea
\tilde\Phi^{j,w}_{m,\bar m}(\xi,\bar \xi)= \Omega~ (-)^{j+m}
\frac{\Gamma\left(1+j-m\right)\Gamma\left(1+j+m
\right)}{\Gamma\left(1+2j\right)}~\Phi^{(P)}
\left(\left|j,w,m,\bar m\right\rangle;\xi,\bar \xi\right)\, ,\label{norm}
\eea 
where $\Omega$ is the normalization of $\Phi^{j,w}_{-j,-j}$. We find perfect 
agreement between the expressions 
(\ref{resul}) and (\ref{final}) for one-point functions, as long as 
$~~\Omega=-\sqrt{\frac{-ik(k-2)}{16}}$ .

\subsection{{\it Cardy structure} in $H_2$ branes}

In section \ref{sect1pf} we reviewed the results for the one-point functions in maximally symmetric D-branes obtained by applying the method of \cite{israel}.
From the one for fields in discrete representations on $H_2$ branes we find the following Ishibashi coefficients (see (\ref{H2dis}) and (\ref{norm}))
\bea
{\cal A}_{(j,w)}^{\sigma'\equiv(j',w')} =
\frac{\pi}{\sqrt k}\left(\frac{2}{k-2}\right)^{\frac34}~ \frac{(-)^w e^{\frac{4\pi i}{k-2}(j'+\frac12-w'\frac{k-2}{2})(j+\frac12-w
\frac{k-2}{2})}}{\sqrt{\sin\left[\frac{\pi}{k-2}(2j+1)\right]}}\sim(-)^{w'}\frac{S_{j,w}{}^{j',w'}}{\sqrt{S_0{}^{j,w}}}\, ,
\eea 
which exhibit a clear compromise between microscopic and modular data. It gives the following degeneracy in the open string spectrum
\bea
{\cal N}_{j_1,w_1~j_2,w_2}{}^{j_3,w_3}\sim \delta_{{\cal D}_j^{+,w}}\sum_{m=-\infty}^{\infty}\delta\left(j_2+j_3-j_1-\left(w_2+w_3-w_1\right)\frac{k-2}{2}+m\right)\, ,\nonumber
\eea
where the divergent integral $\int_0^1 d\lambda \frac{e^{-2\pi i \left(m+\frac12\right)\lambda }}{2i \sin(\pi \lambda)}$ has been replaced by its principal value, $\frac12$. $\delta_{{\cal D}_j^{+,w}}$ is one when $j$ is within the unitary bound and vanishes if it is not. $\sim$ means up to the factor $\frac{-i\, 2\,\pi^2(-)^{w_3}}{k(k-2)}$ which would imply that either the one point function computed in \cite{israel} has a subtle modification to ensure integer coefficients in the boundary spectrum or such a factor should be supplemented by the contribution of other representations appearing in the boundary spectrum (associated to open string) not contained in the $AdS_3$ WZNW spectrum (associated to closed strings) which transform under the modular group in terms of the spectral flow images of the principal lowest weight representations. 

Contrary to what happens in RCFT with maximally symmetric $D$-branes, the Boundary CFT spectrum does not coincide with the fusion rules of the Bulk CFT. The consequence of this observation is the failure of the Verlinde Theorem.

\subsection{Coherent states for twined gluing conditions}

The gluing conditions defining the coherent states $|j,w\gg$
 for $AdS_2$ branes \cite{RR}, 
frequently called  twisted boundary conditions, are 
\bea
\left(J_n^3+\bar J_{-n}^3\right)\left|j,w\gg\right.=0, 
\qquad
\left(J_n^{\pm}+\bar J_{-n}^{\pm}\right)\left|j,w\gg\right.=0
\, .\label{bcads2}
\eea
These constraints are highly restrictive. As we show below, 
coherent states satisfying these conditions can only be found for  
representations where the holomorphic and  antiholomorphic sectors are 
conjugate of each other, $i.e.$ only for
 $w=0$, $\alpha=0,\frac12$  continuous representations   
in the $AdS_3$
model. 

Let us assume $|j,w\gg$ is an Ishibashi state 
 associated to the spectral flow image of a 
 discrete or  continuous 
representation. 
The spectral flow transformation (\ref{wtrans}) 
allows  to translate the problem of solving 
(\ref{bcads2}) 
to that of solving 
\bea
\left(J_n^3+\bar J_{-n}^3+kw\delta_{n,0}\right)\left|j\right\rangle^{w}=0\, 
,\qquad
\left(J_n^{\pm}+\bar J_{-n\mp2w}^{\pm}\right)\left|j\right\rangle^{w}=0
\, ,\label{gcads2}
\eea
where $\left|j\right\rangle^{w}=U_{-w}\bar U_{-w}|j,w\gg$ is 
in an unflowed representation\footnote{Notice that in the case $w=-\bar w$ 
discussed in \cite{RR} for the single covering of SL(2,$\mathbb R$),
one gets (\ref{bcads2}) with the unflowed $|j>^{w,\bar w}$ 
state replacing $|j,w\gg$ 
instead of (\ref{gcads2}).
 Then, once an Ishibashi state is found for $w=-\bar w=0$, 
the solutions for
 generic  representations with $w=-\bar w$ are trivially obtained applying the
spectral flow operation, and
coherent states in arbitrary spectral flow sectors are found.
This fails in $AdS_3$ and thus
 the discussion in $loc.~cit.$ does not apply here, except for
 $w=0$ discrete or $w=0$, $\alpha=0,
\frac12$ continuous representations.}.

The special case $n=0$  in (\ref{gcads2}) 
implies 
$2\alpha+kw~\in{\mathbb Z}$ and $ -2j+kw ~\in{\mathbb Z}$ 
for continuous and discrete representations, respectively.
In particular, for $w=0$ continuous representations there are two solutions 
with $\alpha=0,\frac12$, given by 
\bea
\left|j,0,\alpha\right.\gg= \sum_{\rm n} \epsilon_n\bar U \left|j,w,
\alpha;{\rm n,n}\right\rangle, 
\eea
where the antilinear operator $U$ is defined by
\bea
U\prod_I J_{n_I}^{a_I}\left|j,w=0,m=\alpha\right\rangle=
\prod_I -J^{a_I}_{n_I}\left|j,w=0,m=-\alpha\right\rangle\, .
\eea 
It can be  easily verified that this defines an antiunitary operator and 
it is exactly the same Ishibashi state found in SU(2) \cite{Ishibashi}.

To understand why there are no solutions in other modules, let us expand 
the hypothetical Ishibashi state in the orthonormal base 
$|j,w,\zeta>\{|{\rm n}>\otimes\overline{|{\rm m}>}\}$, with $\zeta=\alpha$ 
or $+$ and $|{\rm n}>,$ $\overline{|{\rm m}>}$ 
eigenvectors of $J_0^3,L_0$ and $\bar J_0^3,\bar L_0$ respectively. 
The constraint that Ishibashi states are annihilated by $L_0-\bar L_0$ 
forces  $|{\rm n}>,\overline{|{\rm m}>}$ to be at the same level.  
But taking into account that 
all modules at a given level are highest or lowest weight representations 
of the zero modes of the currents (with the only exception of $w=0$ 
continuous representations) and the fact that the eigenvalues of the highest 
(lowest) weight operators
 decrease (increase) after descending a finite number of levels, 
the first equation in (\ref{bcads2}) with $n=0$ 
has no solution
below certain level. This implies that below that level there 
are no contributions to the Ishibashi states and so, using for instance the 
constraint $(J_1^a+\bar J_{-1}^a)|j,w\gg=0$, 
it is easy to show by induction that 
no level contributes to the coherent states. 

The 
coherent states defined above are normalized as
\begin{equation}
\ll j,0,\alpha|e^{\pi i\tau\left(L_0+\bar L_0-\frac{c}{12}\right)}
e^{\pi i\theta(J_0^3-\bar J_0^3)}|j',0,\alpha'\gg ~= ~\delta (s-s')
\delta(\alpha-\alpha')~\chi_j^{\alpha,0}(\tau,\theta),\label{IshNormTwis}
\end{equation}
for $\alpha=0,\frac12$. The fact that it is only possible to construct
Ishibashi states associated to $w=0$ continuous representations 
is again in agreement with the one-point functions found in \cite{israel} 
and the conjecture in \cite{D} 
that only states in these representations couple to $AdS_2$ branes.

\chapter{Conclusions and perspectives}\label{chapConcl}

Now it is time to summarize and to discuss the results obtained during the doctoral activity and to indicate possible directions to continue these investigations. 
\bigskip
 
String theory in anti de Sitter spaces has occupied a central place in the last twenty years. The initial motivation was to learn more about string theory with Lorentzian target space with a non trivial timelike direction. Indeed this is one of the simplest theories, if not the simplest, with this property and it took not too much time to realize the complexity and richness of the model still under study. A few years later from the initial investigations by O'Raifeartaigh {\it et al} \cite{BRFW}, the {\it Maldacena conjecture} taught us that string theory with target $AdS$ spaces are not only interesting models to learn more about string theory with the aim of applying it to more realistic scenarios, but it is also an incredible tool which allows us to analyze non perturbative regimes of gauge theories, with applications from quark gluon plasma and quantum gravity of black holes to condensed matter systems. 

In this thesis we have concentrated on some aspects of the worldsheet theory of strings in $AdS_3$. This is described by a WZNW model with non compact group which implies that the Conformal Field Theory is non Rational. RCFT was extensively studied in the literature and was completely solved but unfortunately the situation is much more involved in the case of non RCFT where many of the properties which helped in the resolution of the former do not hold, $e.g.$ the conformal blocks cannot be solved from purely algebraic methods, because the fact that three point functions do not vanish does not necessarily imply that the conjugate of the third field will be in the fusion rules of the first two. The proofs of the factorization and crossing symmetry of four point functions are of considerable complexity. The spectrum contains sectors where $L_0$ is not bounded from below, and some properties like the Cardy formula, determining the one point functions of the Boundary CFT, and the Verlinde Theorem, which determines the fusion rules of the Bulk CFT and the Open string spectrum of the Boundary CFT, do not work any more.   

Some non RCFTs, like Liouville theory and the $H_3^+$ model were successfully solved, both the Bulk and the Boundary CFTs. Instead, even though much is known today about the $AdS_3$ WZNW model, its spectrum was completely determined, there is great confidence on the one point functions of maximally symmetric and some symmetry breaking $D$-branes and on the structure constants, it has not been completely solved as the crossing symmetry has not yet been established. 
\medskip

We made the following contributions to the the resolution of the $AdS_3$ WZNW model. 
 
We determined the operator algebra as presented in chapter \ref{chapOPE}. Performing the analytic continuation of the expressions in the Euclidean $H_3^+$ model proposed in \cite{tesch1, tesch2} and adding spectral flow structure constants, we obtained the OPE of spectral flow images of primary fields in the Lorentzian theory. We have argued that the spectral flow symmetry forces a truncation in order to avoid contradictions and we have shown that the consistent cut amounts to the closure of the operator algebra on the Hilbert space of the theory as only operators outside the physical spectrum must be discarded and moreover, every physical state contributing to a given OPE is also found to appear in  all possible equivalent operator products. The fusion rules obtained in this way are consistent with results in \cite{mo3}, deduced from the factorization of four-point functions of $w=0$ short strings in the boundary conformal field theory, and contain in addition operator products involving states in continuous representations. A discussion of the relation between our results and some conclusions in \cite{mo3} can also be found in chapter \ref{chapOPE}. Several consistency checks have been performed in this chapter and the OPE displayed in items 1. to 3. of  section \ref{secttrunc} can then be taken to stand on solid foundations.

Given that scattering amplitudes of states of string theory on $AdS_3$ should be obtained from correlation functions in the $AdS_3$ WZNW model, our results constitute a step forward towards the construction of the scattering S-matrix in string theory on Lorentzian $AdS_3$ and to learn more about the dual conformal field theory on the boundary through AdS/CFT, in the spirit of \cite{mo3}. Indeed an important application of our results would be to construct the S-matrix of long strings in $AdS_3$ which describes scatterings in the CFT defined on the Lorentzian two-dimensional boundary. In particular, the OPE of fields in spectral flow non trivial continuous representations obtained here sustains the expectations in \cite{mo3} that short and long strings should appear as poles in the scattering of asymptotic states of long strings. 

The full consistency of the fusion rules should follow from a proof of factorization and crossing symmetry of the four-point functions. An analysis of the factorization of amplitudes involving states in different sectors of the theory was presented in chapter \ref{chapFact}. As an interesting check we have shown that using the factorization ansatz and OPE obtained in the previous chapter we can reproduce the spectral flow selection rules of four and higher point functions determined in \cite{mo3}. We illustrated in one example that the amplitudes must factorize as expected in order to avoid inconsistencies, $i.e.$ only states according to the fusion rules determined in chapter \ref{chapOPE} must propagate in the intermediate channels.

In chapter \ref{chapChar} we have computed the characters of the relevant representations of the $AdS_3$ model on the Lorentzian torus and we showed that contrary to other proposals these seem to preserve full information on the spectrum. The price to pay is that characters have now a subtler definition as distributional objects. In the following chapter we studied their modular transformations. We fully determined the generalized $S$ matrix, which depends on the sign of $\tau$, and showed that real modular parameters are crucial to find the modular maps, which implies that if one employs the standard method for Lorentzian models and Wick rotate the characters, these not only loose information on the spectrum but also do not transform properly under modular transformations. 

We have seen that the characters of continuous representations transform among themselves under $S$ while both kinds of characters appear in the $S$ transformation of the characters of discrete representations. An important consequence of this fact is that the Lorentzian partition function is not  modular  invariant (and the departure from modular invariance is not just the sign appearing in (\ref{ZDmod})). The analytic continuation to obtain the Euclidean partition function (which must be invariant) is not fully satisfactory. Following the road of \cite{mo1} and simply discarding the contact terms, one recovers the partition function of the $H_3^+$ model obtained in \cite{gawe}. But even though modular invariant, this expression has poor information about the spectrum. Not only the characters of the continuous representations vanish in all spectral flow sectors but also those of the discrete representations are only well defined in different regions of the moduli space, depending on the spectral flow sector, so that it makes no mathematical sense to sum them in order to find the modular $S$ transformation. An alternative approach was followed in \cite{kounnas}, where an expression for the partition function was found  starting from that of the SL(2,$\mathbb R$)/U(1) coset computed in \cite{hpt} and using path integral techniques. Although formally divergent, it is modular invariant and allows to read  the spectrum of the model\footnote{The spectrum was also obtained from a computation of the Free Energy in \cite{mo2}.}. It was shown that the partition function obtained in \cite{gawe,mo1} is recovered after some formal manipulations. 

In the following chapter we computed the full set of Ishibashi states for all the gluing maps leading to maximally symmetric $D$-branes and to all the representations of the spectrum. The treatment of the boundary states presented here differs from previous works in related models. While we have expressed them as a sum over Ishibashi states, in other related models such as $H_3^+$ \cite{PST}, Liouville \cite{FZZ} or the Euclidean black hole \cite{RS}, the boundary states have been expanded, instead, in terms of primary states and their descendants. The coefficients in the latter expansions directly give the one-point functions of the primary fields. For instance, in the $H_3^+$ model, the gluing conditions were imposed in \cite{PST} not over the Ishibashi states but over the one-point functions. One of the reasons why this approach seems more suitable for $H_3^+$ is the observation that the expectation values used to fix the normalization of the Ishibashi states diverge in the hyperbolic model. As we have seen, this is not the case in $AdS_3$. Another difficulty in applying the standard techniques to $H_3^+$ models comes form the fact that the spectrum does not factorize as holomorphic times antiholomorphic sectors.

We found that there is a one to one map between the representation of the models and the family of maximally symmetric $D$-branes. Thanks to this observation we then showed that the one-point functions of fields in discrete representations coupled to point-like and to $H_2$ branes are determined by the generalized modular $S$ matrix, as usual in RCFT. We also found a generalized Verlinde formula which gives the fusion rules of the degenerate representations of $sl(2)$ appearing in the spectrum of open strings attached to the point-like D-branes of the model.
\medskip

There are some natural directions to follow in the future. The most important and most difficult one is to prove the crossing symmetry of four point functions. 
We hope that this will resolve the problem of the breaking of analyticity and that it can be used to put the fusion rules on a firmer mathematical ground. 

Another interesting line of investigation is to better understand how the information is lost in the procedure implemented in \cite{kounnas} when reproducing the $H_3^+$ partition function and to explore if it is possible to find an analytic continuation of the Lorentzian partition function computed here leading to the integral expression obtained in $loc. cit.$ (or an equivalent one), in a controlled way in which the knowledge on the spectrum is not removed.

The Verlinde like formula was shown to hold for generic $\theta, \tau$  far from $\theta +n\tau\in{\mathbb Z}$. It would be interesting  to study the extension to generic $\theta,\tau$ which requires to consider the $S$ matrix block (\ref{degs}). Furthermore, one could also study the modular transformations of the characters of other degenerate representations and their spectral flow images and explore the validity of generalized Verlinde formulas in these cases.
 It will be very interesting to also compute the boundary spectrum of $H_2,\,dS_2$, light cone and point like $D$-branes independently of the $H_3^+$ and the coset model. This can shed light on the puzzle which arises when considering the open/closed duality which gives negative degeneracies in the open string spectrum of the $H_2$ branes. 
\bigskip

The original contributions discussed along this thesis are based on the author's publications \cite{wc,wc2}. Other investigations performed during the doctoral period were published in \cite{BL,ABMN}. In \cite{BL} we discussed the stability of certain systems of branes with non vanishing background gauge fields in a flat target space. In \cite{ABMN} we performed a dimensional reduction of Double Field Theory (DFT) \cite{HZ,HHZ,HHZ2} and we found that it reproduces the four dimensional bosonic electric sector of gauged $N=4$ supergravity \cite{JW}. We showed that the standard NS-NS (non)-geometric fluxes ($H,\,\omega,\,Q,\,R$) can be identified with the gaugings of the effective action, and that the string d-dimensional background can be decoded from the double twisted 2d-torus. The fluxes obey the standard T-duality chain and satisfy Jacobi identities reproducing the results of \cite{STW}. In this way, the higher dimensional origin of the string fluxes can be traced to the new degrees of freedom of DFT. Thus, the formalism of DFT allows to describe effective field theories in an intermediate stage between supergravity and the full quantum string theory description. Even though this mechanism was successfully implemented in WZNW models \cite{Schulz}, it seems unnatural to introduce dual coordinates to winding because there are no non trivial cycles in $AdS_3$.

\addcontentsline{toc}{chapter}{Appendices}

\appendix

\chapter{Basic facts on CFTs}\label{A:bfcft}

\section{Correlation functions}\label{seccorr}

Correlation functions play a central role in any Quantum Field Theory, as they are the physical objects that connect the theory with measured quantities. Computing correlators in non trivial curved spaces usually involve arduous calculations, but fortunately in many situations the symmetries of the specific model one is analyzing can be exploited to help in the computations. For instance, the global conformal symmetry completely fixes the coordinate dependence of two and three point functions and it determines that of four point functions up to functions of the anharmonic ratios, which are coordinate combinations invariant under the global conformal group. There are two linearly independent ratios in generic spacetime dimension, $e.g.$ we may consider 
\bea
x=\frac{{\bf x}_{12}\cdot{\bf x}_{34}}{{\bf x}_{13}\cdot{\bf x}_{24}}~,~~~~~~ 
y=\frac{{\bf x}_{12}\cdot{\bf x}_{34}}{{\bf x}_{23}\cdot{\bf x}_{14}}, \label{anratios}
\eea
where the notation ${\bf x}_{ij}={\bf x}_i-{\bf x}_j$ was used.

As is well known, in the particular case of two dimensional CFT there is an enhancement of the symmetry group. The conformal algebra is now a local symmetry and it is generated by all Virasoro generators $L_n$ ($n\in{\mathbb Z}$) in contrast with the global conformal symmetry of higher dimensional theories generated by $\{L_{-1},L_0,L_1\}$. If one replaces the two dimensional vectors ${\bf x}_i$  by complex coordinates $x_i,\bar x_i$ one rapidly realizes that the conformal group is nothing else than the group of holomorphic and antiholomorphic maps. In this basis the global invariant combinations are the analogous of (\ref{anratios}) replacing the vectors by the complex variables and the dot product by complex multiplication. An important difference holding in the two dimensional case is that only one of the anharmonic ratios is linearly independent because four points require an extra constraint to lie in the same plane. Indeed it is easily found that $y=\frac{x}{1-x}$ and $\bar y=\frac{\bar x}{1-\bar x}$.   

Even in the simplest non trivial correlators, the three point functions, their coefficients $C_{mnp}$ (see (\ref{3pf})) are not fixed by conformal symmetry. Dynamical inputs are required, like {\it crossing symmetry} as well as the full local conformal symmetry, not just the global one, to completely determine them, as will be clear in section \ref{bootapp}. Genus zero correlation functions of arbitrary higher order can be written in a factorized form when the operator algebra and the three point function coefficients are known and so the theory is said to be completely solved \cite{BPZ}.

Let us consider an arbitrary two point function of primary fields $\phi_m$. As is well known, global conformal symmetry fixes it to be
\bea
<\phi_m(z_1,\bar z_1)\phi_n(z_2,\bar z_2)> = \left\{\begin{array}{lcr} ~~~~~ 0~~~, ~~~~~~\Delta_m\neq \Delta_n~~or~~ \bar\Delta_m\neq \bar\Delta_n,\cr
                                             ~\frac{C_{mn}}{z_{12}^{2\Delta}\bar z_{12}^{2\bar \Delta}}~,~~~\Delta_m=\Delta_n=\Delta,                                                                           \bar\Delta_m=\bar\Delta_n=\bar\Delta.\end{array} \right.
\eea 

$C_{mn}$ is simply the normalization of the primary states defined as the asymptotic states generated by the primary fields, $i.e$
\bea
|\Delta_n>&=&\phi_n(0,0)|0>~,\cr
<\Delta_n|&=&\lim_{z,\bar z\rightarrow\infty} z^{2\Delta_n}\bar z^{2\bar\Delta_n}<0|\phi_n(z,\bar z)~.
\eea

Since correlators are invariant under field permutations, $C_{mn}$ must be symmetric and so it is always possible to find a basis where $C_{mn}$ is diagonalizable and, if desired, simply a Kronecker delta \footnote{Or a combination of Kronecker and Dirac delta functions in the case of continuous conformal families.} when the fields are properly normalized. In order to avoid confusion, the reader has to bear in mind that (for historical reasons) this will not be the convention we will be using when we refer to the $AdS_3$ WZNW model. In order to change to a self conjugate basis we need the field redefinition, $\phi^{\pm}_m=(\delta_m{}^n\pm {\cal C}_m{}^n)\phi_n$, ${\cal C}$ being the charge conjugation matrix, followed by the appropriate  normalization.
\bigskip

Concerning higher point correlation functions, global conformal symmetry forces three point functions of primary fields to be
\bea
<\phi_m(z_1,\bar z_1)\phi_n(z_2,\bar z_2)\phi_p(z_3,\bar z_3)> =\frac{C_{mnp}}{\prod_{i<j}z_{ij}^{\Delta_{ij}}~\bar z_{ij}^{\bar \Delta_{ij}}},\label{3pf}
\eea 
where $\Delta_{ij}=\Delta_i+\Delta_j-\Delta_k,\, k\neq i,j$. $C_{mnp}$, not determined by global conformal symmetry, is the so called three point function coefficient determining the structure constants. As commented above the coordinate dependence of the four point function is not completely fixed. These are given by
\bea
<\phi_{m_1}(z_1,\bar z_1)\dots \phi_{m_4}(z_4,\bar z_4)> =\frac{G^{21}_{34}(z,\bar z)}{\prod_{i<j}z_{ij}^{\Delta_i+\Delta_j-\frac\Delta 3}~\bar z_{ij}^{\bar \Delta_i+\bar \Delta_j-\frac{\bar\Delta}{3} }},
\eea 
where $z$ is the anharmonic ratio defined as $x$ in (\ref{anratios}) and $\Delta=\sum_{i=1}^4 \Delta_i$.

It is interesting to note that global conformal transformations can be used to fixed $z_1\rightarrow\infty,z_2=1,z_4=0$ and so $z_3=z$. So that the knowledge of the correlation function in these points completely fixes the four point function with primary fields inserted in any other points. Indeed $f(z,\bar z)$, which completely determines the correlator, can be alternatively defined as
\bea
G^{21}_{34}(z,\bar z)=\lim_{z'_1,\bar z'_1\rightarrow\infty} z'_1{}^{2\Delta_1}\bar z'_1{}^{2\bar\Delta_1} <\phi_{m_1}(z'_1,\bar z'_1)\phi_{m_2}(1,1)\phi_{m_3}(z,\bar z)\phi_{m_4}(0,0)>.
\eea 

\section{Operator Product Expansion}\label{opesection}

Scale invariance requires that the Operator Product Expansion (OPE) must have the following structure
\bea
\phi_{m_1}(z,\bar z)\phi_{m_2}(0,0)=\sum_p\sum_{\{k,\bar k\}} C^{p\{k,\bar k\}}_{1 2} z^{\Delta_p-\Delta_1-\Delta_2+K} 
\bar z^{\bar \Delta_p - \bar \Delta_1 - \bar \Delta_2 + \bar K} \phi_p^{\{k,\bar k\}}(0,0)\label{generalope}
\eea
where $K=\sum k_i$ and $\{k_i\}$ denotes an arbitrary collection of non negative integers $\{k_1,k_2,\dots k_N\}$, such that $\phi_p^{\{k,\bar k\}}(z,\bar z)$ denotes the descendant field 
\bea
\phi_p^{\{k,\bar k\}}(z,\bar z)=L_{-k_1}\dots L_{-k_N}\bar L_{-\bar k_1}\dots\bar L_{-\bar k_{\bar N}}\phi_p(z,\bar z)
\eea 

So the operator algebra is determined when all the conformal weights, $\Delta_p$, and OPE coefficients, $C^{p\{k,\bar k\}}_{1 2}$, are known.  

In order to get more information on the later, let us consider the asymptotic behavior of the following particular three point function of primary fields
\bea
<\Delta_r|\phi_{m_1}(z,\bar z)|\Delta_{m_2}>&=& lim_{w,\bar w \rightarrow\infty} w^{2\Delta_r} 
\bar w^{2\bar \Delta_r} <\phi_{r}(w,\bar w) \phi_{m_1}(z,\bar z)\phi_{m_2}(0,0)>\cr
&=& \frac{C_{rm_1 m_2}}{z^{\Delta_1+\Delta_2-\Delta_r} \bar z^{\bar \Delta_1+\bar \Delta_2-\bar \Delta_r}}
\eea 

Then, after inserting the OPE (\ref{generalope}) and using the orthogonality in the Verma module we conclude 
\bea
C_{12}^p\equiv C_{12}^{p,\{0,0\}}=C_{pm_1m_2}\label{struccte}
\eea
so that the leading contribution in the operator algebra is simply given by the three point function of primary fields.\footnote{In the convention we will use for primary fields of the $AdS_3$ WZNW model, the fields will not be self conjugate and so (\ref{struccte}) must be replaced by
\bea
C_{12}^p\equiv C_{12}^{p,\{0,0\}}=C_{p^*m_1m_2},
\eea
where $p^*$ denotes the representation conjugate to the one labeled by $p$.}

Following similar steps with insertion of descendant fields, one can determine the other coefficients. The correlations with descendant fields are obtained form those of the primaries by repeatedly applying Virasoro generators, and they are non vanishing only if the correlator with highest weight is not vanishing. So, it is natural to propose the factorization ansatz
\bea
C_{12}^{p,\{k,\bar k\}}=C_{12}^p~ \beta_{12}^{p\{k\}}~ \bar \beta_{12}^{p\{\bar k\}}.
\eea

The coefficients $\beta^{p\{k\}}_{12}$ can be recursively determined as functions of conformal weights and the central charge. Let us show how it works.

Of course, by definition, $\beta_{12}^{\{0\}}=1$. To find the other coefficients, notice that 
\bea
\phi_1(z,\bar z)|\Delta_2,\bar \Delta_2>&=& \phi_1(z,\bar z)\phi_2(0,0)|0>\cr
                              &=& \sum_p z^{\Delta_p-\Delta_1-\Delta_2}\bar z^{\bar \Delta_p-\bar \Delta_1-\bar \Delta_2} |z,\Delta_p>|\bar z,\bar \Delta_p>,\label{expansion}
\eea
where 
\bea
|z,\Delta_p>&=&\sum_{\{k\}}z^k\beta_{12}^{p\{k\}}L_{-k_1}\dots L_{-k_N}|\Delta_p>\cr
       &\equiv&\sum_{N=0}^{\infty}z^N |N,\Delta_p>
\eea
Applying the Virasoro mode $L_n$ ($n>0$) on (\ref{expansion}) and using the commutator of Virasoro modes with primary fields (\ref{cVirPhi})
\bea
L_n\phi_1(z,\bar z)|\Delta_2,\bar \Delta_2>&=& [L_n,\phi_1(z,\bar z)]|\Delta_2,\bar \Delta_2>\cr
                                 &=& \left[z^{n+1}\partial_z+z^n(n+1)\Delta_1\right]\phi_1(z,\bar z)|\Delta_2,\bar \Delta_2>\label{opevir}
\eea
one obtains
\bea
L_n|N,\Delta_p> &=& ~~~~~~~0~~~,~~~~~~~~~~~~~~~~~~~~~~~~~~~~~~~~~~~~~~~~~~~~ N=1,2,\dots,n-1\cr
L_n|N,\Delta_p> &=& (n\Delta_1+\Delta_p-\Delta_2+N-n)|N-n,\Delta_p>, ~~~N=n,n+1,\dots~~. 
\eea

The first, non trivial, case is $|N=1,\Delta_p>=\beta_{12}^{p\{1\}}L_{-1}|\Delta_p>$, satisfying
\bea
L_1|1,\Delta_p>&=& (\Delta_p+\Delta_1-\Delta_2)|\Delta_p>\cr
          &=& \beta_{12}^{p\{1\}}[L_1,L_{-1}]|\Delta_p>\cr
          &=&  2\Delta_p \beta_{12}^{p\{1\}}|\Delta_p>
\eea
from which 
\bea
\beta_{12}^{p\{1\}}=\frac{\Delta_p+\Delta_1-\Delta_2}{2\Delta_p}.\label{beta1}
\eea

At the next level, $|2,\Delta_p>=\beta_{12}^{p\{1,1\}}L_{-1}^2|\Delta_p>+\beta_{12}^{p\{2\}}L_{-2}|\Delta_p>$. Now, acting with $L_1,$ and $L_2$ on $|2,\Delta_p>$ and using the relations
\bea
[L_1,L_{-1}^2]&=&4L_{-1}L_0 + 2L_{-1}~,\cr
[L_1,L_{-2}]&=&3L_{-1}~,\cr
[L_2,L_{-1}^2]&=&6L_{-1}L_1 + 6L_{0}~,\cr
[L_2,L_{-2}]&=&4L_0 + \frac c{12}~,
\eea   
one finds
\bea
\beta_{12}^{p\{1,1\}}&=&\frac{18 \Delta_p(\Delta_p+2\Delta_1-\Delta_2)-3(\Delta_p+\Delta_1-\Delta_2+1)(\Delta_p+\Delta_1-\Delta_2)(4\Delta_p+\frac c2)}
                           {6 \Delta_p\left[18 \Delta_p+(4\Delta_p+\frac c2)(2+4\Delta_p)\right]},\cr\cr
\beta_{12}^{p\{2\}}&=&\frac{(2+4\Delta_p)(\Delta_p+2\Delta_1-\Delta_2)+3(\Delta_p+\Delta_1-\Delta_2+1)(\Delta_p+\Delta_1-\Delta_2)}
                           {18 \Delta_p+(4\Delta_p+\frac c2)(2+4\Delta_p)}.
\eea
\medskip

At level $N$, one has $P(N)$ different coefficients $\beta_{12}^{p\{k\}}$, but this is exactly the number of ways $|N,\Delta_p>$ can be brought to zero level by acting with Virasoro modes, $L_{n},~n=1,2,\dots,N$. The number of unknowns equals the number of equations and so the problem admits solution.

Notice that knowledge of $C_{mnp}$ and the $\beta^{p\{k\}}_{12}$ coefficients determines the three point functions containing arbitrary descendant fields.

The knowledge of the coefficients of the three point functions of primary fields, $C_{mnp}$, as well as the knowledge of the conformal weights, $\Delta_n$, and the central charge, $c$, completely determines the operator algebra, and with it any higher order correlation function can be written, at least in a factorized form, in terms of three point functions. In this sense the theory is completely determined once the central charge, conformal weights and three point function coefficients of primary fields are known. We will come back to this issue in section \ref{bootapp}.
\bigskip

In the case of a WZNW model, the coefficients of the correlation functions also depend on parameters labeling states of representations of the algebra. Nevertheless, new constraints must be satisfied. For instance, from the invariance of the correlation functions under the action of the group G of the WZNW model the following Ward identity must be satisfied
\bea
\sum_{i=1}^n t^a_i<\phi_{m_1}(z_1)\dots\phi_{m_n}(z_n)>=0,
\eea
where the index $i$ in $t^a_i$ denotes action on $\phi_{m_i}$. $E.g.$ invariance under the action of $t^3$ in WZNW models with $sl(2)$ symmetry requires the sum of the eigenvalues of $J_0^3$ to vanish in any correlation function.  

The combination of the Ward identities with the Sugawara construction gives rise to a partial differential equation to be satisfied by the correlators. It plays a fundamental role in WZNW models and is usually referred to as the Knizhnik-Zamolodchikov (KZ) equation: 
\bea
\left(\partial_{z_i}+\frac1{k+{\mathfrak g}_c} \sum_{j\neq i} \frac{\sum_a t_i^a\otimes t_j^a}{z_{ij}}\right) <\phi_{m_1}(z_1)\dots \phi_{m_n}(z_n)>=0.
\eea

We will use it when considering the factorization of four point functions (see section \ref{kz} ). The reason why this is particularly important in the case of four point functions is because in this case the partial differential equation becomes an ordinary differential equation, since only one degree of freedom (the anharmonic ratio) is not fixed by global conformal symmetry.

Extra constraints follow with the introduction of null states in the correlators. Let us suppose that the descendant field generated by the action of certain chain of generators on a primary field is a null vector. The correlation function with such a null field vanishes but by using the Wick theorem such a correlator can be expressed in terms of correlators where the currents act on the other primary fields giving rise to differential equations for the correlators. In fact, the usage of the {\it fusion relations} in four point functions with degenerate fields ($i.e$, primary fields with null descendants) was the key to solve the structure constants of the Liouville theory \cite{DO,ZZ} as well as the ones of the $H_3^+$ model \cite{tesch1}.

\section{The bootstrap approach}\label{bootapp}

As we commented in section \ref{seccorr}, coordinates in 4-point functions can be set to $(\infty,1,x,0)$, $x$ being the anharmonic ratio defined in (\ref{anratios}). Notice that $G^{21}_{34}(x,\bar x)$ defined in last section can also be written as
\bea
G^{21}_{34}(x,\bar x)=<\Delta_1,\bar \Delta_1|\phi_2(1,1)\phi_3(x,\bar x)|\Delta_4,\bar \Delta_4>
\eea

After inserting the operator algebra (\ref{generalope}) one ends with 
\bea
G^{21}_{34}(x,\bar x)=\sum_p C^p_{34}C^p_{12}A^{21}_{34}(p|x,\bar x),\label{4pfact}
\eea
where we introduced the {\it partial waves}, $A^{21}_{34}(p|x,\bar x)$ 
\bea
A^{21}_{34}(p|x,\bar x) &=& \sum_p (C^p_{12})^{-1} x^{\Delta_p-\Delta_3-\Delta_4}\bar x^{\bar \Delta_p-\bar \Delta_3-\bar \Delta_4}<\Delta_1,\bar \Delta_1|\phi_2(1,1)\psi_p(x,\bar x|0,0)|0>\cr
                        &=& {\cal F}^{21}_{34}(p|x) ~\bar{\cal F}^{21}_{34}(p|\bar x), 
\eea
wherein 
\bea
\psi_p(x,\bar x|0,0)=\sum_{\{k,\bar k\}}\beta^{p\{k\}}_{34}~\bar \beta^{p\{\bar k\}}_{34}~x^K \bar x^{\bar K} \phi_p^{\{k, \bar k\}}(0,0),
\eea
the {\it conformal blocks} are given by 
\bea
{\cal F}^{21}_{34}(p|x)=x^{\Delta_p-\Delta_3-\Delta_4}\sum_{\{k\}}\beta_{34}^{p\{k\}} x^K \frac{<\Delta_1|\phi_2(1)L_{-k_1}\dots L_{-k_N}|\Delta_p>}{<\Delta_1|\phi_2(1)|\Delta_p>}
\eea
and similarly for the antiholomorphic ones.

As will be clear below the purpose of factorizing four point functions as in (\ref{4pfact}) is because the {\it conformal blocks} are completely determined when conformal weights and the central charge are known.

The general explicit expression for {\it conformal blocks} is not known. Just in a few cases, ($e.g.$ minimal models) one has a closed expression. The strategy is to expand it in power series 
\bea
{\cal F}^{21}_{34}(p|x)=x^{\Delta_p-\Delta_3-\Delta_4}\sum_{\ell=0}^{\infty}{\cal F}_{\ell} x^\ell,
\eea 
where obviously ${\cal F}_0=1$. The next coefficient, ${\cal F}_1$ is
\bea
{\cal F}_1&=& \beta_{34}^{p\{1\}}\frac{<\Delta_1|\phi_2(1)L_{-1}|\Delta_p>}{<\Delta_1|\phi_2(1)|\Delta_p>}\cr\cr
          &=& \frac{(\Delta_p+\Delta_2-\Delta_1)(\Delta_p+\Delta_3-\Delta_4)}{2\Delta_p},
\eea
where $\beta^{p\{1\}}_{12}$ was computed in  (\ref{beta1}) and the commutator of a primary with a Virasoro mode (see (\ref{cVirPhi}) ) was used. 

The situation becomes rapidly tedious for higher coefficients. For instance ${\cal F}_2$ is an involved combination of conformal weights and the central charge, which is three lines long.

The invariance of correlation functions under field permutations imposes a series of relations over {\it conformal blocks}. For instance, after the conformal transformation $z\rightarrow z^{-1}$ one ends with 
\bea
\sum_p C^p_{21}~ C^p_{34}~ {\cal F}^{21}_{34}(p|x)~\bar {\cal F}^{21}_{34}(p|\bar x)=
\sum_q C^q_{24}~ C^q_{31}~ {\cal F}^{24}_{31}\left(q|x^{-1}\right)~ \bar {\cal F}^{24}_{31}\left(q|\bar x^{-1}\right).
\eea 

On the other hand, after $z\rightarrow 1-z$, one finds 
\bea
\sum_p C^p_{21}~ C^p_{34}~ {\cal F}^{21}_{34}(p|x)~\bar {\cal F}^{21}_{34}(p|\bar x)=
\sum_q C^q_{41}~ C^q_{32}~ {\cal F}^{41}_{32}(p|1-x)~\bar {\cal F}^{41}_{32}(p|1-\bar x).
\eea 

These conditions are a realization of the {\it crossing symmetry} and are usually referred to as the {\it bootstrap equations}, being the sole dynamical input required to solve the theory, as they can, in principle, be exploited to find the three point function coefficients, $C_{mnp}$ and the conformal dimensions, $\Delta_n$. 

Indeed, let us suppose that the {\it conformal blocks} are known for generic conformal weights, and suppose that the theory we are interested in has $N$ conformal families, $e.g.$ a RCFT, then we have $N^3$ ($C_{mnp}$) $+$  $N$ ($\Delta_n$) unknowns. These have to be  contrasted with the $N^4$ conditions which follow from a naive counting. 
There is no proof that the problem admits solution in a generic case, but in many situations, $e.g$ the minimal models, the bootstrap equations were completely solved. This road to solve the three point function coefficients and the conformal weights is usually denoted as the {\it bootstrap aproach}. This method was developed in the seminal work \cite{BPZ} and the interested reader will find there the application to some interesting examples.

\chapter{Analytic structure of $W_1$}\label{A:asw}

The purpose of this appendix is to study the analytic structure of
 $W_1$.
 In particular, we are specially
interested in possible
 zeros appearing in $W_1$ which are not evident in the expression
(\ref{w1}),
 but are very important in our definition of the OPE.

Let us recall some useful identities relating
different expressions for $G\left[\begin{matrix} a,b,c\cr
e,f\cr\end{matrix}\right]$ \cite{slater},
\bea
G\left[\begin{matrix} a,b,c\cr e,f\cr\end{matrix}\right]=
\frac{\Gamma(b)\Gamma(c)}{\Gamma(e-a)\Gamma(f-a)}
G\left[\begin{matrix} e-a,f-a,u\cr
u+b,u+c\cr\end{matrix}\right],\label{idG1}
\eea
\bea
G\left[\begin{matrix} a,b,c\cr e,f\cr\end{matrix}\right]=
\frac{\Gamma(b)\Gamma(c)\Gamma(u)}{\Gamma(f-a)\Gamma(e-b)\Gamma(e-c)}
G\left[\begin{matrix} a,e-b,e-c\cr
e,a+u\cr\end{matrix}\right]\, ,\label{idG2}
\eea
where $u$ is defined
as $u=e+f-a-b-c$. Using the permutation symmetry among $a,b,c$ and
$e,f$, which is evident from the series representation of the
hypergeometric function ${}_3F_2$, seven new
identities may be generated. In what follows we use these
identities in order to obtain the greatest possible amount of
information on $W_1$. 

Consider for instance $C^{12}$
defined in (\ref{Wfull}). Using
(\ref{idG1}), it can be rewritten for $j_1=-m_1+n_1$, with $n_1$ a non
negative 
integer, as
\bea
C^{12}&=&\frac{\Gamma(-N)\Gamma(-j_{13})\Gamma(-j_{12})\Gamma(1+j_2+m_2)}
{\Gamma(-j_3-m_3)}\cr
&\times&\sum_{n=0}^{n_1} \left(\begin{matrix}
n_1\cr n\cr\end{matrix}\right) \frac{(-)^n}{\Gamma(n-2j_1)}
\frac{\Gamma(n-j_{12})}{\Gamma(-j_{12})}\frac{\Gamma(n+1+j_{23})}{\Gamma(1+j_{23})}
\frac{\Gamma(1+j_3-m_3)}{\Gamma(1+j_3-m_3-n_1+n)}.
\label{c12-1}
\eea

Using (\ref{idG2}) instead of (\ref{idG1}), one finds an expression
for $C^{12}$ equal to (\ref{c12-1}) with  $j_3\rightarrow-1-j_3$.

There is a third expression in which 
$C^{12}$ can be written as a finite sum for generic $j_2,j_3$. This follows
from (\ref{Wfull}), using the identity obtained from (\ref{idG2})
with $(e\leftrightarrow f)$. This expression is explicitly
invariant under $j_3\rightarrow-1-j_3$.

Consider for instance 
(\ref{c12-1}). All quotients inside the sum are such that the arguments
in the $\Gamma-$functions of the denominator
equal those in the numerator up to a positive integer,
except for the one with
$\Gamma(n-2j_1)$ which is regular and non vanishing for ${\rm
Re}\,j_1<-\frac 12$.
Then, each
quotient is separately regular. Eventually, some
of them may vanish, but not for all values of $n$. In particular, for
$n=0$ the first two quotients equal one. The last factor may vanish for
$n=0$, but for $n=n_1$ it equals  one.
However, particular configurations of $j_i$, $m_i$
may occur such that one of
the first two quotients vanishes for certain  values of $n$,
namely $n=n_{min}, n_{min}+1,\dots,n_1$, and the last one vanishes
 for other special values, namely $n=0,1,\dots,n_{max}$.
Thus, if $n_{max}\geq n_{min}$, all terms in the sum cancel and
$C^{12}$ vanishes as a simple zero.
   In fact, let us consider for instance  both $1+j_{23}=-p_3$ and
   $1+j_3-m_3=1+n_3$,
with $p_3,n_3$  non negative integers. This requires
$\Phi^{j_2,w_2}_{m_2,\overline m_2}\in {\cal D}^{-,w_2}_{j_2}$
and $j_3=j_1-j_2-1-p_3=m_3+n_1-n_2-1-p_3$, which impose $p_3<n_1$ and
allow to rewrite
the sum in (\ref{c12-1}) as
\bea
\sum_{n=0}^{p_3} \frac{1}{n!}\frac{n_1!}{(n_1-n)!}\frac{p_3!}{(p_3-n)!}
\frac{\Gamma(n-j_{12})}{\Gamma(-j_{12})}\frac{1}{\Gamma(n-2j_1)}
\frac{n_3!}
{\Gamma(1+n_3-n_1+n)}\, .\label{c4}
\eea

Finally, taking into account that $1+n_3-n_1+n=-n_2-(p_3-n)\leq0$,
for $n=0,1,\dots,p_3$, the sum vanishes as a simple zero.
A similar analysis for $j_{12}=p_3\geq0$ and
$1+j_3-m_3=1+n_3\geq1$ shows that
no zeros appear in this case when $\Phi^{j_2,w_2}_{m_2,\overline m_2}$
is the spectral flow image of a primary field.

From the expression obtained for $C^{12}$ by changing $j_3\rightarrow-1-j_3$, 
one finds zeros again for
$\Phi^{j_2,w_2}_{m_2,\overline m_2}\in {\cal D}^{-,w_2}_{j_2}$. These  appear
when both
$j_{3}=j_2-j_1+p_3$ and $j_3=-m_3-1-n_3$ hold simultaneously.

Finally, repeating the analysis for the sum in the third expression
for $C^{12}$, $i.e.$ that explicitly symmetric under $j_3\rightarrow
-1-j_3$,
 one finds the same zeros as in the
previous cases.

Let us now consider the analytic structure of $W_1:=D_1C^{12}\overline
C{}^{12}$.
Expression (\ref{w1}) together with the  discussions above allow to rewrite  $W_1$ as
\bea
W_1(j_i;m_i,\overline m_i )=
\frac{(-)^{m_3-\overline m_3+\overline n_1}\pi^2 \gamma(-N)
}{\gamma(-2j_1)\gamma(1+j_{12})\gamma(1+j_{13})}\frac{\Gamma(1+j_2+m_2)}
{\Gamma(-j_2-\overline m_2)}
\frac{\Gamma(1+j_3+m_3)}{\Gamma(-j_3-\overline m_3)}
E_{12}\overline E{}_{12}\, ,
\eea
where
$E_{12}$ is given by $\Gamma(-2j_1)$ times (\ref{c4}).
$E_{12}$ has
no poles but it may vanish for certain
special configurations if
$\Phi^{j_2,w_2}_{m_2,\overline m_2}\in {\cal D}^{-, w_2}_{j_2}$,
namely $n_2<n_1-p_3$ and $j_3=m_3+n_3$ or
$j_3=-m_3-1-n_3$, with $n_3=0,1,2,\dots$, where $p_3=-1-j_{23}$ 
in the former and $p_3=j_{13}$ in the latter. The same result 
applies to  $\overline E{}_{12}$,  changing
$n_i$ by $\overline n_i$. Obviously one might find, using
other identities, new zeros for special configurations. This
could be a difficult task, because the series does not 
reduce to a finite sum in general. Fortunately,
it is not necessary for our purposes.

\chapter{The Lorentzian torus}\label{A:tlt}

In this appendix we present a description of
 the moduli space of the torus with  Lorentzian  metric\footnote{Tori in $1+1$ dimensions have 
been considered previously in 
\cite{moore} - \cite{russo} in the context of string propagation 
in time dependent 
backgrounds.}. Although it can be easily obtained from the 
Euclidean case, we include it here for completeness. 

Consider the two dimensional torus with 
worldsheet coordinates $\sigma^1, \sigma^2$ 
obeying the identifications
\begin{equation}
(\sigma^1,\sigma^2)\cong(\sigma^1+2\pi n,\sigma^2+2\pi m),~~~~~~n,m
\in{\mathbb Z}. \label{pertoro}
\end{equation}
By diffeomorphisms and Weyl transformations that leave invariant the 
periodicity, 
a general two dimensional Lorentzian metric can be taken to the form
\begin{equation}
ds^2=(d\sigma^1+\tau_+d\sigma^2)(d\sigma^1+\tau_-d\sigma^2),\label{gToroLor}
\end{equation}
where
$\tau_+,\tau_-$ are two real independent parameters.
Recall that the metric of the Euclidean torus, 
namely $ds^2=|d\sigma^1+\tau d\sigma^2|^2$, is
degenerate for $\tau\in\mathbb R$ since $det~g=(\tau-\tau^*)^2$. 
In contrast, here it is degenerate for $\tau_-=\tau_+$.

The linear transformation
\begin{eqnarray}
\tilde\sigma^1=\sigma^1+\tau^+\sigma^2,\quad
\tilde\sigma^2=\tau^-\sigma^2\, , \qquad \tau^{\pm}=\frac{\tau_-\pm\tau_+}{2},
\end{eqnarray}
takes (\ref{gToroLor}) to the Minkowski metric.
The new coordinates obey the periodicity conditions
\begin{equation}
(\tilde\sigma^1,\tilde\sigma^2)
\cong(\tilde\sigma^1+2\pi n+2\pi m\tau^+,\tilde\sigma^2+2\pi 
\tau^-m)\, ,~~~~~~n,m\in{\mathbb Z}\, ,\label{pertoroMink}
\end{equation}
while the light-cone coordinates
$\tilde \sigma_{\pm}=\tilde\sigma^1\pm\tilde\sigma^2$, obey
\begin{equation}
\tilde \sigma_{\pm}\cong\tilde \sigma_{\pm}+2\pi n+2\pi m\tau_{\mp}\, .
\end{equation}

 In the Euclidean case, 
there are in addition
 global transformations that cannot be smoothly connected to the 
identity, generated by Dehn twists. 
A twist along the $a$ cycle of a Lorentzian torus
preserves the metric (\ref{gToroLor}) but
changes the periodicity  to
\begin{equation}
(\tilde\sigma^1,\tilde\sigma^2)\cong(\tilde\sigma^1+2\pi n+2
\pi m(1+\tau^+),\tilde\sigma^2+2\pi m~\tau^-)\, ,~~~~~~n,m\in{\mathbb Z},
\end{equation}
or
\begin{equation}
\tilde \sigma_{\pm}
\cong \tilde \sigma_{\pm}+2\pi n+2\pi m~(\tau_{\mp}+1)\, .
\end{equation}
Thus it gives a torus with modular parameters
 $(\tau_+',\tau_-')=(\tau_++1,\tau_-+1)$.
A twist along the $b$ cycle leads to the following periodicity conditions
\begin{equation}
(\tilde\sigma^1,\tilde\sigma^2)\cong(\tilde\sigma^1+2\pi n(1+\tau^+)+2\pi 
m~\tau^+,\tilde\sigma^2+2\pi n~\tau^-+2\pi m~\tau^-),~~~~~~n,m\in{\mathbb Z},
\end{equation}
or
\begin{equation}
\tilde \sigma_{\pm}\cong\tilde \sigma_{\pm}+2\pi 
n(1+\tau_{\mp})+2\pi m~\tau_{\mp}\, .
\end{equation}
As in the Euclidean case, this is equivalent to a torus with 
$(\tau_+',\tau_-')=(\frac{\tau_+}{\tau_++1},\frac{\tau_-}{\tau_-+1})$ 
and conformally flat metric. But there is a crucial difference. In the 
Euclidean case, the overall conformal factor multiplying the flat metric 
is positive definite, namely $\frac{1}{(1+\tau)(1+\tau^*)}$. 
On the contrary, in the 
Lorentzian torus, the conformal factor $\frac{1}{(1+\tau_-)(1+\tau_+)}$ is not 
positive definite and so, it cannot be generically eliminated through a
Weyl transformation.

Defining the modular $S$ transformation as 
$S\tau_{\pm}=-\frac{1}{\tau_{\pm}}$, we can write
$\tau'_{\pm}=\frac{\tau_\pm}{1+\tau_\pm}=TST~ \tau_\pm$, and then
the problem can be reformulated in the following way. 
The  $T$ transformation works as in the Euclidean case. 
Instead, under a modular $S$ transformation, the torus 
defined by (\ref{pertoro}) and (\ref{gToroLor}) 
is equivalent to a torus with the same 
periodicities but with the following metric (after diffeomorphisms and
Weyl rescaling)   
\bea
ds^2=sgn(\tau_-\tau_+)\left(d\sigma '^1+\tau_+d\sigma'^2\right)
\left(d\sigma '^1+\tau_-d\sigma '^2\right)\, .
\eea

\medskip

\chapter{The mixing block of the $S$ matrix}\label{A:mb}

In this appendix we sketch the computation of the off-diagonal
 block of the $S$ matrix mixing the characters of 
continuous and discrete representations.

\medskip
{\textbf{\textit{ A useful identity}}}
\medskip
 
It is convenient to begin displaying a useful identity.

Let 
$h(x;\epsilon_0)=\frac{1}{1-e^{2\pi i (x+i\epsilon_0)}}$, with
$x\in {\mathbb R}$,  be 
the distribution defined as the weak limit $\epsilon_0\rightarrow0$ 
and  $G(x;\epsilon_1,\epsilon_2,\epsilon_3,\dots)$ a generalized function 
having simple poles outside of the real line\footnote{
$G(x;0,0,0,\dots)$ not necessarily has only simple poles. 
In the most general case, it will have poles of arbitrary order.},
defined as the weak limit $\epsilon_i\rightarrow0, i=1,2,3,...$.  
The non vanishing infinitesimals 
 $\epsilon_i$ are allowed to depend on the $x$ coordinate
and they all differ from each other
in an open set around each simple pole. Then, the following identity holds 
(in a distributional sense):
\bea
&&\frac{1}{1-e^{2\pi i (x+i\epsilon_0)}} G(x;\epsilon_1,\epsilon_2,
\epsilon_3,\dots)~~=~~
\frac{1}{1-e^{2\pi i (x+i\tilde\epsilon_0)}} G(x;\epsilon_1,\epsilon_2,
\epsilon_3,\dots)\cr\cr
&&\ \ \ \ \ \ \ \ \ \ \ \ \ \ \ \ \ \ \ \ \ \ \ 
+~~\sum_{x_i^\downarrow}\delta(x-x_i^\downarrow) G(x;\epsilon_1-\epsilon_0,
\epsilon_2-\epsilon_0,\epsilon_3-\epsilon_0,\dots)\cr
&& \ \ \ \ \ \ \ \ \ \ \ \ \ \ \ \ \ \ \ \ \ \ \ 
-~~\sum_{x_i^\uparrow}\delta(x-x_i^\uparrow) G(x;\epsilon_1-\epsilon_0,
\epsilon_2-\epsilon_0,\epsilon_3-\epsilon_0,\dots),\label{identdeltas}
\eea
where $\tilde\epsilon_0$ is a new infinitesimal parameter, 
$x_i^\downarrow$ ($x_i^\uparrow$) is the real part of the pulled down (up)
poles, $i.e.$ those poles where 
$\epsilon_0(x_i^\downarrow)<0<\tilde\epsilon_0(x_i^\downarrow)$ 
($\tilde\epsilon_0(x_i^\uparrow)<0<\epsilon_0(x_i^\uparrow)$). 
Of course, here $x_i^\downarrow,\,x_i^\uparrow\in{\mathbb Z}$, 
but 
(\ref{identdeltas}) can be trivially generalized to other 
functionals 
 having simple poles, the only change being that the residue
has to multiply each 
delta function. 

The proof of this identity 
follows from multiplying (\ref{identdeltas}) by an arbitrary test 
function ($f(x)\in C_0^{\infty}$) and integrating  over the real line.

As an example, let us consider the simplest case $G=1$, $\epsilon_0=0^+$, 
$\tilde\epsilon_0=0^-$, where one recovers the well known formula
\bea
\frac{1}{1-e^{2\pi i(x+i0^+)}}=\frac{1}{1-e^{2\pi i(x+i0^-)}}
-\sum_{m=-\infty}^{\infty}\delta(x+m)\,.
\eea

{\textbf{\textit{The mixing   
block}}}
\medskip

Let us first consider the modular transformation of the
 elliptic theta function
\bea
\frac{1}{i\vartheta_{11} (\theta+i\epsilon_2^w,\tau+i\epsilon_1)}
&\rightarrow &
\frac{1}{i\vartheta_{11} (\frac\theta\tau+i\epsilon_2^w,
-\frac1\tau+i\epsilon_1)}\equiv
\frac{1}{i\vartheta_{11} (\frac{\theta+i\epsilon'_2{}^w}{\tau+i\epsilon'_1},
-\frac{1}{\tau+i\epsilon'_1})}\cr
&=&\frac{-sgn(\tau)e^{-\pi i \frac{\theta^2}{\tau}}e^{-sgn(\tau)i\frac\pi4}}
{\sqrt{|\tau|}}
\frac{1}{\vartheta_{11} (\theta+i\epsilon'_2{}^w,\tau+i\epsilon'_1)}, 
\label{thetatrans}\\
&&\left\{\begin{array}{lcr}
~\epsilon'_1=\tau^2\epsilon_1~,\cr
\epsilon'_2{}^w=\tau\left(\epsilon_2^w+\theta\epsilon_1\right)~,
\end{array}\right.
\eea
and $\epsilon_1,\epsilon_2^w$ satisfy (\ref{epsiloncond}).
The identity (\ref{stheta}) was used in the last line of (\ref{thetatrans})
and the limits $\epsilon'_1,\epsilon'_2{}^w\rightarrow0$ 
were taken where  it is allowed.  

Let us now concentrate on the last term in (\ref{thetatrans}). 
It is explicitly given by  (\ref{theta11b}), 
where now the $\epsilon$'s are replaced by $\epsilon'_1,\epsilon'_3{}^{n,w},
\epsilon'_4{}^{n,w}$ satisfying $\epsilon'_1>0$,
\bea
\epsilon'_3{}^{n,w}\left\{\begin{array}{lcr}
<0\,~,\theta-n\tau\leq-1-w\cr
>0\,,~\theta-n\tau\geq-w
\end{array}\right.,~~~~~
\epsilon'_4{}^{n,w}\left\{\begin{array}{lcr}
<0\,~,\theta+n\tau\geq-w\cr
>0\,,~\theta+n\tau\leq-1-w
\end{array}\right. \, ,\ \ \ \tau<0 ,\\
\epsilon'_3{}^{n,w}\left\{\begin{array}{lcr}
<0\,~,\theta-n\tau\geq-w\cr
>0\,,~\theta-n\tau\leq-1-w
\end{array}\right. ,~~~~~
\epsilon'_4{}^{n,w}\left\{\begin{array}{lcr}
<0\,~,\theta+n\tau\geq-w\cr
>0\,,~\theta+n\tau\leq-1-w
\end{array}\right.\, , \ \ \ \ \tau>0\, .
\eea

By comparing with  (\ref{epsilon34})
and using (\ref{identdeltas}), one finds, for instance in the case $w<0,
\tau<0$, after a straightforward but tedious computation, the following 
identity:
\bea
&&\frac{1}{i\vartheta_{11} (\theta + i\epsilon '_2{}^w, \tau +i \epsilon '_1)}
~=~
\frac{1}{i\vartheta_{11} (\theta + i\epsilon_2{}^w, \tau +i \epsilon_1)}\cr
&&\ \ \ \ \
-\frac{1}{\eta^3(\tau+i\epsilon_1)}\left[e^{-i\pi\theta}\sum_{n=0}^{\infty}
\sum_{m=-\infty}^{-w-1}(-)^n e^{\pi i\tau n(1+n)}\delta(-\theta+n\tau+m)
\right. \nonumber\\
&& \ \ \ \ \ \ \ \left.\qquad \qquad~~
+~e^{i\pi\theta}\left(\sum_{n=1}^{-w-1}\sum_{m=w+1}^{\infty}-
\sum_{n=-w}^{\infty}\sum_{m=-\infty}^{w}\right)(-)^n e^{\pi i\tau n(1+n)}
\delta(\theta+n\tau+m)\right]\, .\nonumber
\eea
Repeating the same analysis for the other cases one finds, for arbitrary 
$w$,
\bea
\frac{1}{i\vartheta_{11} (\theta+ i\epsilon'_2{}^w, \tau+i \epsilon'_1)}
&=&\frac{1}{i\vartheta_{11} (\theta+ i\epsilon_2{}^w, \tau+i \epsilon_1)}\cr
&+&
\left[~\sum_{n=-\infty}^{w}\left\{\begin{array}{lcr}
\sum_{m=-\infty}^{w}\delta(\theta-n\tau+m),\tau<0\cr
\sum_{m=1+w}^{\infty}\delta(\theta-n\tau+m),\tau>0
\end{array}\right. \right.\cr
&&-
\left.
\sum_{n=1+w}^{\infty}\left\{\begin{array}{lcr}
\sum_{m=1+w}^{\infty}\delta(\theta-n\tau+m),\tau<0\cr
\sum_{m=-\infty}^{w}\delta(\theta-n\tau+m),\tau>0
\end{array}\right.\right]\frac{(-)^{n+m}e^{2i\pi \tau \frac{n^2}{2}}}
{\eta^3(\tau+i\epsilon_1)}\nonumber
\eea

Using (\ref{gaussian}) and summing or subtracting delta function terms 
like in (\ref{chiw<w'}) and (\ref{chiw>w'}), 
in order to construct the characters of discrete representations,
one finds 
\bea
&&\chi_j^{+,w}(\frac\theta\tau,-\frac1\tau,0)~=~
e^{-2\pi i\frac k4\frac{\theta^2}{\tau}}sgn(\tau)\cr\cr
&&~~~~~~\times~\left\{\sum_{w'=-\infty}^{\infty}
\int^{-\frac{1}{2}}_{-\frac{k-1}{2}}dj'
\sqrt{\frac{2}{k-2}}(-)^{w+w'+1} e^{\frac{4\pi i}{k-2} 
\left(j'+\frac12-w'\frac{k-2}{2}\right)\left(j+\frac12-
w\frac{k-2}{2}\right)} \chi_{j'}^{+,w'}(\theta,\tau,0)\right.\cr\cr
&&~~~~~~
+\sum_{w',n,m~\in~ {\cal I(\tau)}}~ \int^{-\frac{1}{2}}_{-\frac{k-1}{2}}dj'
\sqrt{\frac{2}{k-2}}(-)^{w+1} e^{\frac{4\pi i}{k-2} \left(j'+\frac12-w'\frac{k-2}{2}\right)\left(j+\frac12-
w\frac{k-2}{2}\right)}  \cr\cr
&&~~~~~~~~~~~~~~
\times ~\left.\frac{e^{-\frac{2\pi i}{k-2}\tau(j'+\frac12-w'
\frac{k-2}{2})^2} 
e^{-2\pi i\theta(j'+\frac12-w'\frac{k-2}{2})}}{\eta^3(\tau+i\epsilon_1)} 
(-)^{n+m} e^{2\pi i\tau \frac{n^2}{2}}
\delta(\theta-n\tau+m)\right\}\, ,\nonumber
\eea
where $\sum_{w',n,m~\in~ {\cal I(\tau)}}$ is expected to reproduce 
the contribution from the 
continuous representations and is explicitly given by
\bea
\sum_{w',n,m~\in~ {\cal I(\tau)}} &\equiv&-\sum_{w'=-\infty}^{w-1}
\sum_{n=1+w'}^{w}\sum_{m=-\infty}^{\infty}+\sum_{w'=1+w}^{\infty}
\sum_{n=1+w}^{w'}\sum_{m=-\infty}^{\infty}\cr\cr
&&+~\sum_{w'=-\infty}^{\infty}\left(
\sum_{n=-\infty}^{w}\left\{\begin{array}{lcr}
\sum_{m=-\infty}^w\cr
\sum_{m=1+w}^\infty
\end{array}\right. -
\sum_{n=1+w}^{\infty}\left\{\begin{array}{lcr}
\sum_{m=1+w}^\infty\cr
\sum_{m=-\infty}^w
\end{array}\right.\right)\cr\cr
&=&\sum_{w'=-\infty}^\infty\left(
\sum_{n=-\infty}^{w'}\left\{\begin{array}{lcr}
\sum_{m=-\infty}^w\cr
\sum_{m=1+w}^\infty
\end{array}\right. -
\sum_{n=1+w'}^{\infty}\left\{\begin{array}{lcr}
\sum_{m=1+w}^\infty\cr
\sum_{m=-\infty}^w
\end{array}\right.\right)\cr\cr
&=&\sum_{n=-\infty}^\infty~\left(
\sum_{w'=n}^{\infty}~\left\{\begin{array}{lcr}
\sum_{m=-\infty}^w\cr
\sum_{m=1+w}^\infty
\end{array}\right. -
\sum_{w'=-\infty}^{n-1}\left\{\begin{array}{lcr}
\sum_{m=1+w}^\infty\cr
\sum_{m=-\infty}^w
\end{array}\right.\right)\, ,\nonumber
\eea
where the upper lines inside the brackets hold
 for $\tau <0$ and the lower ones
for $\tau >0$.
In the last line we have exchanged the order of summations. 
The sum over $w'$ together with the integral over $j'$, the spin of the states
 in discrete representations, match together 
to give, after analytic continuation, the integral over $s'$,
the imaginary part of the spin of the states in the principal continuous 
representations: 
\bea
&&\sum_{w'=n}^{\infty} \int^{-\frac{1}{2}}_{-\frac{k-1}{2}}dj'
e^{\frac{4\pi i}{k-2} \left(j'+\frac12-w'\frac{k-2}{2}\right)\left(j+\frac12-
w\frac{k-2}{2}\right)} e^{-\frac{2\pi i}{k-2}\tau(j'+\frac12-w'
\frac{k-2}{2})^2} 
e^{-2\pi i\theta(j'+\frac12-w'\frac{k-2}{2})}\cr
&&~~~=~\int_{-\infty}^{0}d\lambda
e^{\frac{4\pi i}{k-2} \left(\lambda-n\frac{k-2}{2}\right)\left(j+\frac12-
w\frac{k-2}{2}\right)} e^{-\frac{2\pi i}{k-2}\tau(\lambda-n\frac{k-2}{2})^2} 
e^{-2\pi i\theta(\lambda-n\frac{k-2}{2})}\cr\cr
&&~~~=\left\{\begin{array}{lcr}
~i\int_0^\infty ds' ~e^{\frac{4\pi i}{k-2} \left(-is'-n\frac{k-2}{2}\right)
\left(j+\frac12-
w\frac{k-2}{2}\right)} e^{-\frac{2\pi i}{k-2}\tau(-is'-n\frac{k-2}{2})^2} 
e^{-2\pi i\theta(-is'-n\frac{k-2}{2})} ,~\tau<0~,\cr\cr
-i\int_0^\infty ds' ~e^{\frac{4\pi i}{k-2} \left(is'-n\frac{k-2}{2}\right)
\left(j+\frac12-
w\frac{k-2}{2}\right)}~ e^{-\frac{2\pi i}{k-2}\tau(is'-n\frac{k-2}{2})^2} 
~e^{-2\pi i\theta(is'-n\frac{k-2}{2})}~ ,~\tau>0~.
\end{array}\right.\nonumber
\eea 
After a similar analysis for the terms 
in the sum $\sum_{w'=-\infty}^{n-1}$ and 
relabeling the dummy index $n\rightarrow w'$, one finds the 
following contribution from the continuous series    
\bea
\sum_{w'=-\infty}^{\infty}~~ i\sqrt{\frac{2}{k-2}}\int_{0}^{\infty} ds'
(-)^{w+w'+1} \left[\sum_{m=-\infty}^{w}e^{\frac{4\pi i}{k-2} \left(-is'-w'
\frac{k-2}{2}\right)\left(j+\frac12-
w\frac{k-2}{2}\right)}e^{-2\pi i m\left(\frac12+is'+w'\frac{k-2}{2}\right)}
\right.  \cr\cr
-\left. \sum_{m=1+w}^{\infty}e^{\frac{4\pi i}{k-2} \left(is'-w'\frac{k-2}{2}
\right)\left(j+\frac12-
w\frac{k-2}{2}\right)}e^{-2\pi i m\left(\frac12-is'+w'\frac{k-2}{2}\right)}
\right]               
\frac{e^{2\pi i\tau\left(\frac{s'^{2}}{k-2}+\frac k4 w'{}^2\right)}}
{\eta^3(\tau+i\epsilon_1)}\delta(\theta-w'\tau+m)\nonumber
\eea

Finally, using (\ref{conttrick}), with the appropriate relabeling and 
performing the sum over $m$ (which then simply  reduces 
to a geometric series)
one gets
\bea
\sum_{w'=-\infty}^{\infty}\int_0^\infty ds'\int_0^1d\alpha' 
{\cal S}_{j,w}{}^{s',\alpha',w'} \chi_{s'}^{\alpha',w'}(\theta,\tau,0),
\eea
with
\bea
{\cal S}_{j,w}{}^{s',\alpha',w'}=-i\sqrt{\frac{2}{k-2}}e^{-2\pi i\left(w'j-w
\alpha'-ww'\frac k2\right)} \left[\frac{e^{\frac{4\pi}{k-2}s'\left
(j+\frac12\right)}}{1+e^{-2\pi i(\alpha'-is')}} + \frac{e^{-\frac{4\pi}{k-2}s'
\left(j+\frac12\right)}}{1+e^{-2\pi i(\alpha'+is')}}\right]\, .~~~~
\eea
\medskip

It is interesting to note that (repeated indices denote implicit sum)
\bea
{\cal S}_{j,w}{}^{s_1,\alpha_1,w_1} ~~
{\cal S}_{s_1,\alpha_1,w_1}{}^{s',\alpha',w'}&=&
-~{\cal S}_{j,w}{}^{j_1,w_1}~~{\cal S}_{j_1,w_1}{}^{s',\alpha',w'}\cr
&=&\frac{(-)^{w+w'+1}}{2\pi}\sum_{m=-\infty}^{\infty}\left[\frac{1}
{\frac12+\alpha '-is'-m}+ \frac{1}{\frac12+\alpha'+is'-m} \right]\cr 
&&\times ~ \delta\left(j-\alpha'-(w+w')\frac{k-2}{2}+m\right)\, .
\label{S2}
\eea

The first line implies ${\cal S}^2_{~j,w}{}^{s',\alpha',w'}=0$.

To show that $({\cal S}T)^3_{~j,w}{}^{s',\alpha',w'}=0$ is a bit more involved. This block is explicitly given by
\bea
{\cal S}_{j,w}{}^{s_1,\alpha_1,w_1}\left[  \left(T{\cal S} T{\cal S} T\right)_{s_1,\alpha_1,w_1} {}^{s',\alpha',w'}\right] +
 {\cal S}_{j,w}{}^{j_1,w_1}\left[ \left(T{\cal S} T {\cal S} T\right)_{j_1,w_1}{}^{s',\alpha',w'} \right]\, .
\eea
The first term above coincides with the first one in (\ref{S2}). This is a consequence of (\ref{STcontprop}), which implies $ \left(T{\cal S} T{\cal S} T\right)_{s_1,\alpha_1,w_1} {}^{s',\alpha',w'}= {\cal S}_{s_1,\alpha_1,w_1} {}^{s',\alpha',w'}$. So, in order for this block to vanish it is sufficient to show that the term inside the second bracket is exactly the ${\cal S}$ matrix mixing block. 

The factor inside the last bracket splits into the sum
\bea
&&T_{j_1,w_1}{}^{j_2,w_2}{\cal S}_{j_2,w_2}{}^{s_3,\alpha_3,w_3} T_{s_3,\alpha_3,w_3}{}^{s_4,\alpha_4,w_4} {\cal S}_{s_4,\alpha_4,w_4}{}^{s_5,\alpha_5,w_5} T_{s_5,\alpha_5,w_5}{}^{s',\alpha',w'} \cr 
&&~+ ~~~~ T_{j_1,w_1}{}^{j_2,w_2}{\cal S}_{j_2,w_2}{}^{j_3,w_3} T_{j_3,w_3}{}^{j_4,w_4} {\cal S}_{j_4,w_4} {}^{s_5,\alpha_5,w_5} T_{s_5,\alpha_5,w_5}{}^{s',\alpha',w'}\, .\label{ST3d-c}
\eea
These terms are very difficult to compute separately because each 
one gives the integral of a Gauss error function. So, we show 
here how  the sums  can be reorganized
in order to cancel all the intricate 
integrals when summing both terms and one ends with the mixing block 
${\mathcal S}_{j_1,w_1}{}^{s',\alpha',w'}$. In fact, after some few steps, 
the first line can be expressed as 
\bea
&&\sqrt{\frac{2}{k-2}}\int_0^\infty ds~ \left\{ \tilde S_{j_1,w_1}{}^{s_2,\alpha_2,w_2} \left[\sum_{w=-\infty}^0 e^{-i\frac\pi4}e^{-\frac{2\pi i}{k-2}\left[-is-w\frac{k-2}{2}-(j_1+\frac12)+is_2\right]^2} e^{2\pi i w\left(\alpha_2+\frac12-is_2\right)}\right.\right.\cr
&&~~~~~~-~~~~\left.\left. \sum_{w=1}^\infty e^{-i\frac\pi4}e^{\frac{2\pi i}{k-2}\left[-is-w\frac{k-2}{2}-(j_1+\frac12)+is_2\right]^2} e^{2\pi i w\left(\alpha_2+\frac12-is_2\right)}\right] + (s_2\rightarrow -s_2) \right\},~~\label{ST3a}
\eea
where we have introduced $~~\displaystyle \tilde S_{j_1,w_1}{}^{s_2,\alpha_2,w_2}=-i\sqrt{\frac{2}{k-2}}~ e^{-2\pi i\,\left(w_2j_1-w_1\alpha_2-w_1w_2\frac k2\right)}~ e^{\frac{4\pi s_2}{k-2}\left(j_1+\frac 12\right)}$.

On the other hand, the second line in (\ref{ST3d-c}) takes the form
\bea
&&\sum_{w=-\infty}^{\infty}\sqrt{\frac{2}{k-2}}\int_{-\frac{k-1}{2}}^{-\frac12} dj~e^{i\frac\pi2} e^{-\frac{2\pi i}{k-2}\left[j+\frac12-w\frac{k-2}{2}-(j_1+\frac12)+is_2\right]^2} e^{2\pi i w\left(\alpha_2+\frac12-is_2\right)} \cr &&~~~~~~~~~~~~~\times~~~~ \frac{\tilde S_{j_1,w_1}{}^{s_2,\alpha_2,w_2}}{1+e^{-2\pi i\left(\alpha_2-is_2\right)}}~~ + ~~~~(s_2\rightarrow -s_2) \, .
\eea    

Now notice that, for $w\leq-1$, the integral over $j$ can be replaced by an integral over $-\frac{k-1}{2}+is$ minus an integral over $-\frac12+is$ with $s
 \in [-\infty,0]$. For $w\geq1$,
 the original integral splits into the same two integrals, but now 
with $s \in [0,\infty ]$. Adding these terms to (\ref{ST3a})
 one ends, after some extra contour deformations in the remaining integrals,
 with 
$~~S_{j_1,w_1}^{s',\alpha',w'}$ 
and we can conclude that $({\mathcal S} T)^3_{~j,w}{}^{s',\alpha',w'}=0$.

\chapter{A generalized Verlinde formula}\label{A:gvf}

As is well known, the Verlinde theorem allows to compute the fusion
coefficients in RCFT as:
\bea
{\cal N}_{{\mu\nu}}{}^{\rho}
=\sum_{\kappa}\frac{S_{\mu}{}^{\kappa} S_{\nu}{}^{\kappa} 
\left(S_{\rho}{}^{\kappa}\right)^{-1}}{S_{0}{}^{\kappa}}\, ,\label{verlinde}
\eea
where the index ``$0$'' refers
to the representation containing the identity field. 
In the
case of the
fractional level admissible representations of the $\widehat{sl(2)}$
 affine Lie algebra, the negative integer
fusion coefficients obtained from (\ref{verlinde})
 in \cite{mat} were interpreted as 
 a consequence
of the identification $j\rightarrow -1-j$ in \cite{ay}\footnote{
Interestingly, it was shown 
in a recent
detailed study of the $\hat{sl}(2)_{k=\frac12}$ model \cite{Ridaut},
that the origin of the negative signs
is the absence of
 spectral flow images of the admissible representations  in the analysis of
 \cite{ay}.},
where it was also
shown that fusions are not allowed by the Verlinde formula
if the fields involved are not highest- or lowest-weight.
Applications to other non RCFT were discussed in 
\cite{TJ}, where generalizations of the theorem were proposed
for 
certain representations in
the Liouville theory, the $H_3^+$ model and the SL(2,$\mathbb R)$/U(1)
 coset.

In order to explore alternative expressions 
in the $AdS_3$ model, let us 
consider the more tractable finite dimensional
 degenerate representations. From the results 
for the characters  obtained in
section \ref{Sdeg}, it is natural to
propose the following generalization of
the Verlinde formula\footnote{ A similar 
expression was 
obtained in \cite{TJ} 
 for the $H_3^+$ model applying the Cardy ansatz.}

\bea
\sum_{J_3}~~{\cal N}_{J_1J_2}{}^{J_3}~~
\chi_{J_3}(\theta,\tau,0)=\sum_{w=-\infty}^{\infty}
\int^{-\frac12}_{-\frac{k-1}{2}}dj~~
\frac{S_{J_1}{}^{j,w}S_{J_2}{}^{j,w}}{S_{0}{}^{j,w}}~~
e^{2\pi i\frac k4 \frac{\theta^2}{\tau}}\chi_j^{+,w}(\frac\theta\tau,
-\frac1\tau,0)\, ,\cr&&\label{verlindedeg}
\eea
which holds for generic $(\theta,\tau)$ far from the 
points $\theta+n\tau\in{\mathbb Z},\forall n\in{\mathbb Z}$.
In order to prove it,
 notice that, in the region of the parameters where we claim it holds, 
one can neglect the
$\epsilon's$ and contact terms on both sides of the equation
and show that the fusion 
coefficients
${\mathcal N}_{J_1J_2}{}^{J_3}$ 
coincide with those 
obtained in the $H_3^+$ model, namely
\bea
{\cal N}_{J_1J_2}{}^{J_3}=\left\{\begin{array}{lcr}
1 & & |J_1-J_2|\leq J_3\leq J_1+J_2,\cr
0 & &{\rm otherwise}\, .~~~~~~~~~~
\end{array}\right.
\label{coef}
\eea
\medskip

Let us denote the $r.h.s.$ of (\ref{verlindedeg}) 
as  $I(J_1,J_2)$ and  rewrite it as
(see (\ref{tmd}))

\bea
I(J_1,J_2)&=&\sqrt{\frac{2}{k-2}}\frac{e^{\frac{2\pi i}{k-2}
\left(\frac{k-2}{2}\right)^2\frac{\theta^2}{\tau}}}{\sqrt{i\tau}~
i\vartheta_{11}(\theta,\tau)}\int_{-\infty}^{\infty}d\lambda~~
\frac{e^{\frac{2\pi i}{k-2}\frac{\lambda^2}{\tau}}e^{2\pi i\frac{\theta}\tau
\lambda}}{e^{\pi i\sqrt{\frac{2}{k-2}}\lambda}-e^{-\pi i\sqrt{\frac{2}{k-2}}
\lambda}}
\cr\cr\cr
&&\times~\left[e^{\frac{2\pi i}{k-2}N_1\lambda}+e^{-\frac{2\pi i}{k-2}N_1
\lambda}-e^{\frac{2\pi i}{k-2}N_2\lambda}-e^{-\frac{2\pi i}{k-2}N_2\lambda}
\right],
\eea
where $N_1=2(J_1+J_2+1)$ and $N_2=2(J_1-J_2)$.
Changing $\lambda\rightarrow-\lambda$ in the second and fourth terms, we get
\bea
I(J_1,J_2)=I(N_1)-I(N_2)~,~~~~~~I(N_i)=\tilde I(N_i,\theta,\tau)+ 
\tilde I(N_i,-\theta,\tau)~, \label{sum}
\eea
with
\bea
\tilde I(N_i,\theta,\tau)=\frac{\sqrt{\frac{2}{k-2}}}{\sqrt{i\tau}~
i\vartheta_{11}(\theta,\tau)}\int_{-\infty}^{\infty}d\lambda~
\frac{e^{\frac{2\pi i}{k-2}
\frac{1}{\tau}\left(\lambda+\theta\frac{k-2}{2}\right)^2}e^{\pi 
i\sqrt{\frac{2}{k-2}}N_i\lambda}}{e^{\pi i\sqrt{\frac{2}{k-2}}
\lambda}-e^{-\pi i\sqrt{\frac{2}{k-2}}\lambda}}\, .\label{sinn}
\eea
The divergent terms in this expression cancel in the sum (\ref{sum}).

Without loss of generality, let us assume 
$J_1\geq J_2$. To perform the $\lambda$-integral in (\ref{sinn}), it is 
convenient to
split the cases with
odd and even $N_i$.
Writing
$N_i+1=2m_i$, $m_i\in\mathbb N$,  in the first case
we get
\bea
\tilde I(N_i,\theta,\tau)=\sum_{L=0}^{m_i-1}\frac{e^{\frac{-2\pi i}{k-2}
\tau L^2}e^{-2\pi i \theta L}}{i\vartheta_{11}(\theta,\tau)}-\frac{e^{\pi i
\frac{k-2}{2}\frac{\theta^2}{\tau}}}{\sqrt{i\tau}~i\vartheta_{11}
(\theta,\tau)}\int_{-\infty}^{\infty}d\lambda~~\frac{e^{\pi i\frac{\lambda^2}
{\tau}}e^{2\pi i\sqrt{\frac{k-2}{2}}\frac{\theta\lambda}{\tau}}}{1-e^{2\pi i
\sqrt{\frac{2}{k-2}}\lambda}}\, ,
\eea
where the second term diverges.
For even $N_i$, take $N_i+2=2n_i$ with $n_i\in\mathbb N$, and then
\bea
\tilde I(N_i,\theta,\tau)&=&\sum_{L=0}^{n_i-1}\frac{e^{\frac{-2\pi i}{k-2}\tau 
\left(L-\frac12\right)^2}e^{-2\pi i \theta \left(L-\frac12\right)}}
{i\vartheta_{11}(\theta,\tau)}\cr\cr
&&-~ \frac{e^{\pi i\frac{k-2}{2}\frac{\theta^2}{\tau}}}{\sqrt{i\tau}~
i\vartheta_{11}(\theta,\tau)}\int_{-\infty}^{\infty}d\lambda~~
\frac{e^{\pi i\frac{\lambda^2}{\tau}}e^{2\pi i\sqrt{\frac{k-2}{2}}
\frac{\theta\lambda}{\tau}}e^{-\pi i\sqrt{\frac{2}{k-2}}\lambda}}
{1-e^{2\pi i\sqrt{\frac{2}{k-2}}\lambda}}\, ,
\eea
where again the second term diverges.

Notice that
$N_1$ and $N_2$ are either both even or odd, and since the divergent term
is the same in 
$I(N_1)$ and $I(N_2)$, it cancels in the sum
$I(J_1,J_2)$. Thus,
putting all together we get
\bea
I(J_1,J_2)= \sum_{J_3=J_1-J_2}^{J_1+J_2}\frac{-e^{\frac{-2\pi i}{4(k-2)}
\tau(2J_3+1) ^2}2~\sin\left(\pi i \theta (2J_3+1)\right)}
{\vartheta_{11}(\theta,\tau)}=\sum_{J_3=J_1-J_2}^{J_1+J_2}\chi_{J_3}
(\theta,\tau,0)\,.
\eea
where we have defined $J_3=L-\frac12$ for odd $N_1$ and $N_2$ and 
$J_3=L-1$ for  even $N_1$ and $N_2$.

From a similar analysis of the case $J_2>J_1$, we obtain
(\ref{verlindedeg}) and (\ref{coef}). 

 In conclusion, consistently 
with the assumption that correlation functions of fields in 
degenerate representations in the $H_3^+$ and $AdS_3$ models
 are 
related by analytic continuation,
the generalized Verlinde formula (\ref{verlindedeg})
reproduces the  fusion rules of degenerate representations
previously obtained in the Euclidean model.
However,  even if it is not expected to reproduce
the fusion rules of  continuous representations \cite{ay},
applying it for discrete representations
 also fails.


\begin{thebibliography}{99}

\bibitem{magoo} O. Aharony, S. S. Gubser, J. M. Maldacena, H. Ooguri and Y. Oz, {\it Large N field theories, 
  string theory and gravity.}, Phys. Rept. {\bf323}, 183 (2000); arXiv: 9905111 [hep-th].
\bibitem{malda} J. Maldacena, {\it The large N limit of superconformal field
  theories and supergravity}, Adv. Theor. Math. Phys. {\bf 2}, 231(1998); arXiv:9711200 [hep-th]
\bibitem{wittenhol} E. Witten, {\it Anti-de Sitter space and holography}, Adv. Theor. Math. Phys. 
  {\bf 2} 253 (1998); arXiv: 9802150 [hep-th]. 
\bibitem{NS}  C. Núñez, I. Y. Park, M. Schvellinger and T. A. Tran, {\it Supergravity duals of gauge theories 
  from F(4) gauged supergravity in six-dimensions.}, J. High Energy Phys. {\bf0104}, (2001) 025; arXiv: 0103080 [hep-th]. 
\bibitem{ST} M. Schvellinger and T. A. Tran, {\it Supergravity duals of gauge field theories from 
  SU(2) x U(1) gauged supergravity in five-dimensions},J. High Energy Phys. {\bf0106}, (2001) 025; arXiv: 0105019 [hep-th].
\bibitem{GNS} U. Gursoy, C. Núñez and M. Schvellinger, {\it RG flows from spin(7), CY 4 fold and HK manifolds to AdS, 
  Penrose limits and pp waves} J. High Energy Phys. {\bf0206}, (2002) 015; arXiv: 0203124 [hep-th].
\bibitem{BMN}D. Berenstein, J. Maldacena and H. Nastase, {\it Strings in flat space and pp waves from N=4 
  superYang-Mills} J. High Energy Phys. {\bf0204} (2002) 013; arXiv: 0202021 [hep-th]. 
\bibitem{GKP} S. S. Gubser, I. R. Klebanov and A. M. Polyakov, {\it A Semiclassical limit of the gauge / string 
  correspondence.},Nucl. Phys. {\bf B636}, 99 (2002); arXiv: 0204051 [hep-th].
\bibitem{S} M. Schvellinger, {\it Spinning and rotating strings for N=1 SYM theory and brane constructions.}, J. 
  High Energy Phys. {\bf0402}, (2004) 066; arXiv: 0309161 [hep-th].
\bibitem{GHK} S. Gubser, C. Herzog, I. Klebanov, {\it Symmetry breaking and axionic strings in the warped deformed 
  conifold}, J. High Energy Phys. {\bf0409} (2004) 036; arXiv: 0405282 [hep-th].
\bibitem{S2} M. Schvellinger, {\it Glueballs, symmetry breaking and axionic strings in non-supersymmetric deformations of 
  the Klebanov-Strassler background}, J. High Energy Phys {\bf0409} (2004) 057; arXiv: 0407152 [hep-th].
\bibitem{HHR}S.W. Hawking, T. Hertog, H.S. Reall, {\it Brane new world} Phys.Rev. {\bf D62}:043501, 2000; arXiv: 0003052 
  [hep-th]. 
\bibitem{AEN}L. Anchordoqui, J. Edelstein, C. Nunez, S. Perez Bergliaffa, M. Schvellinger, M. Trobo, F. Zyserman, {\it Brane 
  worlds, string cosmology, and AdS / CFT.}, Phys.Rev.{\bf D64}:084027, 2001; arXiv: 0106127 [hep-th].
\bibitem{EKSS}J. Erlich, E. Katz, D. Son, M. Stephanov, {\it QCD and a holographic model of hadrons}, Phys.Rev.Lett.{\bf95}:261602, 
  2005; arXiv: 0501128 [hep-th]. 
\bibitem{RP} L. Da Rold, A. Pomarol, {\it Chiral symmetry breaking from five dimensional spaces},  Nucl.Phys.{\bf B721}:79-97, 2005; 
  arXiv: 0501218 [hep-th]. 
\bibitem{HHMS} T. Hambye, B. Hassanain, J. March-Russell, M. Schvellinger, {\it Four-point functions and Kaon decays in 
  a minimal AdS/QCD model}Phys.Rev.{\bf D76}:125017, 2007; arXiv: 0612010 [hep-th]. 
\bibitem{HHMS2} T. Hambye, B. Hassanain, J. March-Russell, M. Schvellinger, {\it On the Delta I = 1/2 rule in holographic 
  QCD}, Phys.Rev.{\bf D74}:026003, 2006; arXiv: 026003 [hep-th]. 
\bibitem{HMS}B. Hassanain, J. March-Russell, M. Schvellinger, {\it Warped Deformed Throats have Faster (Electroweak) Phase 
  Transitions}, JHEP {\bf0710}:089, 2007; arXiv: 0708.2060 [hep-th].
\bibitem{MMT} J. Maldacena, D. Martelli and Y. Tachikawa, {\it Comments on string theory backgrounds with non-relativistic 
  conformal symmetry} JHEP {\bf0810}:072, 2008; arXiv: 0807.1100 [hep-th].
\bibitem{ABM} A. Adams, K. Balasubramanian and J. McGreevy, {\it Hot Spacetimes for Cold Atoms}, JHEP {\bf0811}:059, 2008; arXiv: 
  0807.1111[hep-th].
\bibitem{HRM} C. P. Herzog, M. Rangamani and S. F. Ross, {\it Heating up Galilean holography}, JHEP {\bf0811}:080, 2008; arXiv: 
  0807.1099 [hep-th].
\bibitem{S3} M. Schvellinger, {\it Kerr-AdS black holes and non-relativistic conformal QM theories in diverse dimensions}, 
  JHEP {\bf 0812}:004,2008; arXiv: 0810.3011 [hep-th].
\bibitem{landscape} M. Douglas, {\it The Statistics of string / M theory vacua} JHEP {\bf 0305}:046 (2003); arXiv: 0303194 
  [hep-th].
\bibitem{dif} P. Di Francesco, P. Mathieu and D. S\'en\'echal, {\it Conformal Field Theory},  Springer-Verlag, New York, 
  1997.
\bibitem{PZ}V. B. Petkova and J. B. Zuber, {\it Conformal boundary conditions and what they teach us}, arXiv: 0103007 
  [hep-th].
\bibitem{SA} C. Angelantonj and A. Sagnotti, {\it Open strings}, Phys. Rept. {\bf 371}, 1 (2002),[Erratum-ibid. 376, 339 
  (2003)]; arXiv: 0204089 [hep-th].
\bibitem{SNRat}V. Schomerus, {\it Non-compact string backgrounds and non-rational CFT}, Phys.Rept.{\bf 431}:39 (2006); arXiv: 
  0509155 [hep-th].
\bibitem{M} R. Metsaev, {\it Type IIB Green-Schwarz Superstring in Plane Wave Ramond-Ramond Background}, Nucl. Phys. {\bf B625}, 
  70 (2002); arXiv: 0112044 [hep-th].
\bibitem{MT} R. Metsaev and A. Tseytlin, {\it Exactly Solvable Model of Superstring in Ramond-Ramond Plane Wave Background}, 
  Phys. Rev. {\bf D65}:126004 (2002); arXiv: 0202109 [hep-th].
\bibitem{KG} M. Gaberdiel and I. Kirsch, {\it Worldsheet correlators in AdS(3)/CFT(2)}, JHEP {\bf 0704}, 050 (2007); 
  arXiv:0703001 [hep-th].  
\bibitem{DP}A. Dabholkar and A. Pakman, {\it Exact chiral ring of $AdS_3$/CFT$_2$}, hep-th/0703022.
\bibitem{SP} A. Pakman and A. Server, {\it Exact N=4 correlators of AdS(3)/CFT(2)}, Phys. Lett. {\bf B652}, 60 
 (2007); arXiv:0704.3040 [hep-th]. 
\bibitem{Taylor} M. Taylor, {\it Matching of correlators in $AdS_3$/CFT$_2$}, JHEP {\bf 0806}, 010 (2008); arXiv:
  0709.1838 [hep-th].
\bibitem{cn} C. Cardona, C. Nuñez, {\it Three-point functions in superstring theory on $AdS_3 \times S^3 \times T^4$.},
  JHEP {\bf 0906} 009 (2009); arXiv: 0903.2001 [hep-th].
\bibitem{ck} C. Cardona, I. Kirsch, {\it Worldsheet four-point functions in $AdS_3/CFT_2$}, JHEP {\bf 1101} 015 (2011);
  arXiv: 1007.2720 [hep-th]. 
\bibitem{BRFW} J. Balog, L. O'Raifeartaigh, P. Forgacs, A. Wipf {\it Consistency of String Propagation on Curved 
Space-Times: An SU(1,1) Based Counterexample}, Nucl.Phys. {\bf B325}:225 (1989).
\bibitem{mo1} J. Maldacena and H. Ooguri, {\it Strings in $AdS_3$ and the $SL(2,\mathbb R)$ WZW Model: Part 1: The Spectrum},
  J. Math. Phys. {\bf 42}, 2929 (2001); arXiv:0001053 [hep-th].
\bibitem{EGP} J. Evans, M. Gaberdiel, M. Perry, {\it The no ghost theorem for $AdS(3)$ and the stringy exclusion principle}, 
  Nucl. Phys. {\bf B535}, 152 (1998); arXiv: 9806024 [hep-th].
\bibitem{mo2} J. Maldacena, H. Ooguri and J. Son, {\it Strings in $AdS_3$ and the $SL(2,\mathbb R)$ WZW Model: Part 2: 
  Euclidean black hole},J. Math. Phys. {\bf 42}, 2961 (2001); arXiv:0005183 [hep-th]
\bibitem{mo3} J. Maldacena and H. Ooguri, {\it Strings in $AdS_3$ and the $SL(2,\mathbb R)$ WZW Model: Part 3: Correlation Functions},
  Phys. Rev. {\bf D65}, 106006 (2002); arXiv:0111180 [hep-th].
\bibitem{tesch1} J. Teschner, {\it On Structure constants and fusion rules in the $SL(2,\mathbb C)/SU(2)$ WZNW model}, Nucl. Phys. {\bf
  B546}, 390 (1999); arXiv:9712256 [hep-th]
\bibitem{tesch2} J. Teschner, {\it Operator product expansion and factorization in the $H_3^+$-WZNW model}, Nucl. Phys. {\bf B571},
  555 (2000); arXiv:9906215 [hep-th]
\bibitem{wc} {\bf W. Baron} and C. N\'u\~nez, {\it Fusion rules and four-point functions in the $AdS_3$ WZNW model}, Phys. Rev. {\bf D79}, 
  086004 (2009); arXiv:0810.2768 [hep-th].
\bibitem{verlinde} E. Verlinde, {\it Fusion rules and modular transformationsin conformal field theory}, Nucl. Phys. {\bf B300}, 360 
  (1988).
\bibitem{TJ} C. Jego and J. Troost, {\it Notes on the Verlinde formula in non rational conformal field theories},  Phys. Rev. {\bf D74}, 
  106002 (2006); arXiv: 0601085 [hep-th]. 
\bibitem{stt} A. M. Semikhatov, A. Taormina and I. Yu. Tipunin, {\it Higher-level appell functions, modular transformations and 
  characters}, arXiv: 0311314 [math].
\bibitem{est} T. Eguchi, Y. Sugawara and A. Taormina, {\it Liouville field, modular forms and elliptic genera}, JHEP {\bf 0703}, 119 
  (2007); arXiv:0611338 [hep-th].
\bibitem{t} A. Taormina, {\it Liouville theory and elliptic genera}, Prog. Theor. Phys. Suppl. {\bf 177}, 203 (2009);arXiv:0808.2376 
  [hep-th].
\bibitem{wc2} {\bf W. Baron}, C. Núñez {\it On modular properties of the AdS3 CFT}, Phys.Rev. {\bf D83},106010 (2011); arXiv:1012.2359 [hep-th].
\bibitem{hhrs} M. Henningson, S. Hwang, P. Roberts and B. Sundborg, {\it Modular invariance of SU(1,1) strings}, Phys. Lett. {\bf B267}, 
  350 (1991).
\bibitem{HS} Y. Hikida, Y. Sugawara, {\it Boundary states of D branes in AdS(3) based on discrete series},  Prog. Theor. Phys. {\bf 107}, 
  1245 (2002); arXiv: 0107189 [hep-th]. 
\bibitem{fnp} A. Fotopoulos, V. Niarchos and N. Prezas, {\it D-branes and extended characters in SL(2,R)/U(1)}, Nucl. Phys. {\bf B710}, 
  309 (2005); arXiv: 0406017 [hep-th].
\bibitem{ipt} D. Israel, A. Pakman and J. Troost, {\it Extended SL(2,R)/U(1) characters, or modular properties of a simple non-rational 
  conformal field theory}, JHEP {\bf 0404}, 043 (2004); arXiv: 0402085[hep-th].
\bibitem{bf} J. Bj\"ornsson and J. Fjelstad, {\it Modular invariant partition functions for non-compact $G/Ad(H)$ models}, arXiv:1012.3920 
  [hep-th].
\bibitem{gawe}K. Gawedski, {\it Noncompact WZW conformal field theories},Proceedings of the NATO Advanced Study Institute, {\it New 
  Symmetry Principles in Quantum Field Theory}, Cargese, 1991, p. 247, eds. J. Frolich, G. ´t Hooft, A. Jaffe, G. Mack, P.K. Mitter and R. 
  Stora, Plenum Press 1992; arXiv:9110076 [hep-th].
\bibitem{SH} S. Helgason, {\it Differential Geometry and Symmetric Spaces}. Academic Press, New York, 1962.
\bibitem{teschmssl} J. Teschner, {\it The Minisuperspace limit of the $SL(2,C)/SU(2)$ WZNW model}, Nucl.Phys. {\bf B546}, 369 (1999); arXiv: 
  9712258 [hep-th].
\bibitem{hosrib} K. Hosomichi, S. Ribault, {\it Solution of the $h_3^+$ model on a disc}, JHEP {\bf 0701}, 057 (2007); arXiv: 0610117 [hep-th].
\bibitem{rib} S. Ribault, {\it Minisuperspace limit of the $AdS_3$ WZNW model},
  JHEP {\bf 1004}, 096 (2010); [arXiv:0912.4481 [hep-th]].
\bibitem{basu} D. Basu, {\it The plancherel formula for the universal covering group of $sl(2,r)$ revisited}, arXiv:0710.2224 [hep-th].
\bibitem{Novikov} S. P. Novikov, {\it The hamiltonian formalism and a many valued analog of Morse theory}, Usp. Mat. Nauk {\bf 37N5} 3 (1982).
\bibitem{Witten1} E. Witten, {\it Nonabelian bosonization in two dimensions}, Commun. Math. Phys. {\bf 92} 455 (1984).
\bibitem{GW} D. Gepner and E. Witten, {\it String theory on group manifolds}, Nucl. Phys. {\bf B278} 493 (1986).
\bibitem{Witten2} E. Witten, {\it On holomorphic factorization of WZW and coset models}, Commun. Math. Phys. {\bf 144} 189 (1992).  
\bibitem{Mohammedi} N. Mohammedi, {\it On the unitarity of string propagating on $SU(1, 1)$}, Int. J. Mod. Phys. {\bf A5} 3201 
  (1990).
\bibitem{BN} I. Bars and D. Nemeschansky, {\it String propagation in backgrounds with curved space-time}, Nucl. Phys. {\bf B348} 
  89 (1991).
\bibitem{Hwang1} S. Hwang, {\it No ghost theorem for $SU(1, 1)$ string theories}, Nucl. Phys. {\bf B354} 100 (1991). 
\bibitem{Hwang2} S. Hwang, {\it Unitarity of strings and non-compact hermitian symmetric spaces}, Phys. Lett. {\bf B435} 331 (1998); 
  arXiv: 9806049 [hep-th].
\bibitem{Petropoulos} P. M. S. Petropoulos, {\it String theory on $AdS_3$: some open questions}, Proceedings of TMR European program 
  meeting, Quantum Aspects of Gauge Theories, Supersymmetry and Unification (Paris, France, 1999), hep-th/9908189.
\bibitem{MMS} J. Maldacena, J. Michelson, and A. Strominger, {\it Anti-de Sitter fragmentation}, JHEP {\bf 02} 011 (1999), 9812073 
  [hep-th].
\bibitem{SW} N. Seiberg and E. Witten, {\it The D1/D5 system and singular CFT}, JHEP {\bf 04}, 017 (1999), arXiv: 9903224 [hep-th].
\bibitem{BORT} J. de Boer, H. Ooguri, H. Robins, J. Tannenhauser, {\it String theory in $AdS_3$}, JHEP {\bf 9812}, 026 (1998); 
  arXiv: 9812046 [hep-th].
\bibitem{kounnas} D. Israel, C. Kounnas and P. Petropoulos, {\it Superstrings on NS5 backgrounds, deformed $AdS_3$ and holography},
  JHEP {\bf 0310}, 028 (2003); [arXiv:0306053 [hep-th]].
\bibitem{israel} D. Israel, {\it D-branes in Lorentzian $AdS_3$}, JHEP {\bf 0506}, 008 (2005); [arXiv: 0502159 [hep-th]].
\bibitem{BPZ} A.A. Belavin, Alexander M. Polyakov, A.B. Zamolodchikov, {\it Infinite Conformal Symmetry in Two-Dimensional 
  Quantum Field Theory}, Nucl.Phys. {\bf B241}, 333 (1984).
\bibitem{hs} K. Hosomichi and Y. Satoh, {\it Operator product expansion in SL(2) conformal field theory}, Mod. Phys. Lett. {\bf
  A17}, 683 (2002); arXiv:0105283 [hep-th]
\bibitem{satoh} Y. Satoh, {\it Three-point functions and operator product expansion in the SL(2) conformal field theory}, Nucl. 
  Phys. {\bf B629}, 188 (2002); hep-th/0109059
\bibitem{gk} A. Giveon and D. Kutasov, {\it Little string theory in a double scaling limit}, JHEP {\bf 9910}, 034 (1999); 
  arXiv:9909110 [hep-th]. 
\bibitem{gk2} A. Giveon and D. Kutasov, {\it Comments on double scaled little string theory}, JHEP {\bf 0001}, 023 (2000); arXiv:
  9911039 [hep-th]. 
\bibitem{fukuda} T. Fukuda and K. Hosomichi, {\it Three-point functions in sine-Liouville theory}, JHEP {\bf 0109}, 003 (2001);
  arXiv:0105217 [hep-th].
\bibitem{in} S. Iguri and C. N\'u\~nez, {\it Coulomb integrals for the $SL(2,\mathbb R)$ WZNW model}, Phys. Rev. {\bf D77}, 066015 
  (2008); arXiv:0705.4461[hep-th].
\bibitem{ay} H. Awata and Y. Yamada, {\it Fusion rules for the fractional level $\widehat {sl}_2$ algebra}, Mod. Phys. Lett. {\bf A7} 
  1185 (1992).  
\bibitem{holman} W. J. Holman and L. C. Biedernharn, {\it Complex angular momenta and the groups SL(1,1) and SU(2)}, Ann. Phys. 
  {\bf 39}, 1 (1966); Ann. Phys. {\bf 47}, 205 (1968).
\bibitem{ribault} S. Ribault, {\it Knizhnik-Zamolodchikov equations and spectral flow in $AdS_3$ string theory}, JHEP {\bf 0509}, 045
  (2005); arXiv:0507114 [hep-th]
\bibitem{zf} A. B. Zamolodchikov and V. A. Fateev, {\it Operator algebra and correlation functions in the two-dimensional 
  $SU(2)\times SU(2)$ chiral Wess-Zumino model}, Sov. J. Nucl. Phys. {\bf 43}, 657 (1986).
\bibitem{mn} P. Minces and C. N\'u\~nez, {\it Four-point functions in the $SL(2,\mathbb R)$ WZW model}, Phys. Lett. {\bf B647}, 500
  (2007); arXiv:0701293 [hep-th]
\bibitem{in2} S. Iguri and C. N\'u\~nez, {\it Coulomb integrals and conformal blocks in the $AdS_3$-WZNW model}, JHEP {\bf 0911} 
  090 (2009) 090; arXiv::0908.3460[hep-th]
\bibitem{tesch3} J. Teschner, {\it Crossing symmetry in the $H_3^+$ WZNW model}, Phys. Lett. {\bf B521}, 127-132, 2001;  
  0108121 [hep-th].
\bibitem{GKS} A. Giveon, D. Kutasov, A Shwimmer, {\it Comments on D-branes in $AdS_3$}, Nucl. Phys. {\bf B615}, 133 (2001); arXiv: 
  0106005 [hep-th].
\bibitem{BP} C. Bachas, M. Petropoulos, {\it Anti-de Sitter D-branes}, JHEP {\bf 0102}, 025 (2001); arXiv: 0012234 [hep-th]. 
\bibitem{Stanciu} S. Stanciu, {\it D-branes in an $AdS_3$ background}, JHEP {\bf9909}, 028 (1999); arXiv: 9901122 [hep-th].
\bibitem{RR} A. Rajaraman and M. Rozali, {\it Boundary States for D-branes in $AdS_3$}, Phys. Rev. {\bf D66}, 026006 (2002); 
  arXiv:0108001 [hep-th].
\bibitem{PST} B. Ponsot, V. Schomerus, J. Teschner, {\it Branes in the Euclidean $AdS_3$}, JHEP 0202, 016 (2002); arXiv:0112198 
  [hep-th]. 
\bibitem{FS} J. M. Figueroa-O'Farrill and S. Stanciu, {\it D-branes in $AdS_3\times S_3\times S_3\times S_1$ background}, 
  JHEP {\bf 0004}, 005 (2000); [arXiv:0001199 [hep-th]].
\bibitem{PR} P.M. Petropoulos, S. Ribault, {\it Some comments on Anti-de Sitter D-branes}, JHEP {\bf 0107}, 036 (2001); arXiv: 
  0105252 [hep-th].
\bibitem{LOPT} P. Lee, H. Ooguri, J. Park, J. Tannenhauser, {\it Open strings on $AdS_2$ branes}, Nucl. Phys. {\bf B610}, 3 (2001); 
  arXiv: 0106129 [hep-th].
\bibitem{LOP} P. Lee, H. Ooguri, J. Park, {\it Boundary states for $AdS_2$ branes in $AdS_3$}, Nucl.Phys. {\bf B632}, 283 (2002); 
  arXiv:0112188 [hep-th].
\bibitem{PS} A. Parnachev, D. Sahakyan, {\it Some remarks on D-branes in $AdS_3$}, JHEP {\bf 0110}, 022 (2001); arXiv: 0109150 [hep-th].
\bibitem{D}  C. Deliduman, {\it $AdS_2$ D-branes in Lorentzian $AdS_3$}, Phys. Rev. {\bf D68}, 066006 (2003); arXiv: 0211288 [hep-th].
\bibitem{H} W. H. Huang, {\it Anti-de Sitter D-branes in Curved Backgrounds}, JHEP {\bf 0507}, 031 (2005); arXiv: 0504013 [hep-th].
\bibitem{RS} S. Ribault, V. Schomerus, {\it Branes in the 2-D Euclidean Black hole}, JHEP {\bf 0402}, 019 (2004); arXiv: 0310024 
  [hep-th]. 
\bibitem{kir} G. D'Appollonio, E. Kiritsis, {\it D-branes and BCFT in Hpp-wave backgrounds}, Nucl. Phys. {\bf B712} (2005) 433; 
  arXiv:0410269 [hep-th].
\bibitem{AS} A. Y. Alekseev, V. Schomerus, {\it D-branes in the WZW model}, Phys. Rev. {\bf D60}, 061901 (1999); arXiv: 9812193 [hep-th]. 
\bibitem{KO}M. Kato, T. Okada, {\it D-branes on group manifolds}, Nucl. Phys. {\bf B499}, 583 (1997); arXiv:9612148 [hep-th]. 
\bibitem{Ishibashi} N. Ishibashi, {\it The Boundary and Crosscap States in Conformal Field Theories}, Mod. Phys. Lett. {\bf A4}, 251 (1989). 
\bibitem{Cardy}  J. L. Cardy. {\it Boundary Conditions, Fusion Rules and the Verlinde Formula}, Nucl. Phys. {\bf B324} 581 (1989).
\bibitem{schomerus} Volker Schomerus, {\it Lectures on branes in curved backgrounds}, Class. Quant. Grav. {\bf 19}, 5781 (2002);
  arXiv:0209241 [hep-th].
\bibitem{BL} {\bf W. Baron}, A. Lugo, {\it On supersymmetric $D6$-$\overline{D6}$ 
  systems with magnetic fields}, Phys.Lett. {\bf B656} 243 (2007); arXiv: 0709.3930 [hep-th].
\bibitem{ABMN} G. Aldazabal, {\bf W. Baron}, D. Marques, C. Nunez, {\it The effective action of Double 
  Field Theory}, JHEP {\bf 1111} 052 (2011); arXiv:1109.0290 [hep-th].
\bibitem{HZ} C. Hull, B. Zwiebach, {\it Double Field Theory}, JHEP {\bf 0909}, 090 (2009); arXiv:0904.4664 [hep-th].
\bibitem{HHZ} O. Hohm, C. Hull, B. Zwiebach, {\it Background independent action for double field theory},  JHEP {\bf 1007}, 016 (2010); 
  arXiv:1003.5027 [hep-th].
\bibitem{HHZ2} O. Hohm, C. Hull, B. Zwiebach, {\it Generalized metric formulation of double field theory},  JHEP {\bf 1008}, 008 (2010);
  arXiv:1006.4823 [hep-th].
\bibitem{JW} J. Schon, M. Weidner, {\it Gauged $N=4$ supergravities}, JHEP {\bf 0605}, 034 (2006); arXiv: 0602024[hep-th].
\bibitem{STW} J.~Shelton, W.~Taylor and B.~Wecht, {\it Nongeometric flux compactifications}, JHEP {\bf 0510}, 085 (2005); 
  arXiv: 0508133[hep-th].

\bibitem{Schulz} M. Schulz, {\it T-folds, doubled geometry, and the $SU(2)$ WZW model}, arXiv: 1106.6291 [hep-th].  
\bibitem{hpt}  A. Hanany, N. Prezas and J. Troost, {\it The partition function of the two-dimensional black hole conformal field theory}, 
  JHEP {\bf 0204},  014  (2002) arXiv:0202129 [hep-th].
\bibitem{FZZ} V. Fateev, A. Zamolodchikov, Al. Zamolodchikov, {\it Boundary Liouville field theory. 1. Boundary state and boundary 
  two point function}, arXiv: 0001012 [hep-th].
\bibitem{DO} H. Dorn, H. J. Otto, {\it Two and three point function in Liouville theory}, Nucl. Phys. {\bf B429}, 375 (1994): arXiv: 
  9403141 [hep-th].
\bibitem{ZZ} A. B. Zamolodchikov, Al. B. Zamolodchikov, {\it Structure constants and conformal bootstrap in Liouville field theory}, Nucl. 
  Phys. {\bf B477}, 577 (1996); arXiv: 9506136[hep-th].
\bibitem{slater} L. J. Slater, {\it Generalized hypergeometric functions}, Cambridge Univ. Press, Cambridge (1960).
\bibitem{moore} G. Moore, {\it Finite in all directions}, arXiv:hep-th/9305139.
\bibitem{seiberg} H. Liu, G. Moore and N. Seiberg, {\it Strings in a time dependent orbifold}, JHEP {\bf 0206}, 045 (2002); arXiv:0204168 
  [hep-th].
\bibitem{kutasov} B. Craps, D. Kutasov and G. Rajesh, {\it String propagation in the presence of cosmological singularities}, JHEP 
  {\bf 0206}, 053 (2002); arXiv:0205101 [hep-th].
\bibitem{russo} G. Papadopoulos, J. Russo and A. Tseytlin, {\it Solvable models of strings in a time-dependent plane-wave background}, 
  Class. Quant. Grav. {\bf 20}, 969-1016 (2003); arXiv:0211289 [hep-th].
\bibitem{mat} P. Mathieu and M. Walton, Prog. Theor. Phys. Suppl. {\bf 102}, 229  (1990).
\bibitem{Ridaut} D. Ridout, {\it $\hat{sl}(2)_{-1/2}$: A Case Study}, Nucl. Phys. {\bf B814}, 485 (2009); [arXiv:0810.3532 [hep-th]].





\end{thebibliography}
\end{document}